\documentclass[aps,prd,twocolumn,groupedaddress]{revtex4} 

\usepackage{amsmath}
\usepackage{amsfonts}
\usepackage{amssymb}
\usepackage{color}
\usepackage{graphicx} 
\usepackage{graphics} 
\usepackage{multirow}
\usepackage{setspace}
\usepackage{ragged2e}

\newcommand{\bq}{\begin{equation}}
\newcommand{\eq}{\end{equation}}
\newcommand{\bqn}{\begin{eqnarray}}
\newcommand{\eqn}{\end{eqnarray}}
\newcommand{\nb}{\nonumber}
\newcommand{\lb}{\label}

\begin{document}

\title{Complete Classification of Analytical Models in Einstein-Aether Cosmology}

\author{R. Chan$^{1}$, M. F. A. da Silva$^{2}$ and V. H. Satheeshkumar$^{3}$}
\email[]{chan@on.br, mfasnic@gmail.com, vhsatheeshkumar@gmail.com}
\affiliation{
	$^{1}$Coordena\c{c}\~{a}o de Astronomia e Astrof\'{i}sica,
	Observat\'{o}rio Nacional (ON), Rio de Janeiro, RJ 20921-400, Brazil
	\\
	$^{2}$Departamento de F\'{i}sica Te\'{o}rica, 
	Universidade do Estado do Rio de Janeiro (UERJ), Rio de Janeiro, RJ 20550-900, Brazil
	\\
	$^{3}$Departamento de F\'{\i}sica, Universidade Federal do Estado do Rio de Janeiro (UNIRIO), Rio de Janeiro, RJ 22290-240, Brazil
}

\date{\today}
	
\begin{abstract}

We present all possible analytical solutions of the Friedmann-Lemaître-Robertson-Walker metric in Einstein-aether theory for all values of the cosmological constant and spatial curvature with many reasonable values of the equation-of-state parameter. We analyze the dynamics of each model analytically and also graphically by plotting the geometric radius, Hubble and deceleration parameters along with the effective energy conditions. All our results are compared with the corresponding models in General Relativity. The two key results are (i) the aether does not qualitatively change the dynamics of the cosmological models but merely scales the geometric radius, Hubble, and deceleration parameters, and (ii) we found eight models that are entirely void of any aether, meaning in such a universe aether does not play any role cosmologically, although it affects the solar system dynamics.

\end{abstract}


\maketitle

\section{Introduction}

Although our quest to understand the universe has been as old as humanity, the necessary mathematical apparatus and observational inputs have been available only for a little over a century now. Our current best theory of gravity, General Relativity (GR), was formulated in 1916, while Edwin Hubble only announced that there are galaxies outside our Milky Way on January 1, 1925. However, the first cosmological model based on GR was discovered by Einstein himself in 1917, which was soon followed by a seminal paper by de Sitter. Since then, there have been several new theoretical models of the universe and many impressive observational results in cosmology. Cosmology today is driven by high-precision data from multimessenger astronomical observations. Yet, there is room for theoreticians to construct alternatives to the concordance model of the standard cosmology to explain outstanding problems such as dark matter, dark energy, initial singularity, and, recently, the Hubble and S8 tensions\cite{footnote1}. The interested reader may find the milestone papers in cosmology in reference \cite{Bernstein} and a beautiful summary of key developments in modern cosmology in reference \cite{Peebles}.

In this article, we are interested in finding all possible analytical cosmological solutions in Einstein-aether (EA) theory, firstly because it is still observationally viable for a particular choice of the free parameters of the theory \cite {Foster:2005dk}. Secondly, it is the most general vector-tensor theory of gravity and allows the investigation of the violations of Lorentz Invariance in a cosmological context that can be verified observationally (See \cite{Moore:2013sra} for the black hole context). Thirdly, even though most of the models here do not represent our universe, since we can compare them with the corresponding models in GR, we get an insight on the theoretical behavior of vector fields and their role in cosmology. In our previous articles, we have explored the role of aether in the weak-field limit \cite{Goldoni:2016osp}, spacetime singularities \cite{Chan:2019mdn} \cite{Chan:2020amr}, black hole thermodynamics \cite{Chan:2021ela}, fundamental symmetries \cite{Chan:2022mxd} and gravitational waves \cite{Chan:2023fzw}.

In 2012 Ha and collaborators systematically study the evolution of the Friedmann-Lema{\^{\i}}tre-Robertson-Walker (FLRW) universe coupled with a cosmological constant $\Lambda$ and a perfect fluid
that has the equation of state $p = {\rm w} \rho$, where $p$ and $\rho$ denote, respectively, 
the pressure and energy density of the fluid, and ${\rm w}$ is an arbitrary real constant.
They studied the dynamics of the models using a time analysis of the potential of the system.
Instead of our present work where we solved analytically the field equations and obtain
explicitly solutions, Ha et al. just studied the time behavior of the acceleration universe
using the potential without solving the field equations.

{Einstein-Aether theory of gravity has an aether vector field in addition to the metric tensor that together determines the local spacetime structure. This theory preserves locality and covariance even with the Lorentz-violating vector field, i.e., aether. This helps us investigate the consequences of Lorentz-violating theories of gravity in cosmology. There have been several works on AE cosmology, a subset of which can be found in references \cite{Carroll:2004ai} \cite{Barrow2011} \cite{Alhulaimi:2013sha}, \cite{Coley:2015qva}.} In a previous paper \cite{Campista2020}, we have studied some exact solutions of the with FLRW metric allowed by the EA theory for two particular cases of perfect fluid: a fluid with constant energy density ($p_{m}=-\rho_{0\,m}$) and a fluid with zero energy density ($\rho_{0\,m}=0$), corresponding to the vacuum solution with and without cosmological constant ($\Lambda>0$), respectively. In another previous work \cite{Chan:2019mdn} we have classified all the possible cosmological vacuum solutions in EA theory, which are 
$\Lambda > 0$, $\Lambda = 0$, $\Lambda < 0$ and for $k = -1$, $k= 0$, $k=1$.

The paper is organized as follows. After a short Introduction in Section 1, we briefly outline EA theory in Section 2. In Section 3, we present the solutions of FLRW metric to EA field equations for $\Lambda>0$, $\Lambda=0$, and $\Lambda<0$, each for three values of $k=-1, 0, 1$ with different values of equation-of-state parameter $w =-2, -1, -2/3, -1/3, 0$ and $1/3$. In Section 4, we study the dynamics of each model and analyze their effective energy conditions in Section 5. We end the article with a summary of the results. All analytical solutions, effective energy conditions, and detailed graphs are given in the three Appendices, respectively.

\section{Einstein-Aether theory}

To investigate dynamical preferred frame effects in GR, Jacobson, and Mattingly proposed the EA theory 2001 \cite{Jacobson:2000xp}. It is a generally covariant theory with the local  LI  broken by a dynamical unit timelike vector field $u^a$ dubbed aether. The EA theory belongs to a class of modified gravity theories called Vector-Tensor theories \cite{Satheeshkumar:2021zvl}. More extensive literature on EA theory can be found in \cite{Barrow2011, Bhattacharjee} and references therein. The action of the general EA theory is given by,
\bq 
S = \int \sqrt{-g} (L_{\rm Einstein}+L_{\rm aether}+L_{\rm {matter}}) d^{4}x,
\lb{action} 
\eq
where, 
\bqn 
L_{\rm Einstein} &=&  \frac{1}{16\pi G} \left( R - 2 \Lambda \right), \\ 
L_{\rm aether}   &=&  \frac{1}{16\pi G} [-K^{ab}{}_{mn} \nabla_a u^m \nabla_b u^n \nb \\
&& ~ ~ ~~~~~~ + \lambda(g_{ab}u^a u^b + 1)],
\lb{LEAG}
\eqn
with
\bq {{K^{ab}}_{mn}} = c_1 g^{ab}g_{mn}+c_2\delta^{a}_{m} \delta^{b}_{n}
+c_3\delta^{a}_{n}\delta^{b}_{m}-c_4u^a u^b g_{mn},
\lb{Kab}
\eq
and the $c_i$ are the dimensionless coupling constants and $\lambda$
is a Lagrange multiplier that enforces the unit timelike constraint on the aether vector. The term, $L_{matter}$ is the matter Lagrangian.

The EA theory reduces to Newtonian gravity under the weak-field, slow-motion limit, where Newton's gravitational constant $G_{\rm N}$ is related to the parameter $G$ in the action (\ref{action})  by  \cite{Garfinkle},
\bq
G = G_N\left(1-\frac{c_1+c_4}{2}\right).
\lb{Ge}
\eq
It is interesting to note that $G$ is equal to $G_N$ for $c_1 = -c_4$ without necessarily requiring the dimensionless coupling constants of the theory to identically vanish, i.e., $c_1 = c_4 = 0$. The Newtonian limit is recovered only if $c_1 + c_4 < 2$. Gravity turns repulsive for $c_1 + c_4 > 2$, while for $c_1 + c_4 = 2$, the $G = 0$, which means that gravity does not couple to matter \cite{Jacobson:2008aj}.

By extremizing the action with respect to the independent  variables $\lambda$, $u^a$ and $g_{mn}$, we get the following equations of motion \cite{Garfinkle},
\bq
g_{ab}u^a u^b = -1,
\lb{LagMul}
\eq
\bq
\nabla_a \left( K^{am}_{bn} \nabla_m u^n \right) + c_4 u^m \nabla_m u_n \nabla_b u^n + \lambda u_b = 0,
\eq
\bq
G^{Einstein}_{ab} = T^{aether}_{ab} +8 \pi G \, T^{matter}_{ab},
\lb{EA}
\eq
with
\bqn
G^{Einstein}_{ab} &=& R_{ab} - \frac{1}{2} R\, g_{ab}  + \Lambda \, g_{ab}, \\
T^{aether}_{ab}&=& \nabla_c [ J^c\;_{(a} u_{b)} + u^c J_{(ab)} - J_{(a} \;^c u_{b)}] \nb \\
& &- \frac{1}{2} g_{ab} J^c\;_d \nabla_c u^d+ \lambda u_a u_b  \nb \\
& & + c_1 [\nabla_a u_c \nabla_b u^c - \nabla^c u_a \nabla_c u_b] \nb \\
& & + c_4 u^c \nabla_c u_a u^c \nabla_c u_b, \nb \\
T^{\, matter}_{ab} &=&  \frac{- 2}{\sqrt{-g}} \frac{\delta \left( \sqrt{-g} L_{matter} \right)}{\delta g_{ab}}.
\lb{fieldeqs}
\eqn

We obtain the following constraints on the free parameters $c_i$, by requiring that the tensor, vector, and scalar parts of the linearly perturbed AE action are ghost-free, \cite{Foster:2005dk}
$$c_1+ c_3 < 1,$$
$$c_1+ c_4 > 0,$$
$$c_2 > -1.$$
Thus, mathematical consistency demands $\beta = c_1+3c_2+c_3 > -2$. The observational results including the primordial nucleosynthesis \cite{Carroll:2004ai}, ultra-high energy cosmic rays \cite{Elliott:2005va}, the solar system tests \cite{Eling:2003rd, GrEAsser:2005bg}, binary pulsars \cite{Foster:2007gr, Yagi:2013ava}, and more recently gravitational waves \cite{Gong:2018cgj, Oost:2018tcv} have placed stringent constraints on the free parameters $c_i$.

\section{Einstein-Aether Cosmology}

The Friedmann-Lemaître-Robertson-Walker (FLRW) metric describes the most general isotropic and homogeneous universe,
\bq
ds^2= -dt^2+B(t)^2\left[\frac{dr^2}{1-kr^2} +r^2 d\theta^2 +r^2 \sin^2 \theta d\phi^2\right],
\lb{ds2}
\eq
where $B(t)$ is the scale factor, and $k$ is a Gaussian curvature of the space at a given time. According to our current best cosmological observations, this metric describes our universe as spatially homogeneous and isotropic when averaged over large scales. This leaves us with a choice of
\bq
u^a=(1,0,0,0).
\eq

The time-time component (the Friedmann equation) of the Einstein field equations is given by,
\bq
\left(1+\frac{\beta}{2}\right) \left(\frac{{\dot B}(t)}{B(t)}\right)^2+ \frac{k}{B(t)^2}-\frac{\Lambda}{3} =\frac{8\pi G}{3} \rho_{m}(t),
\lb{G00}
\eq
while the space-space (the Raychaudhuri equation) component is given by,
\bqn
\left(1+\frac{\beta}{2}\right) \left(\frac{2{\ddot B}(t)}{B(t)}+\frac{{\dot B}(t)^2}{B(t)^2}\right) &&+ \frac{k}{B(t)^2}-\Lambda  \nb \\
&& =-8\pi G\,\, p_{m}(t),
\lb{G11}
\eqn
where $\beta=c_1+3c_2+c_3$, $\Lambda$ is the cosmological constant, $p_m(t)$ is the isotropic pressure and $\rho_{m}(t)$ is the density of energy (subscript $m$ means $matter$ from here on). For recent literature on EA cosmology, the reader may consult the references \cite{Coley:2019tyx}, \cite{Leon:2019jnu} and \cite{Campista2020}.

The equation of continuity is given by
\bq
{\dot \rho_{m}}(t) + 3 \frac{{\dot B}(t)}{B(t)}\left [p_{m}(t)+\rho_{m}(t) \right]=0, 
\eq
and the equation of state given by
\bq
p_{m}(t)={\rm w}\, \rho_{m}(t),
\lb{prho}
\eq
where ${\rm w}$ is a constant.

Substituting $p_m(t)$ and $\rho_m(t)$ from the field equations into the equation of state, we get the master equation,
\bqn
\left(\beta+2\right) \frac{{\ddot B}(t)}{B(t)} &&+ \left(1+3 {\rm w} \right) \left[ \left(1+\frac{\beta}{2}\right) \frac{{\dot B}(t)^2}{B(t)^2} + \frac{k}{B(t)^2} \right] \nb \\
&& -\left(1+ {\rm w} \right) \Lambda =0,
\lb{G11a}
\eqn
where we must have $\beta > -2$.

Before we proceed any further, it is interesting to note that, the aether contribution via $\beta$ vanishes completely in the following scenarios,\\
(a) for any ${\rm w}$ when $\Lambda = 0$, $k = 0$\\
(b) for any $\Lambda$ when ${\rm w} = -1$, $k = 0$\\ 
(c) for any $k$ when $\Lambda = 0$, ${\rm w} = -1/3$.

In order to find explicit solutions of the master equation (\ref{G11a}) for the metric function, we used the Maple 16 algebraic software following the algorithm. We have made a survey of all possible analytical solutions for each value of $k=-1,0,1$ and imposing ${\rm w}=m/n$, where $m$ and $n$ are integer constants in the interval $-10<n<10$ and $-10<m<10$ varying in step of one,
and limiting the values of ${\rm w}$ between -1 and 1. We also studied a limit case where
${\rm w = -2}$ for the Big Rip cases \cite{Barrow2009} (we have found 16 cases for $-2<{\rm w}<-1$
but we present only this one
for the sake of simplicity). In the next Section, we present all these explicit analytical solutions and compare them with those found in the literature, whenever possible.

We can notice that our previous work \cite{Campista2020} is not studied in the present paper
since it is a particular case where the energy density is constant.

Before proceeding, let us show in more details the analysis procedure of Ha and collaborators (2012). We would like to make a subtle but important observation on the aether's role in cosmology. Firstly, we can rewrite the Friedmann equation to separate the kinetic and potential terms as the following.
\bq
\frac{1}{2} {{\dot B}(t)}^2+ \frac{1}{2+\beta} \left[ k - \frac{B(t)^2}{3} \left( \Lambda  + 8\pi G \rho_{m}(t) \right) \right] = 0,
\eq
Secondly, by using the equation of state in the equation of continuity and integrating, we get the expression for energy density given by, 
\bq
\rho = \rho_0 \left(\frac{B_0}{B} \right)^{3(1+{\rm w)}}
\eq
where $B_{0}$ and $\rho_{0}$ correspond to the current values of the scale factor and the energy density. Putting these two together, we have an expression of the potential given by,
\bq
V(B) = \frac{1}{2+\beta} \left[ k - \frac{\Lambda B(t)^2}{3}   -  \frac{C}{B(t)^{1+3{\rm w}}} \right],
\eq

where, we have used the conventional value $B_{0} = 1$ and $C = 8 \pi G \rho_0 /3$. 
Thus, the acceleration of the universe can be written as
\bq
{\ddot B(t)} \propto -\frac{dV(B)}{dB},
\eq
without solving the field equations.
Interestingly, the aether contribution via $\beta$ is entirely sitting in the overall common factor. This means the aether does not change the overall behavior of the cosmological models but scales them depending on the value of $\beta$. This is evident in our graphs.


\section{Dynamics of the Cosmological Models}

The analytical solutions of the master equation of EA cosmology, Eq.(\ref{G11a}), are given here under three groups: $\Lambda>0$, $\Lambda=0$, and $\Lambda<0$. For each $\Lambda$, we have considered all three possible values of spatial curvature, $k=-1,0,1$, and for each $k$, we have found analytical solutions for different equation-of-state parameter w: $-2, -1$ $-2/3$, $-1/3$, $0$, $1/3$. All the results are summarized in the following Tables \ref{table0a}, \ref{table0b} and \ref{table0c}.

Now, we analyze the dynamics of the cosmological models presented in the previous section by studying the time evolution of scale factor and its derivatives, represented by the Hubble expansion rate
\begin{equation}
	\label{Ht}
	H(t)=\frac{\dot{B}(t)}{B(t)},
\end{equation}
the deceleration parameter,
\bq
q(t) = -\frac{\ddot{B}(t)}{B(t)H(t)^2}
\lb{qt}
\eq
and the geometrical radius
\bq
R_{g}=r|B(t)|.
\lb{Rg}
\eq
The parameters $B(t)$, $H(t)$, and $q(t)$ are helpful in the comparison of EA solutions with that of GR, allowing us to show explicitly the differences and similarities between the two theories.

The Friedmann equation can be written as,
\bq
H(t)^2 = \frac{1}{1 + \frac{\beta}{2}} \left[ \frac{8\pi G}{3}   \rho_{m}(t) -  \frac{k}{B(t)^2} + \frac{\Lambda}{3} \right],
\lb{G00.1}
\eq
while the Raychaudhuri equation takes the form,
\bq
\frac{{\ddot B}(t)}{B(t)} = \frac{1}{1 + \frac{\beta}{2}} \left[ - \frac{4\pi G}{3}  \left( \rho_{m}(t) + 3 p_{m}(t)   \right) + \frac{2 \Lambda}{3} \right],
\lb{G11.1}
\eq

\begin{figure}[!htp]
\begin{minipage}{175 mm}
	\centering	
	\includegraphics[width=3.4cm]{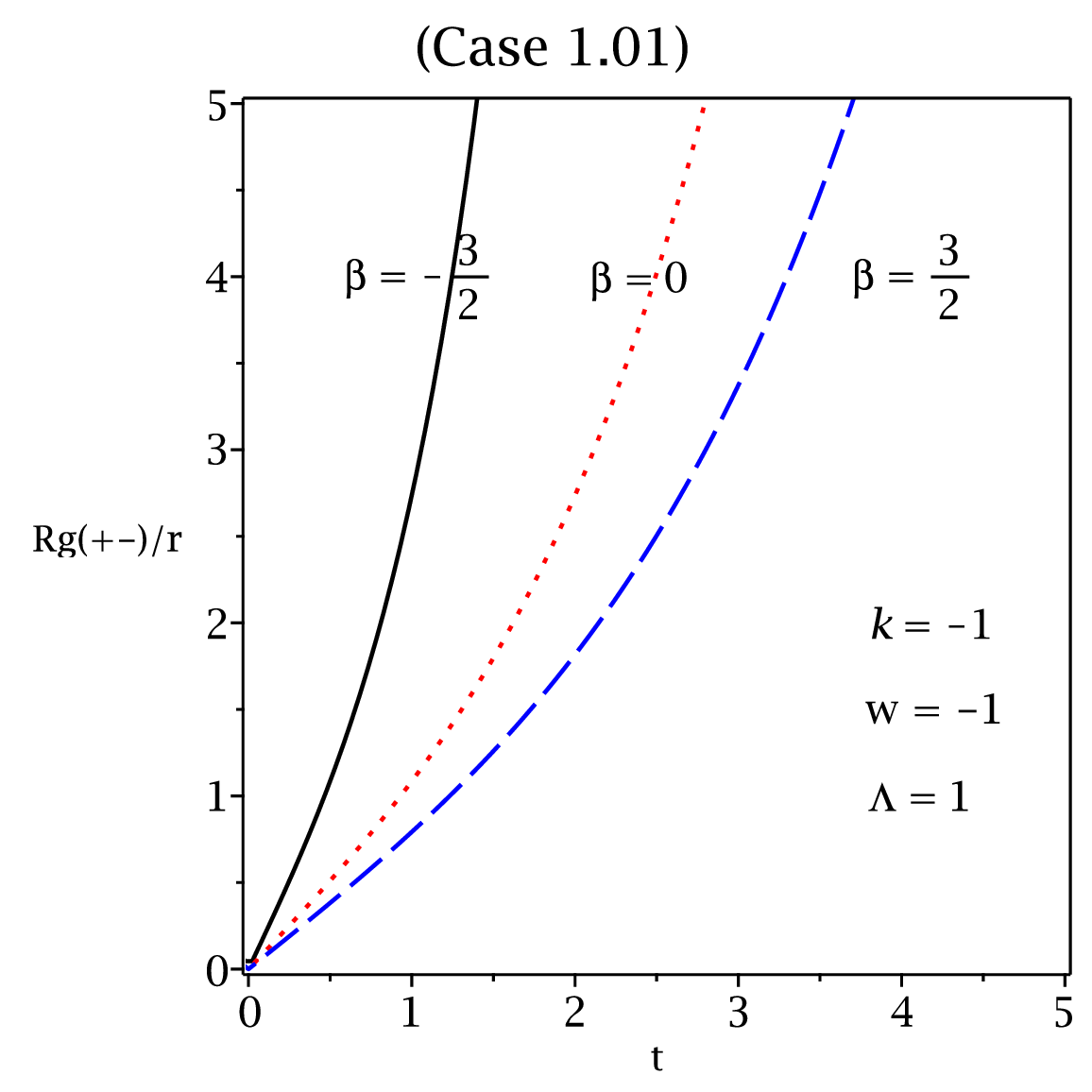}
	\includegraphics[width=3.4cm]{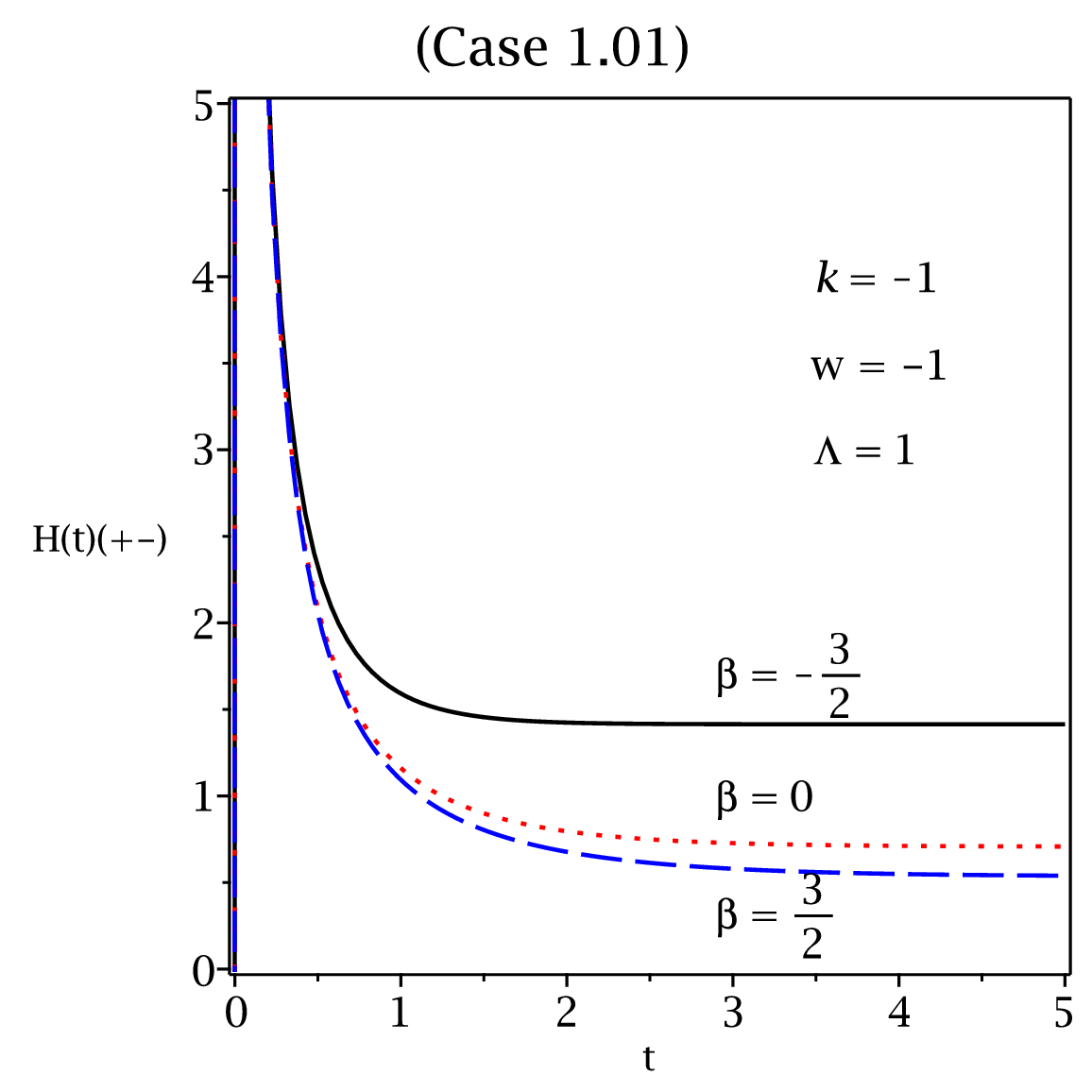}
	\includegraphics[width=3.4cm]{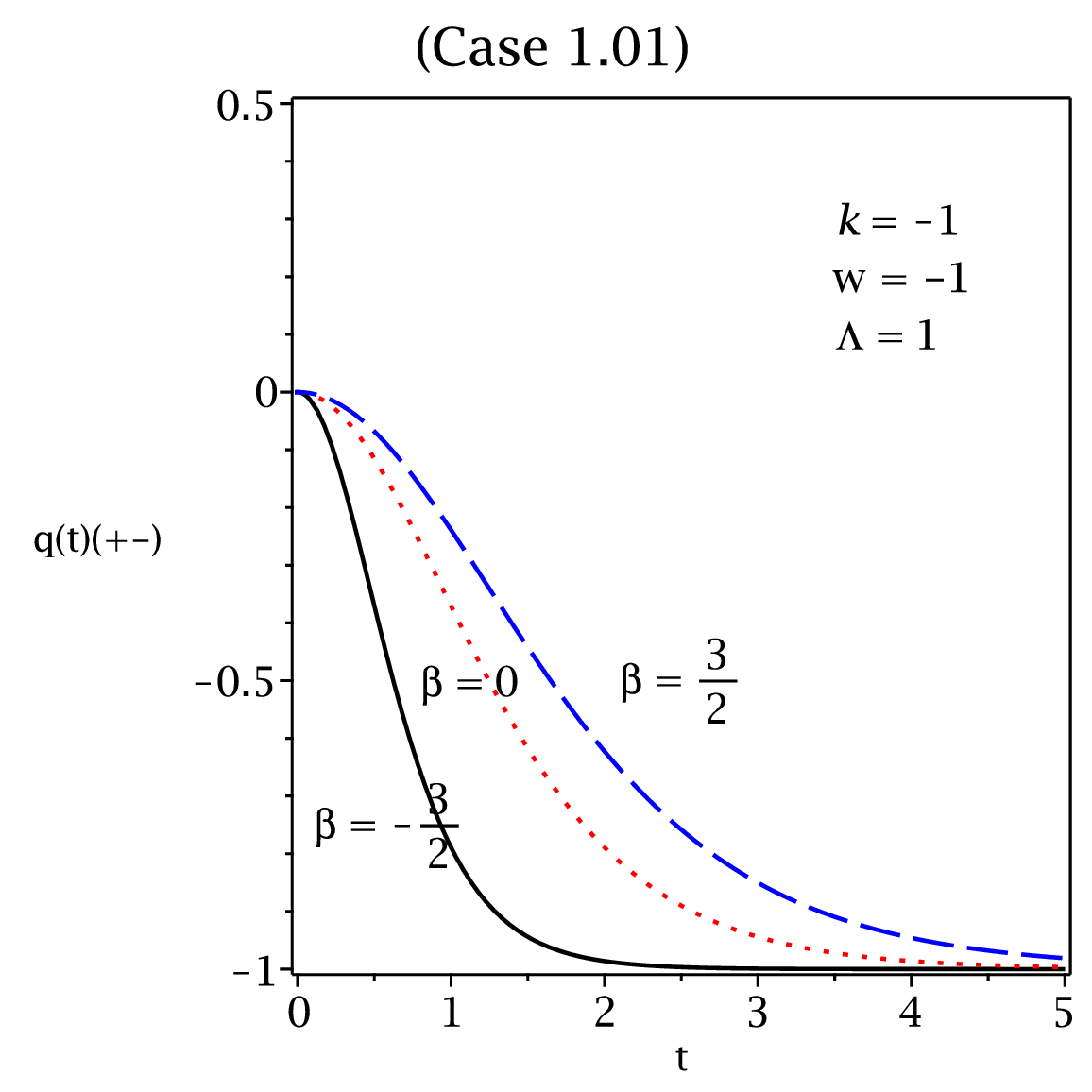}
	\includegraphics[width=3.4cm]{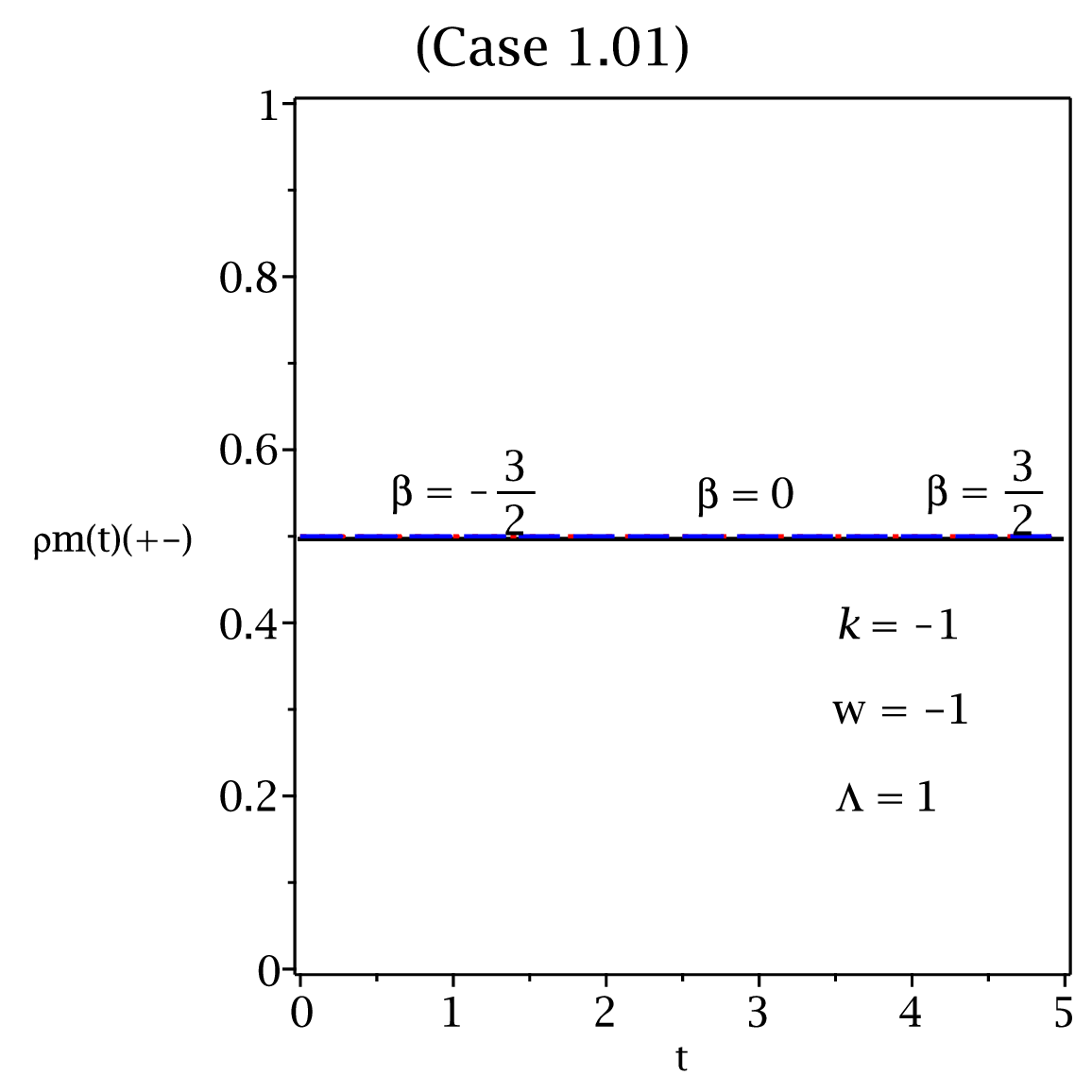}
	\includegraphics[width=3.4cm]{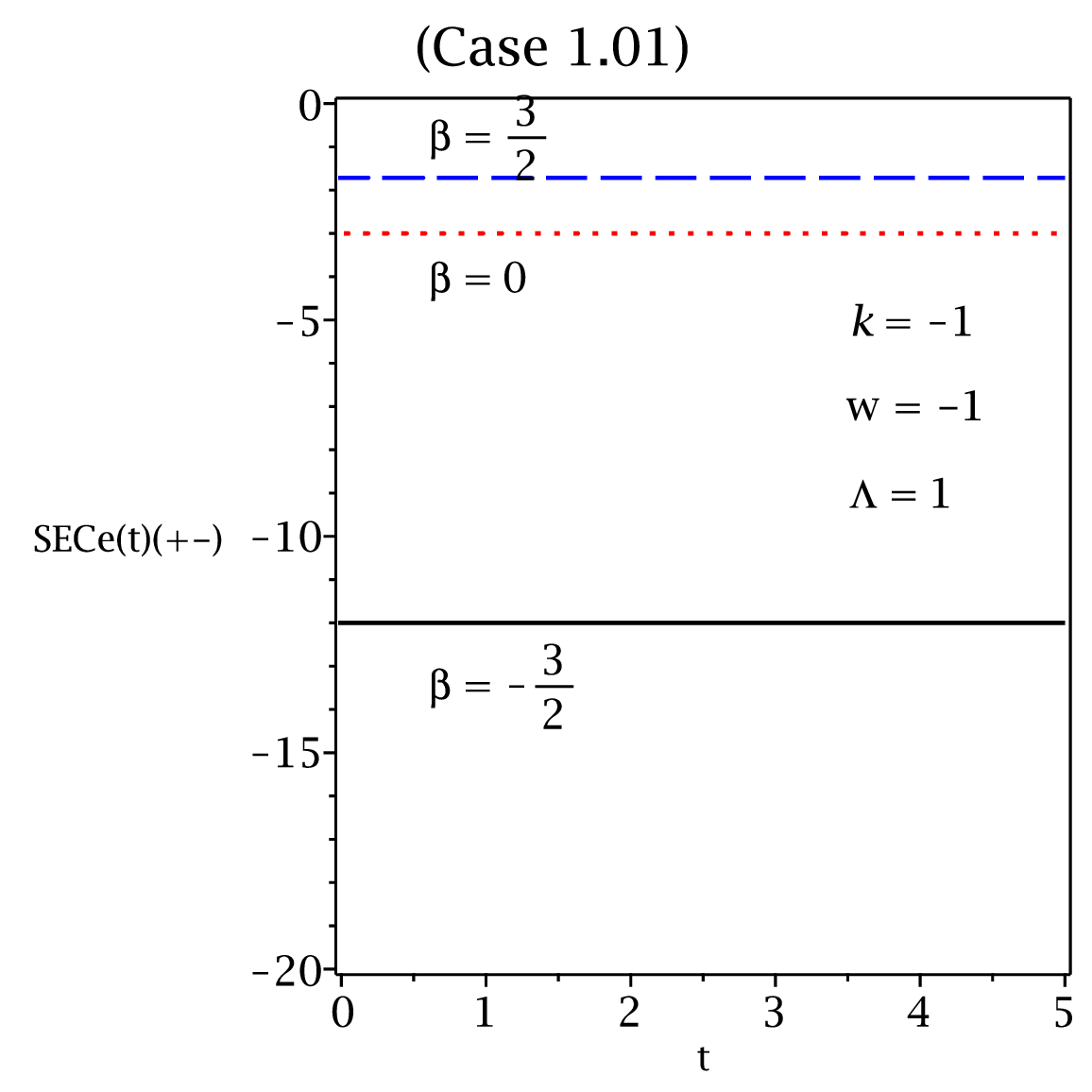}
	\includegraphics[width=3.4cm]{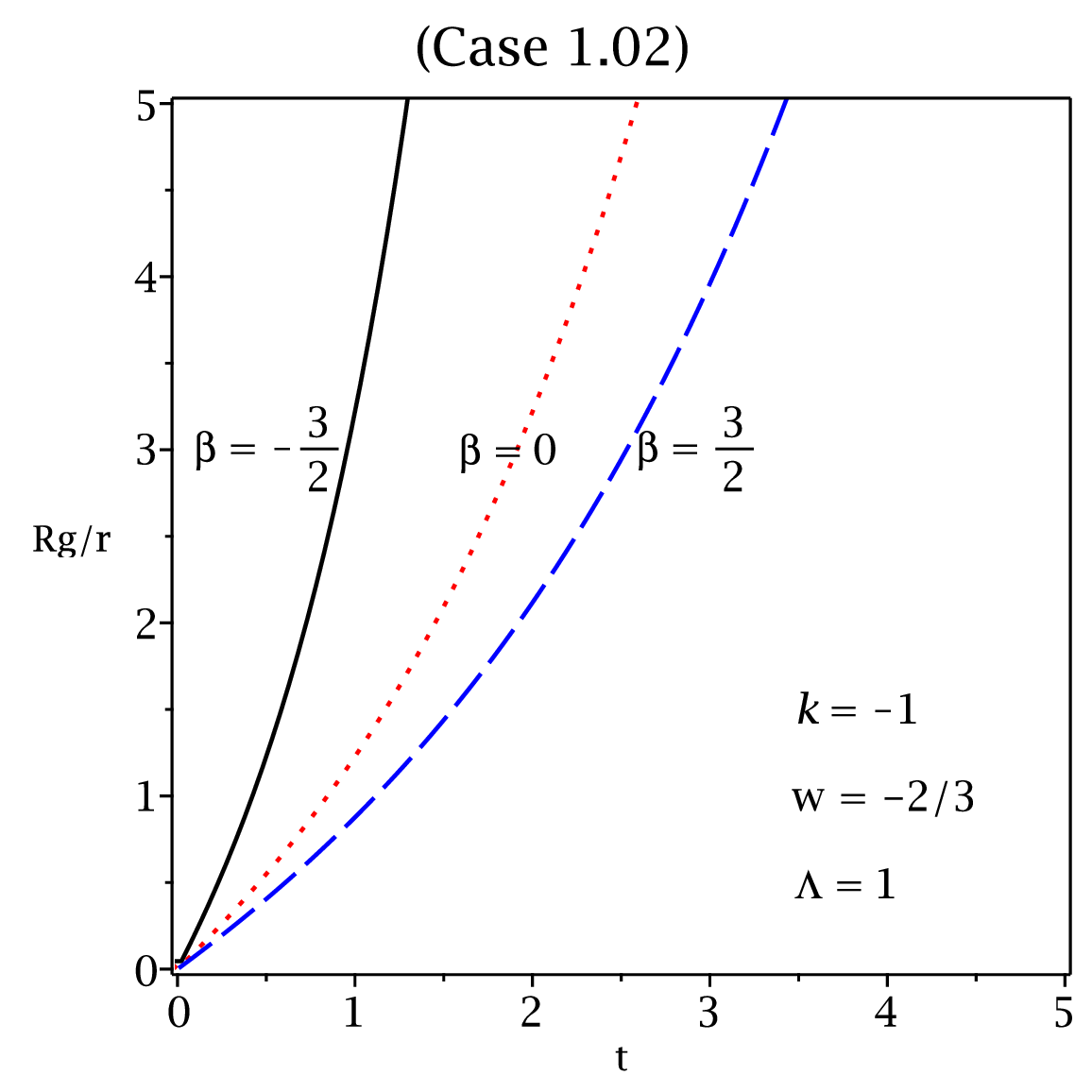}
	\includegraphics[width=3.4cm]{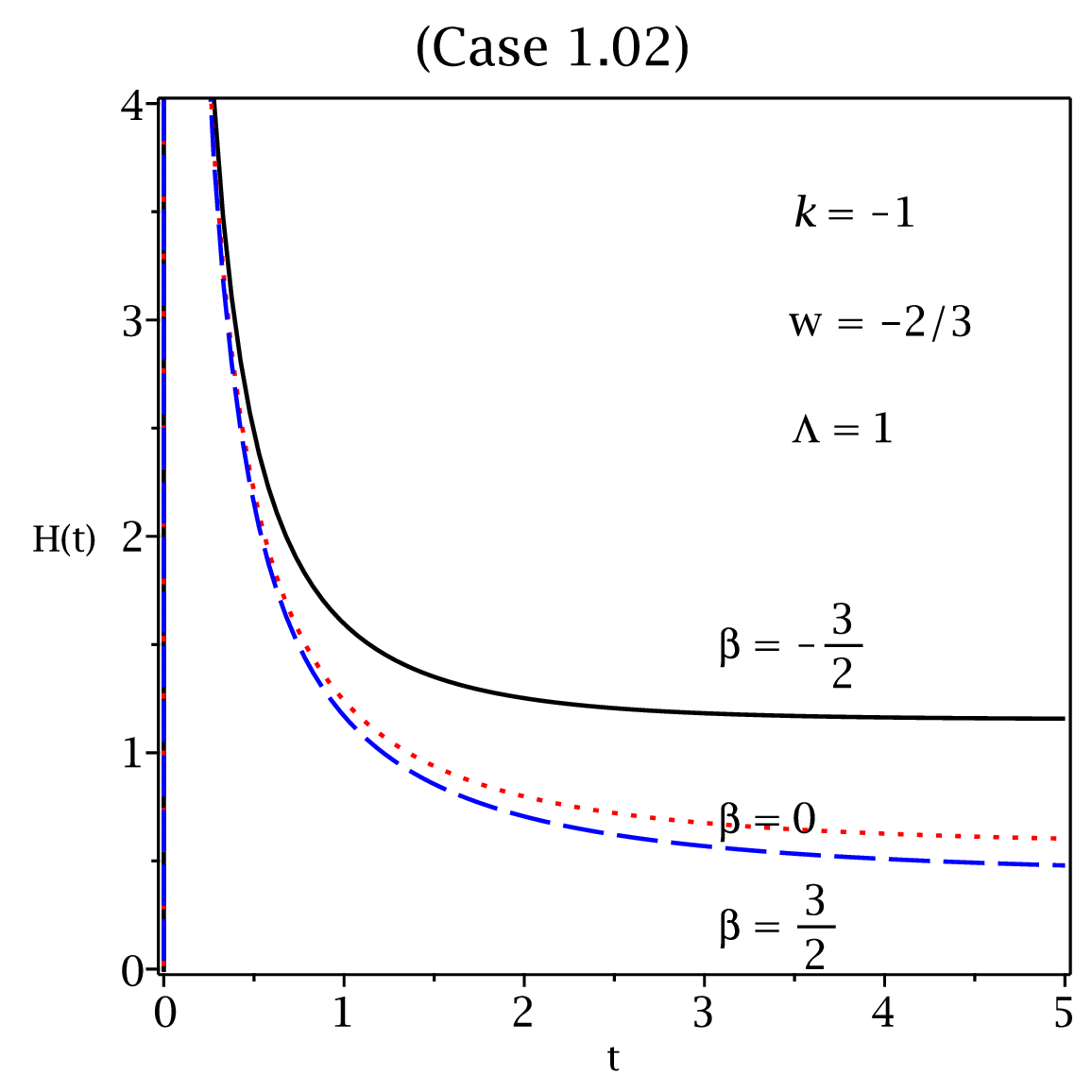}
	\includegraphics[width=3.4cm]{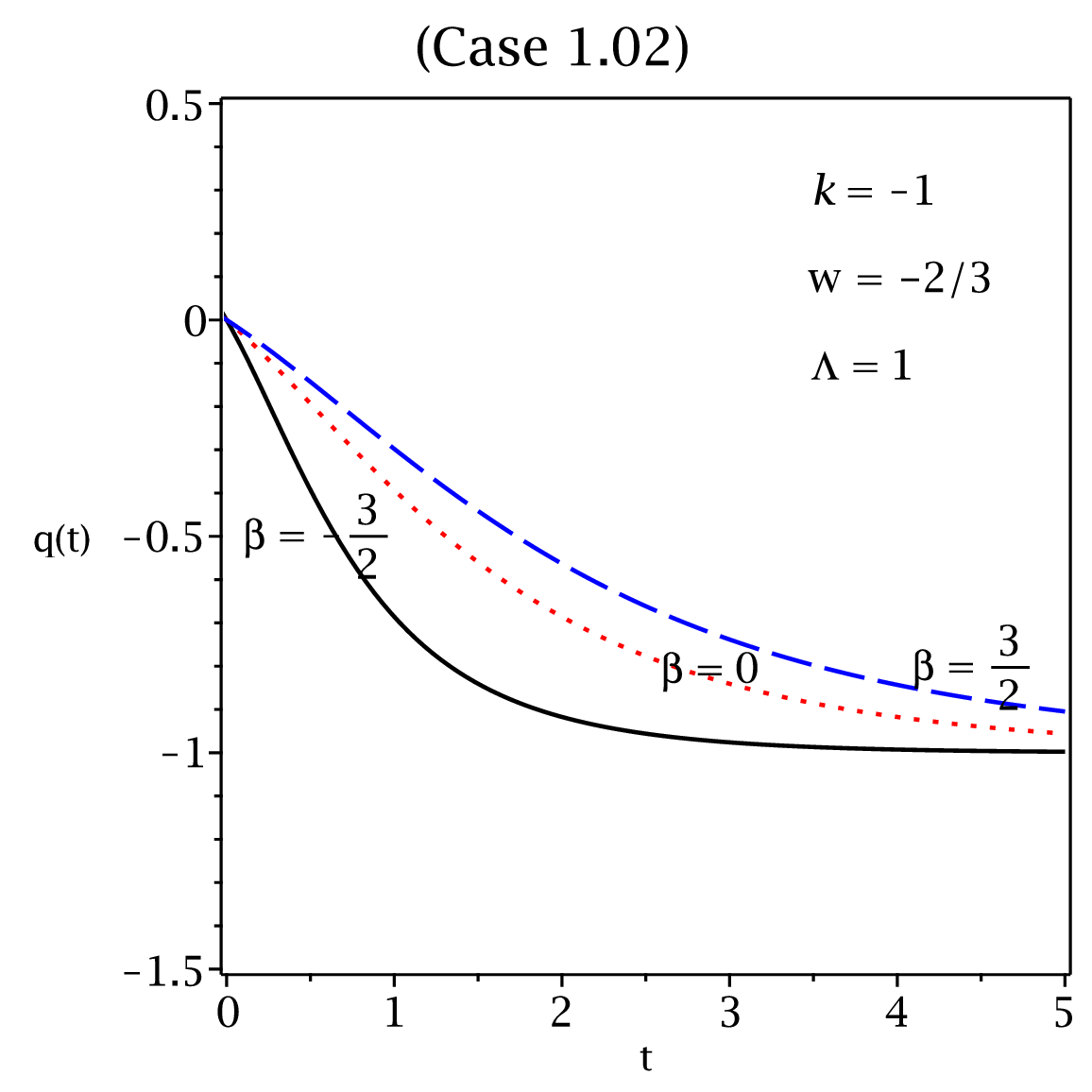}
	\includegraphics[width=3.4cm]{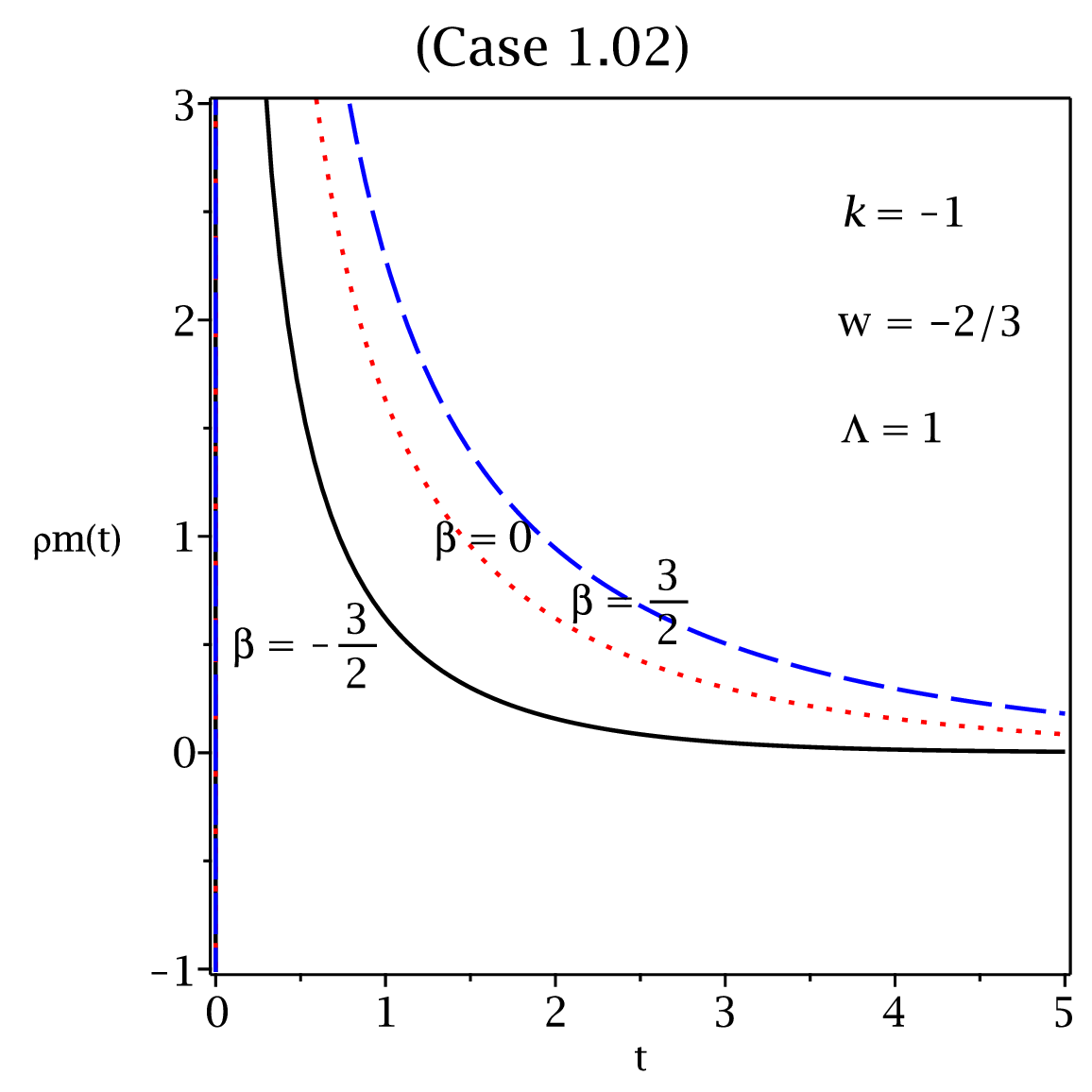}
	\includegraphics[width=3.4cm]{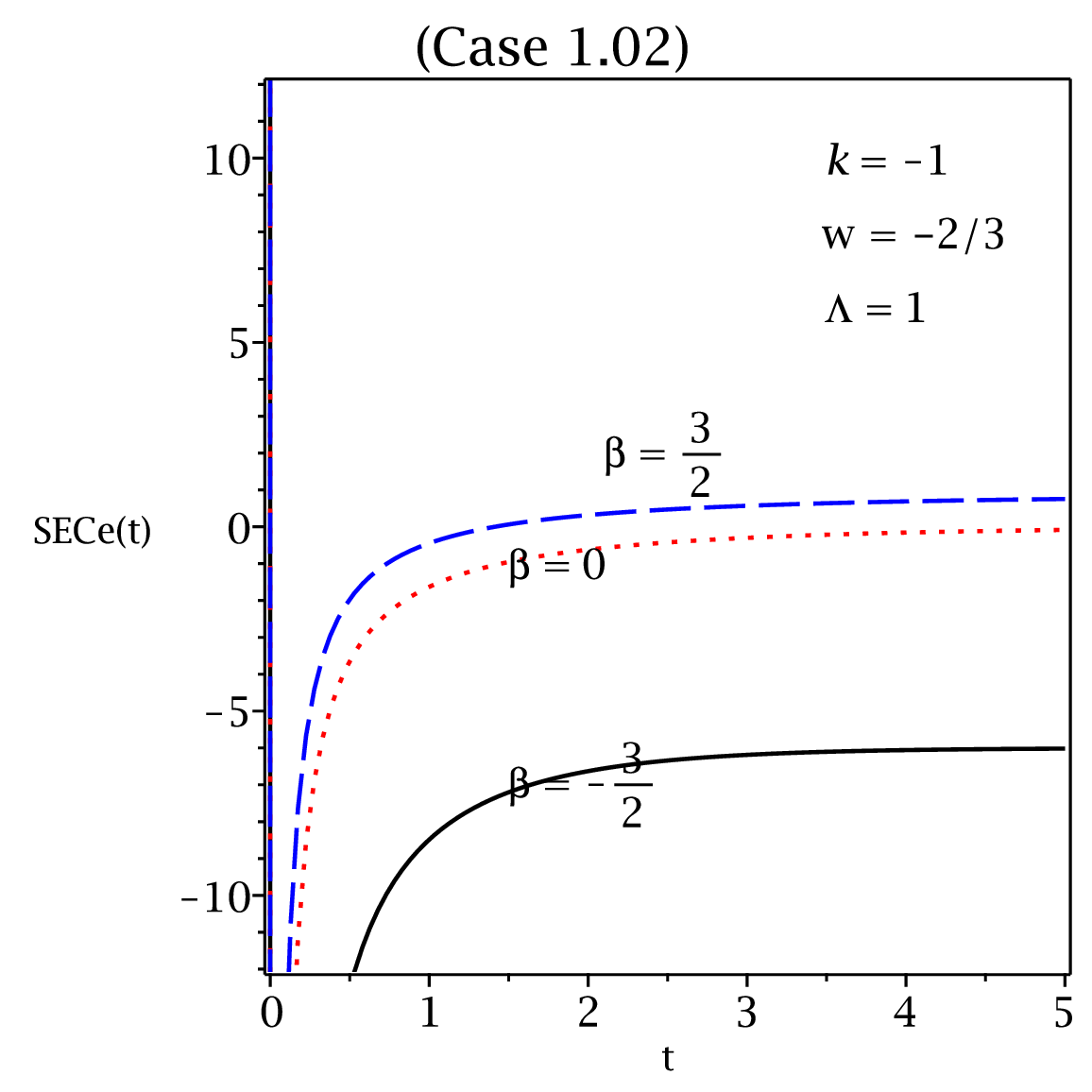}
	\includegraphics[width=3.4cm]{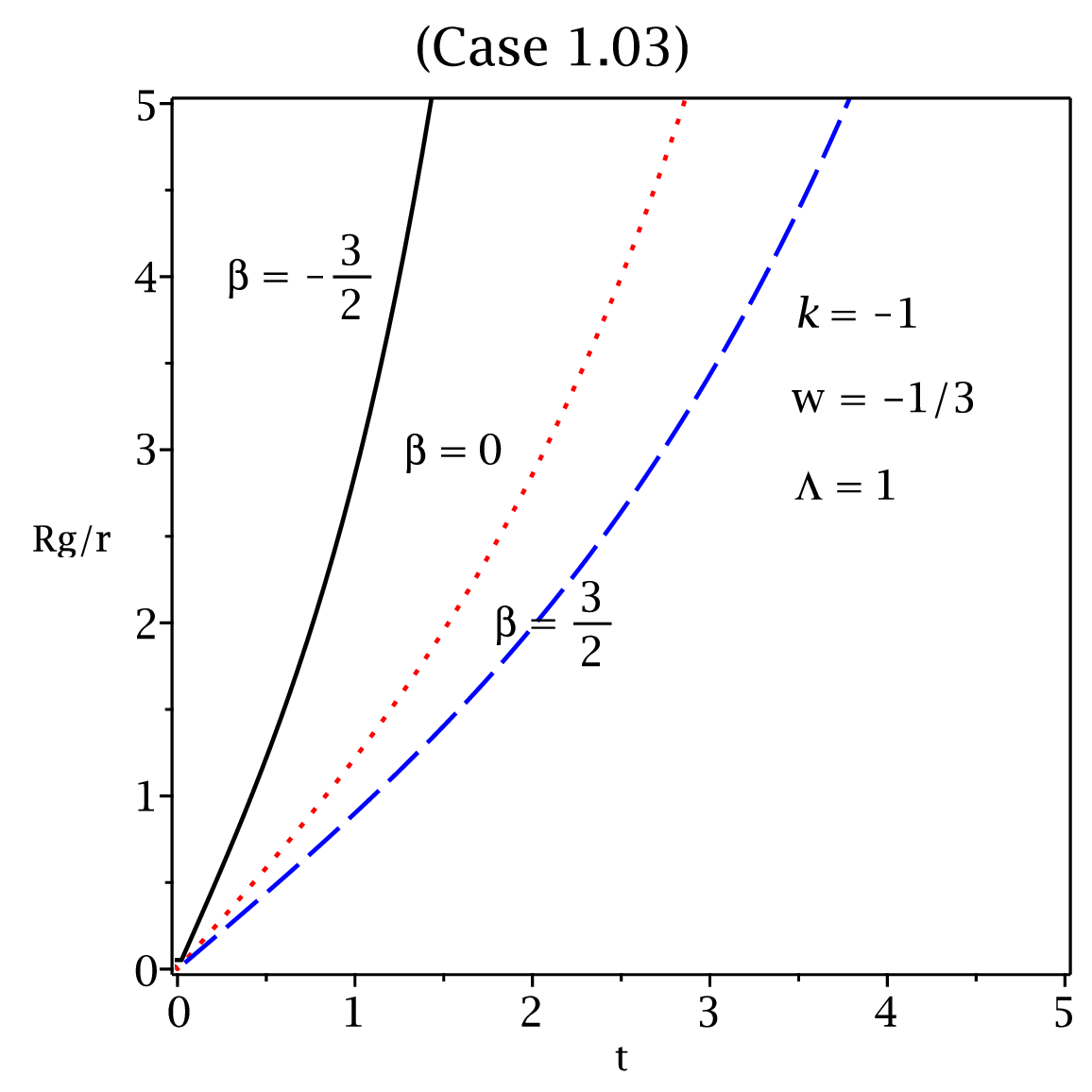}
	\includegraphics[width=3.4cm]{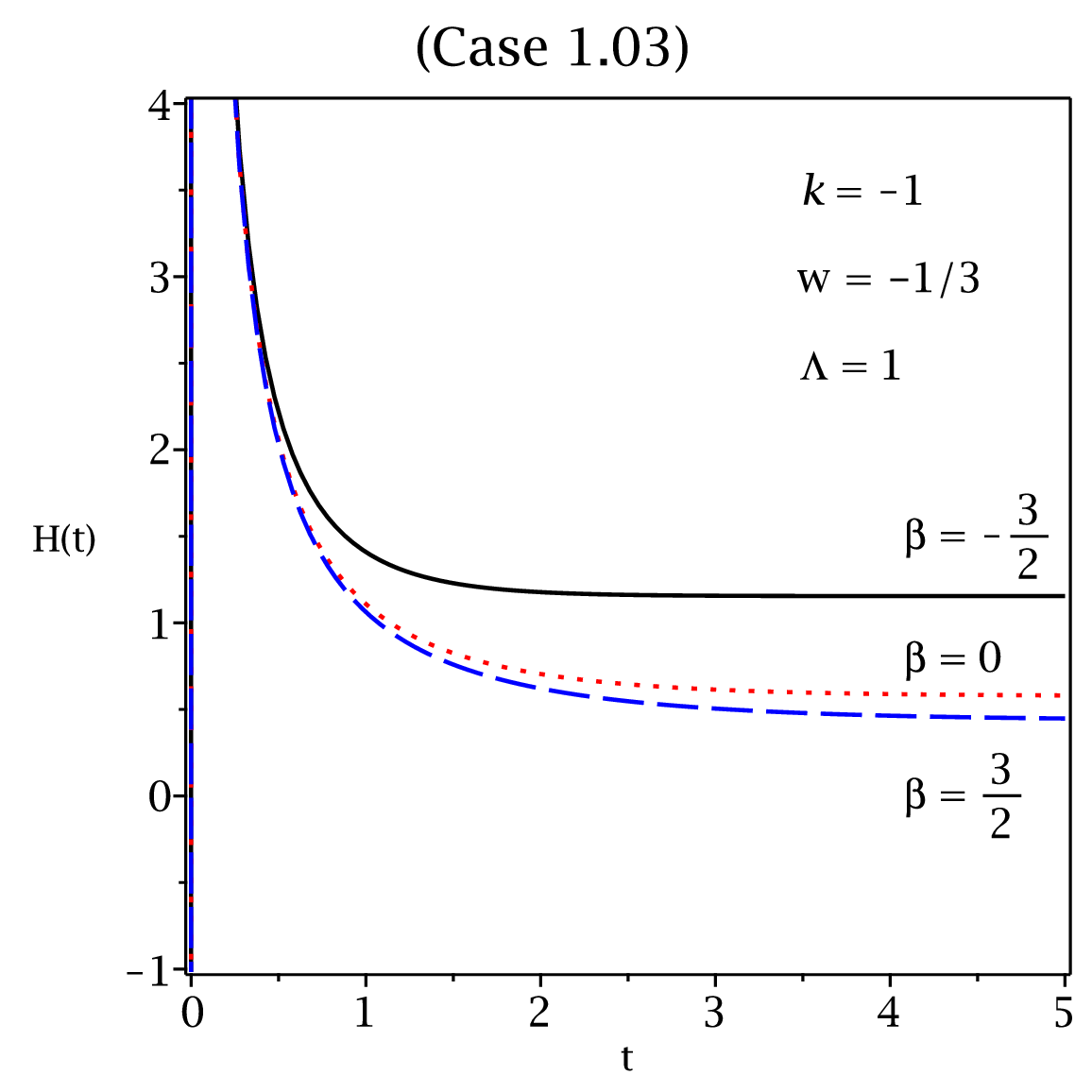}
	\includegraphics[width=3.4cm]{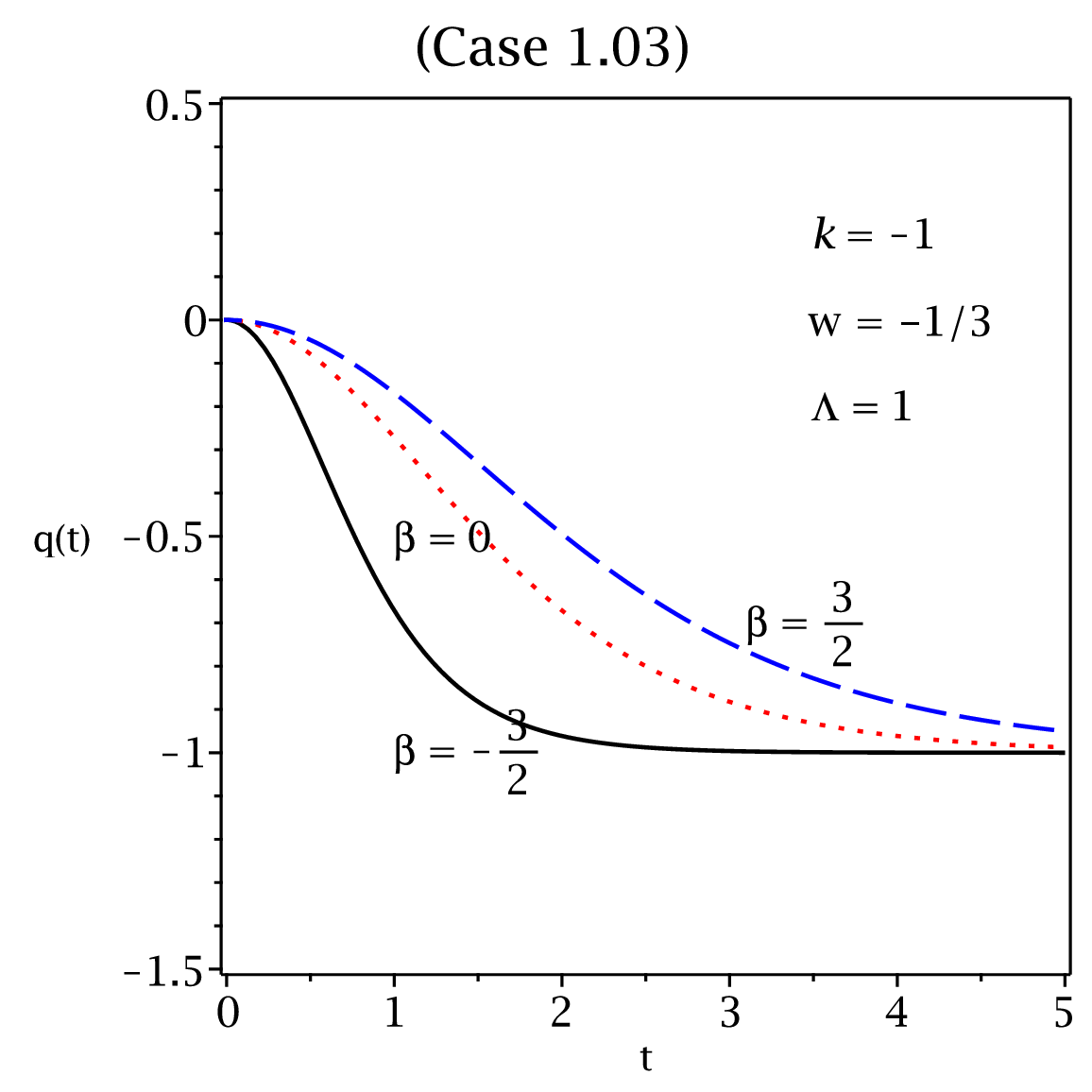}
	\includegraphics[width=3.4cm]{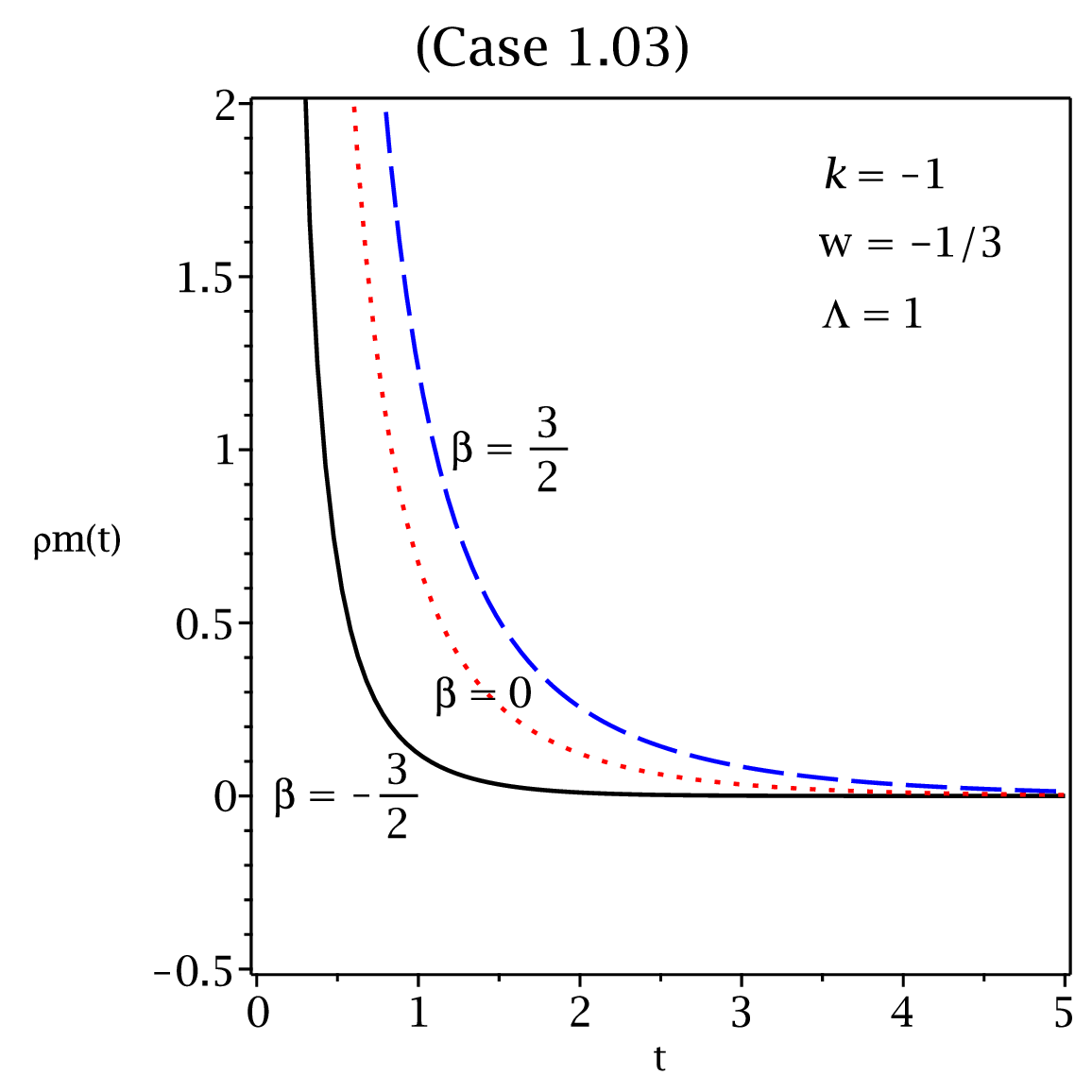}
	\includegraphics[width=3.4cm]{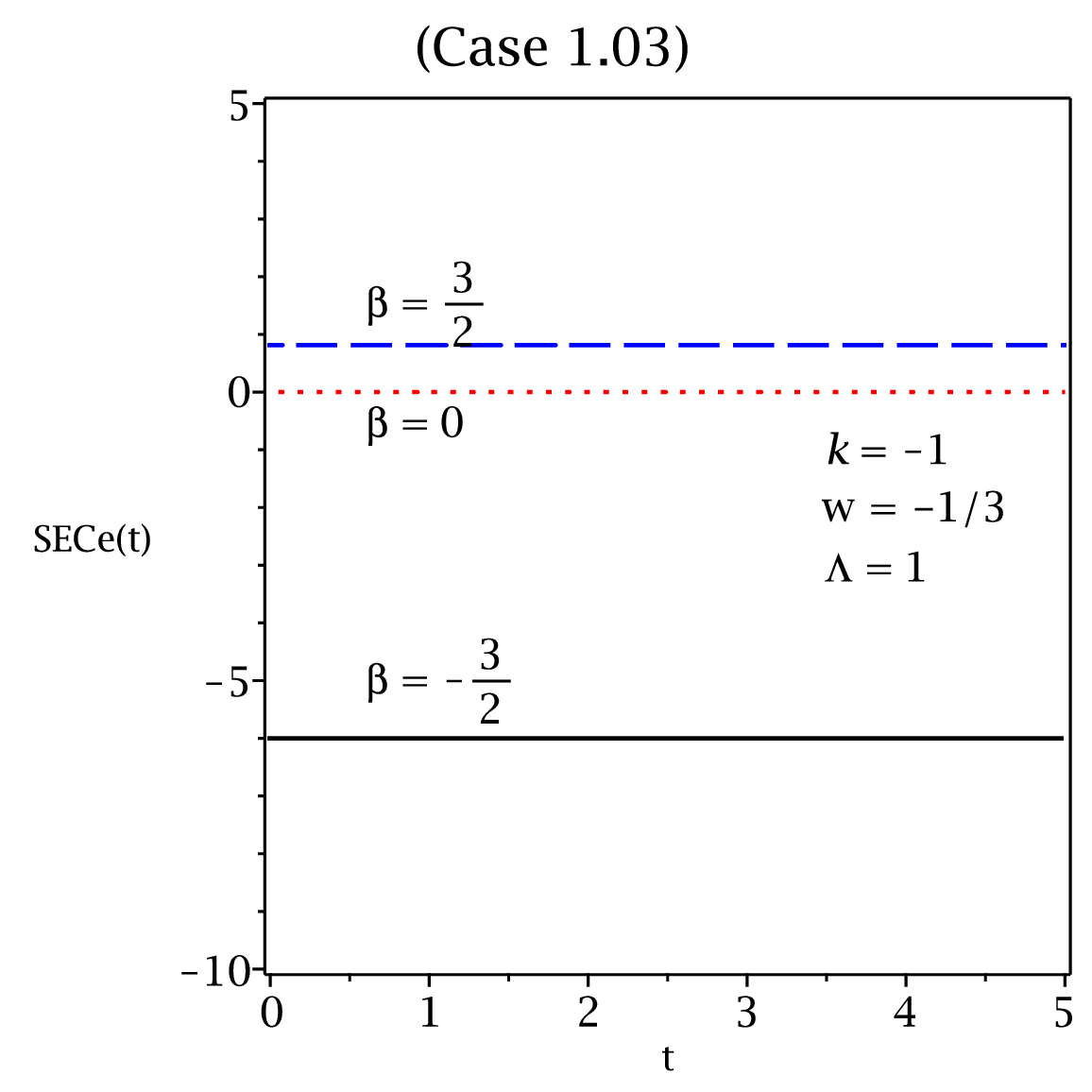}
	\includegraphics[width=3.4cm]{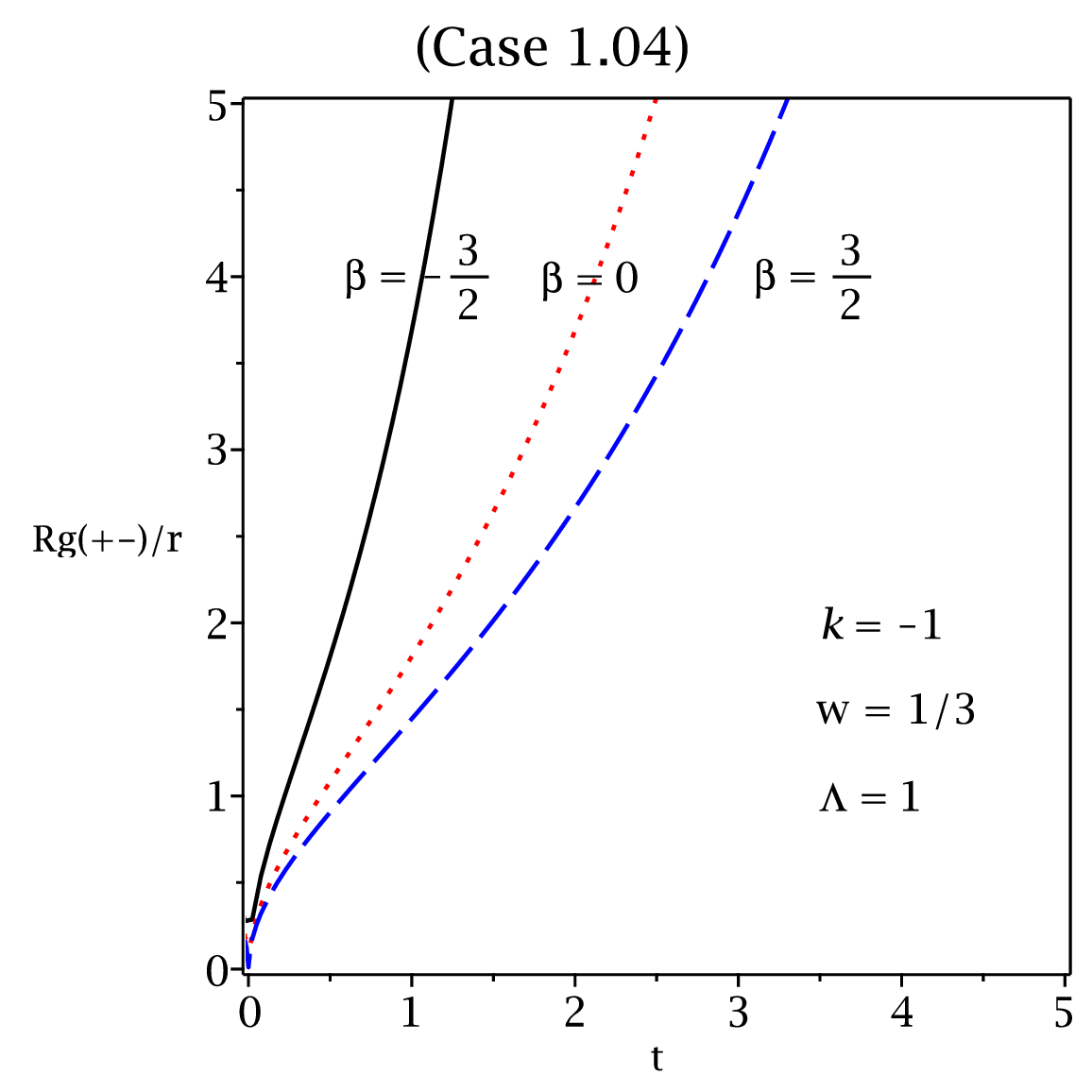}
	\includegraphics[width=3.4cm]{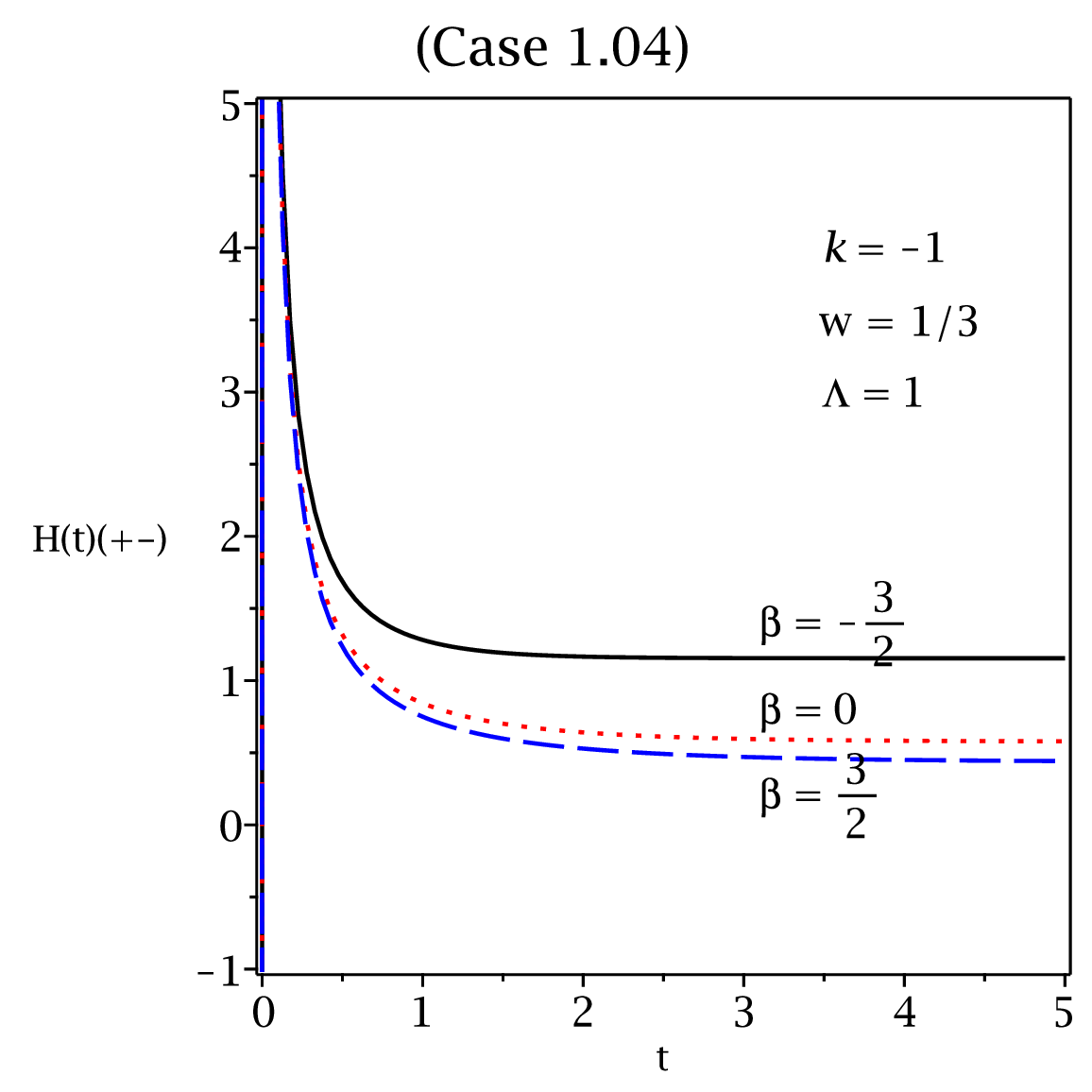}
	\includegraphics[width=3.4cm]{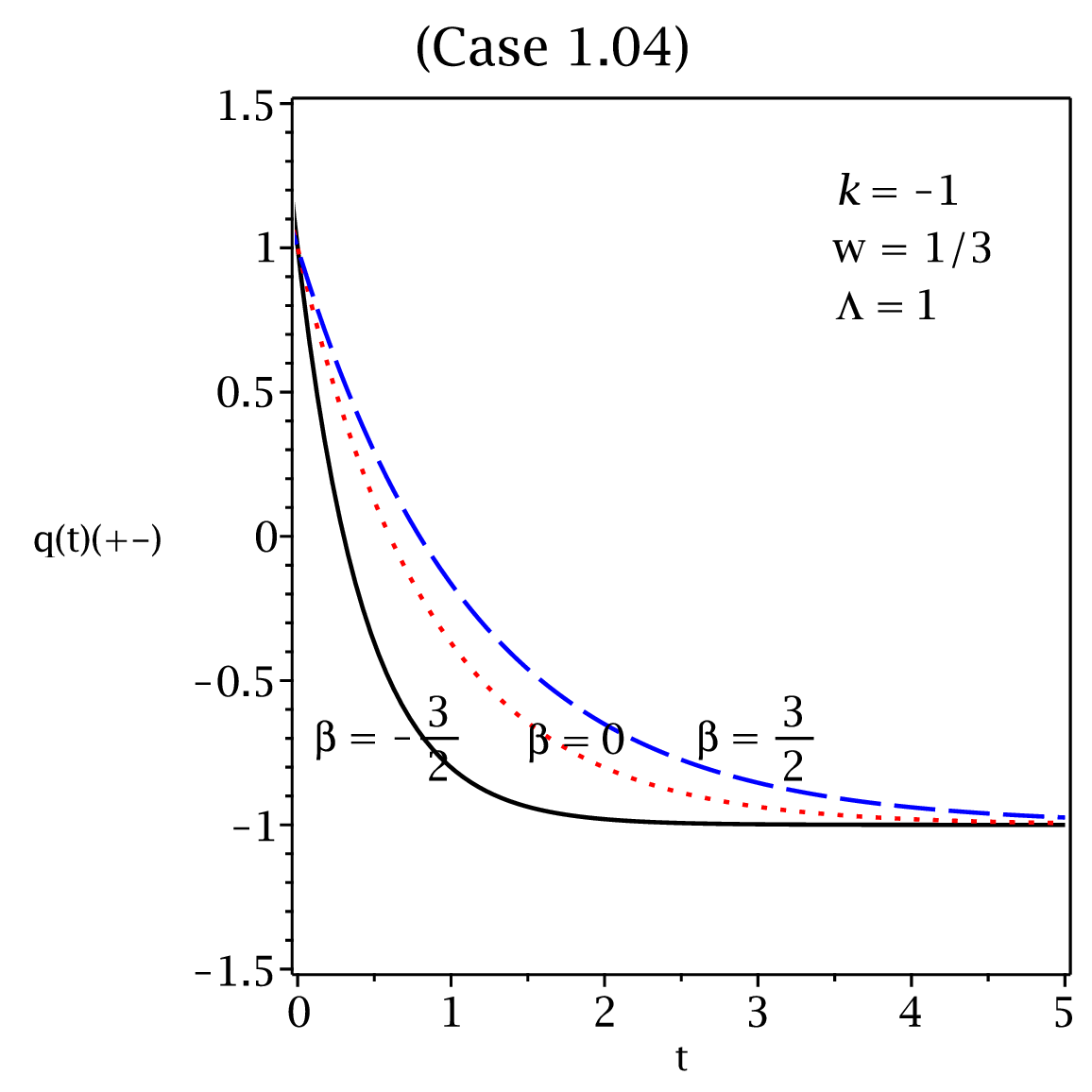}
	\includegraphics[width=3.4cm]{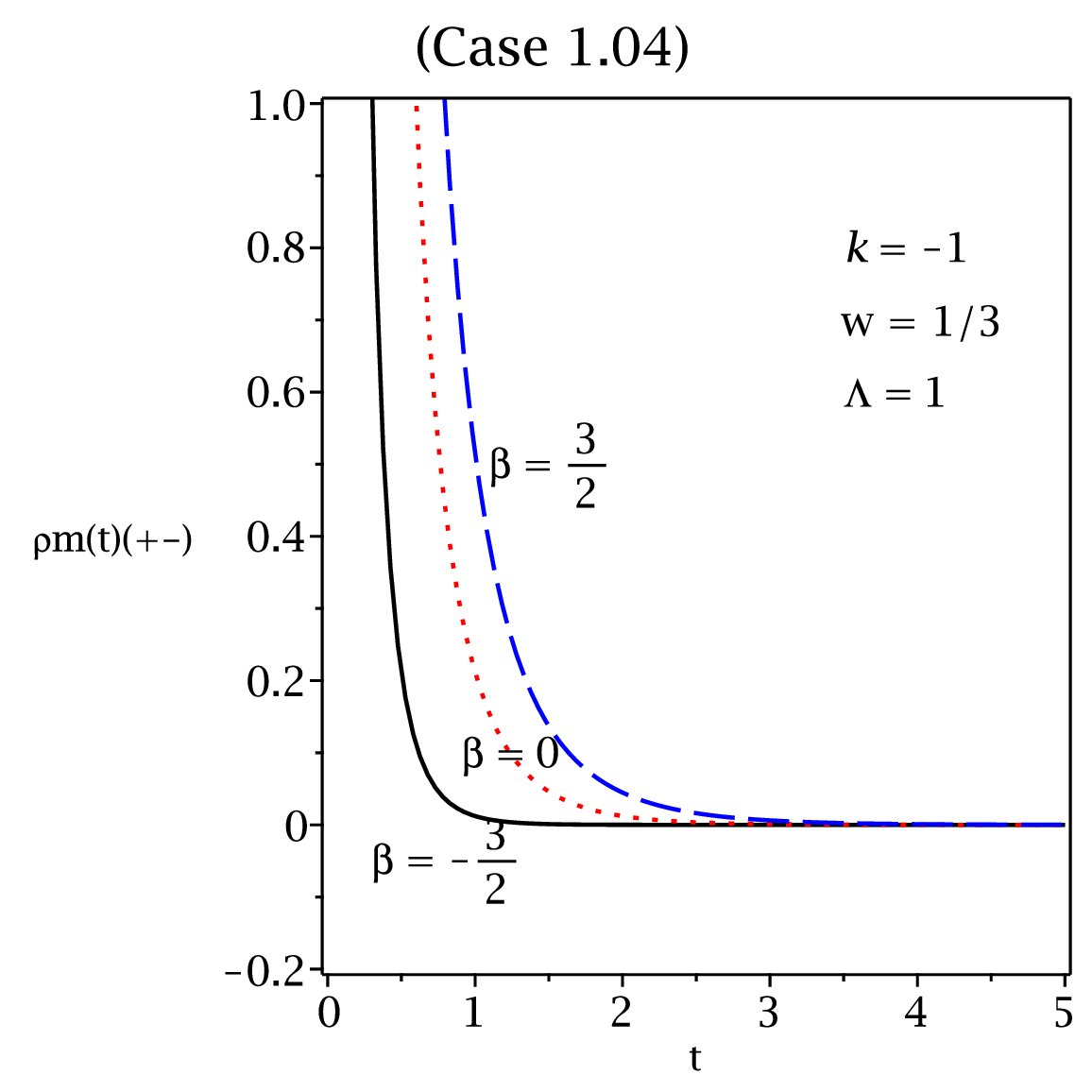}
	\includegraphics[width=3.4cm]{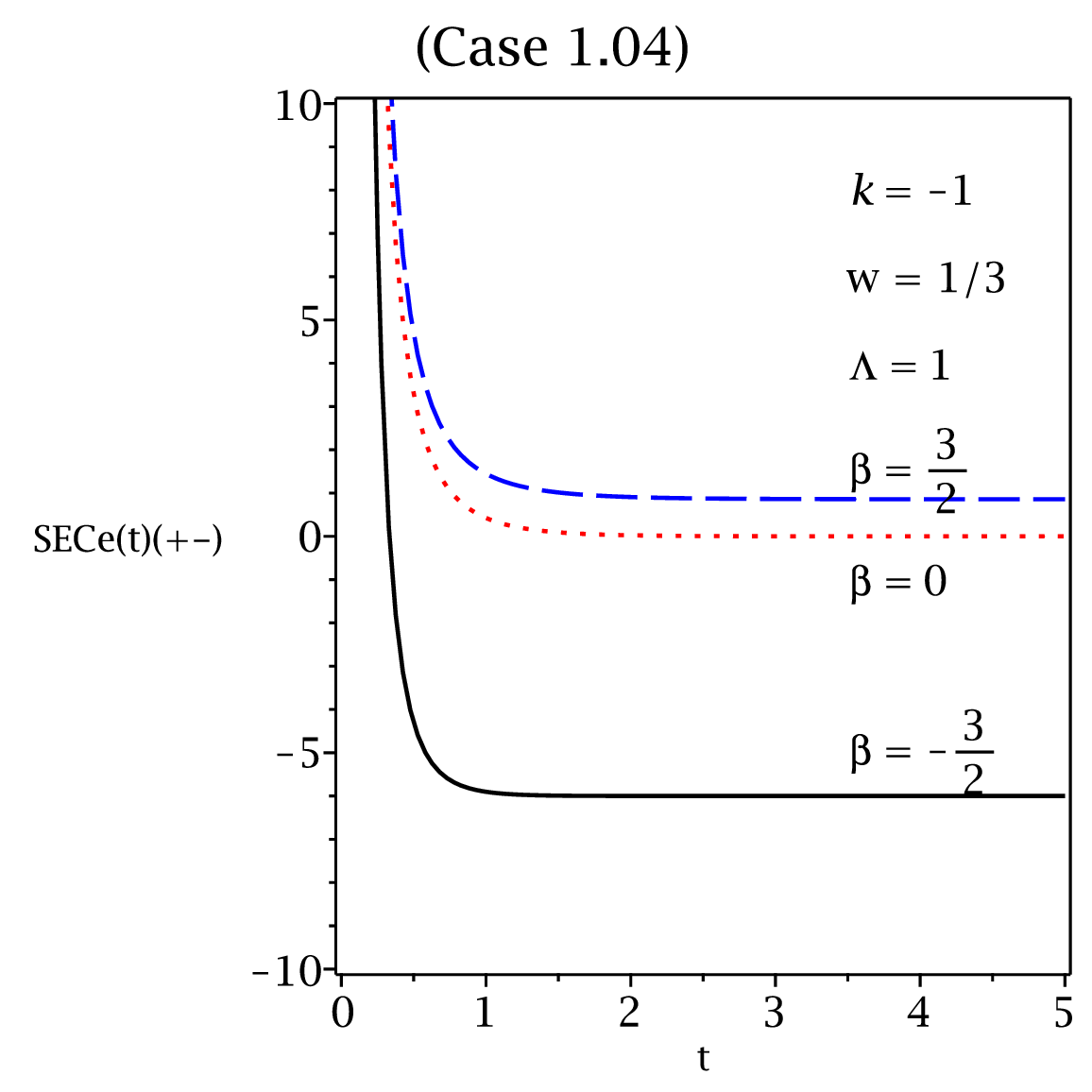}
	\caption{These figures are for $\Lambda>0$ and $k=-1$.
		These figures represent the quantities $R_g$ (geometrical radius), 
		$H(t)$ (Hubble parameter) and $q(t)$ (deceleration parameter) $\rho_m(t)$ 
		(energy density of the aether fluid) and $SEC_{e} \equiv SEC_{\rm eff}$ 
		(strong energy condition for the effective fluid) for the different
		values of $\beta=-3/2$ (black solid line), $\beta=0$ (red dotted line), 
		$\beta=3/2$ (blue dashed line). Assuming that $8 \pi G=1$ and
		$R_g(t=0)=0$. Assuming also that $C_1=1$, $C_2=0$ (Cases 1.01 and 1.04); $C_1=1$, $C_2=1$
		(Case 1.02); $C_1=1$, $C_2=-1$ (Case 1.03).}
	\label{Figure-101-104}
\end{minipage}
\end{figure}

\begin{figure}[!htp]
\begin{minipage}{175 mm}
	\centering	
	\includegraphics[width=3.05cm]{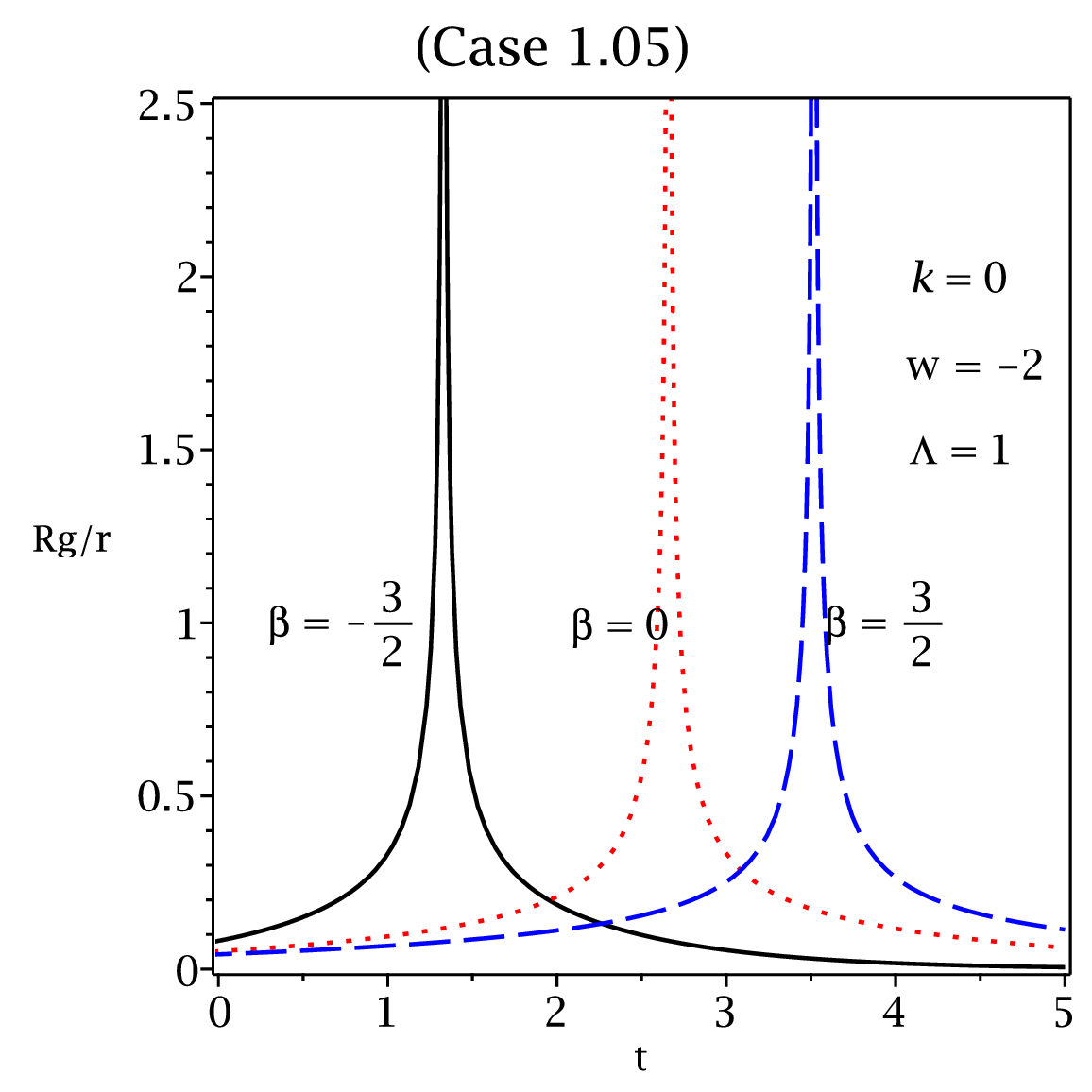}
	\includegraphics[width=3.05cm]{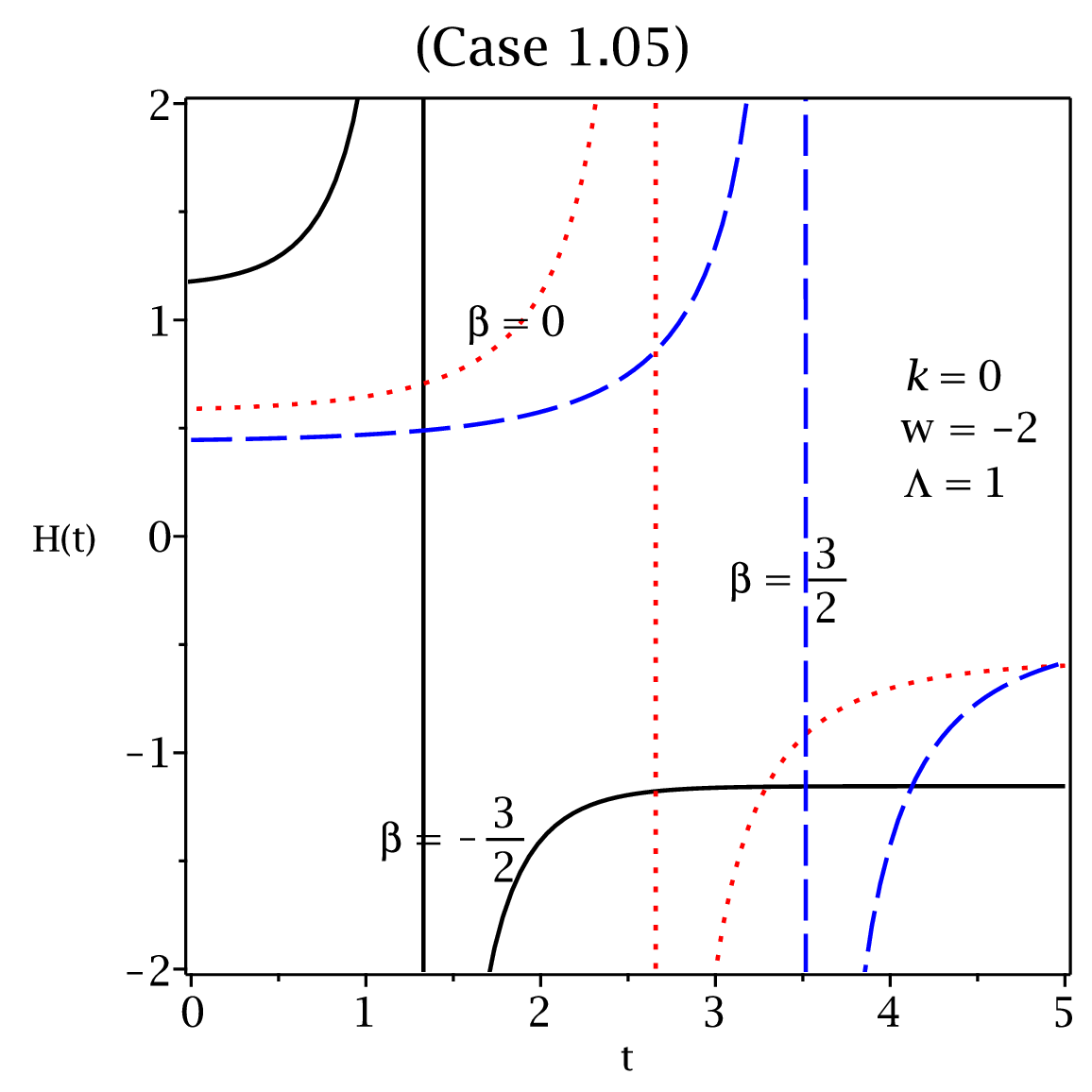}
	\includegraphics[width=3.05cm]{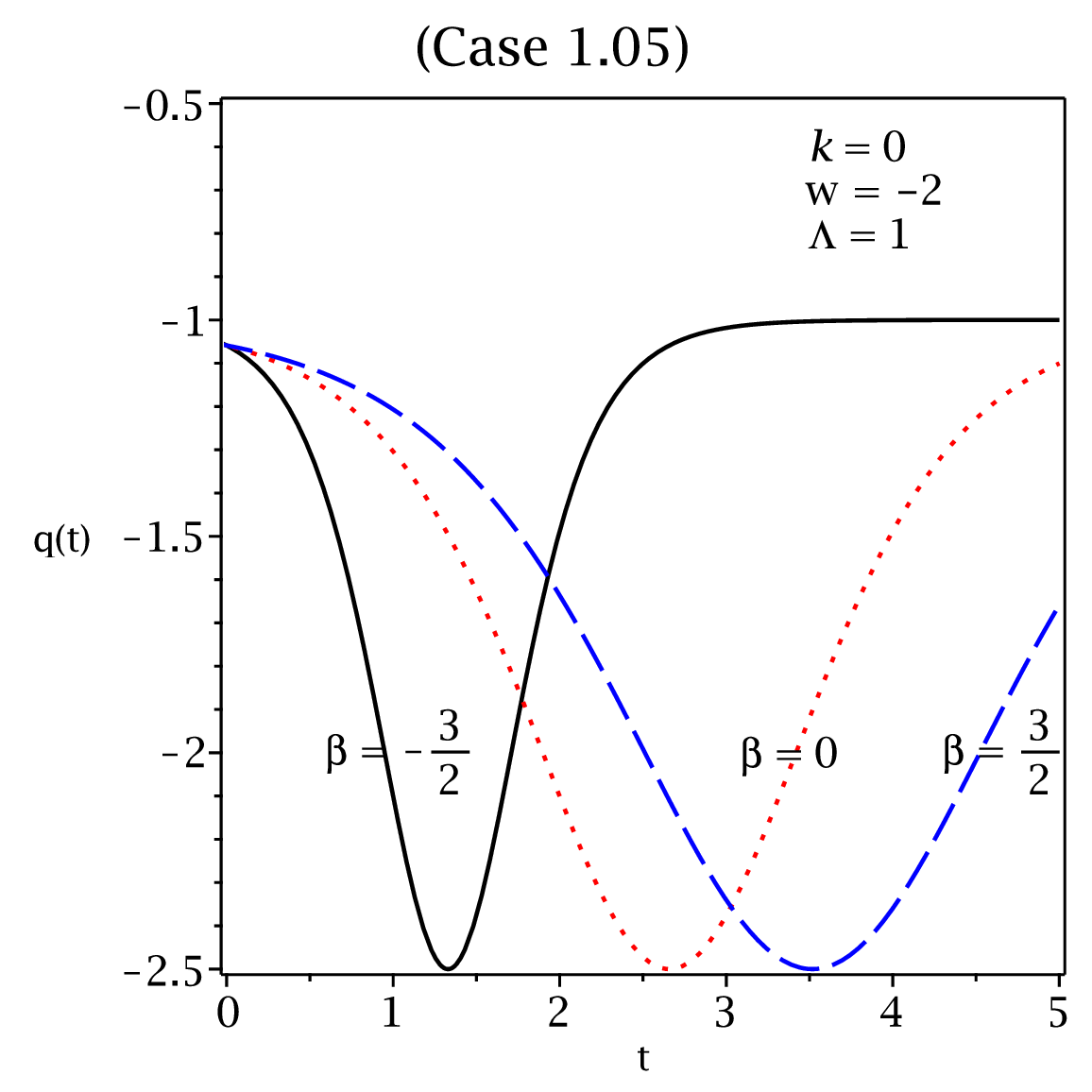}
	\includegraphics[width=3.05cm]{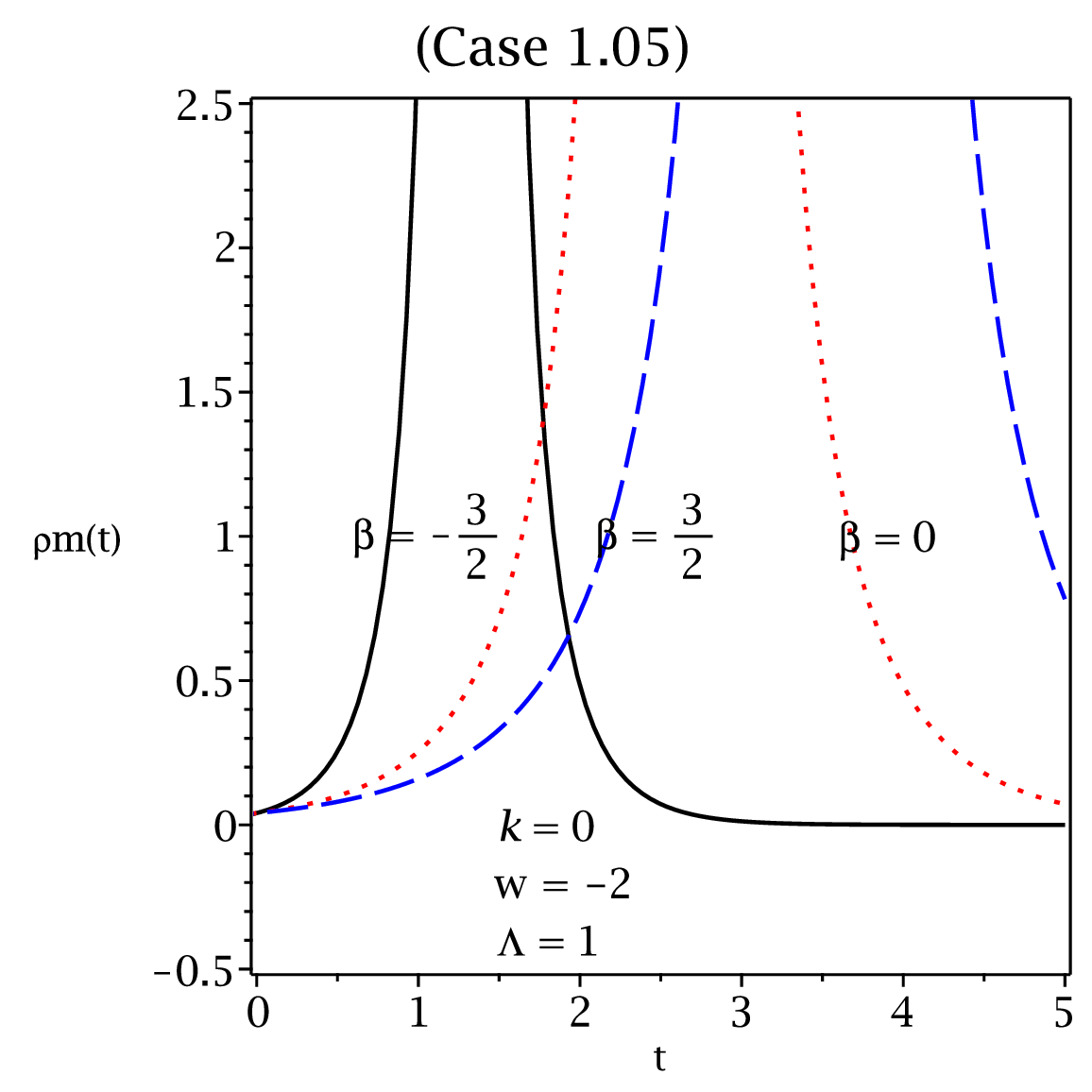}
	\includegraphics[width=3.05cm]{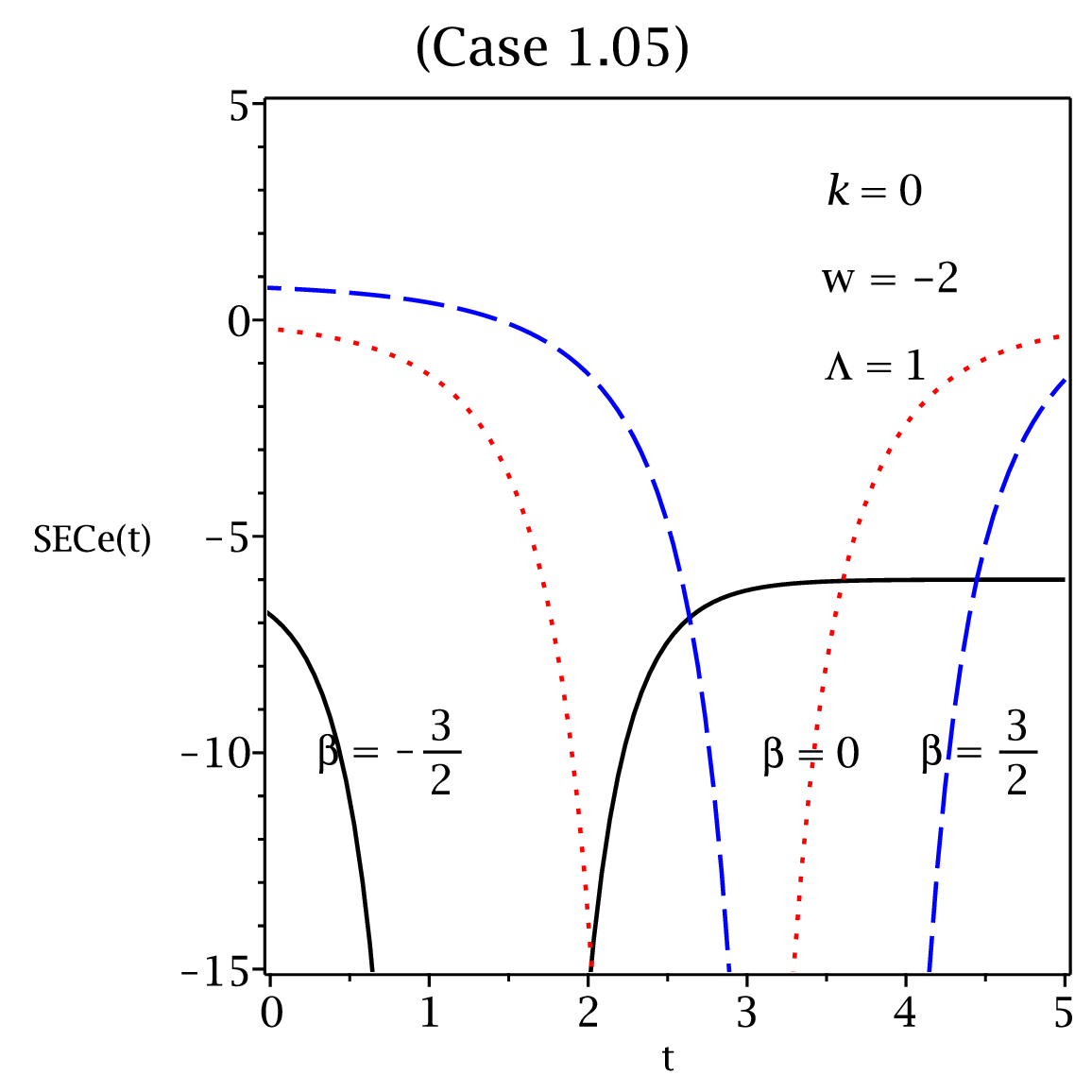}
	\includegraphics[width=3.05cm]{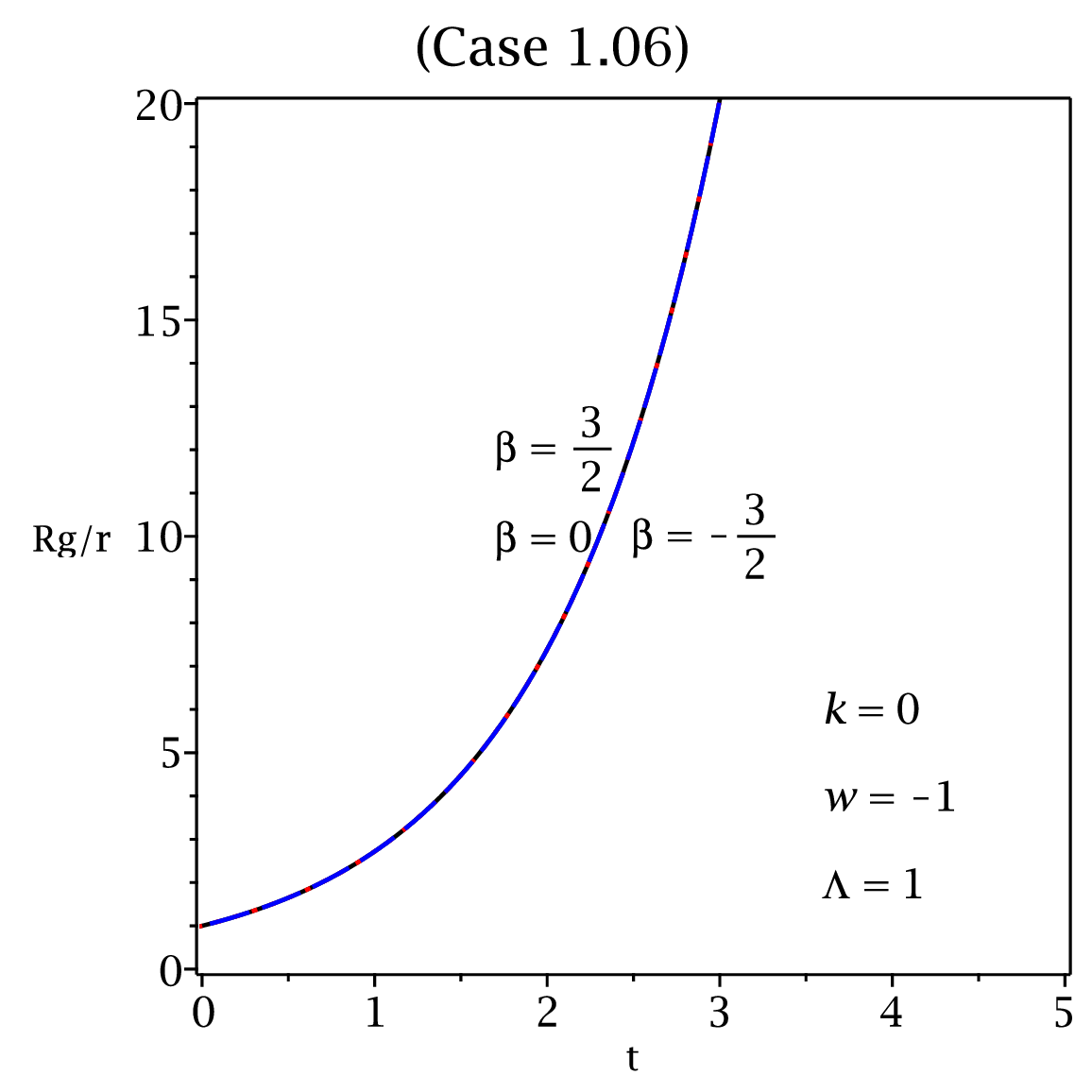}
	\includegraphics[width=3.05cm]{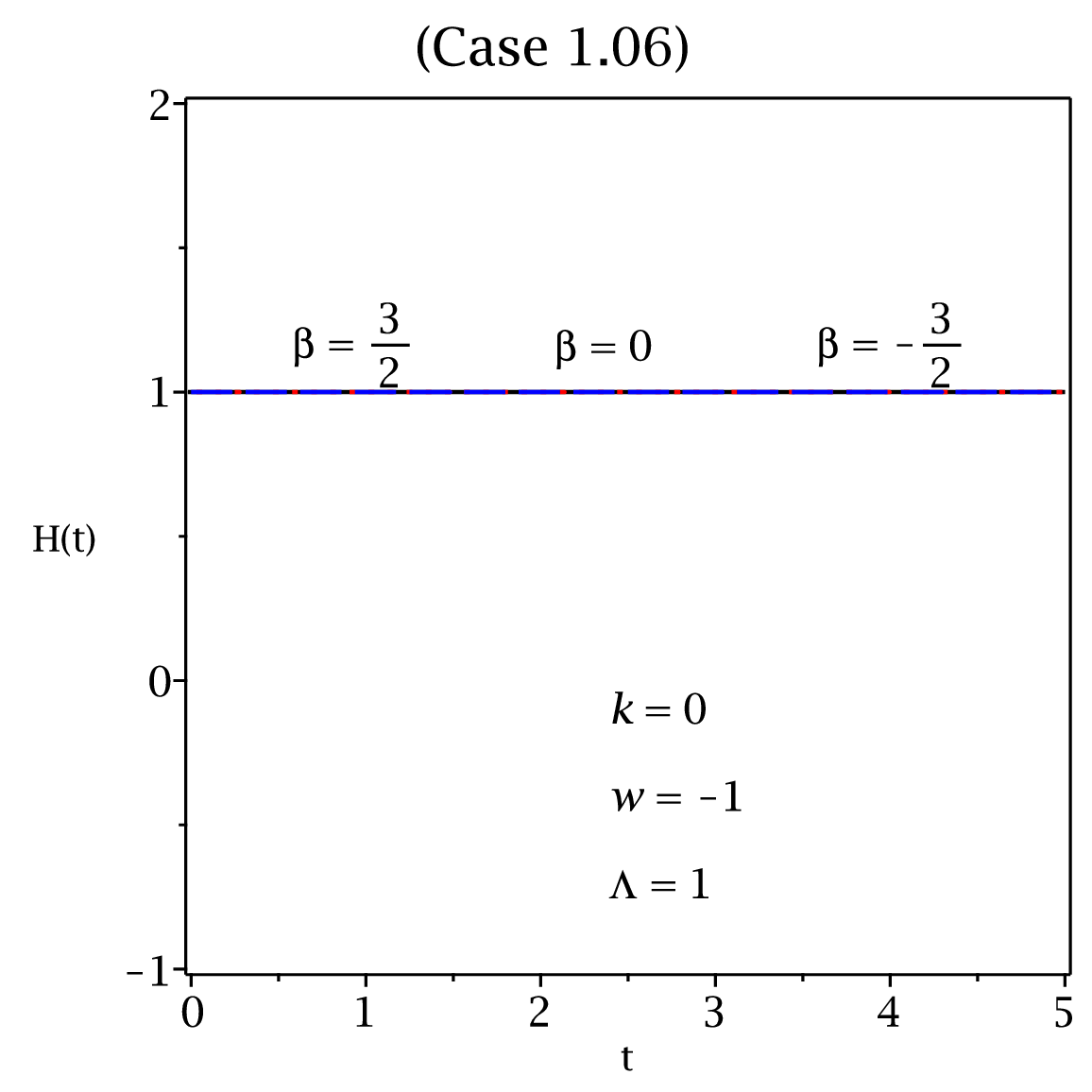}
	\includegraphics[width=3.05cm]{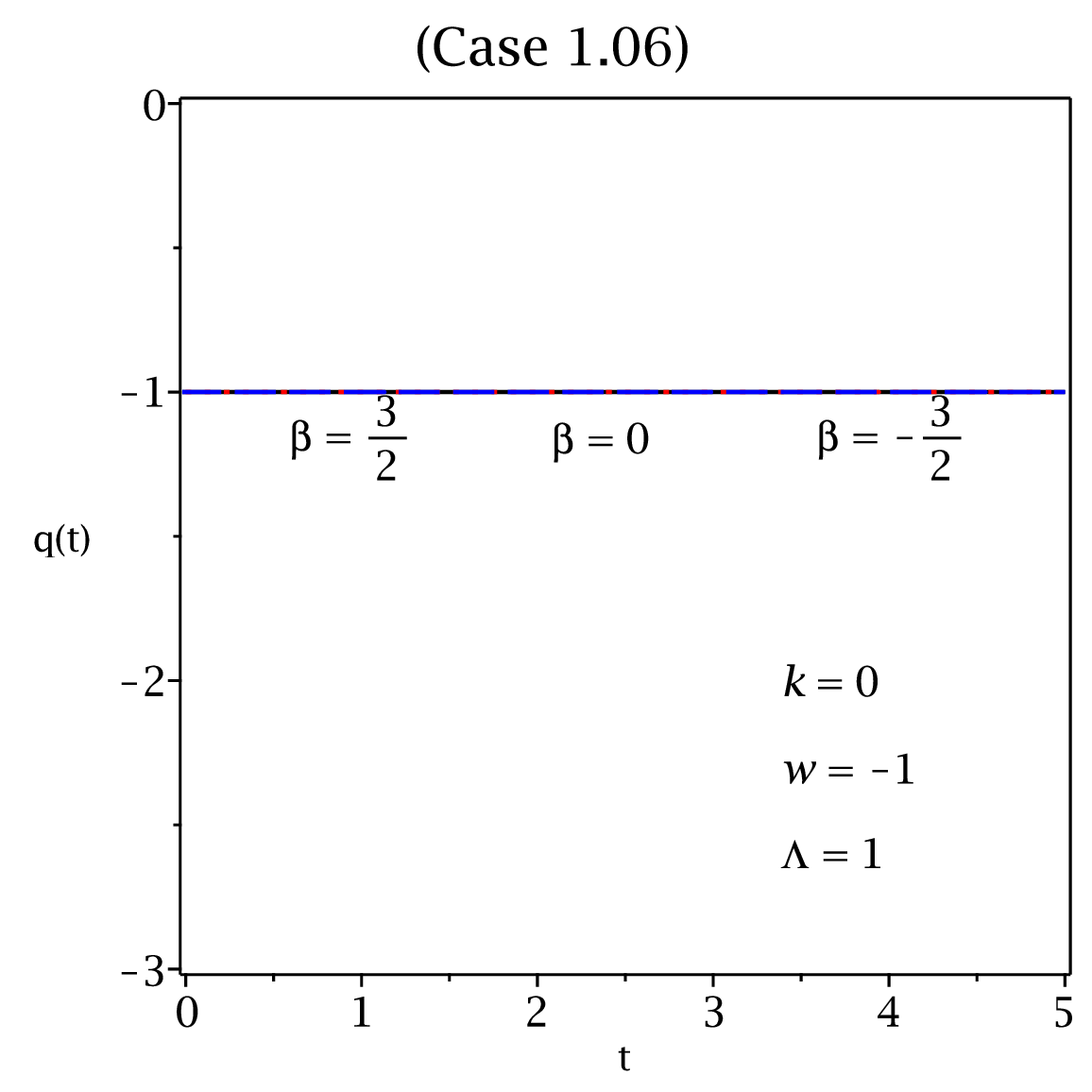}
	\includegraphics[width=3.05cm]{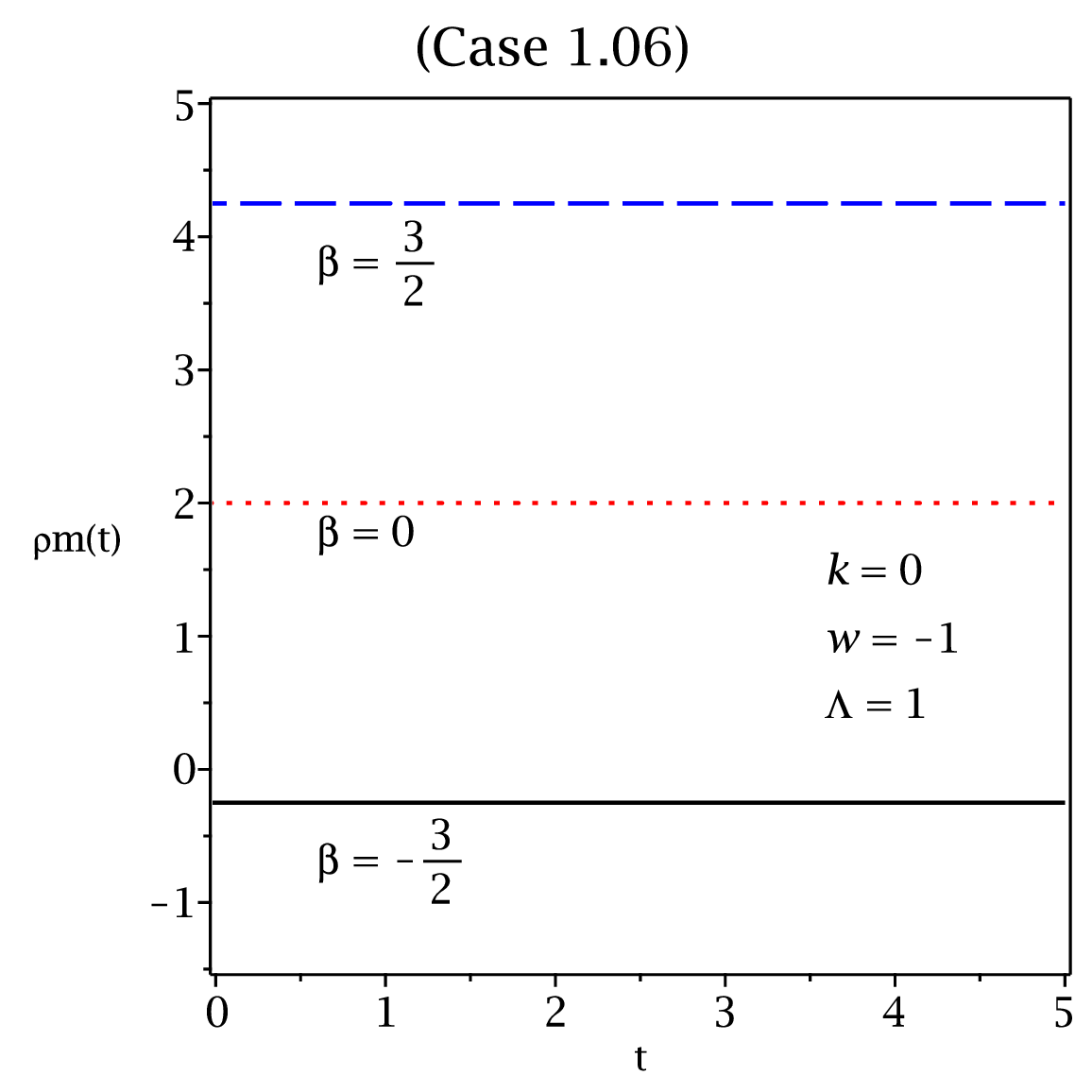}
	\includegraphics[width=3.05cm]{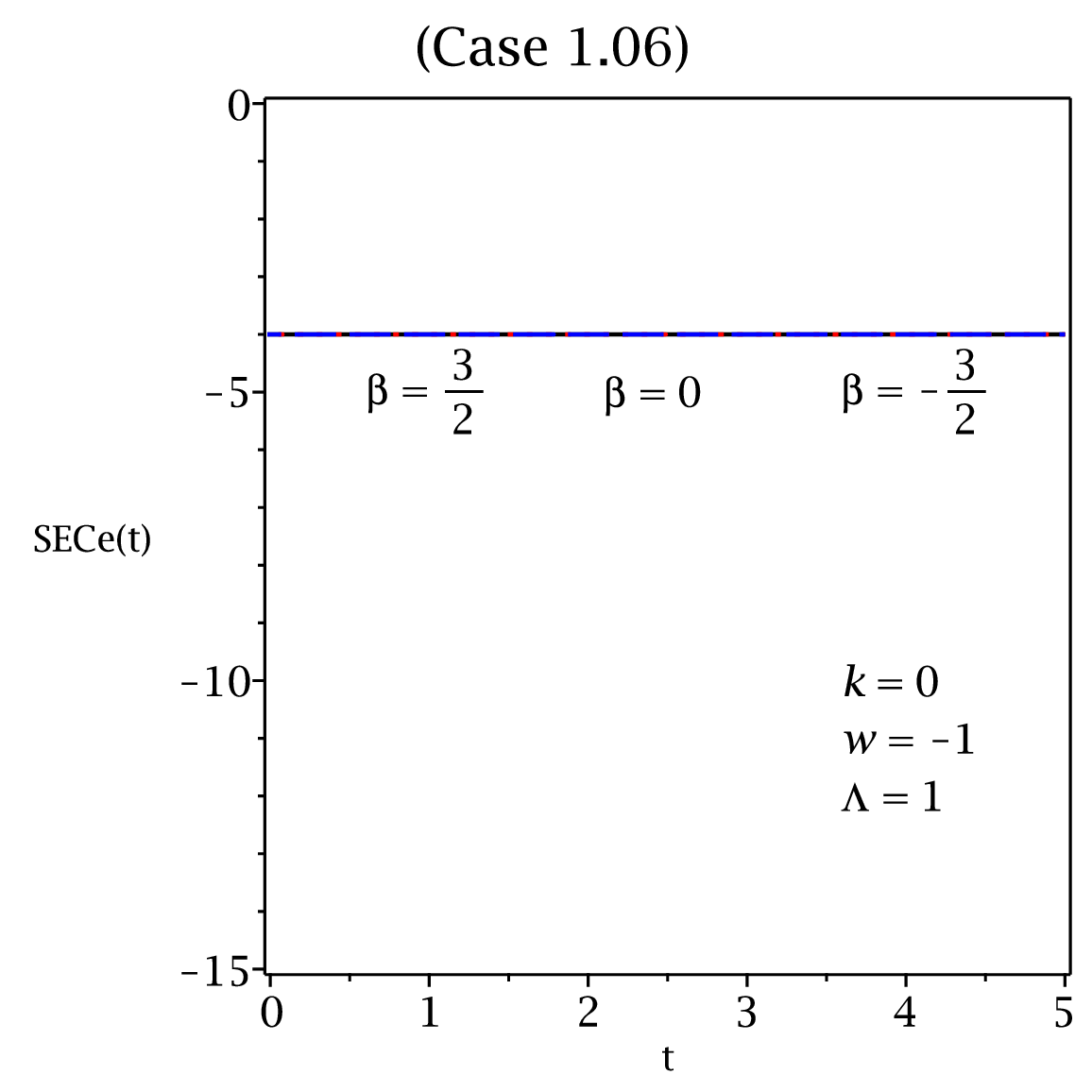}
	\includegraphics[width=3.05cm]{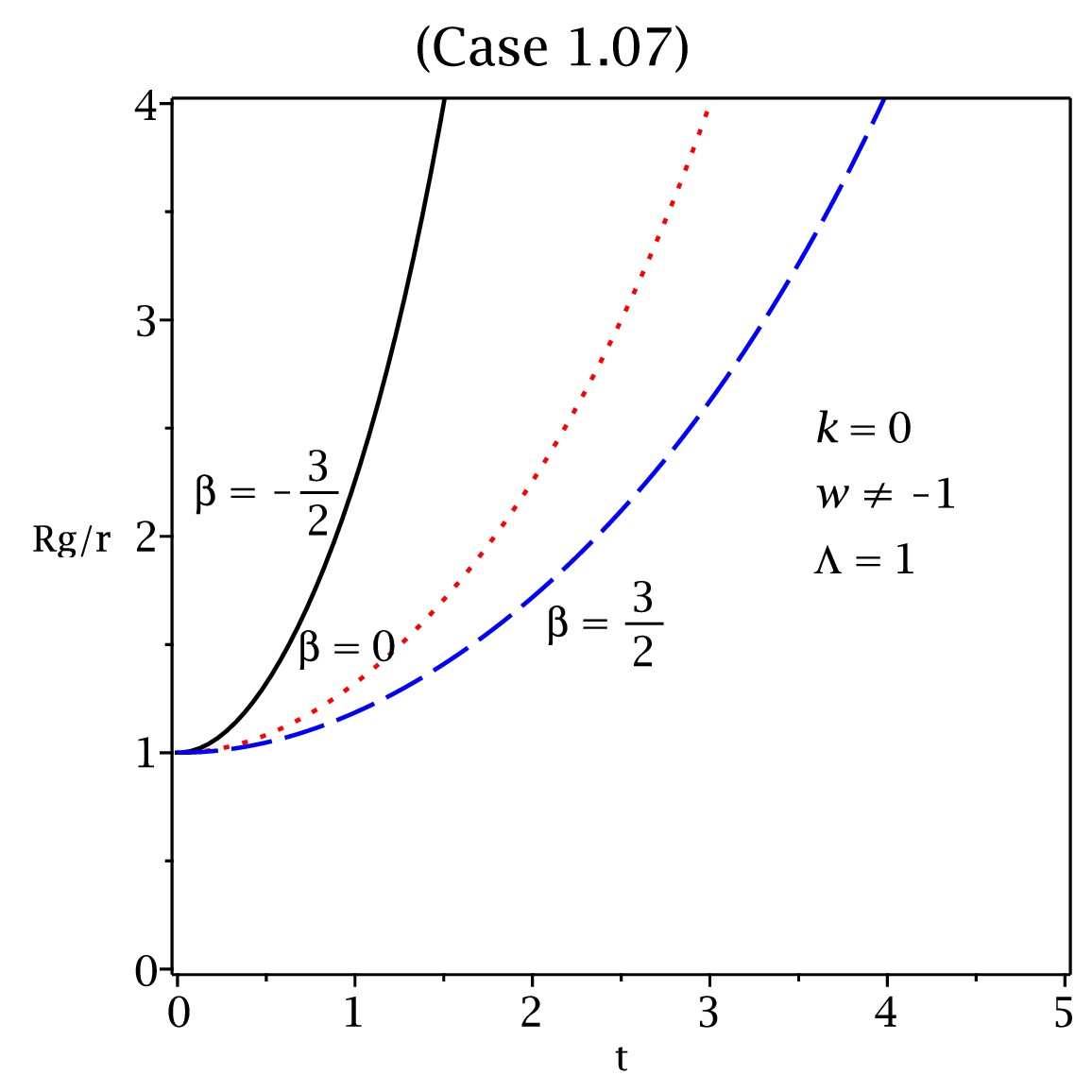}
	\includegraphics[width=3.05cm]{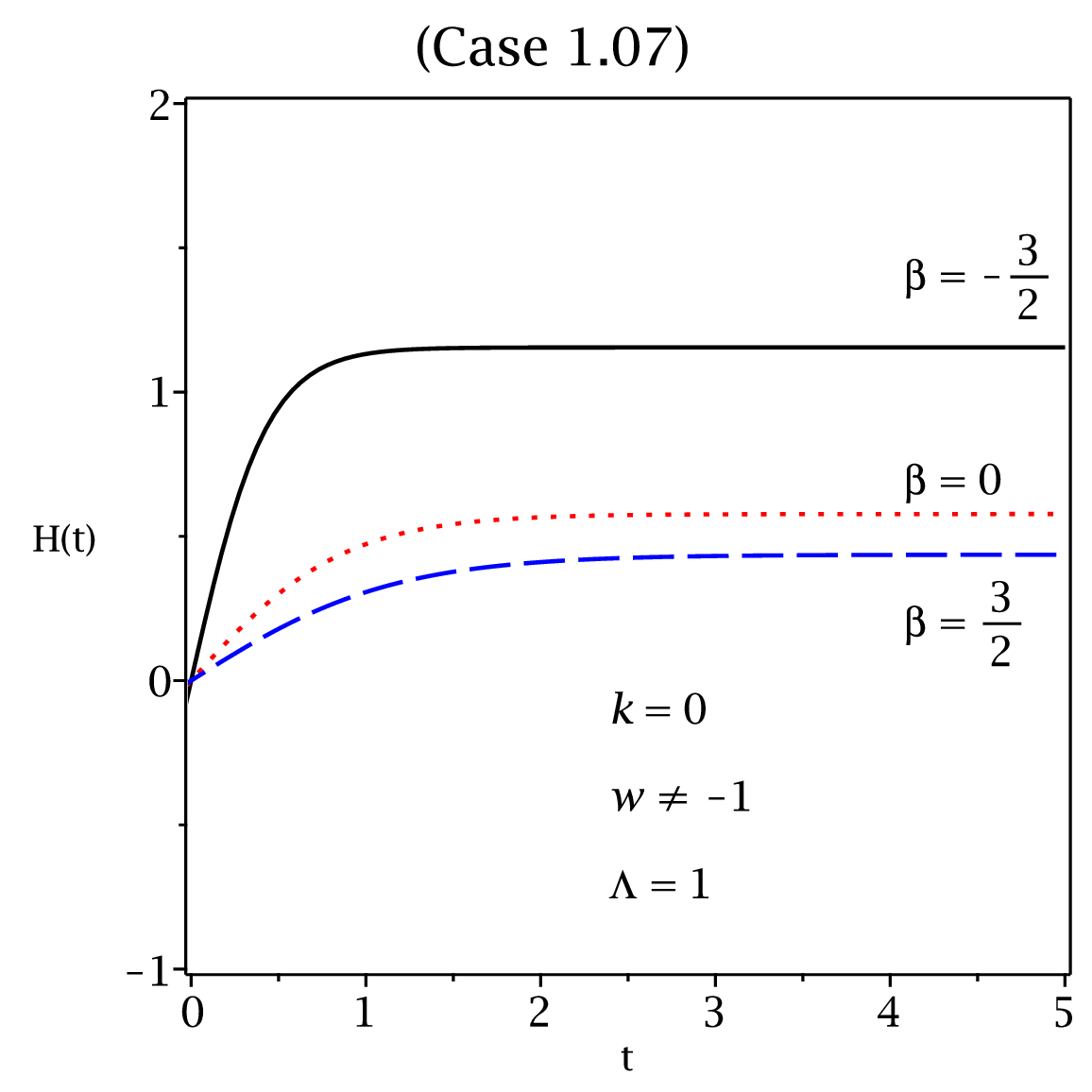}
	\includegraphics[width=3.05cm]{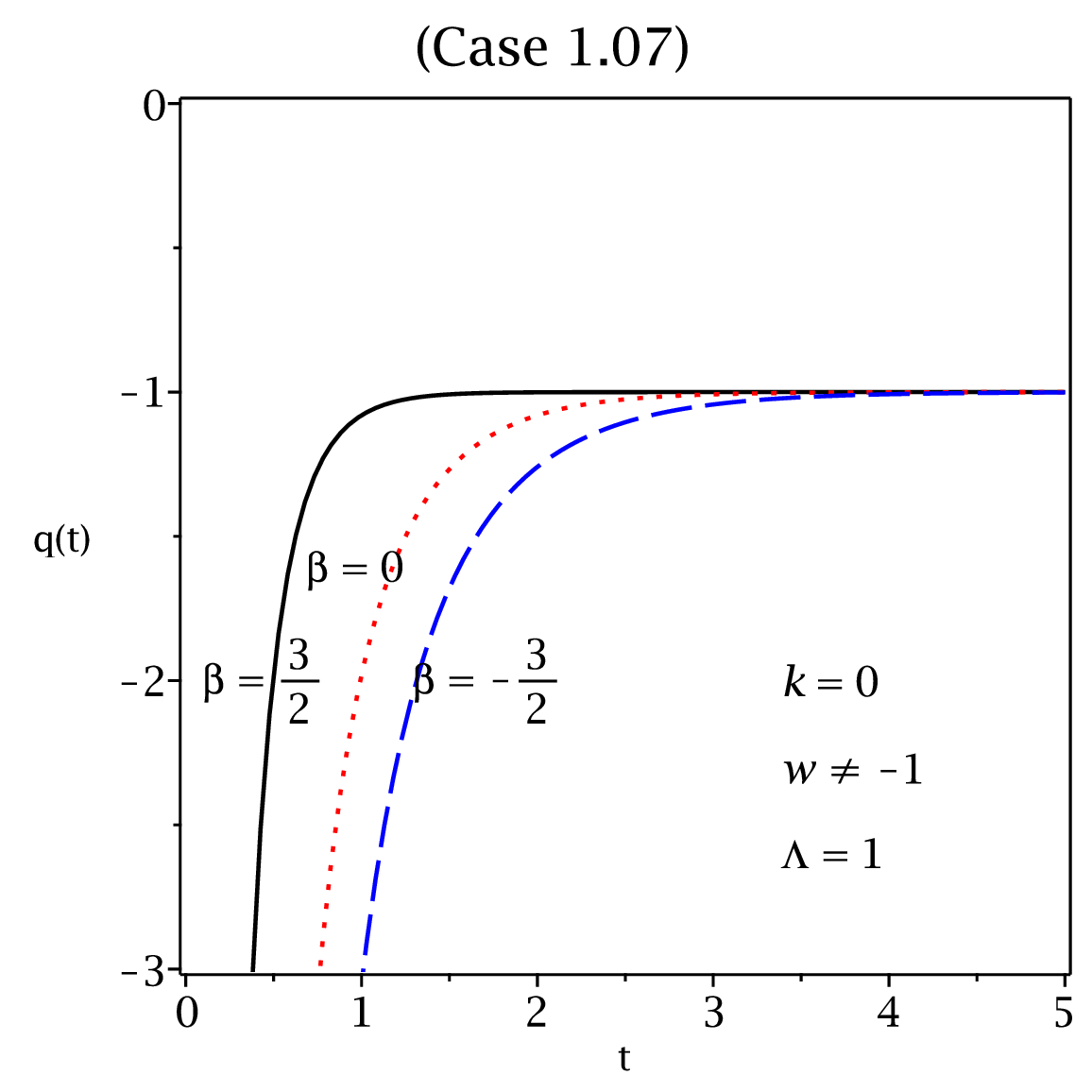}
	\includegraphics[width=3.05cm]{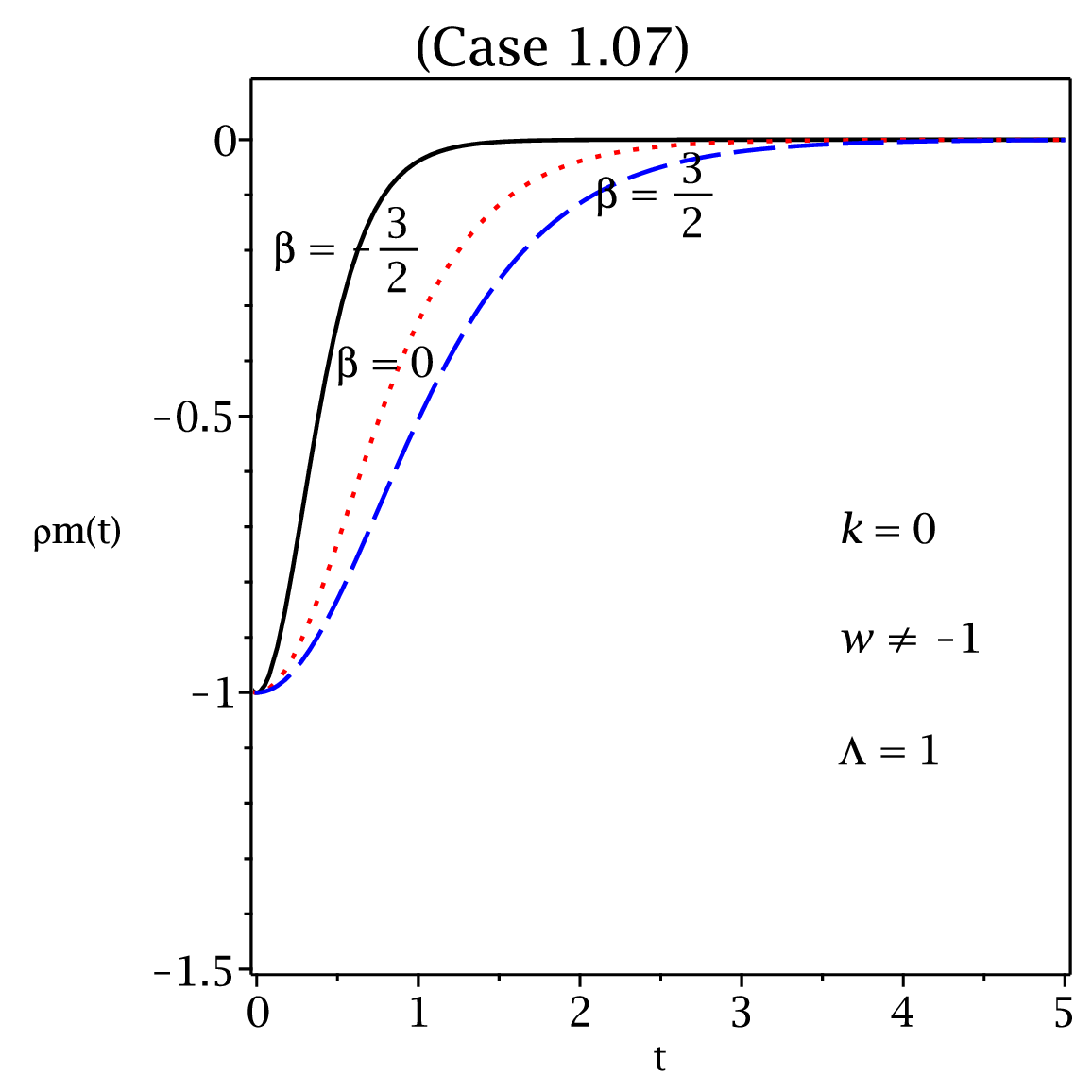}
	\includegraphics[width=3.05cm]{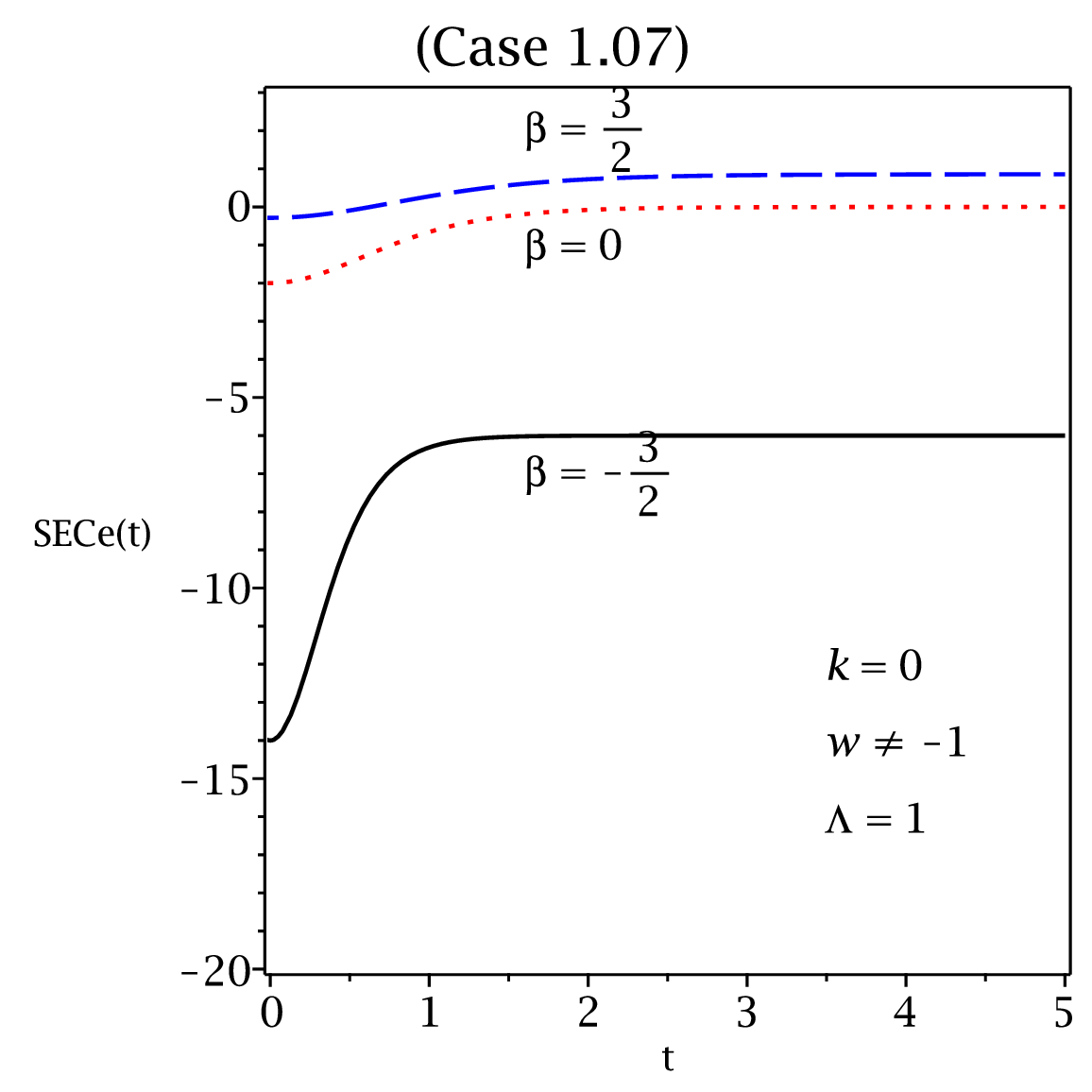}
	\includegraphics[width=3.05cm]{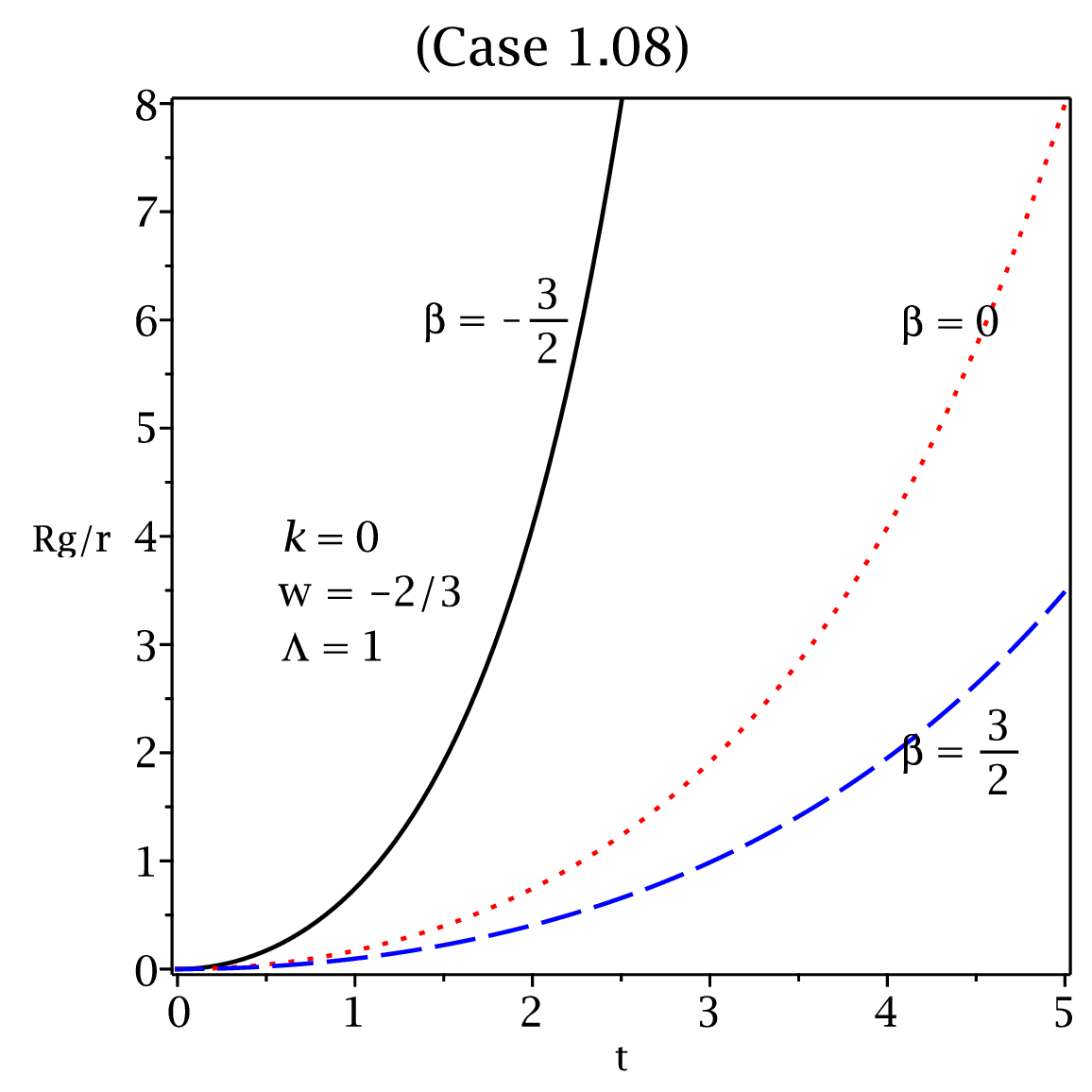}
	\includegraphics[width=3.05cm]{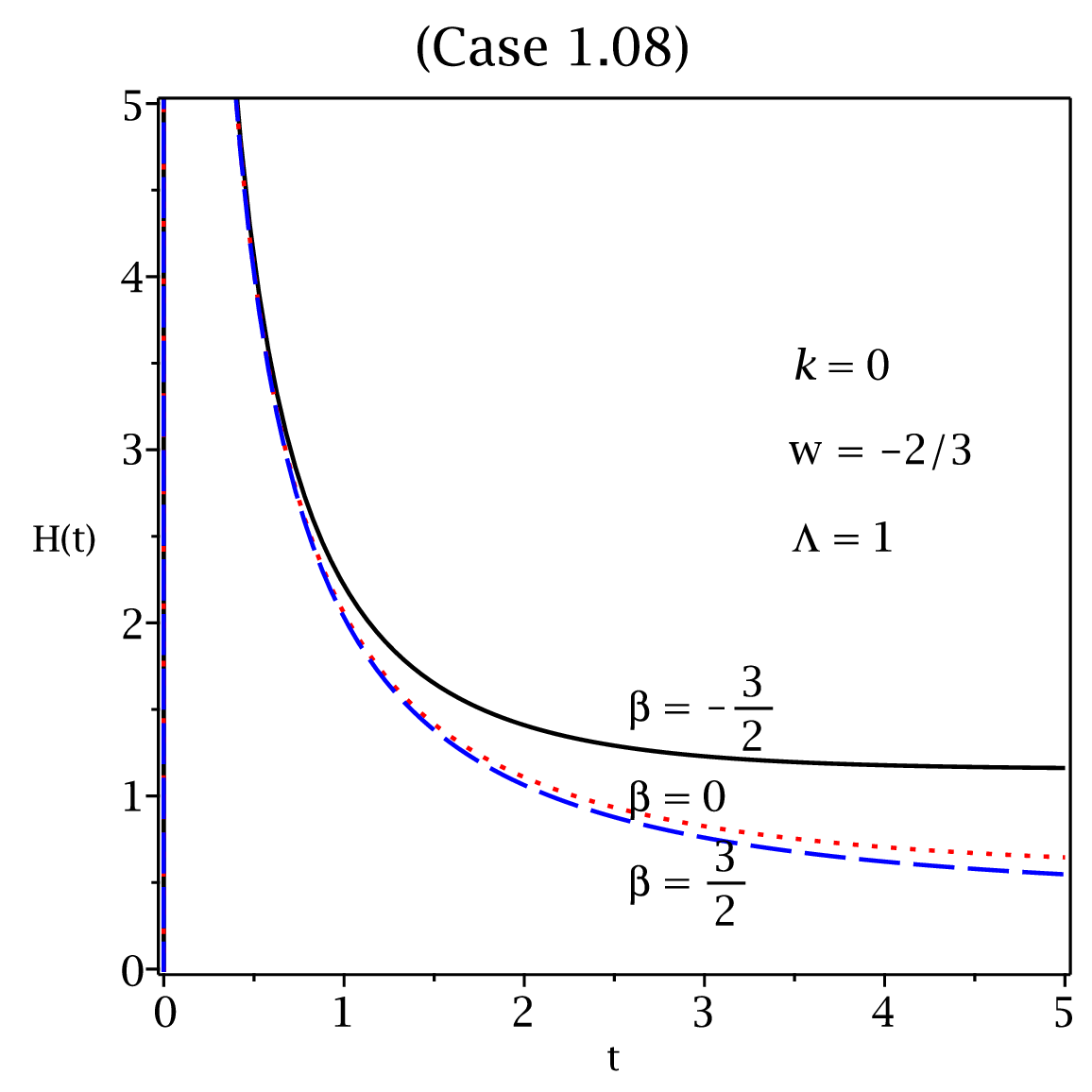}
	\includegraphics[width=3.05cm]{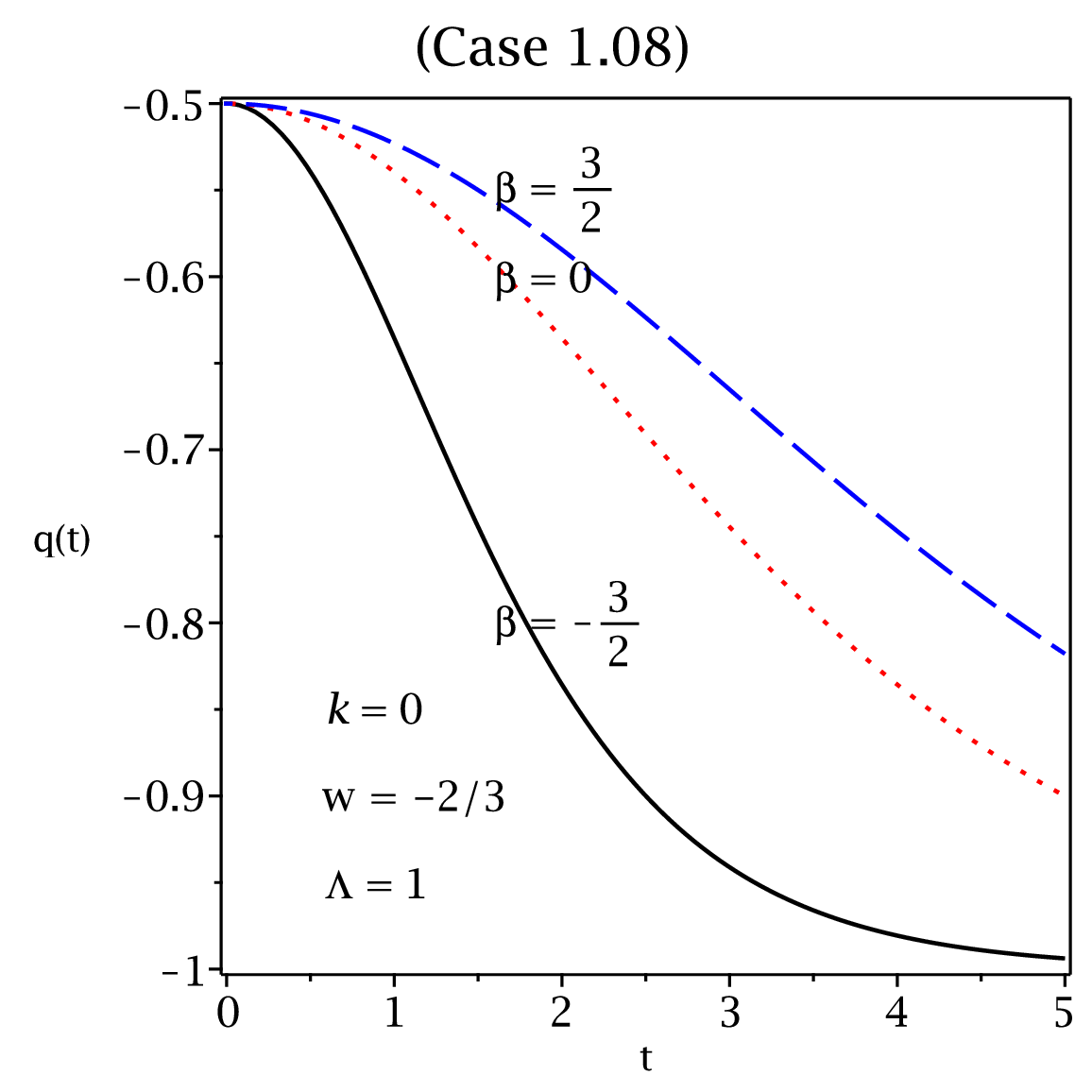}
	\includegraphics[width=3.05cm]{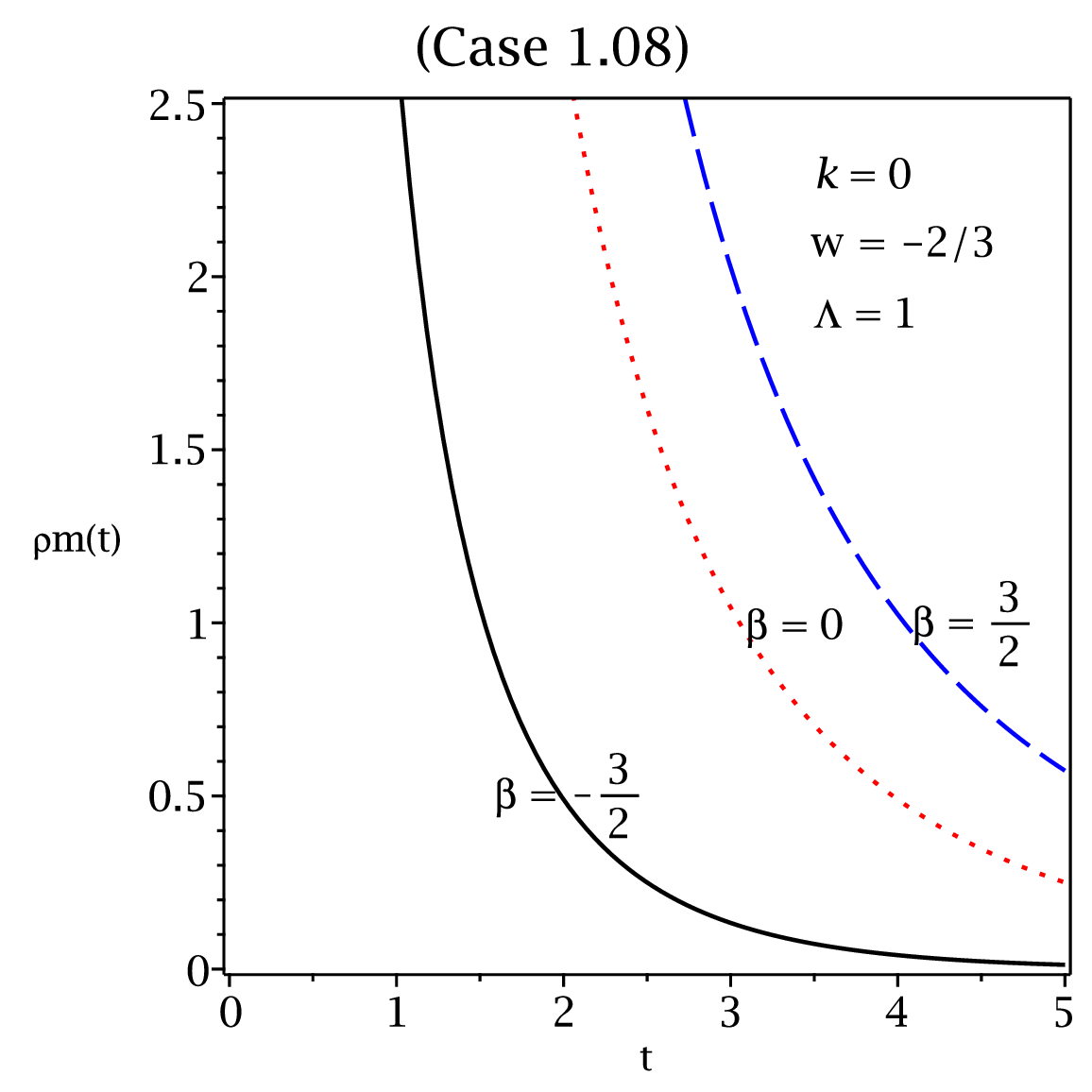}
	\includegraphics[width=3.05cm]{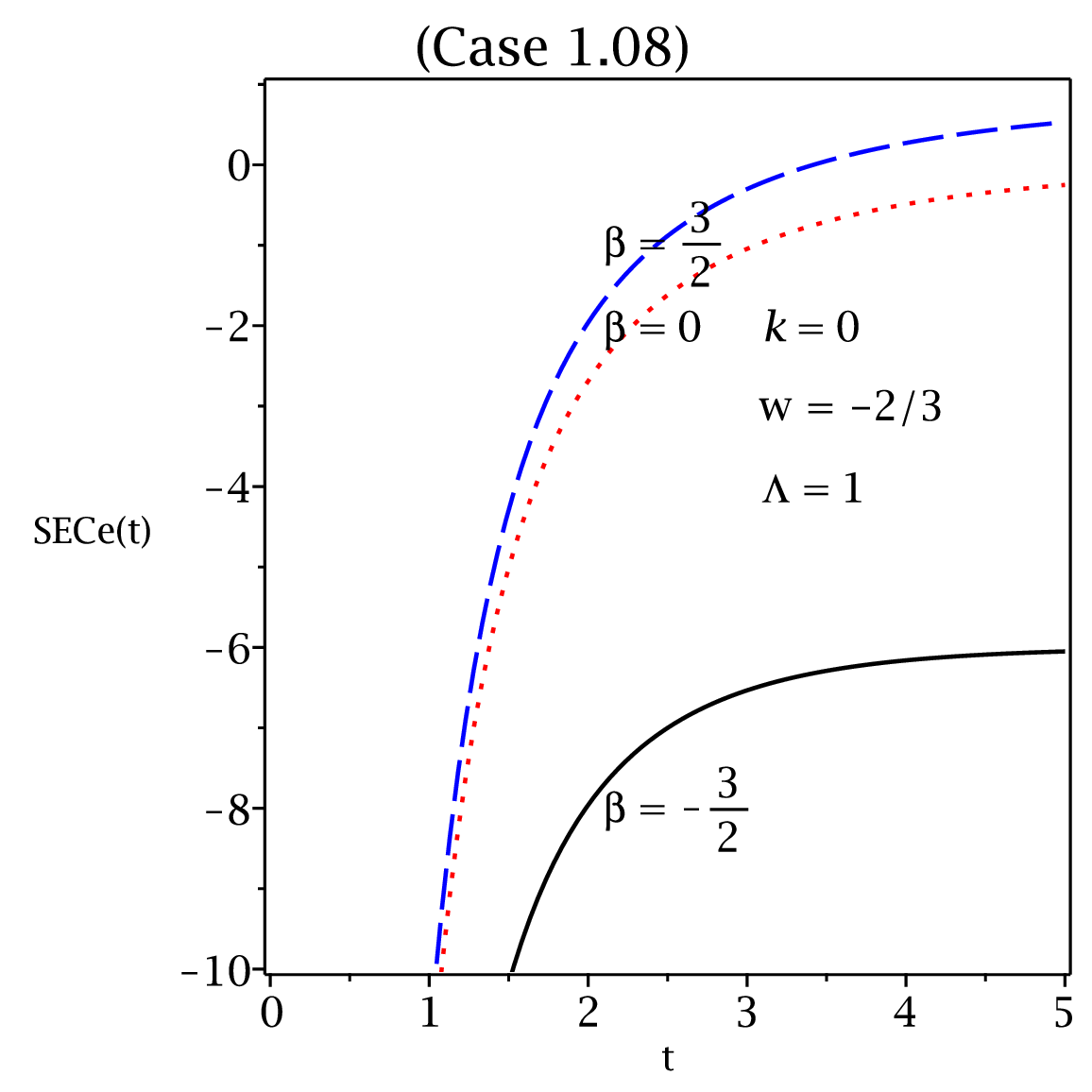}
	\includegraphics[width=3.05cm]{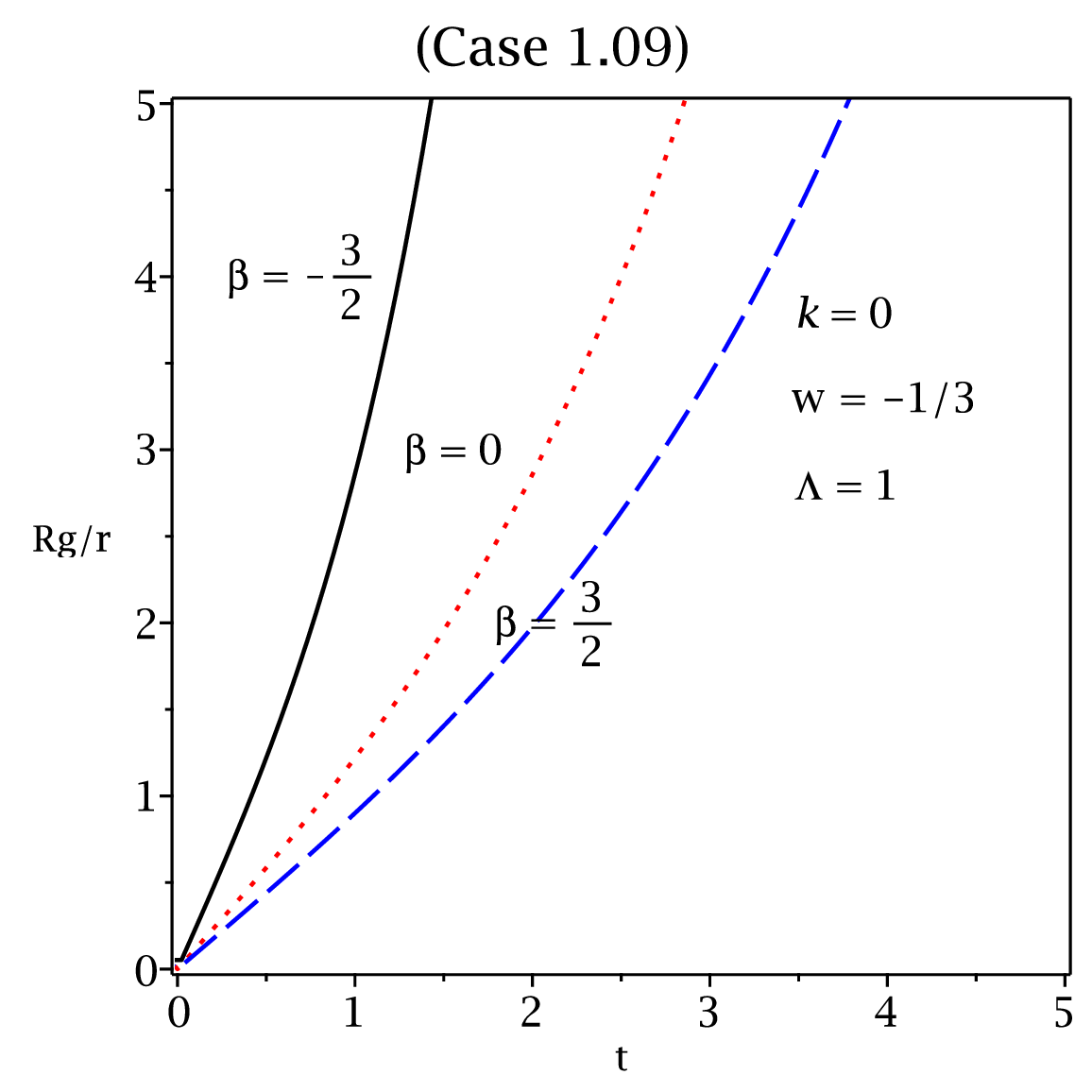}
	\includegraphics[width=3.05cm]{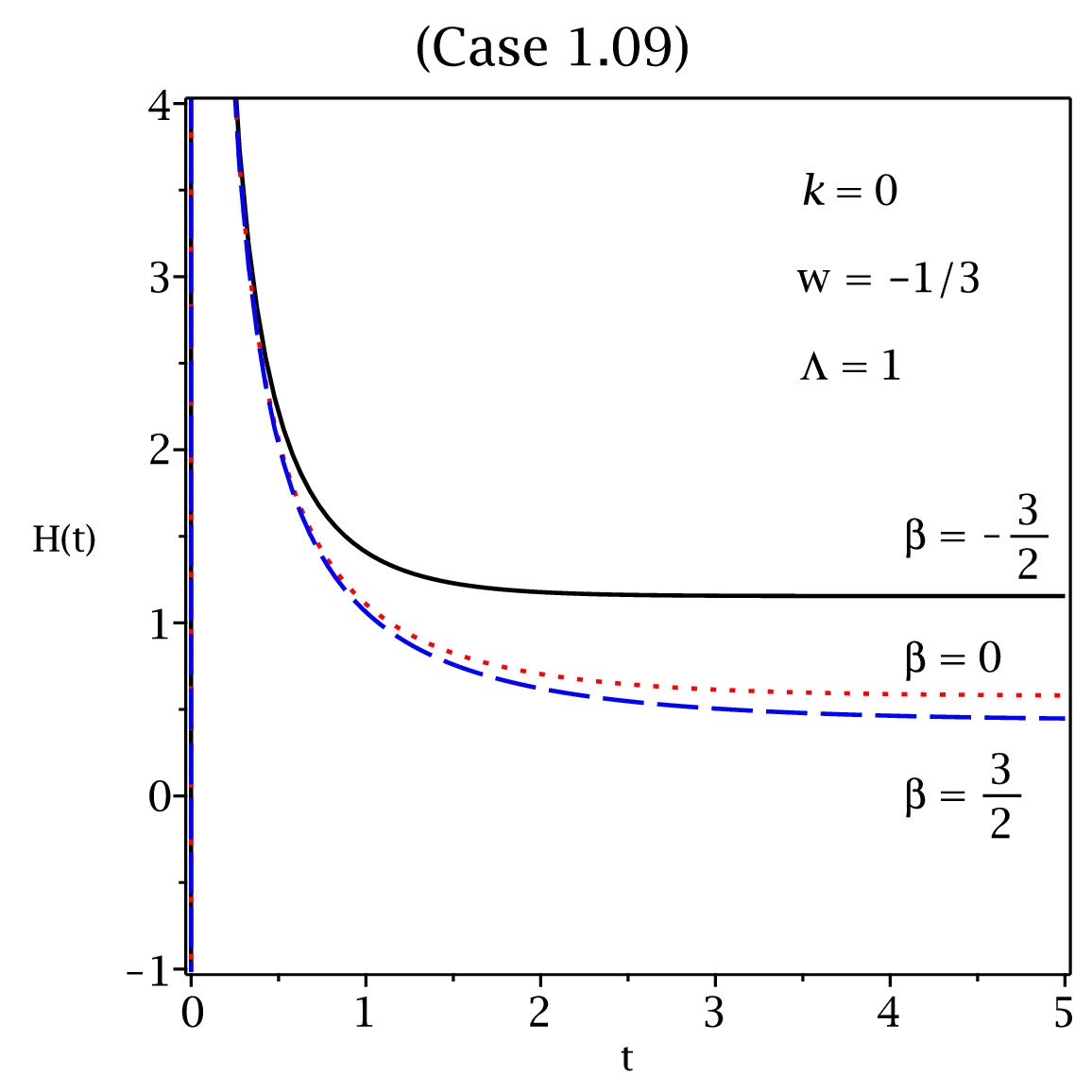}
	\includegraphics[width=3.05cm]{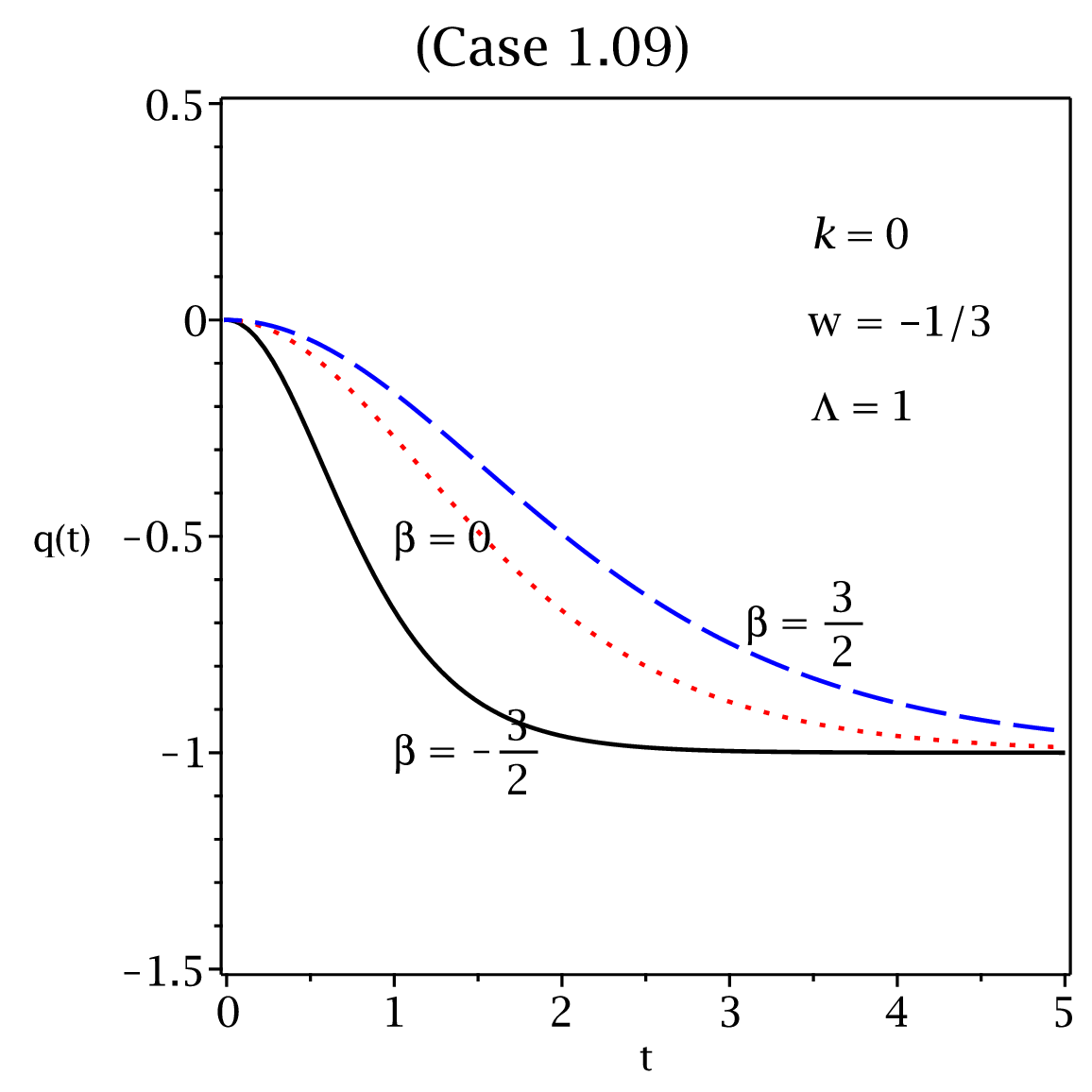}
	\includegraphics[width=3.05cm]{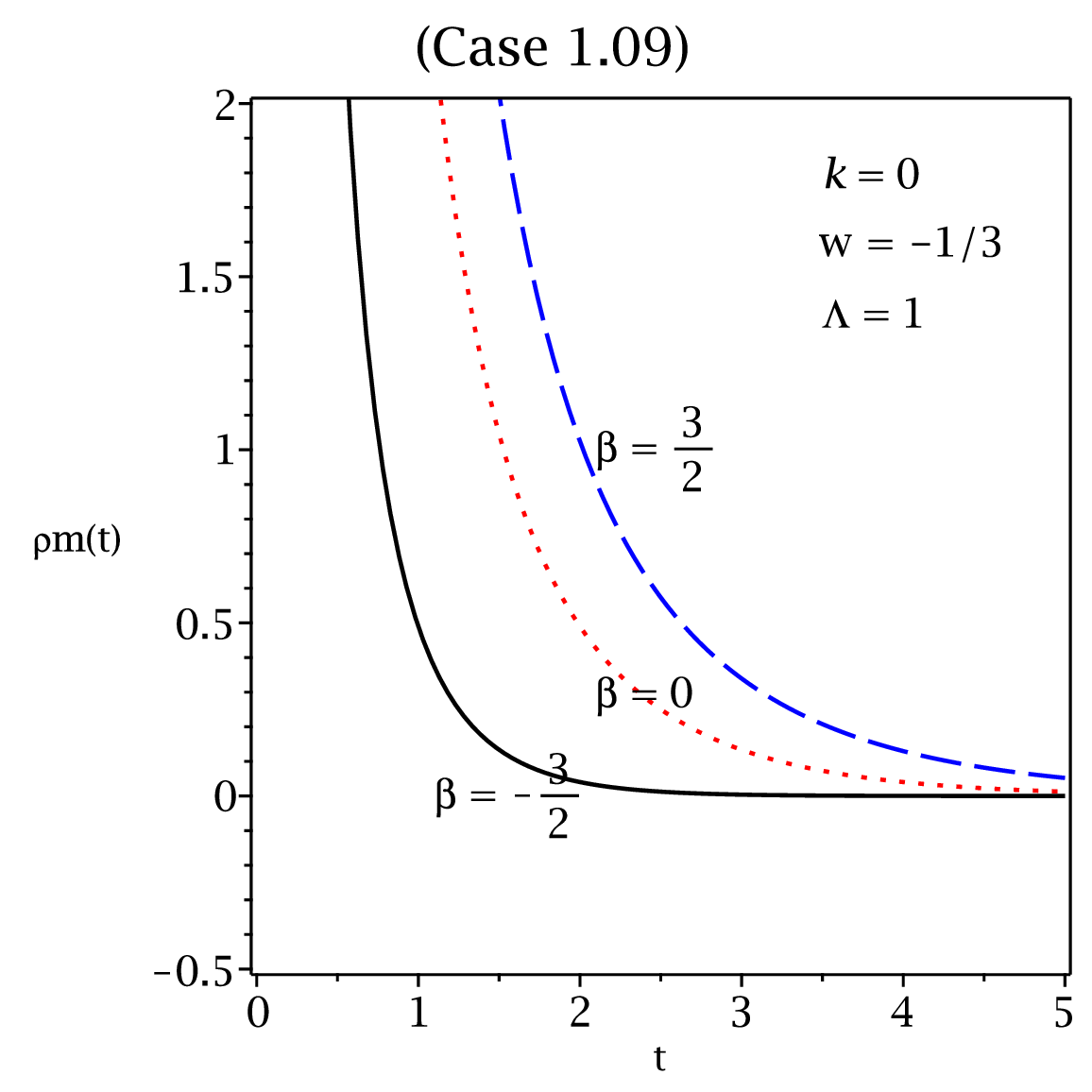}
	\includegraphics[width=3.05cm]{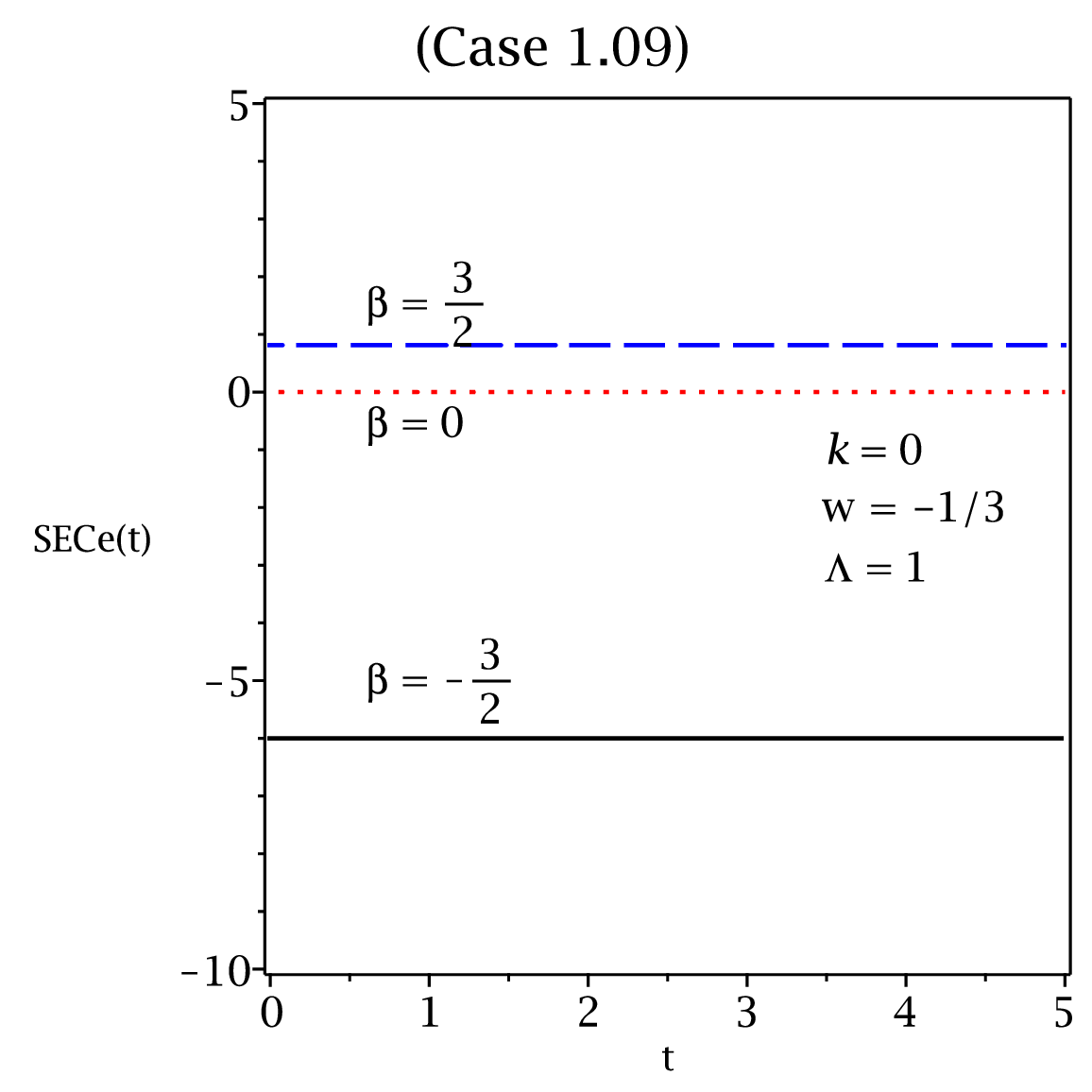}
	\includegraphics[width=3.05cm]{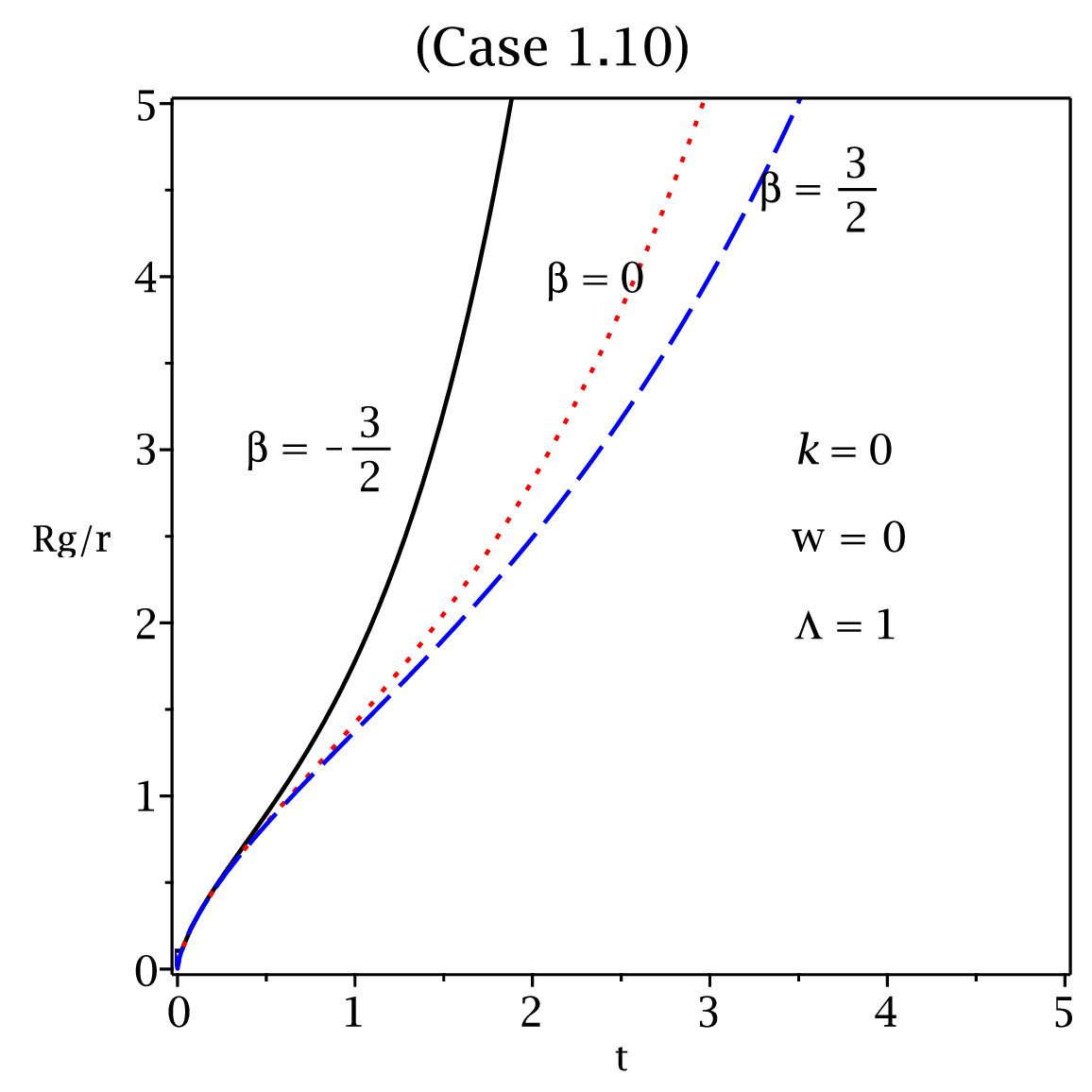}
	\includegraphics[width=3.05cm]{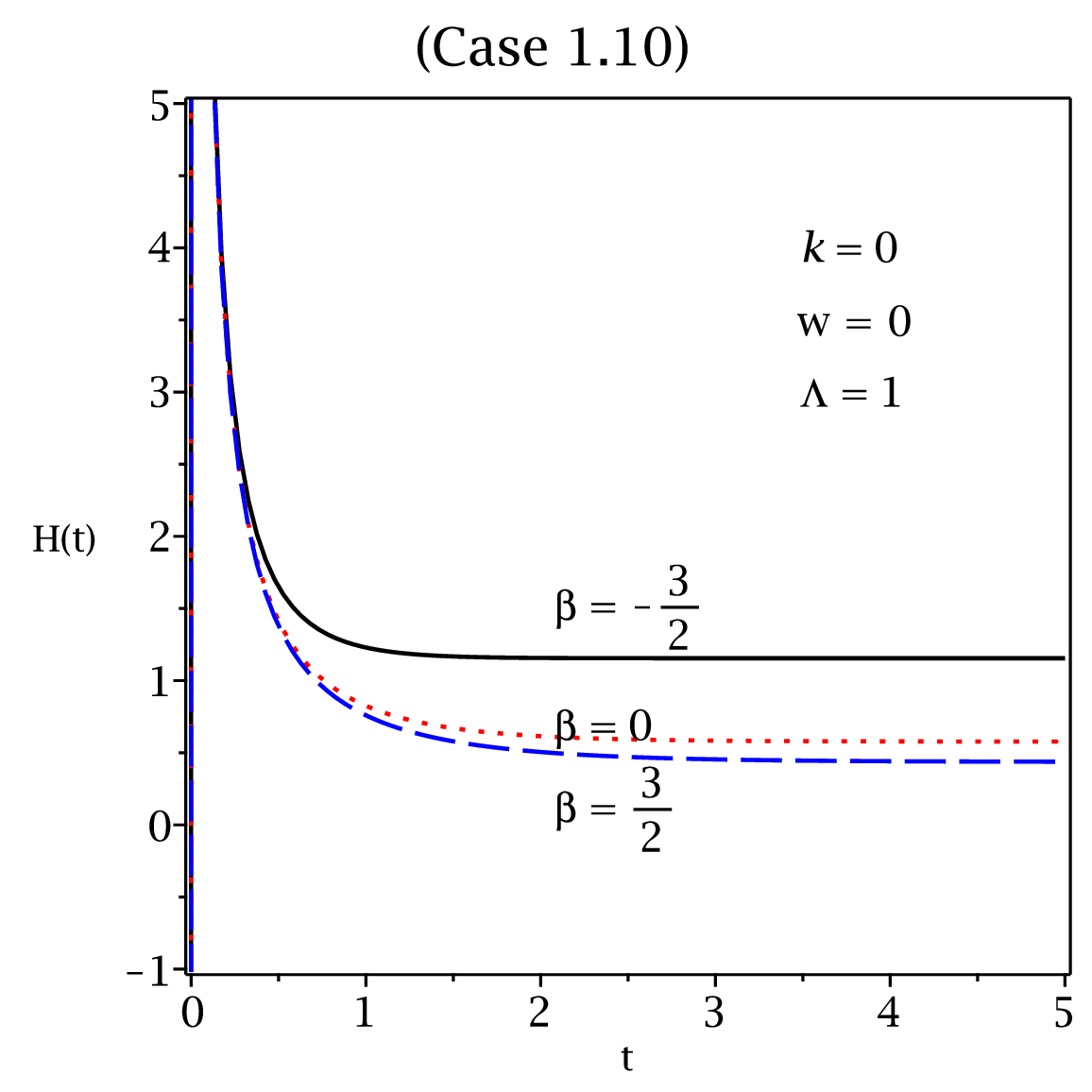}
	\includegraphics[width=3.05cm]{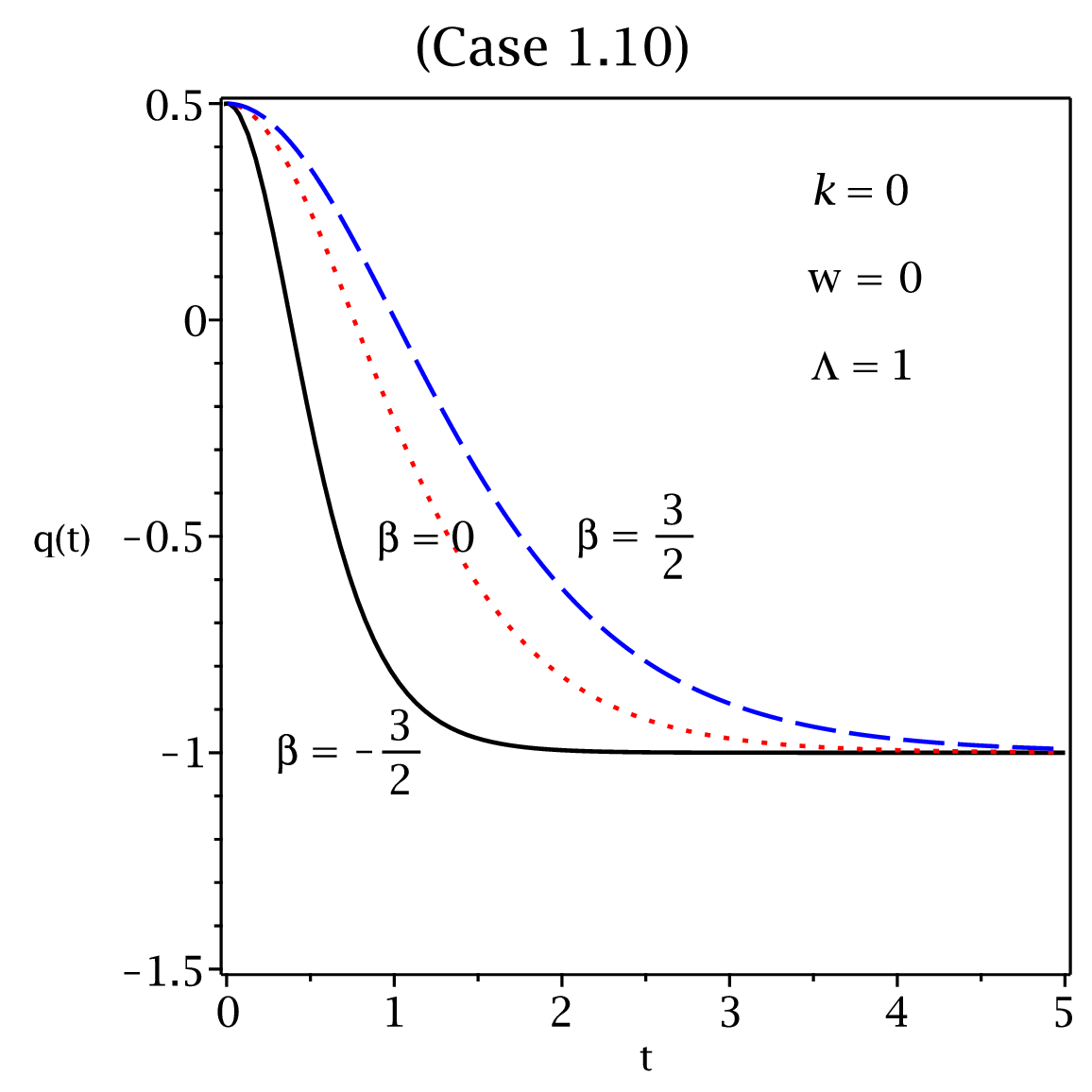}
	\includegraphics[width=3.05cm]{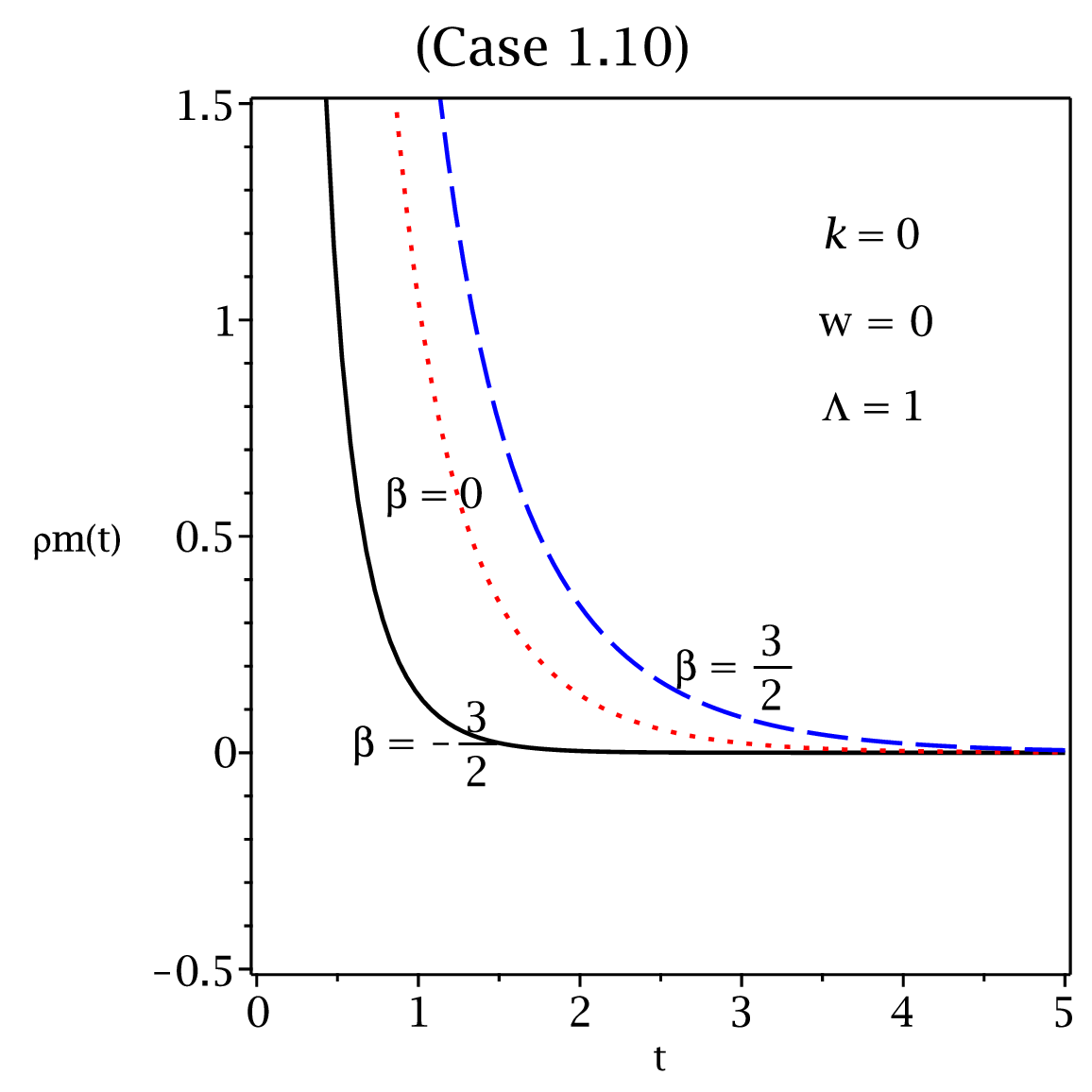}
	\includegraphics[width=3.05cm]{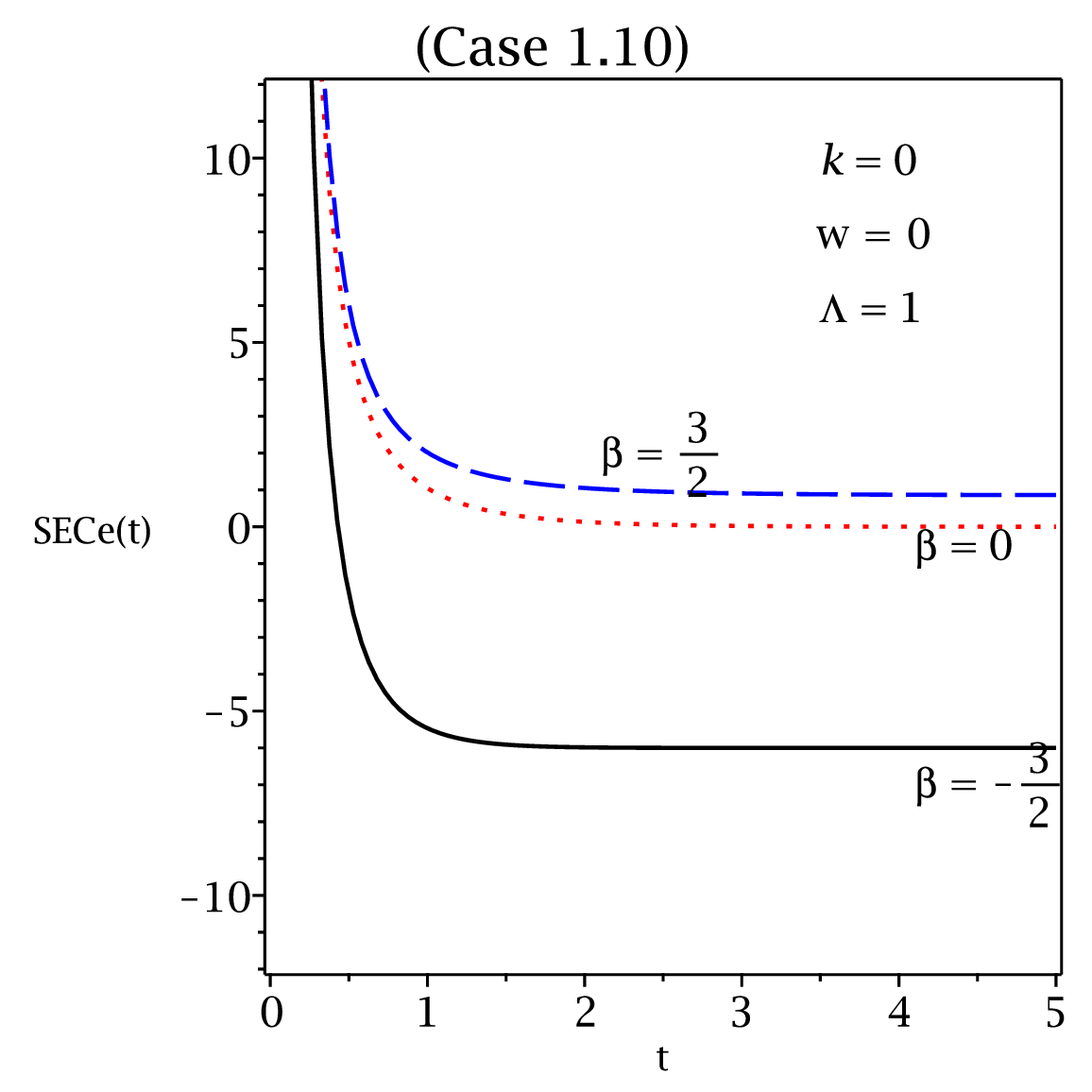}
	\includegraphics[width=3.05cm]{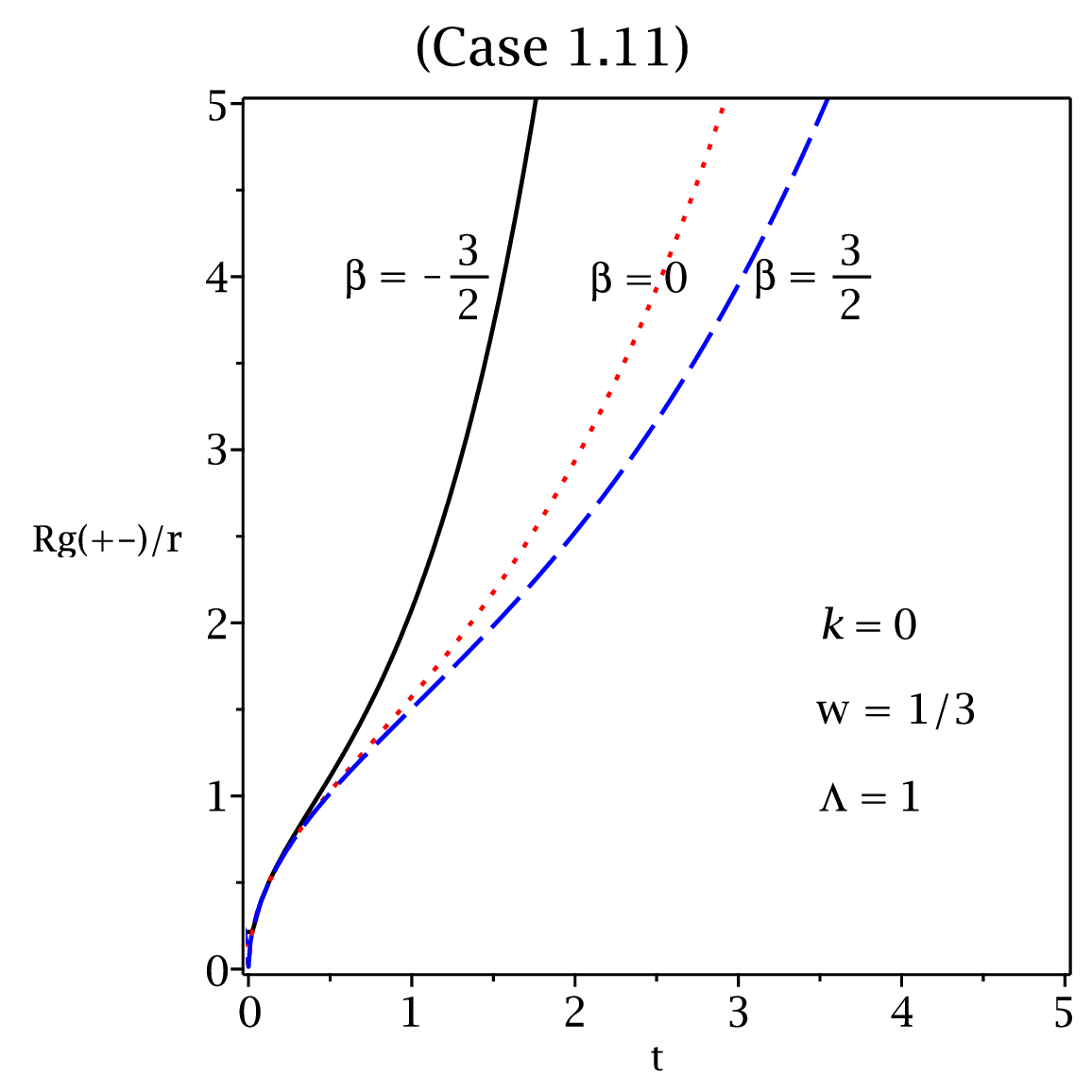}
	\includegraphics[width=3.05cm]{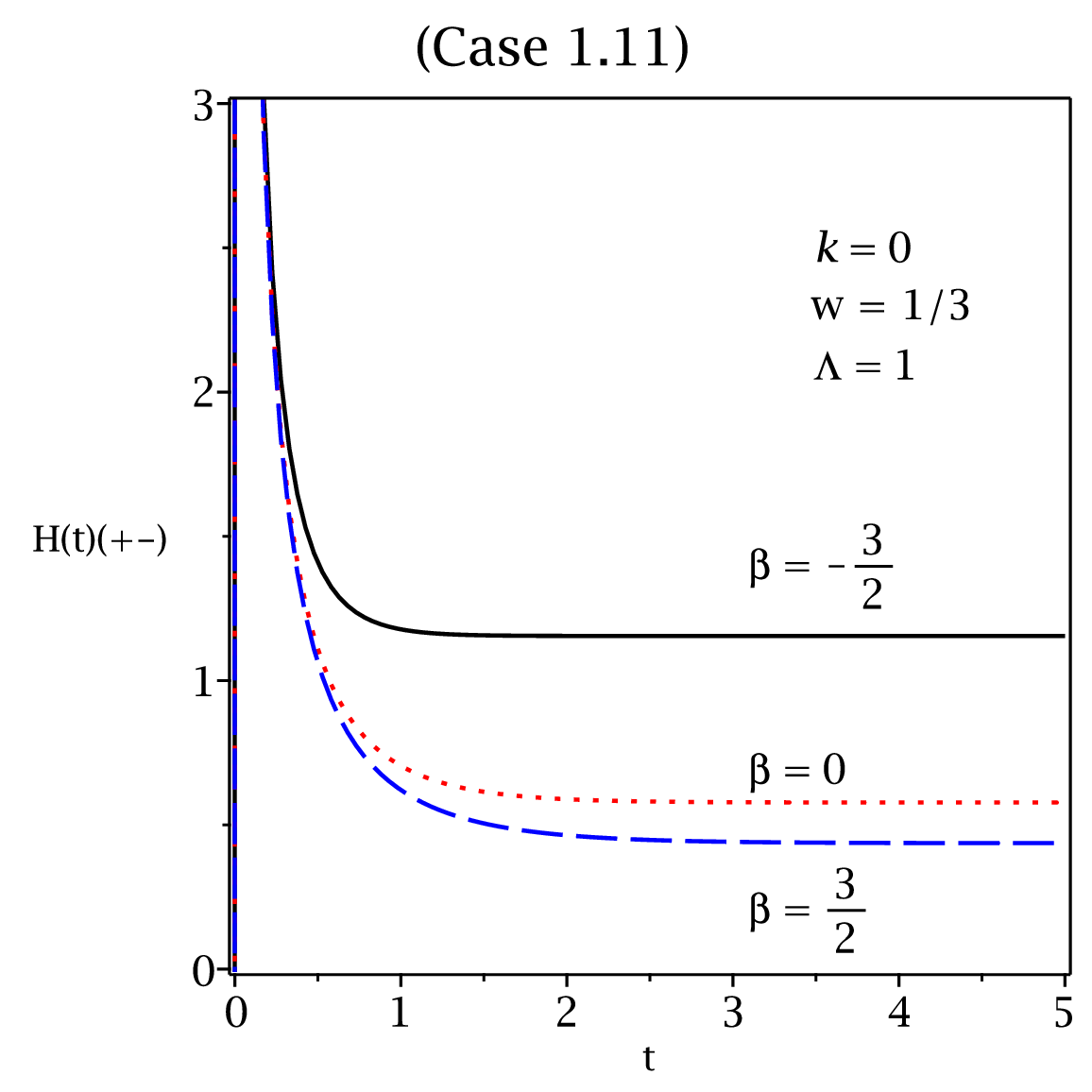}
	\includegraphics[width=3.05cm]{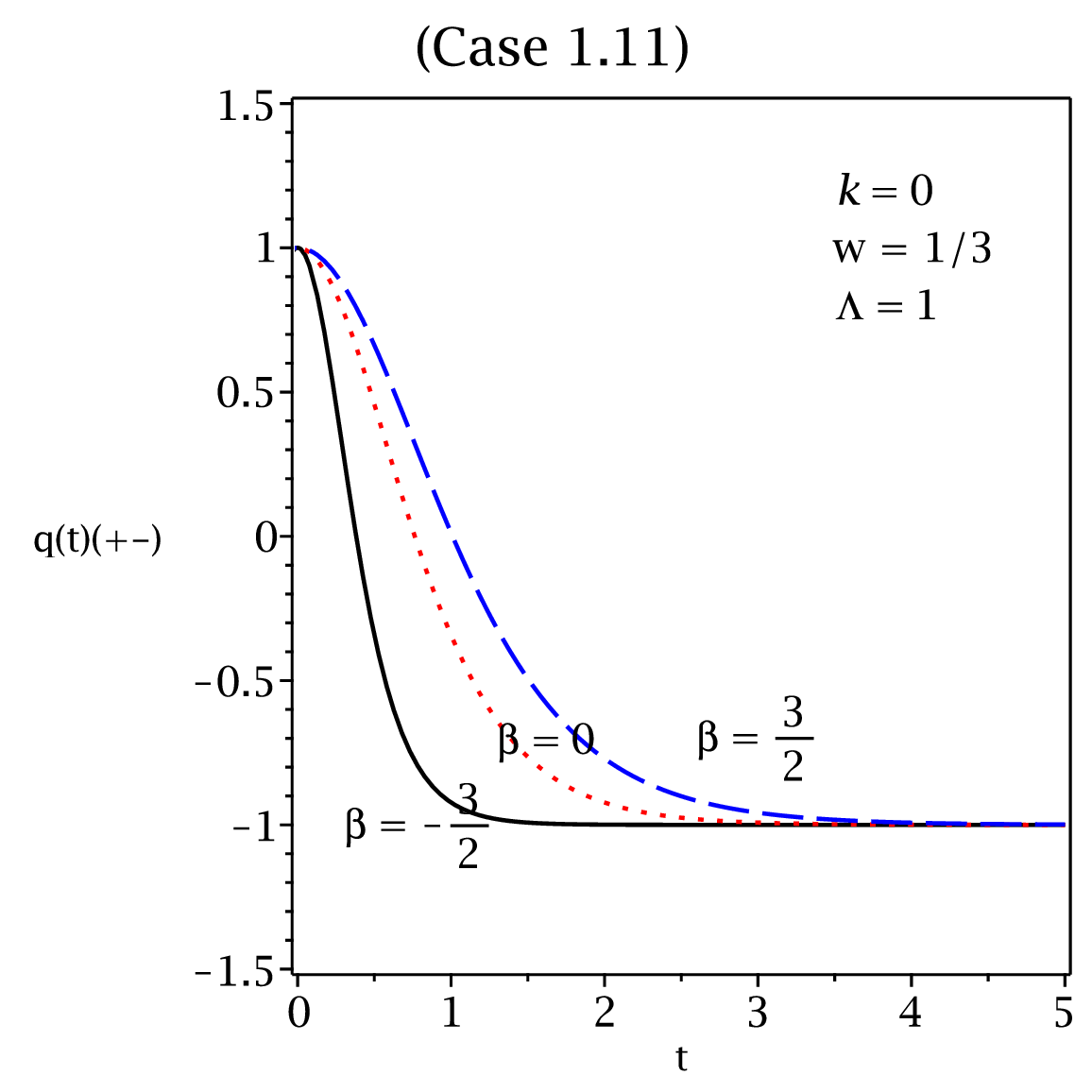}
	\includegraphics[width=3.05cm]{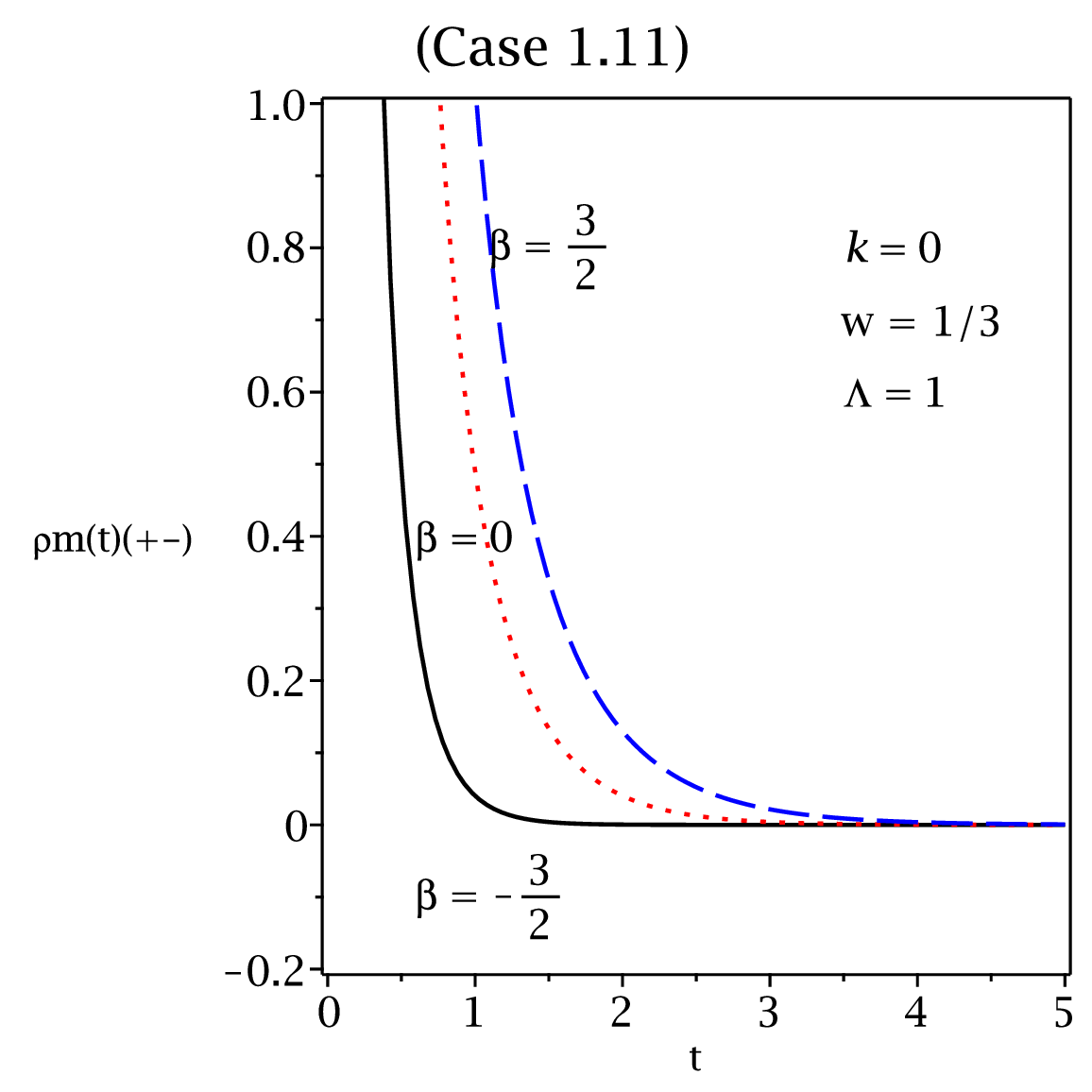}
	\includegraphics[width=3.05cm]{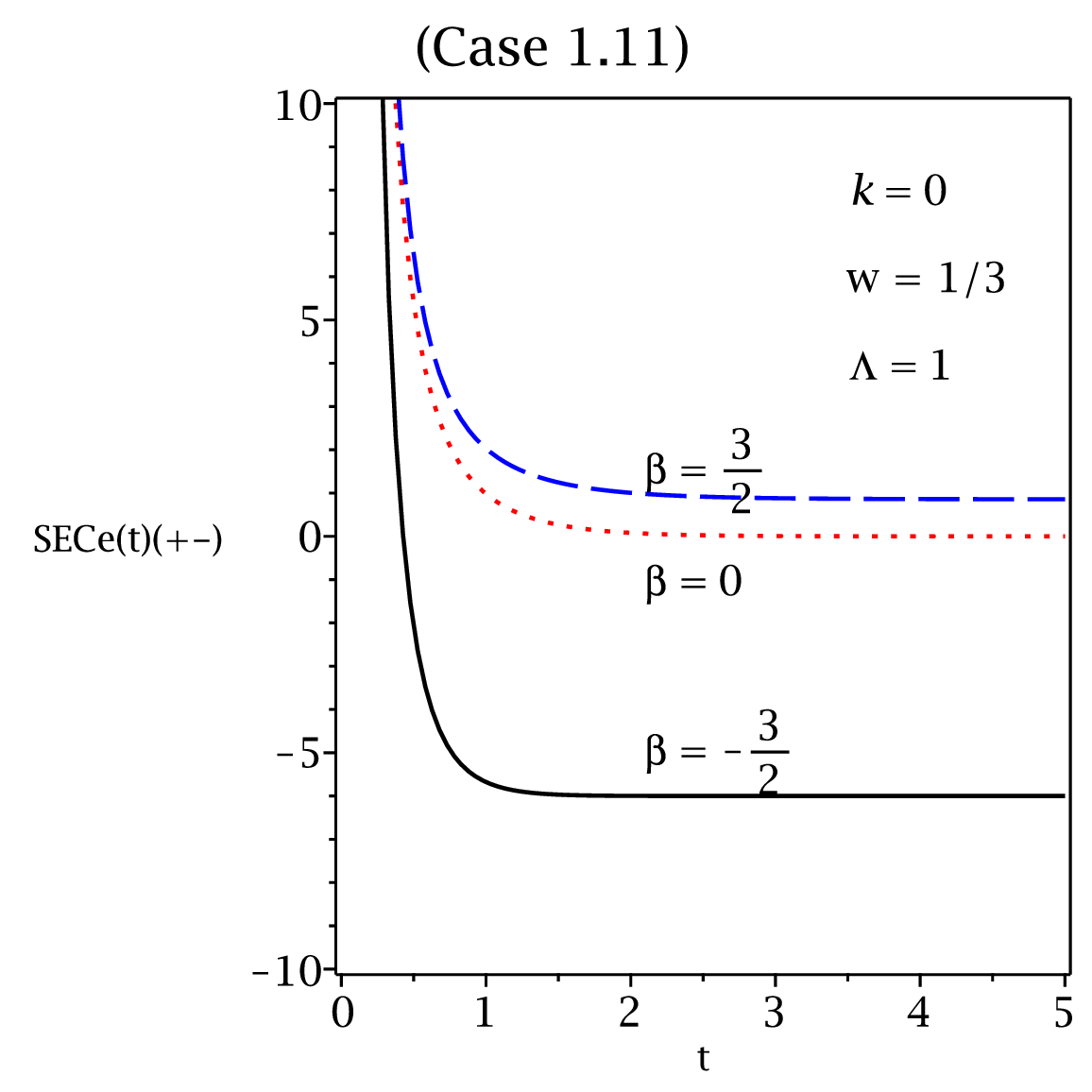}
	\caption{These figures are for $\Lambda>0$ and $k=0$.
		These figures represent the quantities $R_g$ (geometrical radius), 
		$H(t)$ (Hubble parameter) and $q(t)$ (deceleration parameter) $\rho_m(t)$ 
		(energy density of the aether fluid) and $SEC_{e} \equiv SEC_{\rm eff}$ 
		(strong energy condition for the effective fluid) for the different
		values of $\beta=-3/2$ (black solid line), $\beta=0$ (red dotted line), 
		$\beta=3/2$ (blue dashed line). Assuming that $8 \pi G=1$, $R_g(t=0)=0$ 
		and $C_1=1$, $C_2=1$ (Cases 1.10 and 1.06); $C_1=0$, $C_2=1$
		(Case 1.07); $C_1=1$, $C_2=-1$ (Cases 1.08 and 1.09); $C_1=1$, $C_2=1$ 
		(Case 1.11); $C_1=1$, $C_2=100$ (Case 1.05).}
	\label{Figure-105-111}
\end{minipage}	
\end{figure}

\begin{figure}[!htp]
\begin{minipage}{175 mm}
	\centering	
	\includegraphics[width=3.4cm]{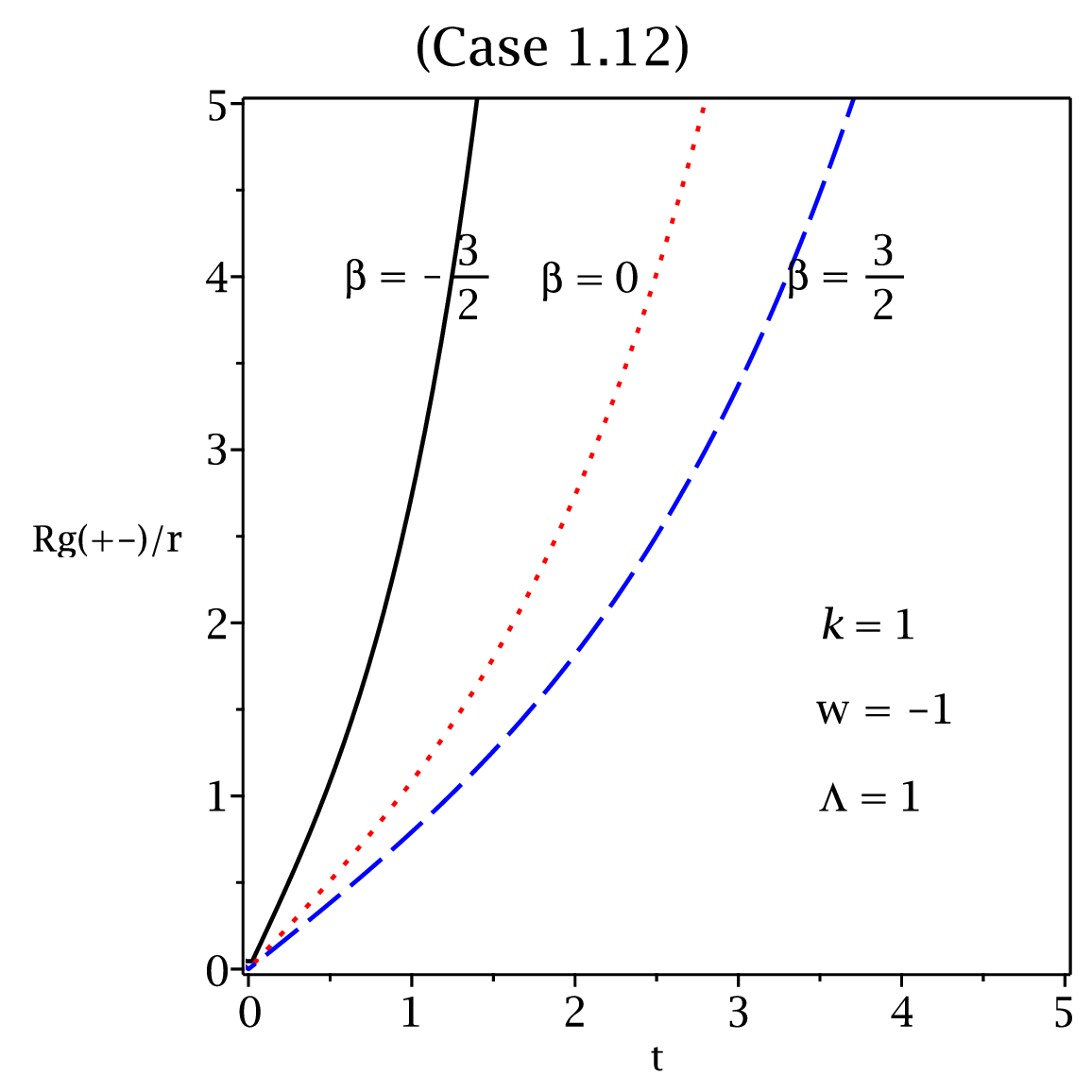}
	\includegraphics[width=3.4cm]{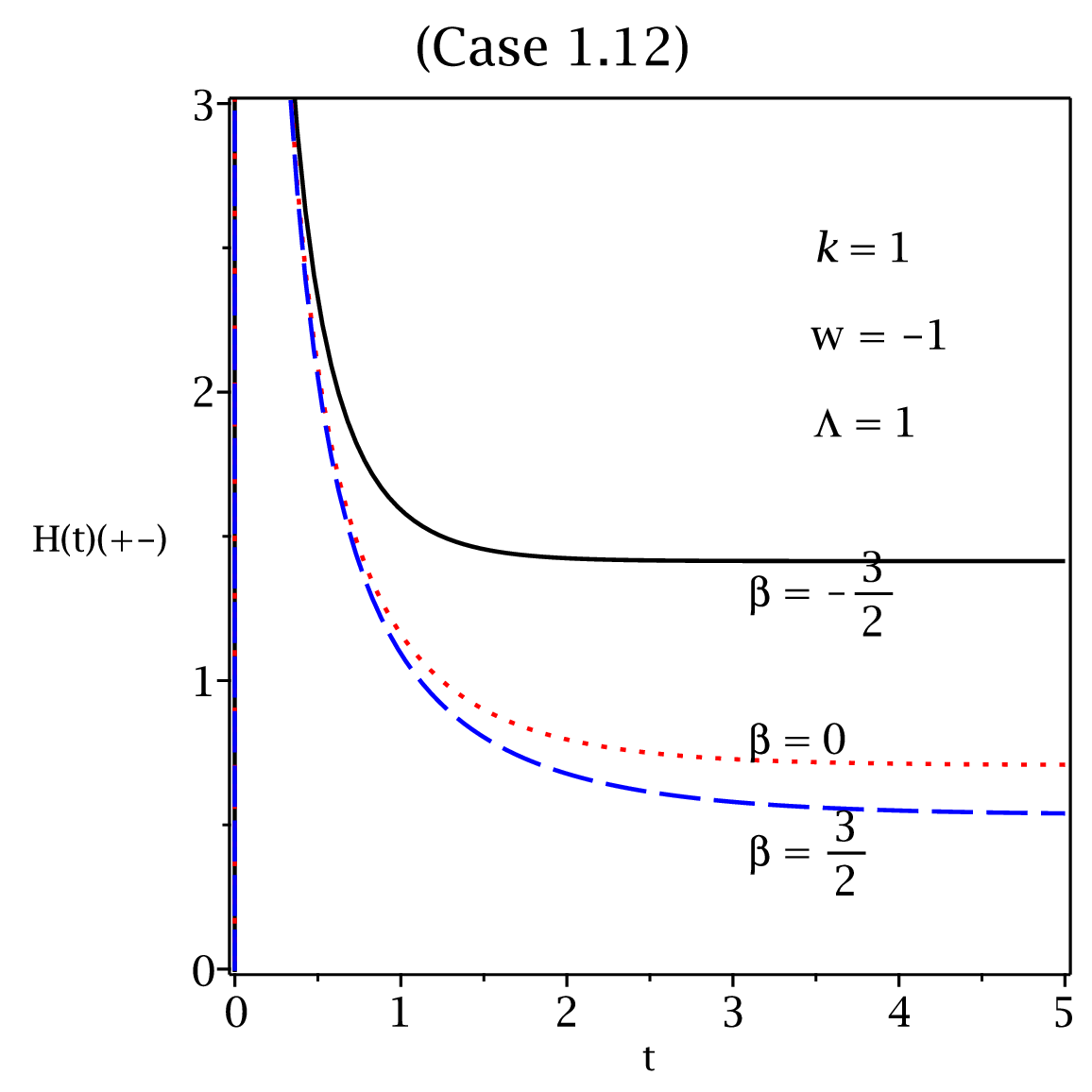}
	\includegraphics[width=3.4cm]{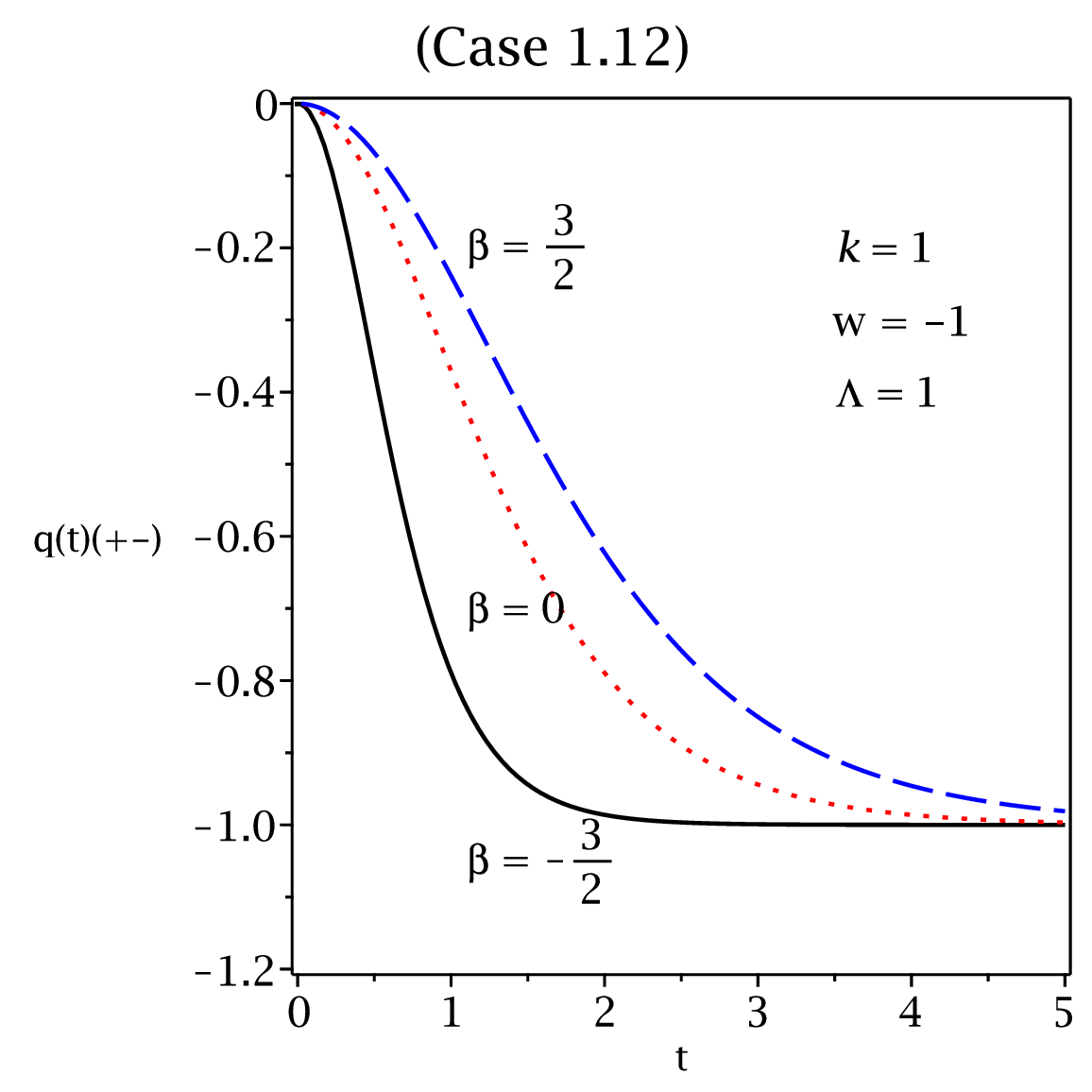}
	\includegraphics[width=3.4cm]{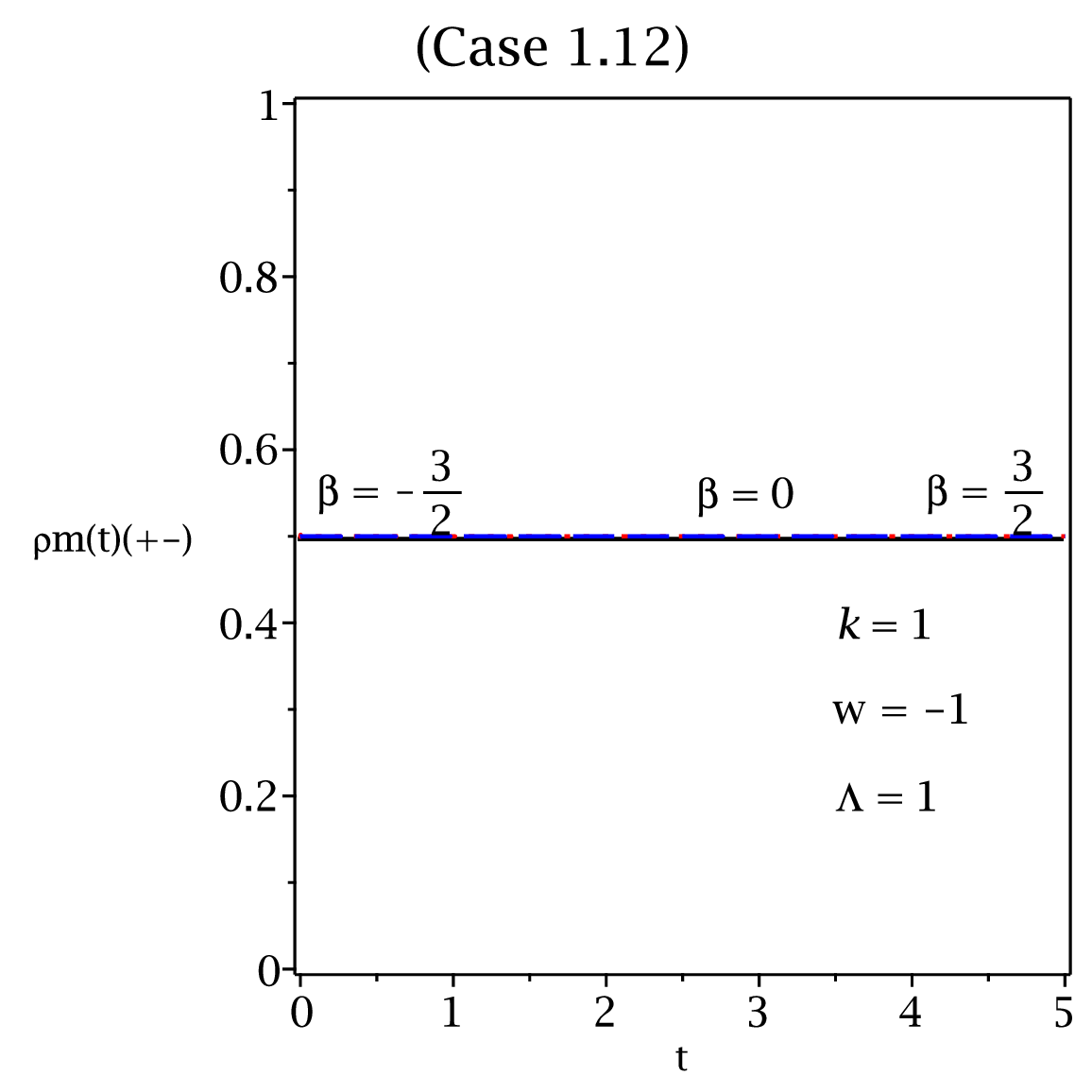}
	\includegraphics[width=3.4cm]{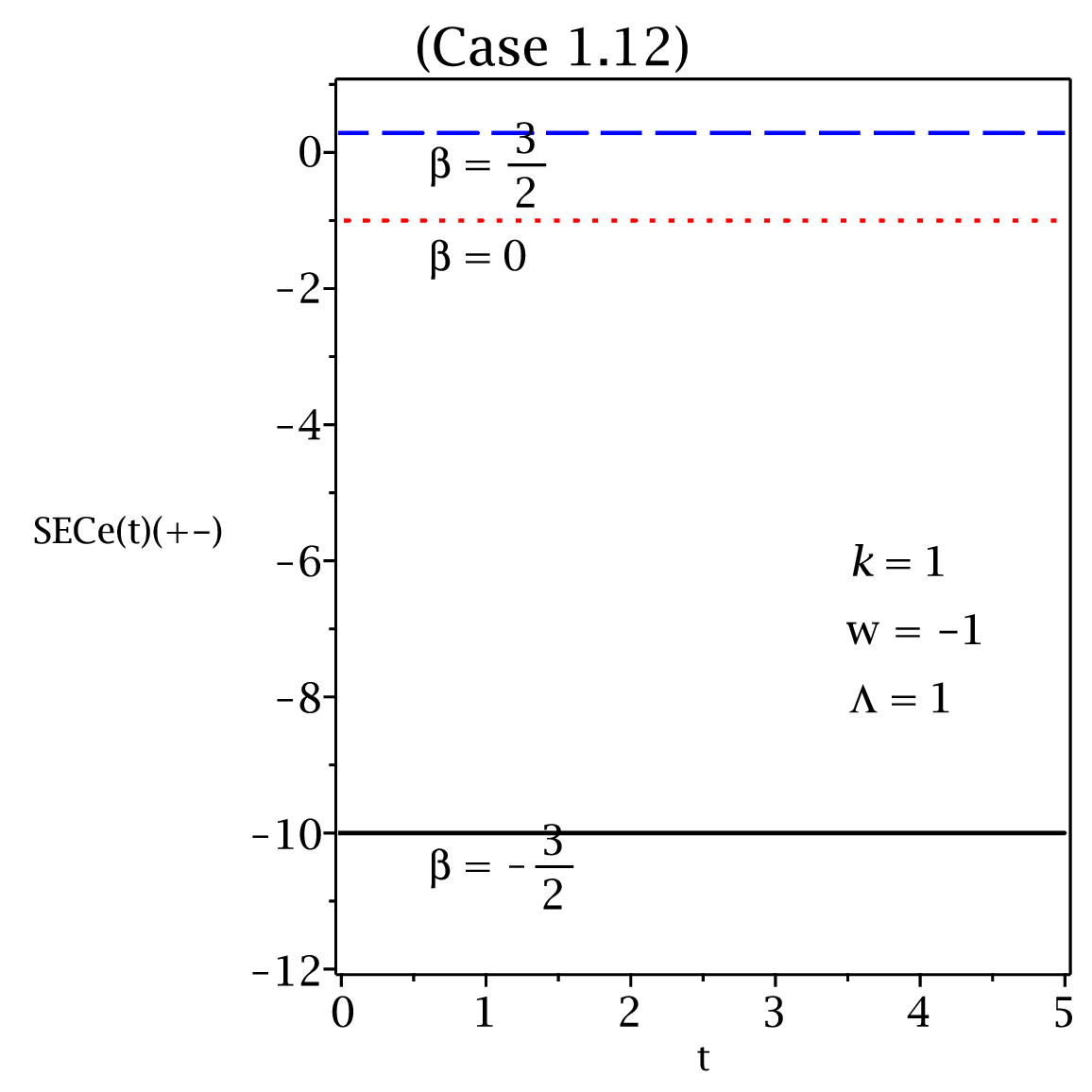}
	\includegraphics[width=3.4cm]{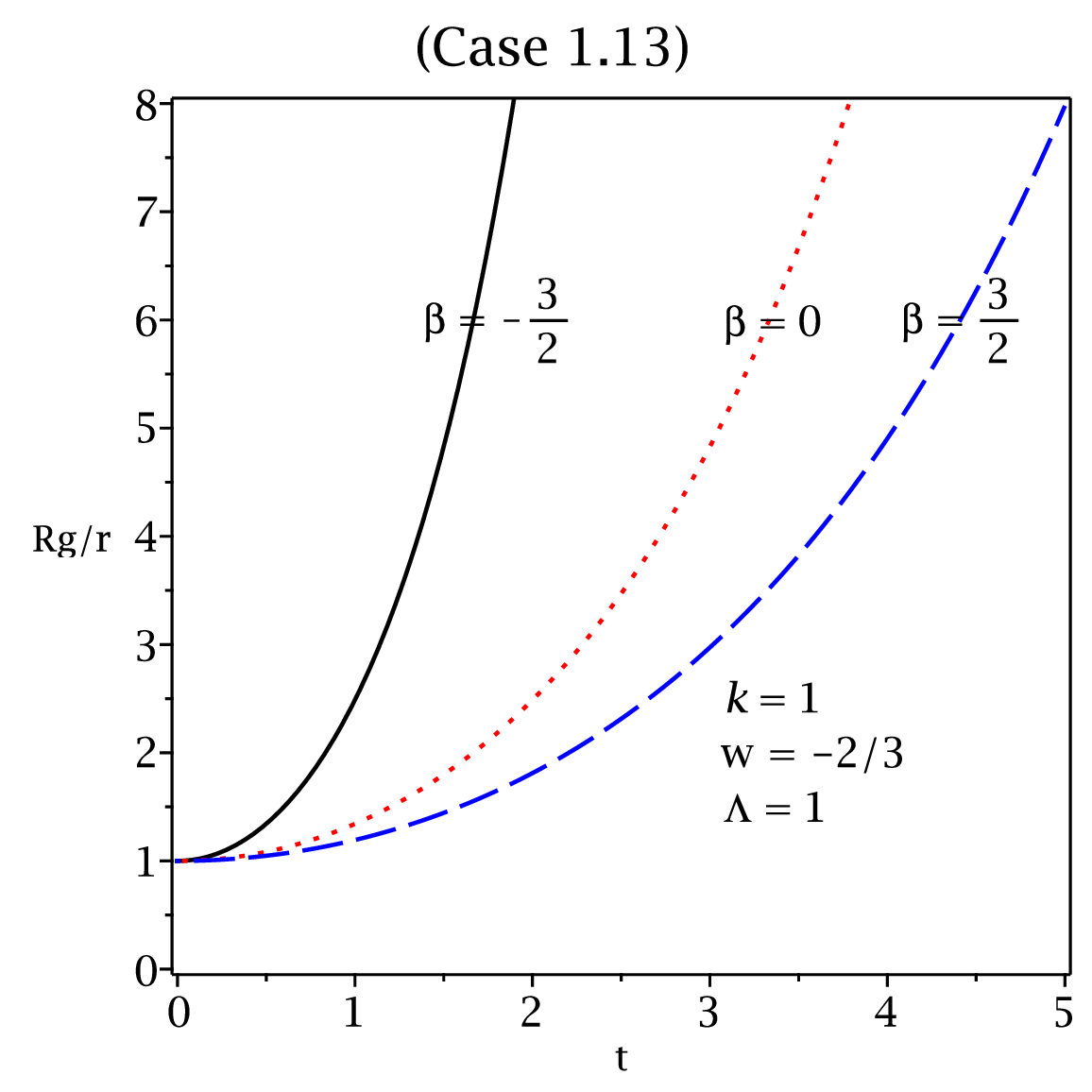}
	\includegraphics[width=3.4cm]{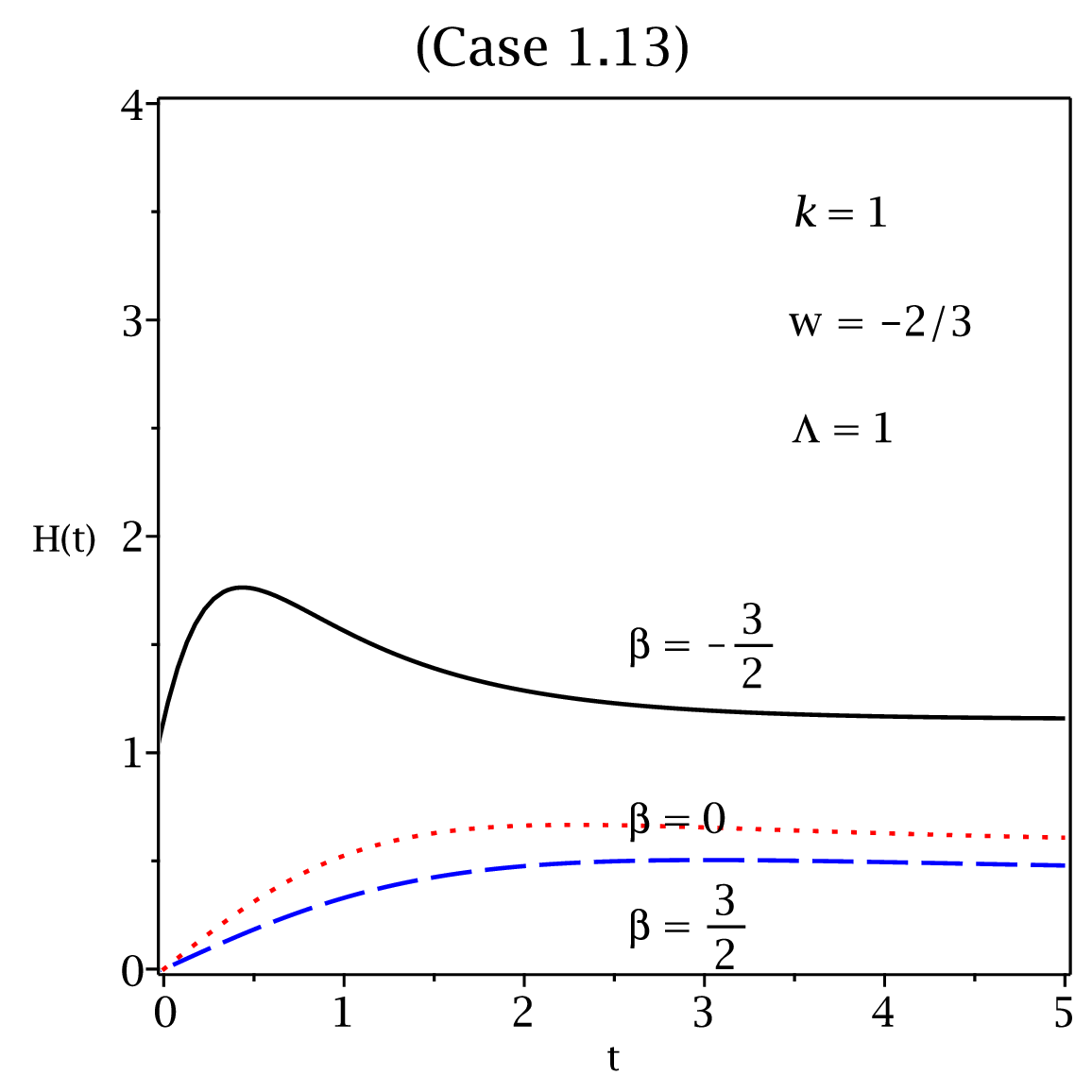}
	\includegraphics[width=3.4cm]{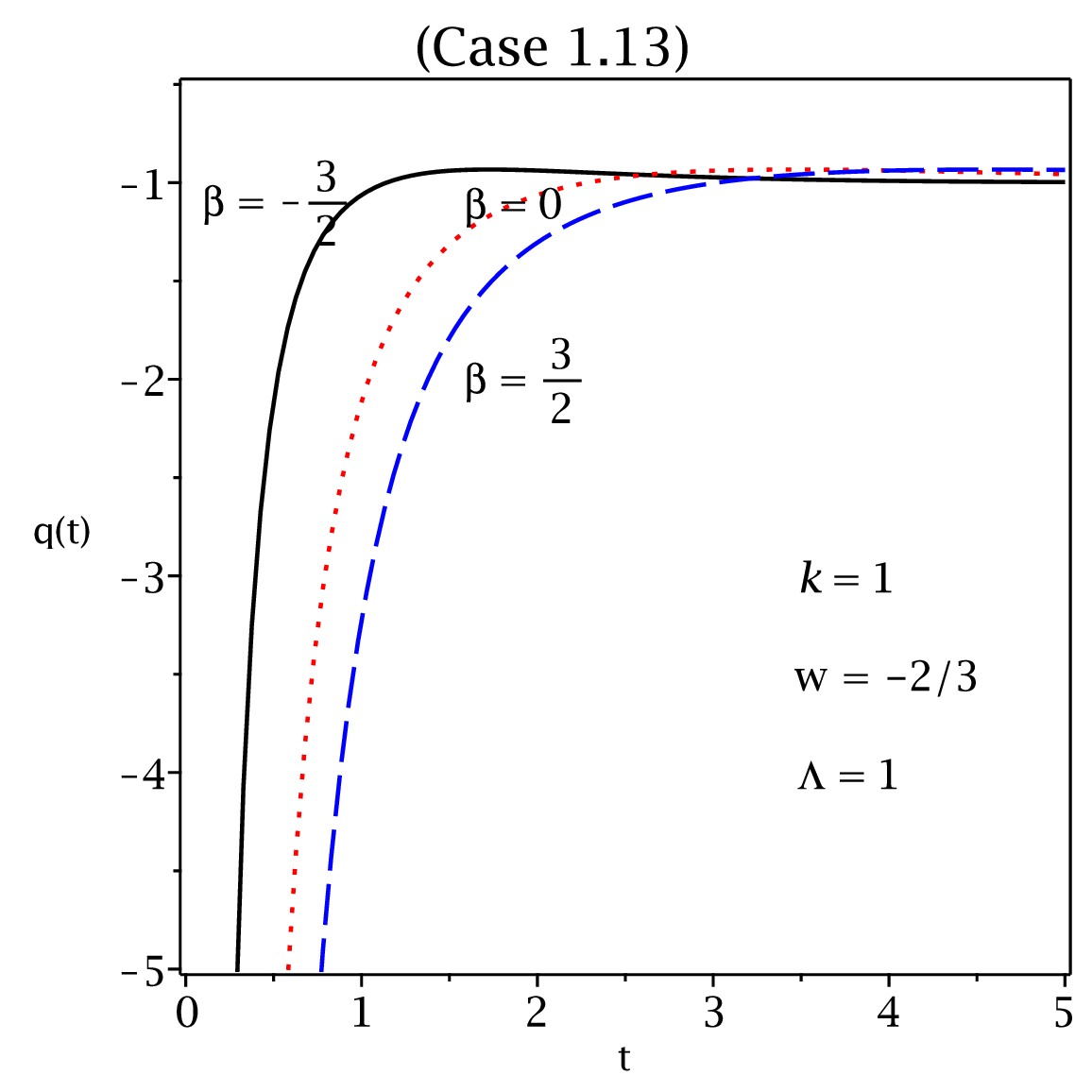}
	\includegraphics[width=3.4cm]{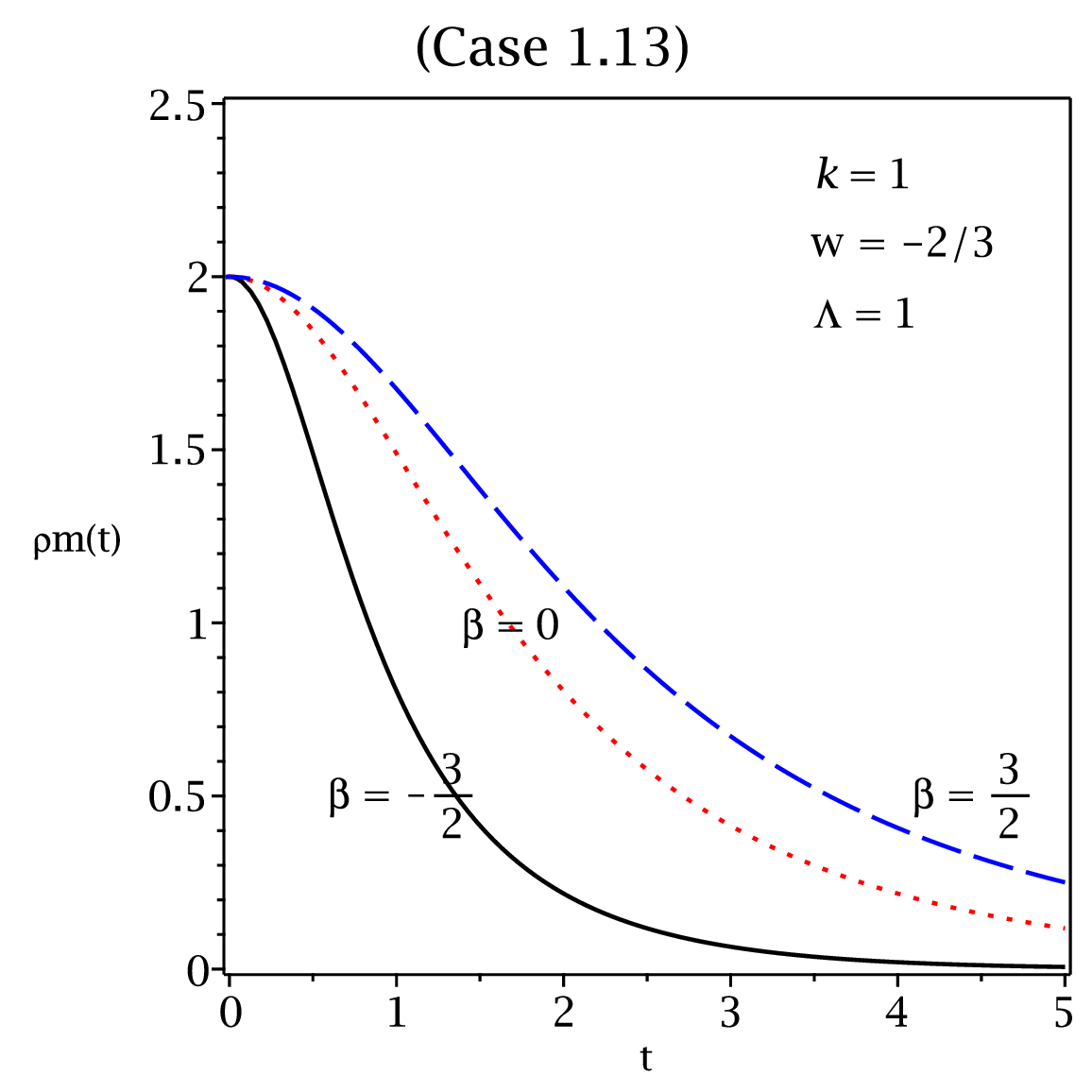}
	\includegraphics[width=3.4cm]{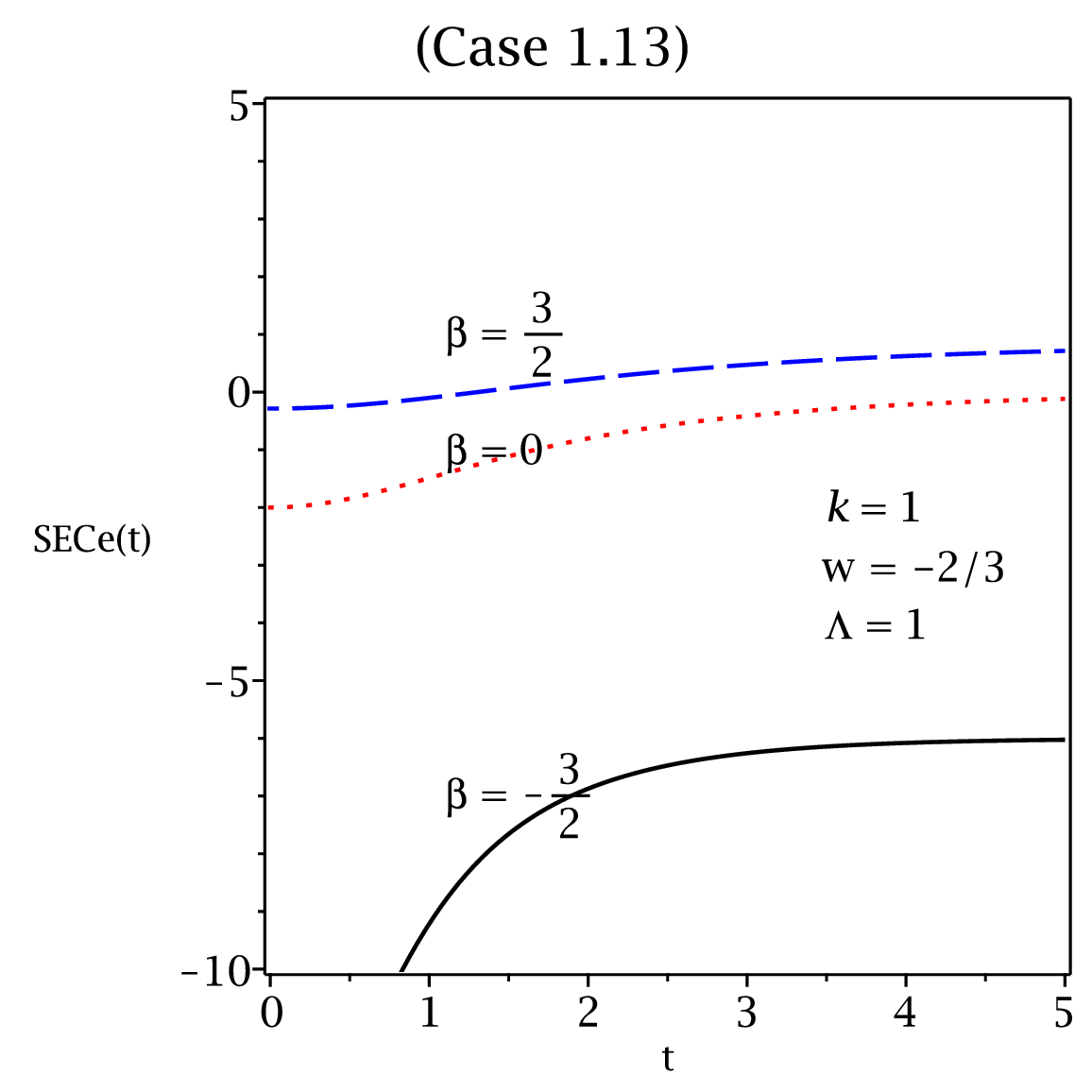}
	\includegraphics[width=3.4cm]{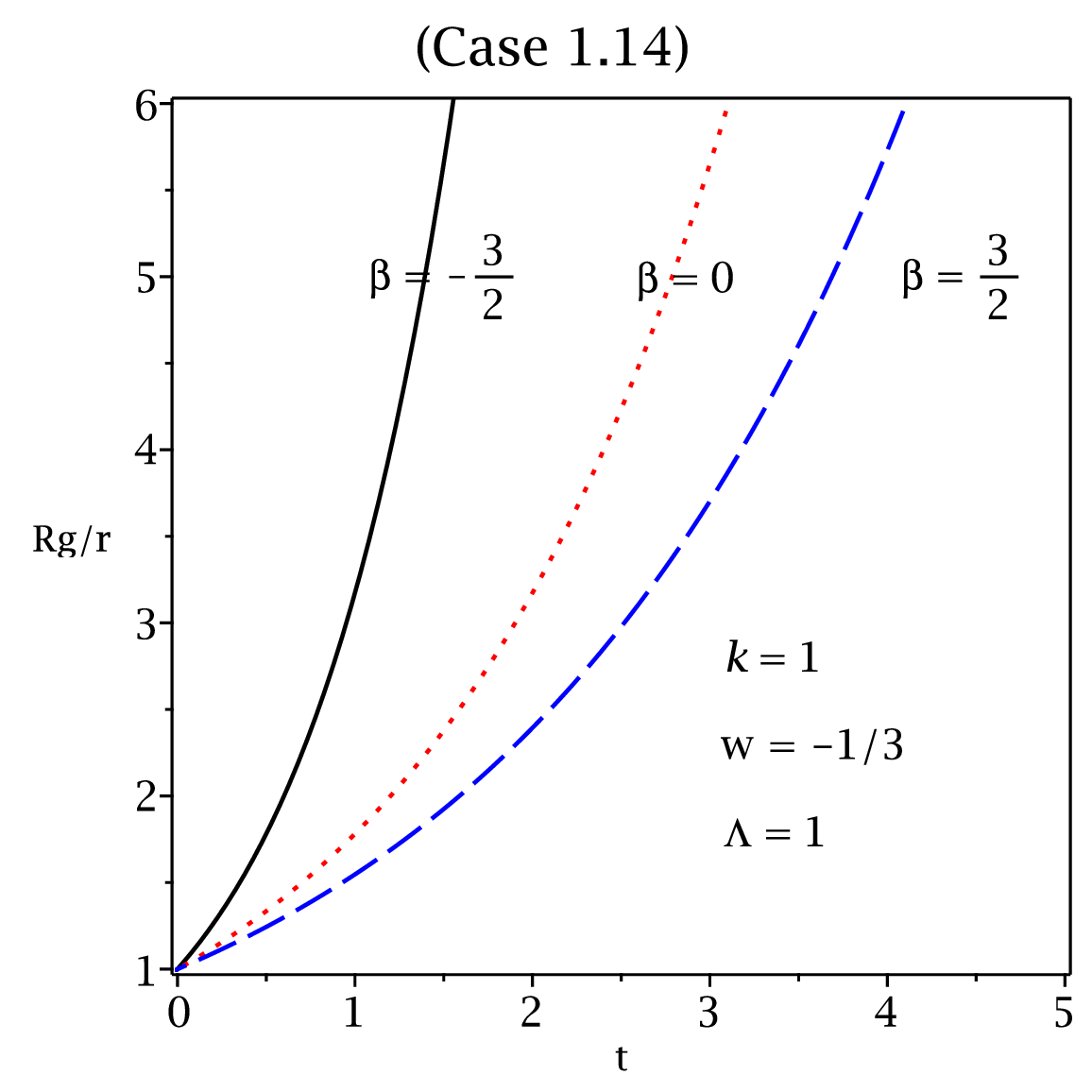}
	\includegraphics[width=3.4cm]{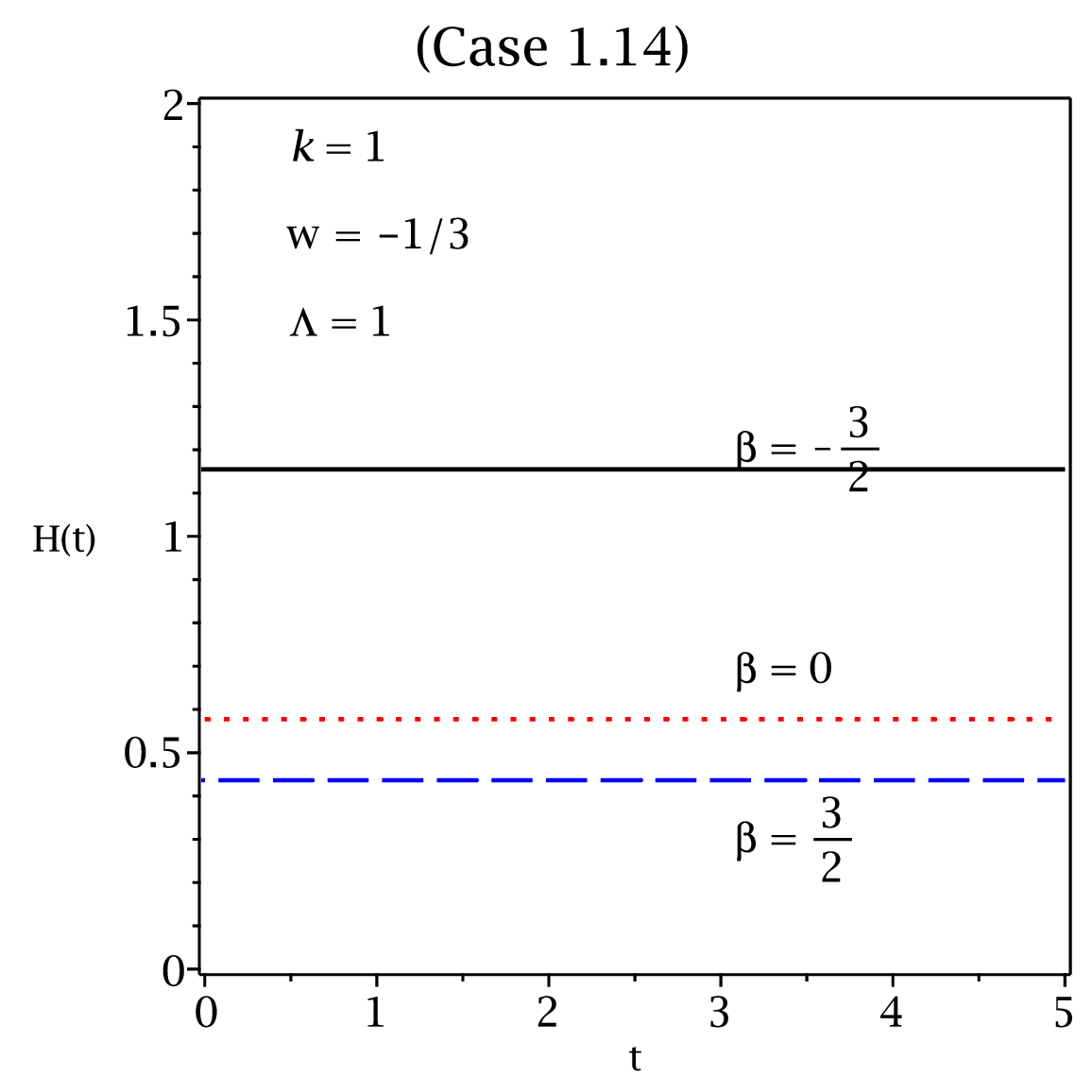}
	\includegraphics[width=3.4cm]{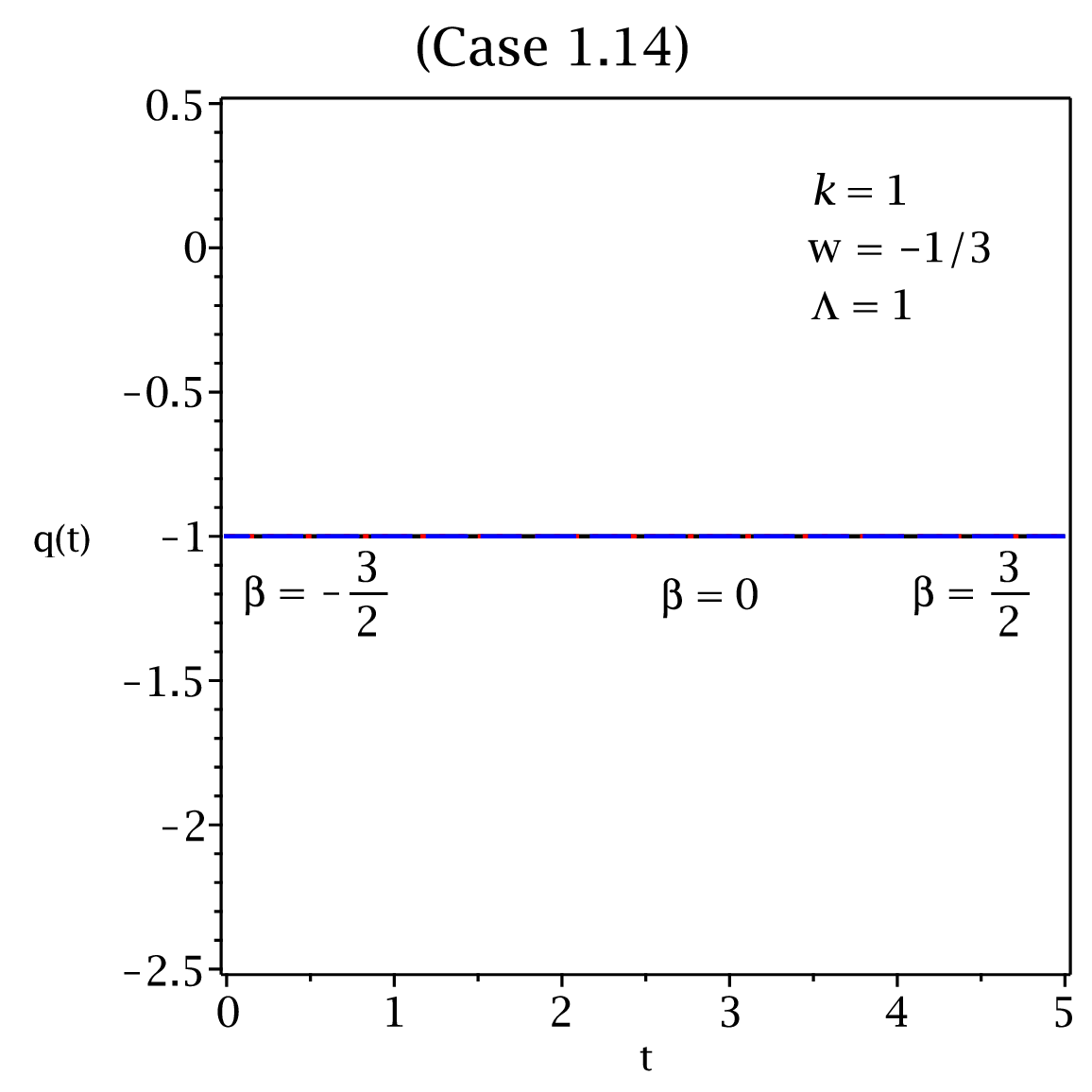}
	\includegraphics[width=3.4cm]{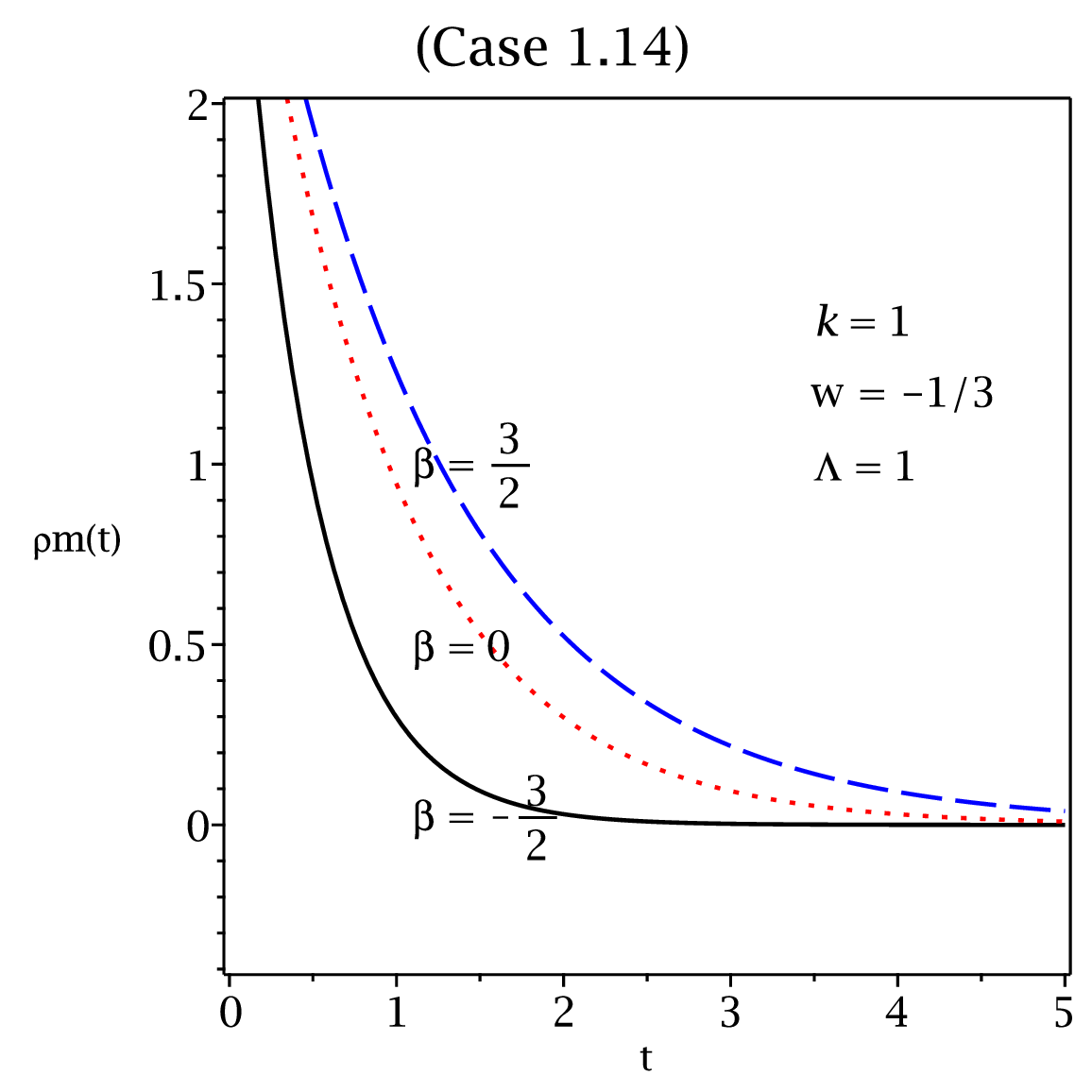}
	\includegraphics[width=3.4cm]{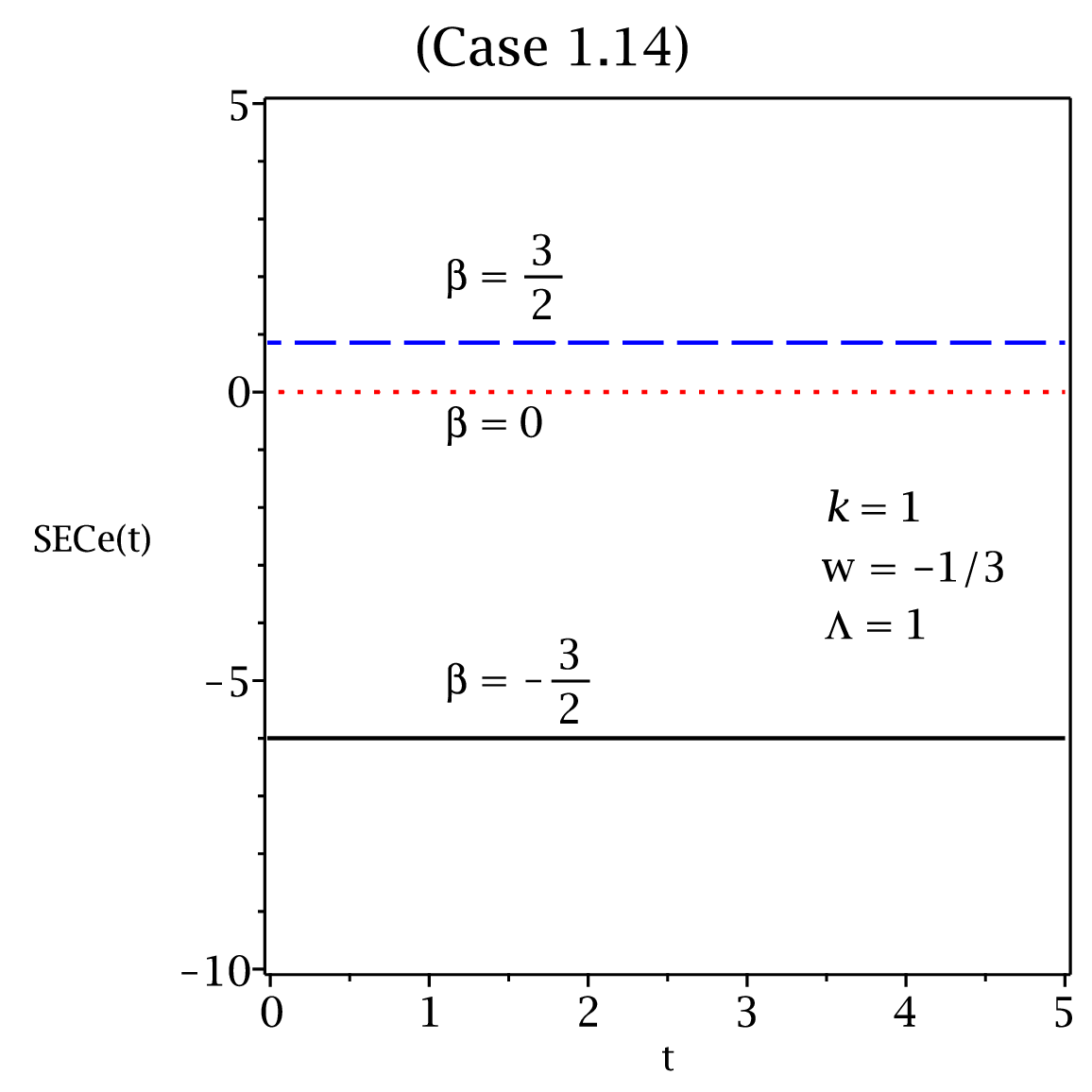}
	\includegraphics[width=3.4cm]{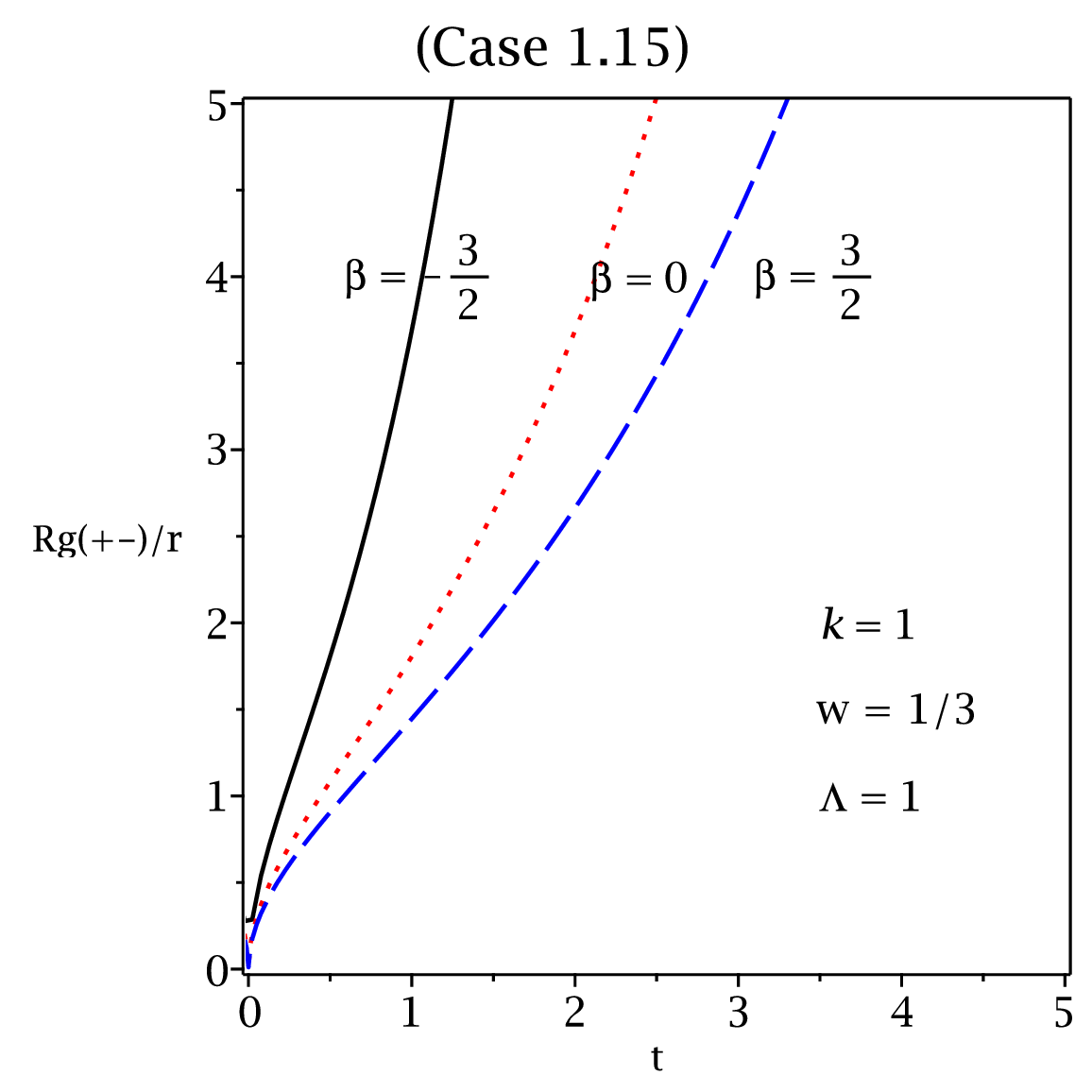}
	\includegraphics[width=3.4cm]{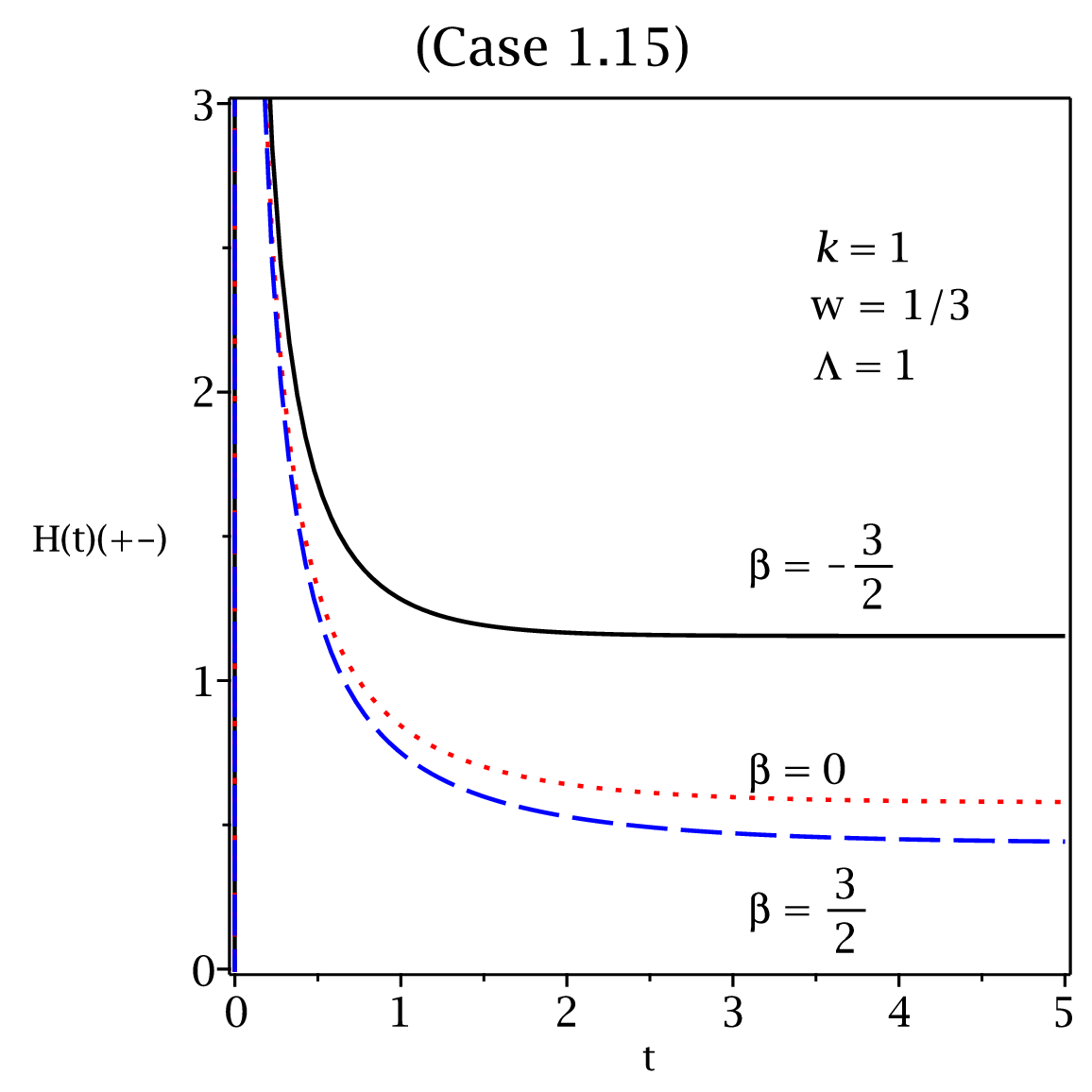}
	\includegraphics[width=3.4cm]{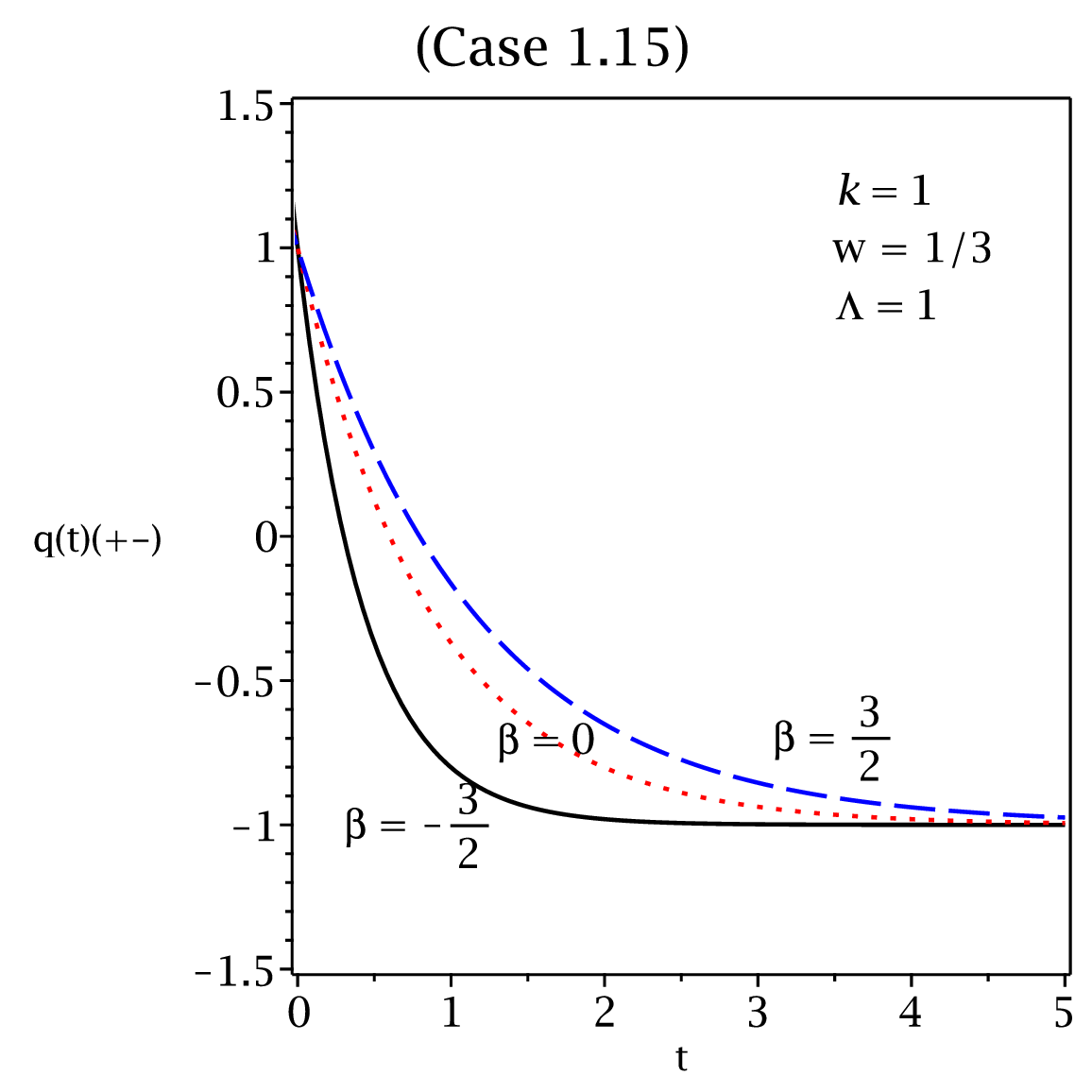}
	\includegraphics[width=3.4cm]{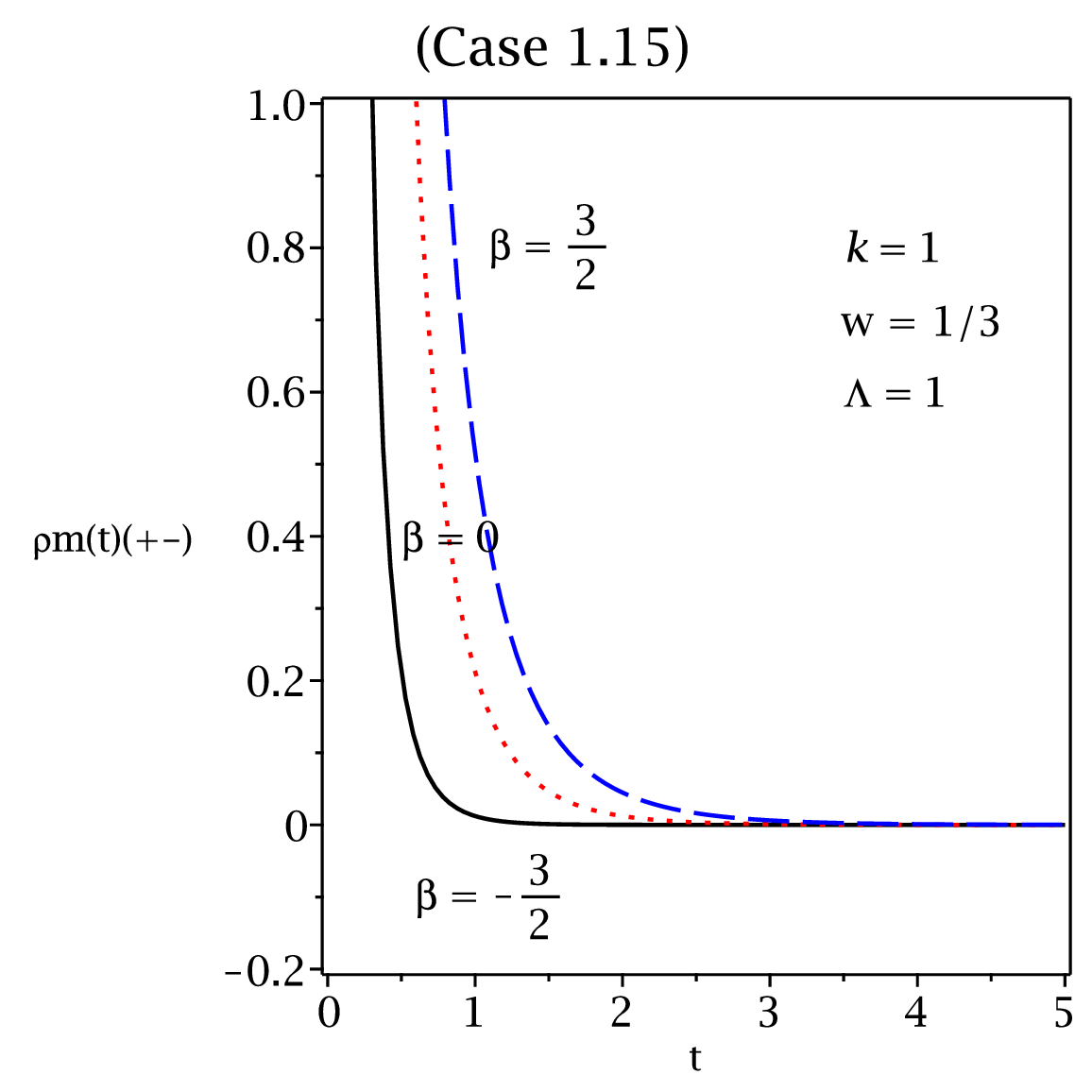}
	\includegraphics[width=3.4cm]{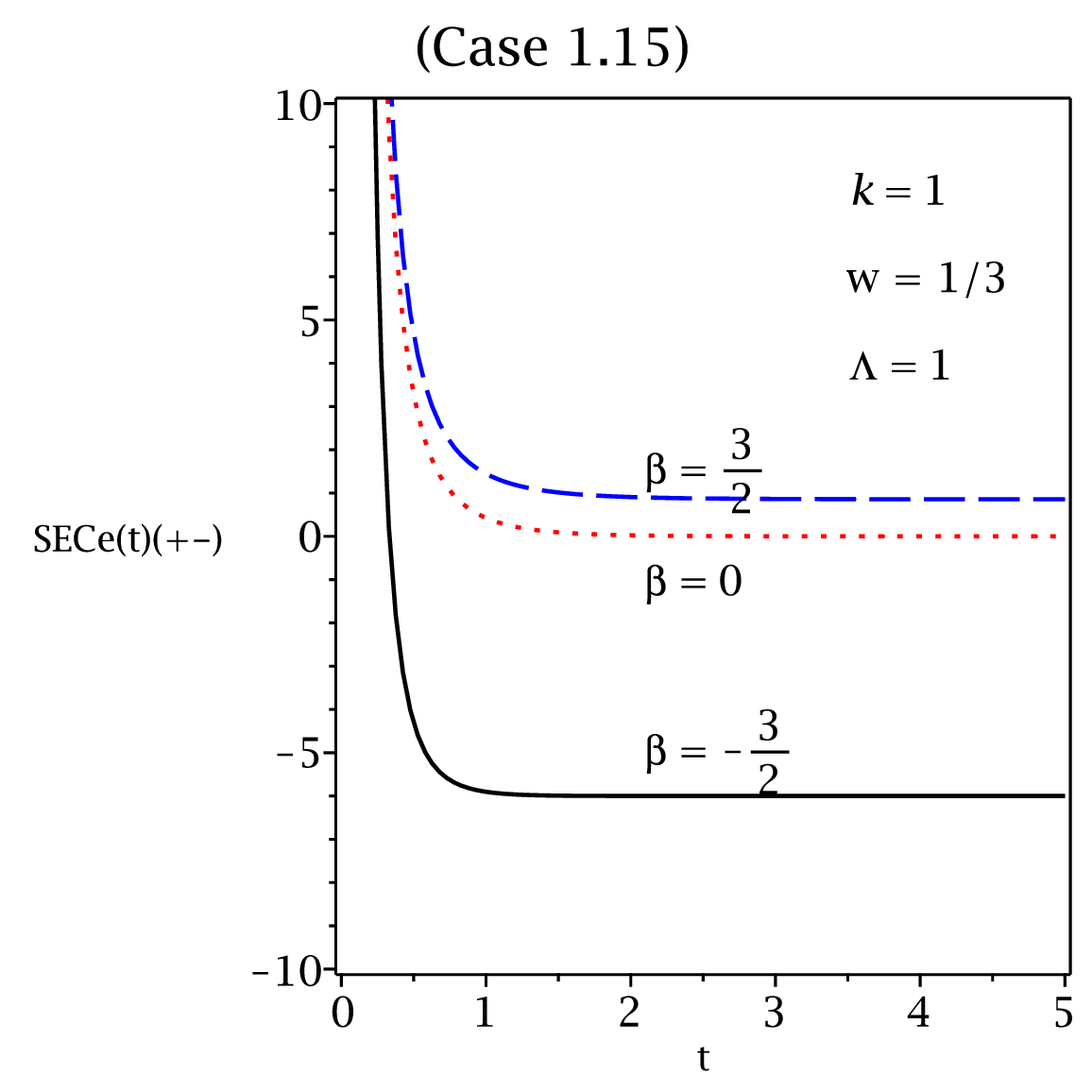}
	\caption{These figures are for $\Lambda>0$ and $k=1$.
		These figures represent the quantities $R_g$ (geometrical radius), 
		$H(t)$ (Hubble parameter) and $q(t)$ (deceleration parameter) $\rho_m(t)$ 
		(energy density of the aether fluid) and $SEC_{e} \equiv SEC_{\rm eff}$ 
		(strong energy condition for the effective fluid) for the different
		values of $\beta=-3/2$ (black solid line), $\beta=0$ (red dotted line), 
		$\beta=3/2$ (blue dashed line). Assuming that $8 \pi G=1$ and
		$R_g(t=0)=0$. Assuming also that $C_1=1$, $C_2=0$ (Cases 1.12, 1.14 and 1.15); 
		$C_1=1$, $C_2=-1$ (Case 1.13). The subscripts $(+)$ and $(-)$, denote the two different 
		solutions for $B(t)$.}
	\label{Figure-112-115}
\end{minipage}	
\end{figure}

\begin{figure}[!htp]
\begin{minipage}{175 mm}
	\centering	
	\includegraphics[width=3.4cm]{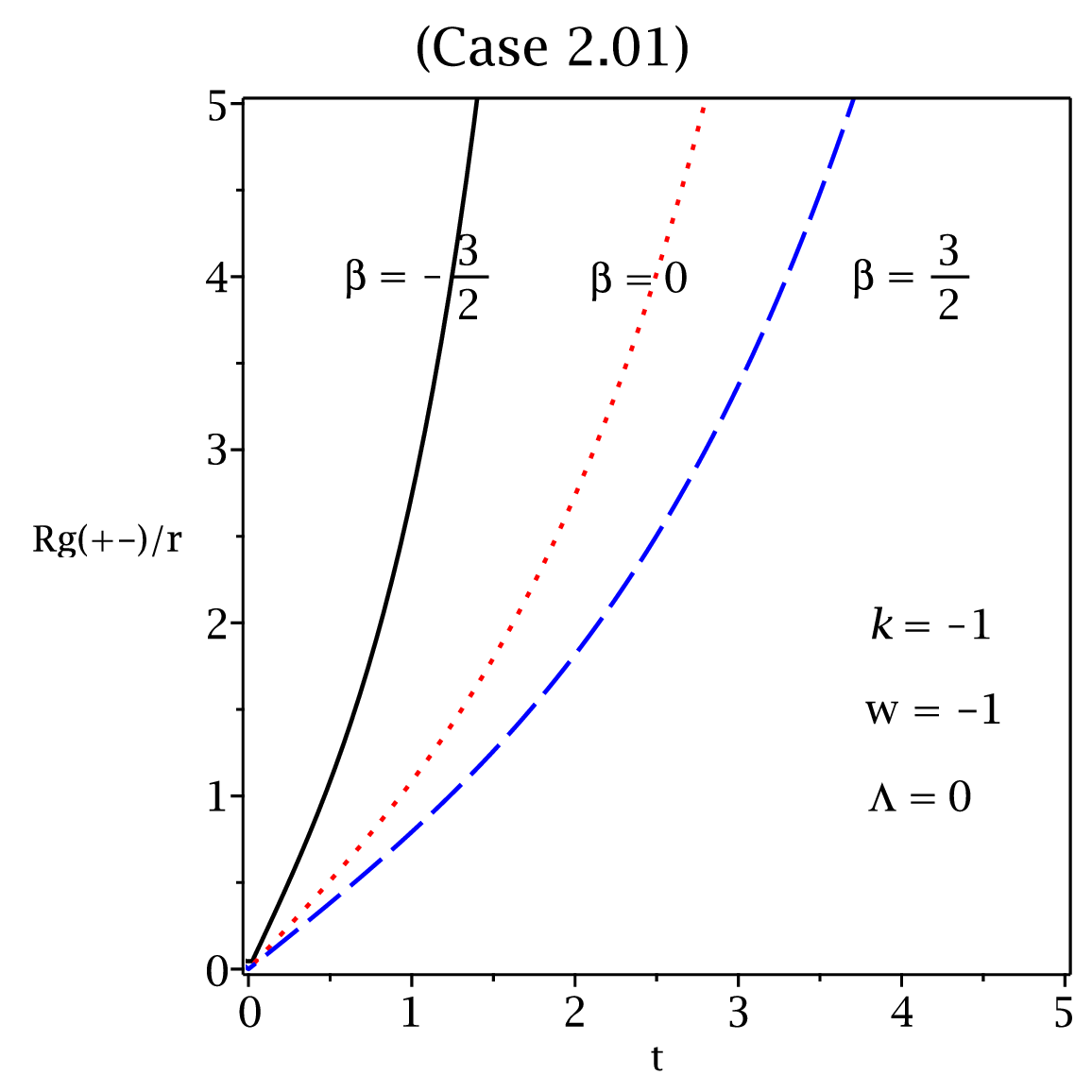}
	\includegraphics[width=3.4cm]{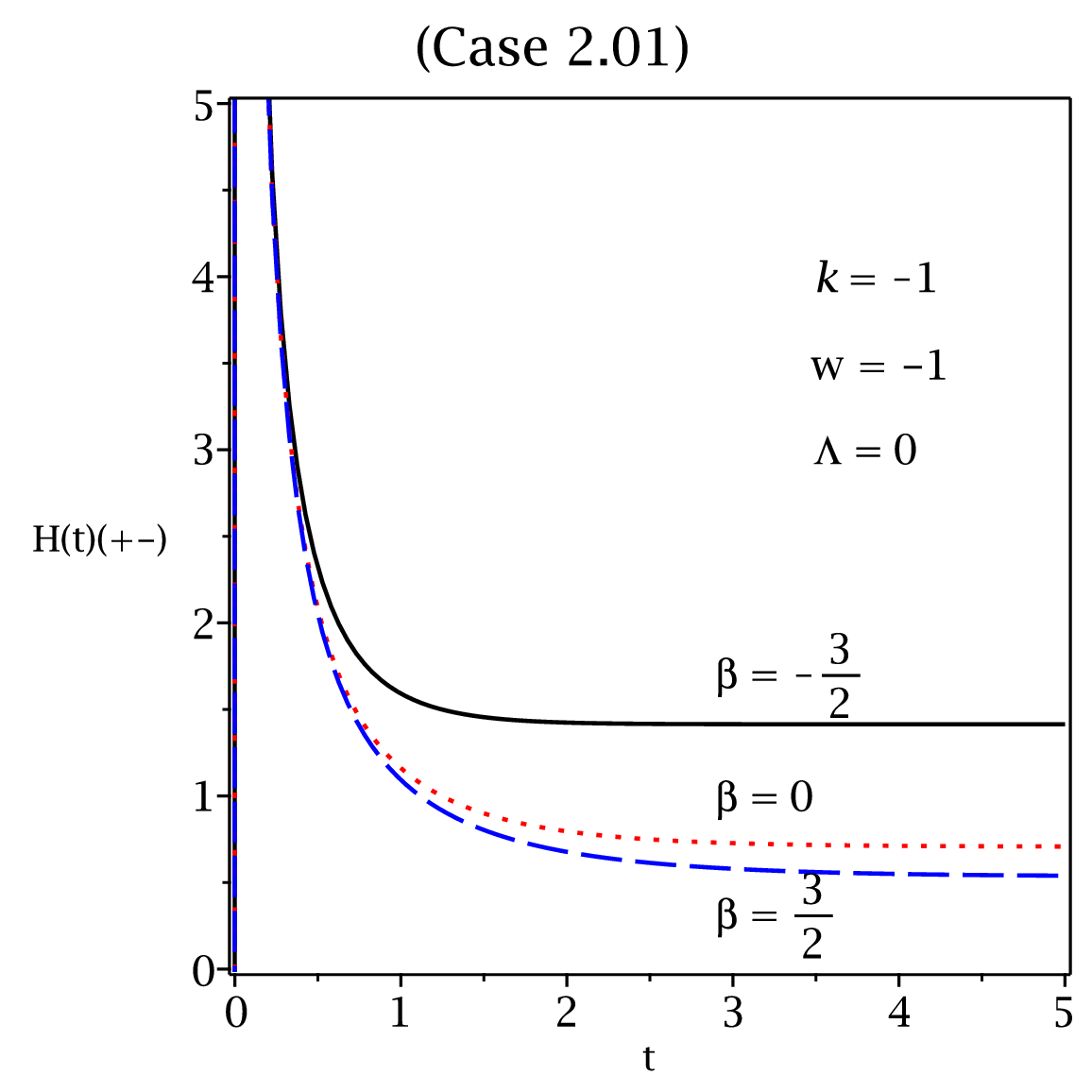}
	\includegraphics[width=3.4cm]{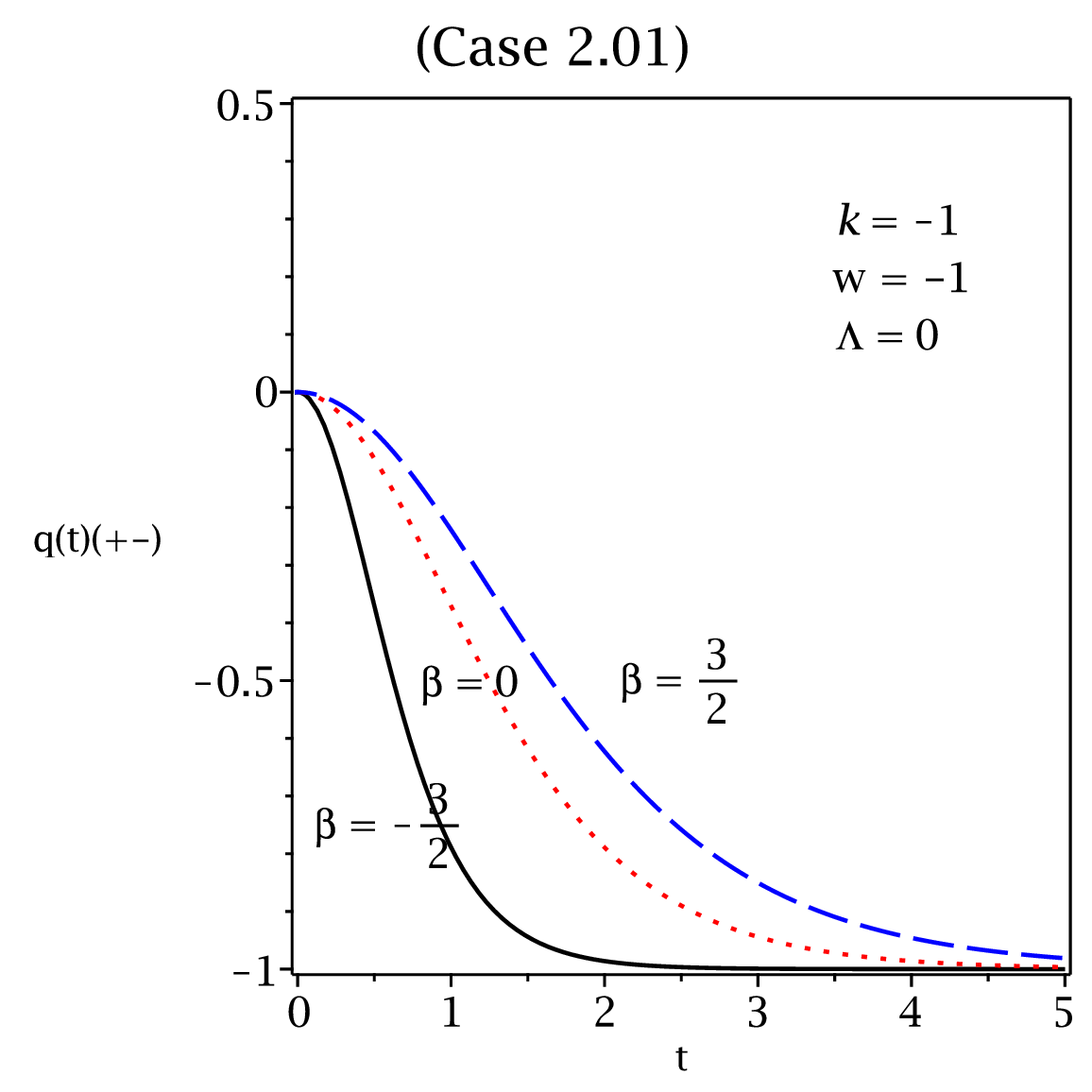}
	\includegraphics[width=3.4cm]{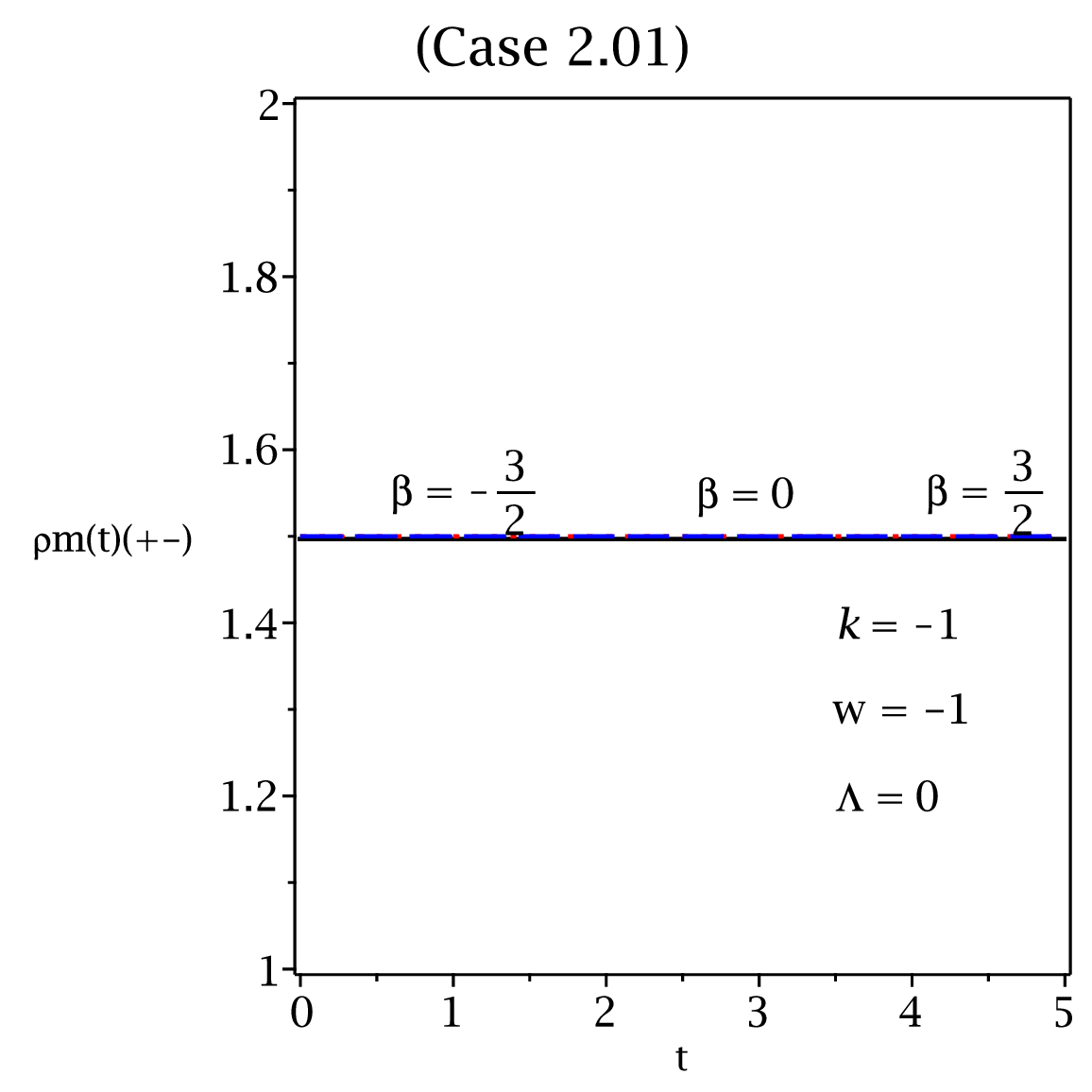}
	\includegraphics[width=3.4cm]{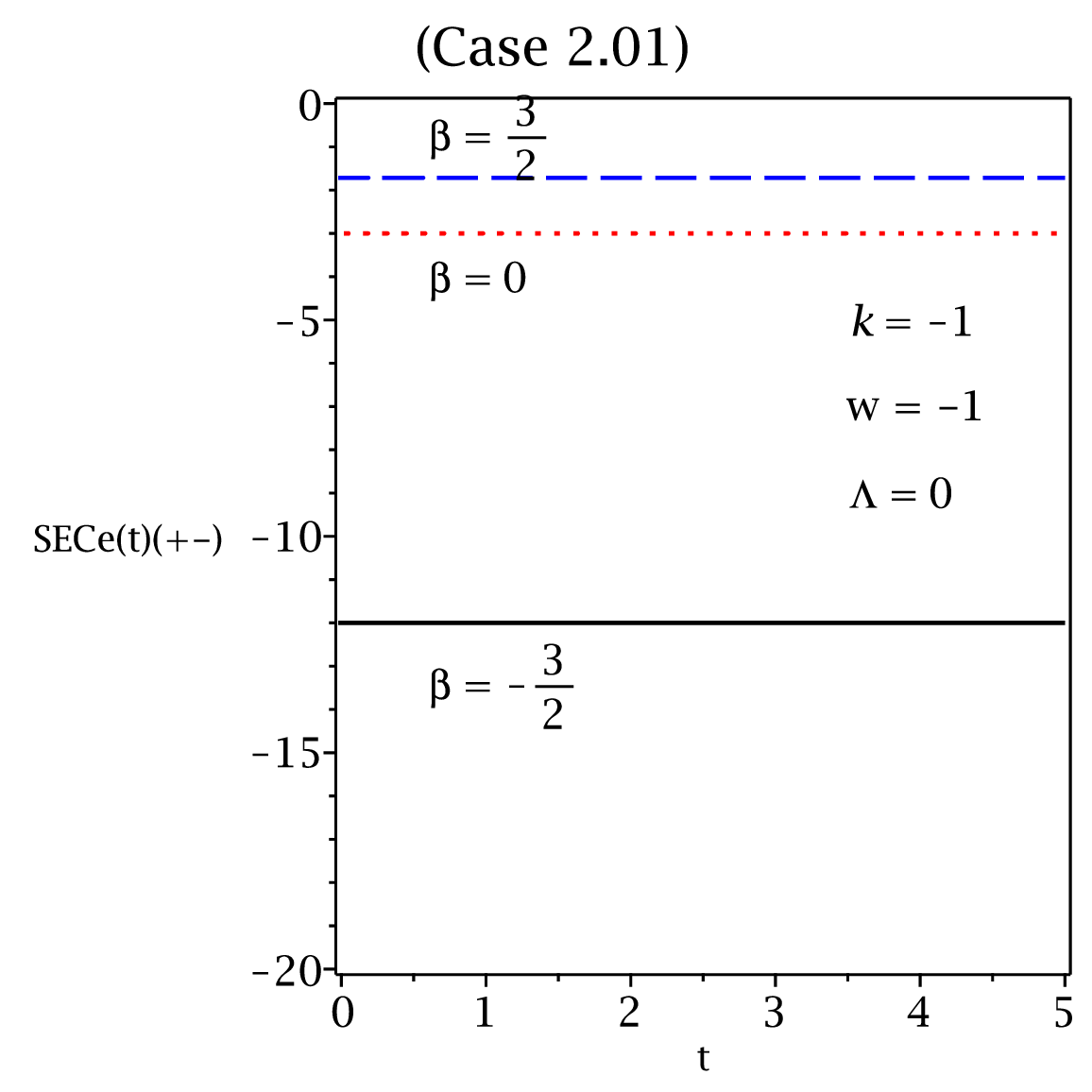}
	\includegraphics[width=3.4cm]{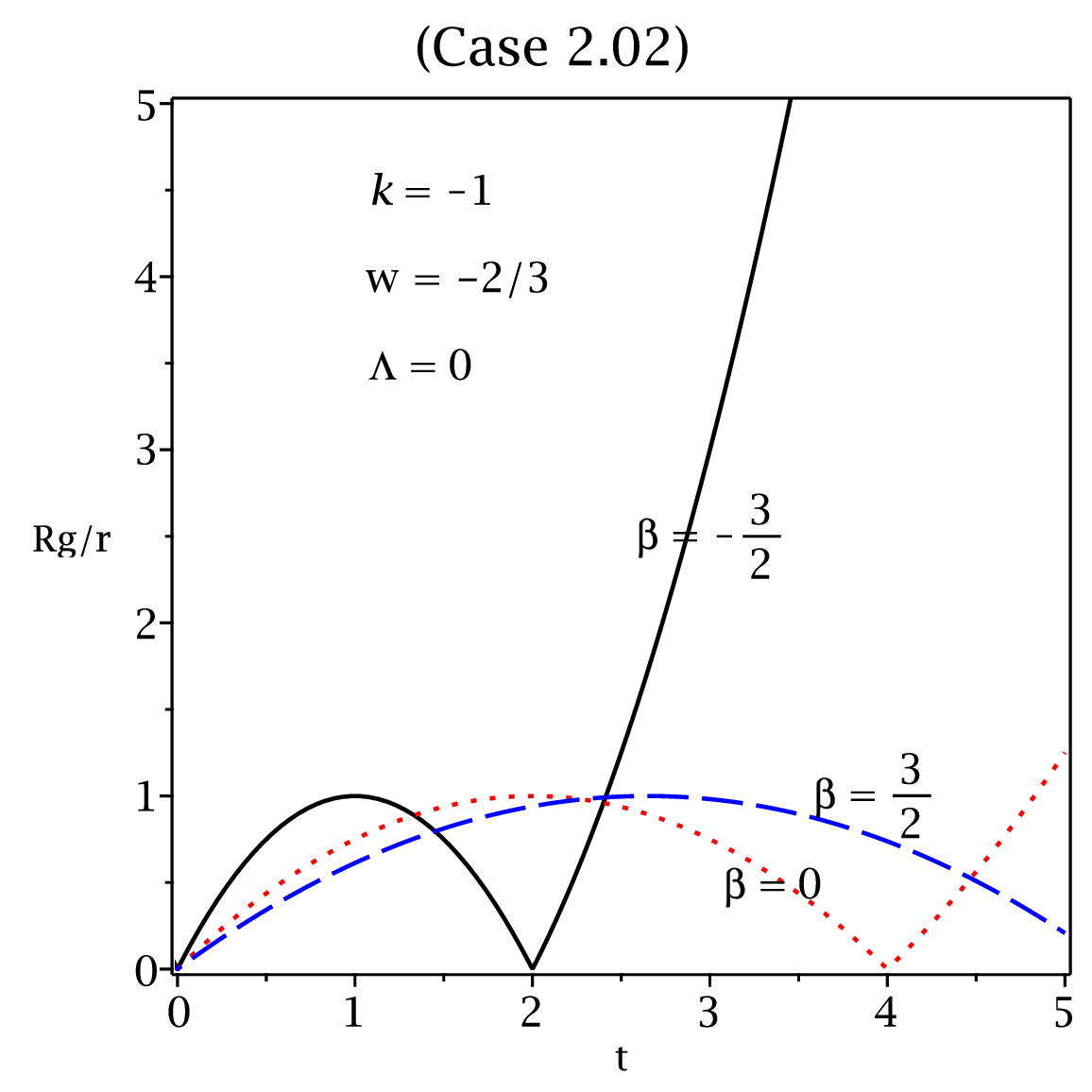}
	\includegraphics[width=3.4cm]{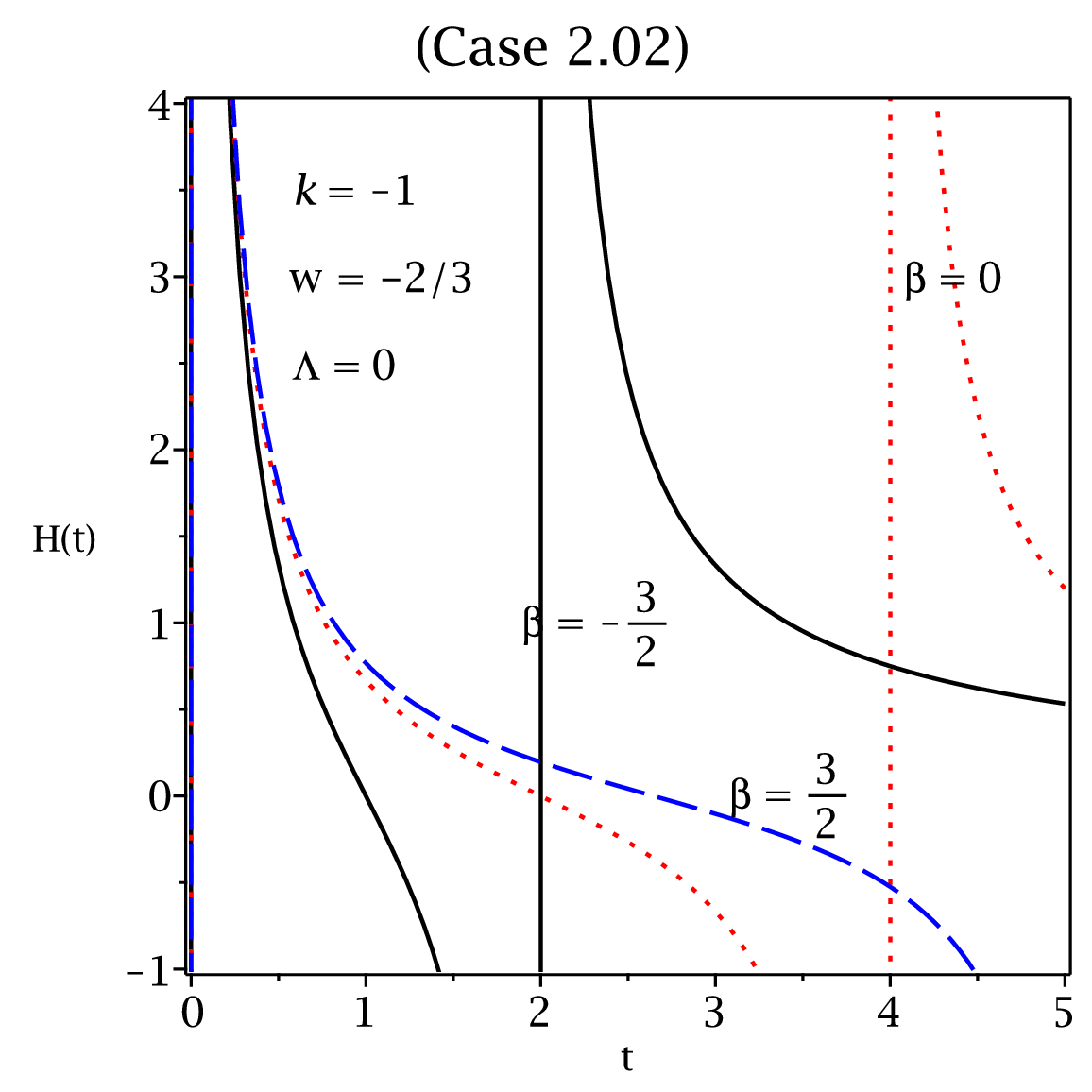}
	\includegraphics[width=3.4cm]{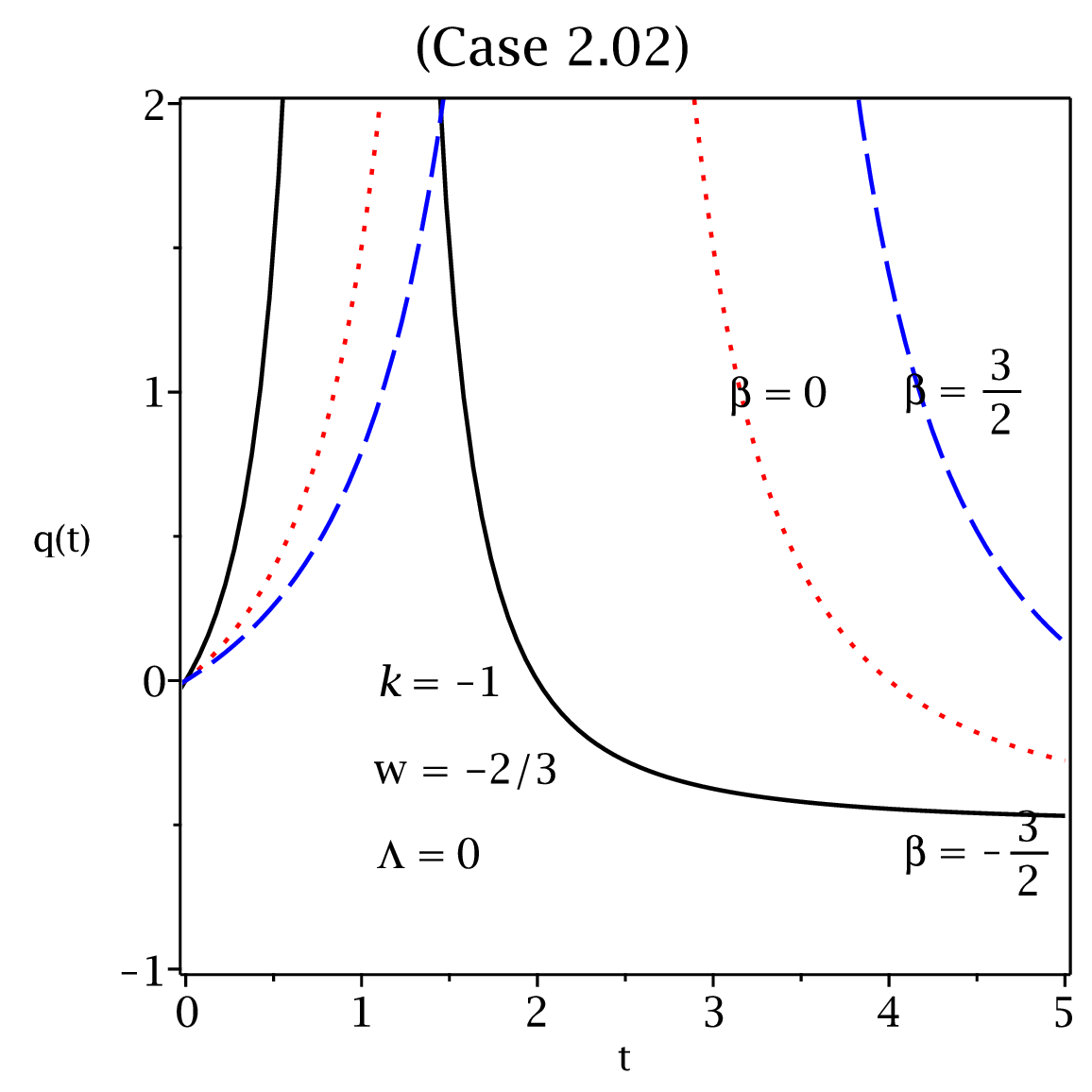}
	\includegraphics[width=3.4cm]{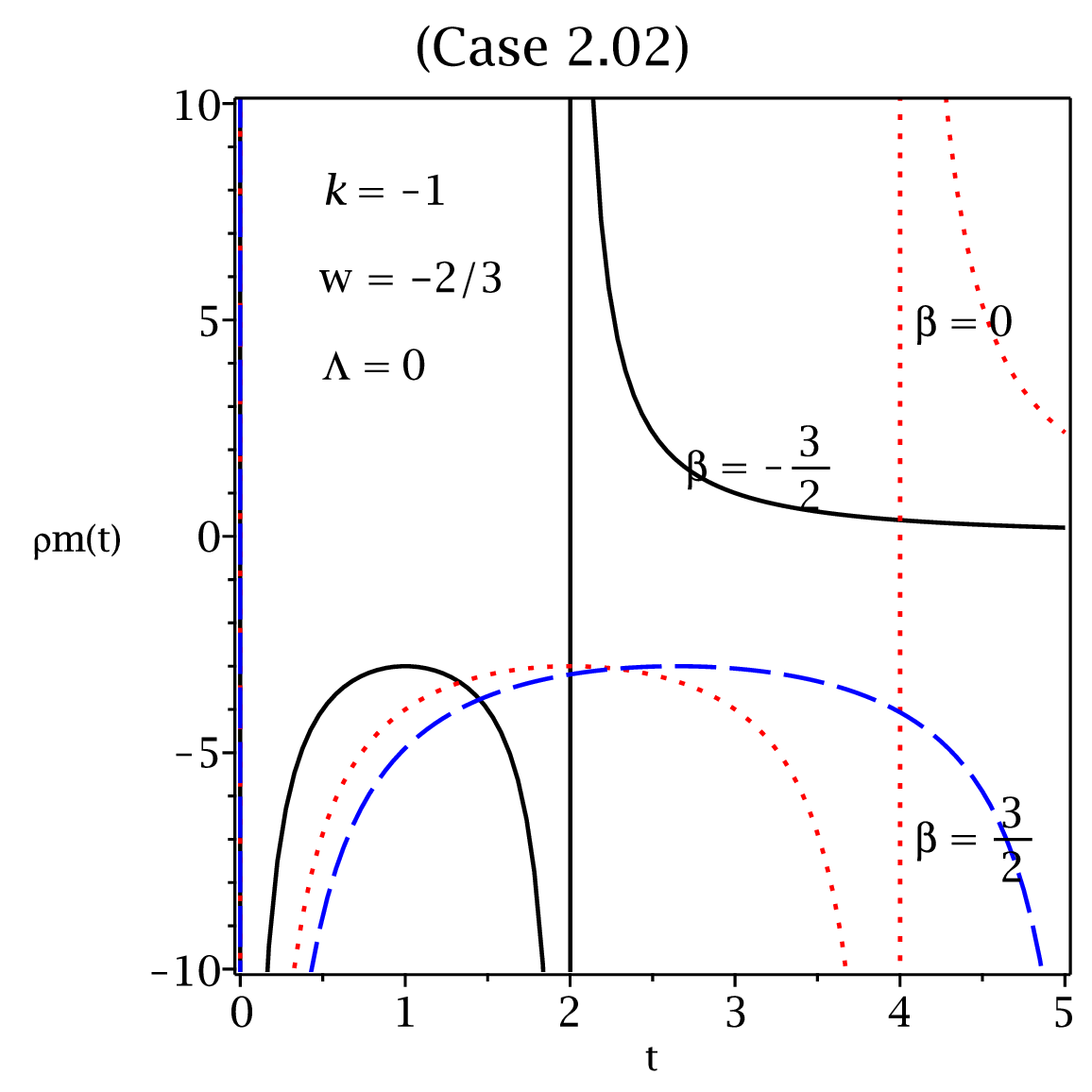}
	\includegraphics[width=3.4cm]{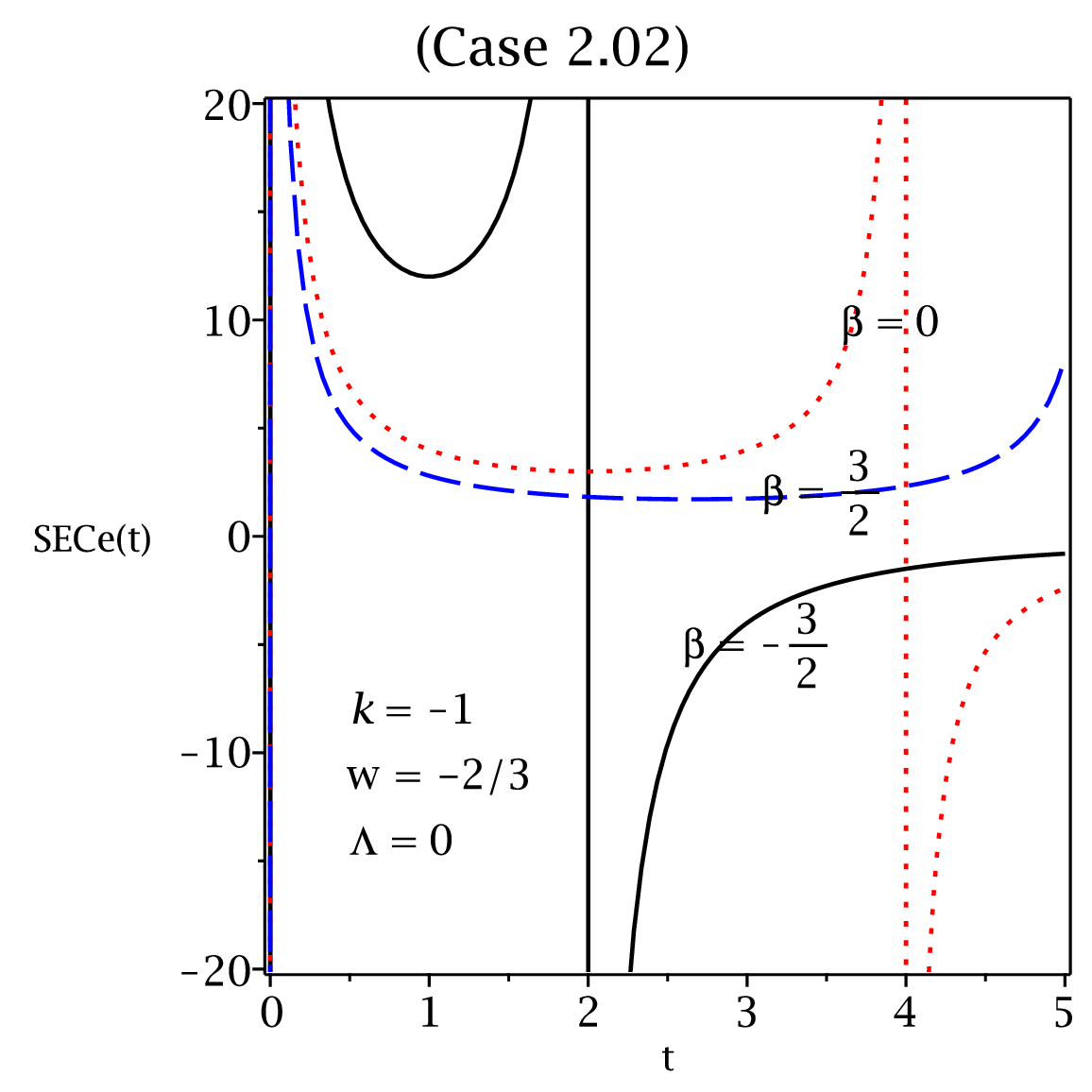}
	\includegraphics[width=3.4cm]{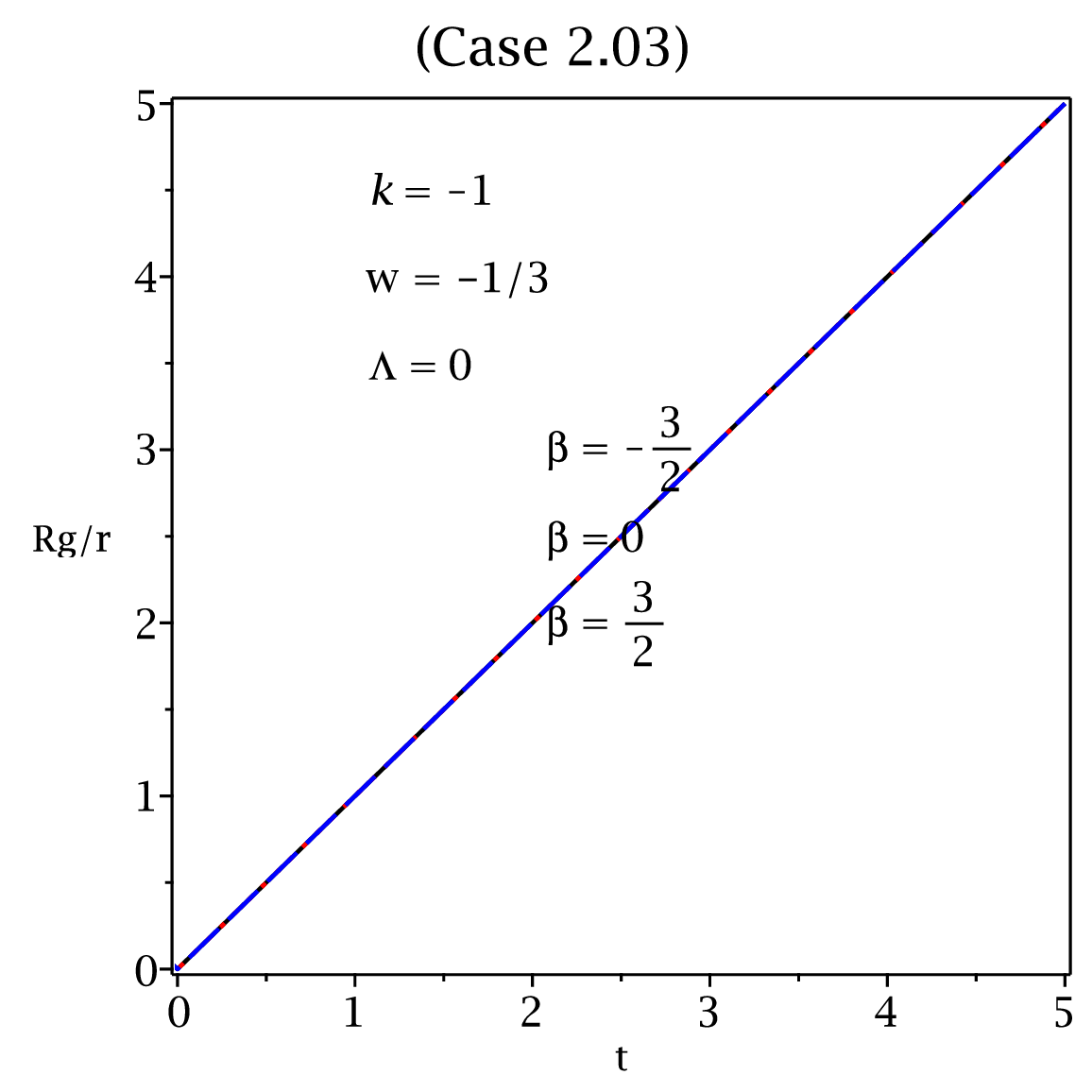}
	\includegraphics[width=3.4cm]{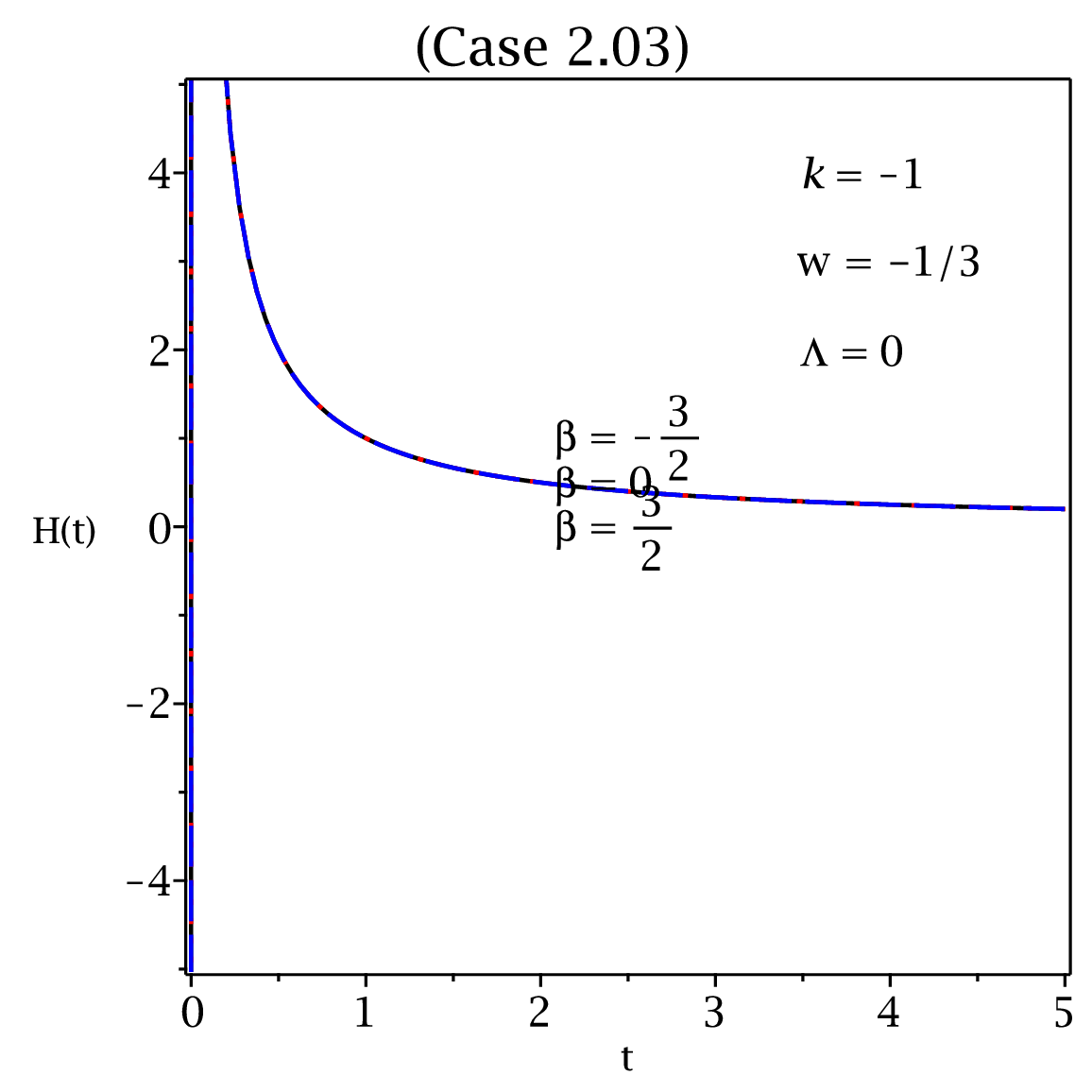}
	\includegraphics[width=3.4cm]{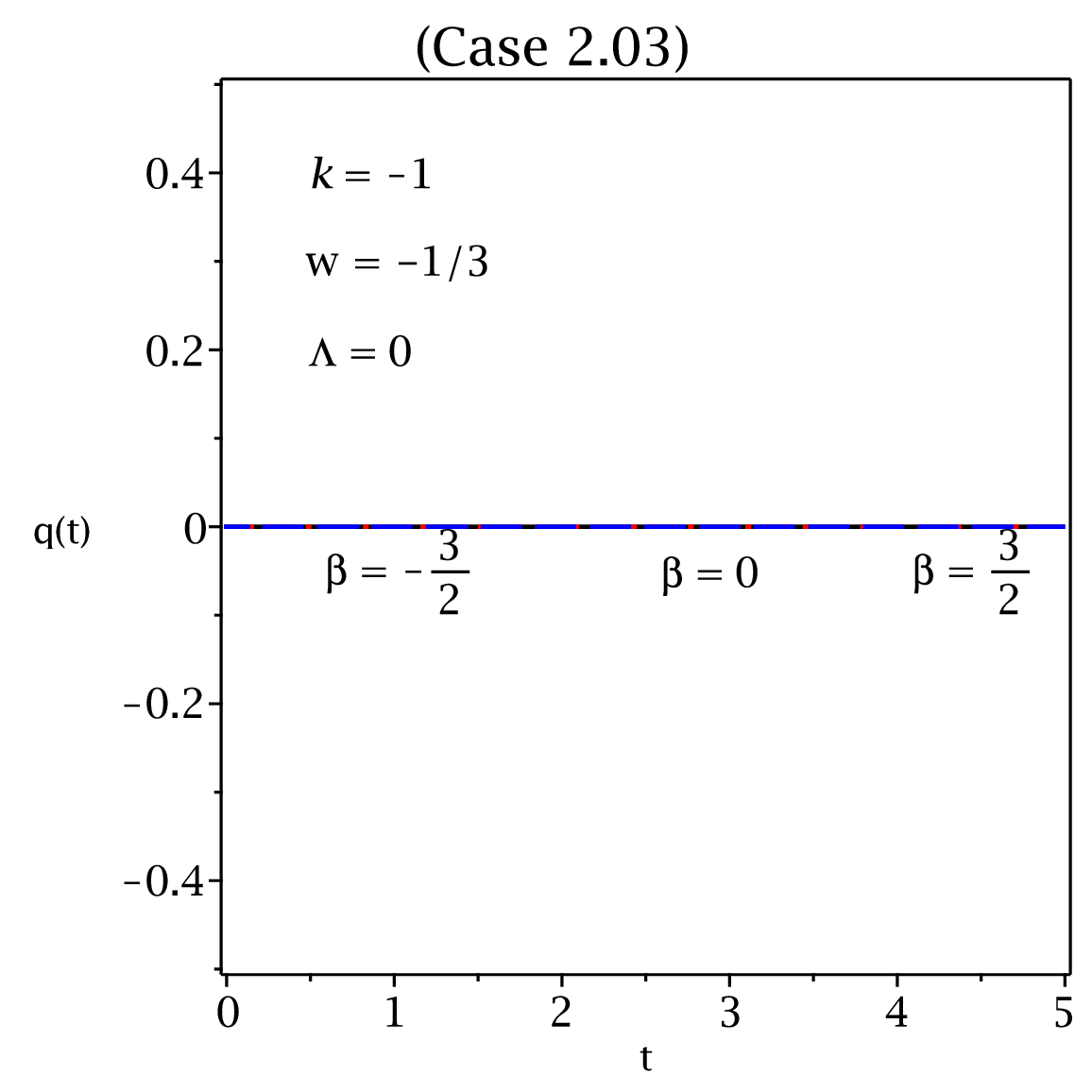}
	\includegraphics[width=3.4cm]{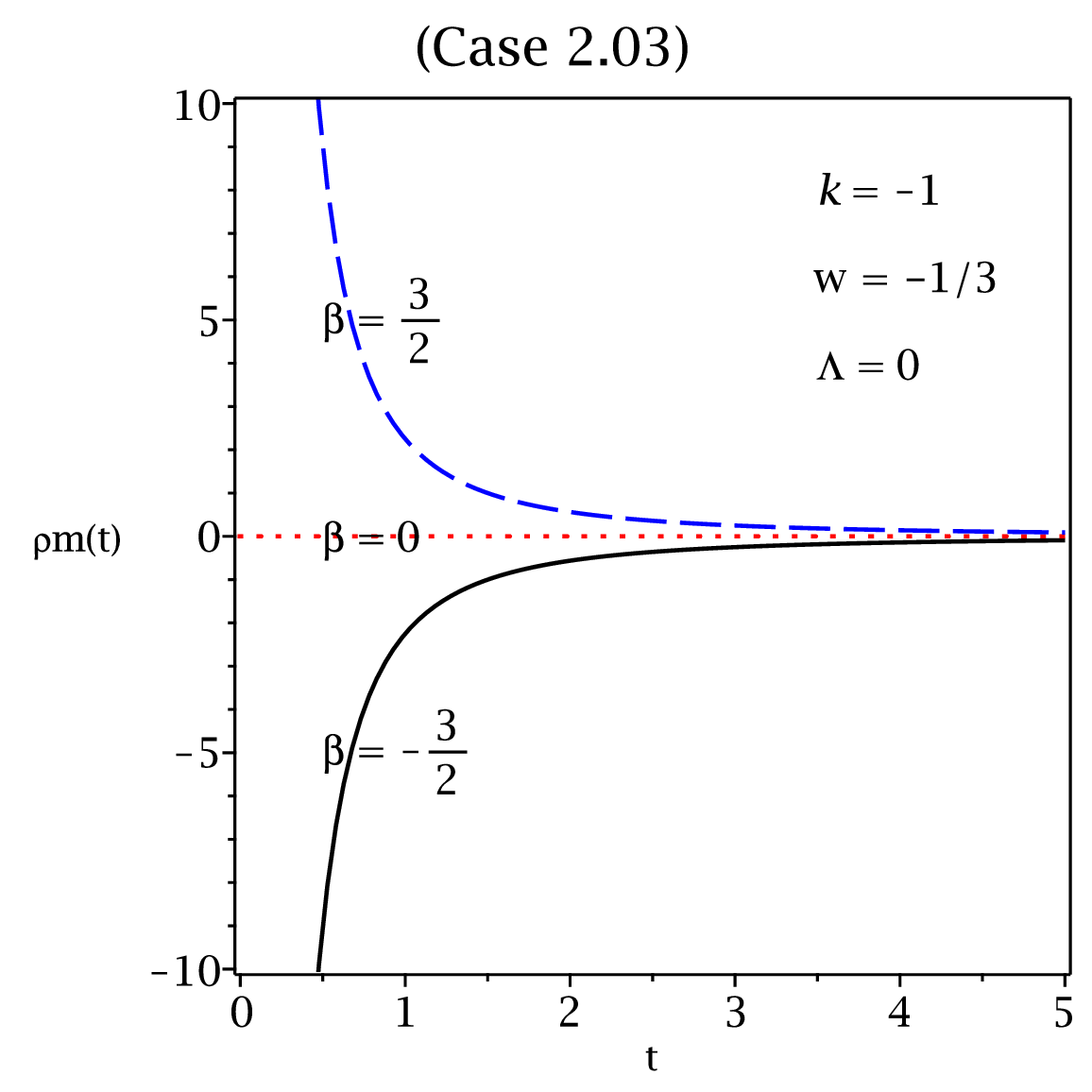}
	\includegraphics[width=3.4cm]{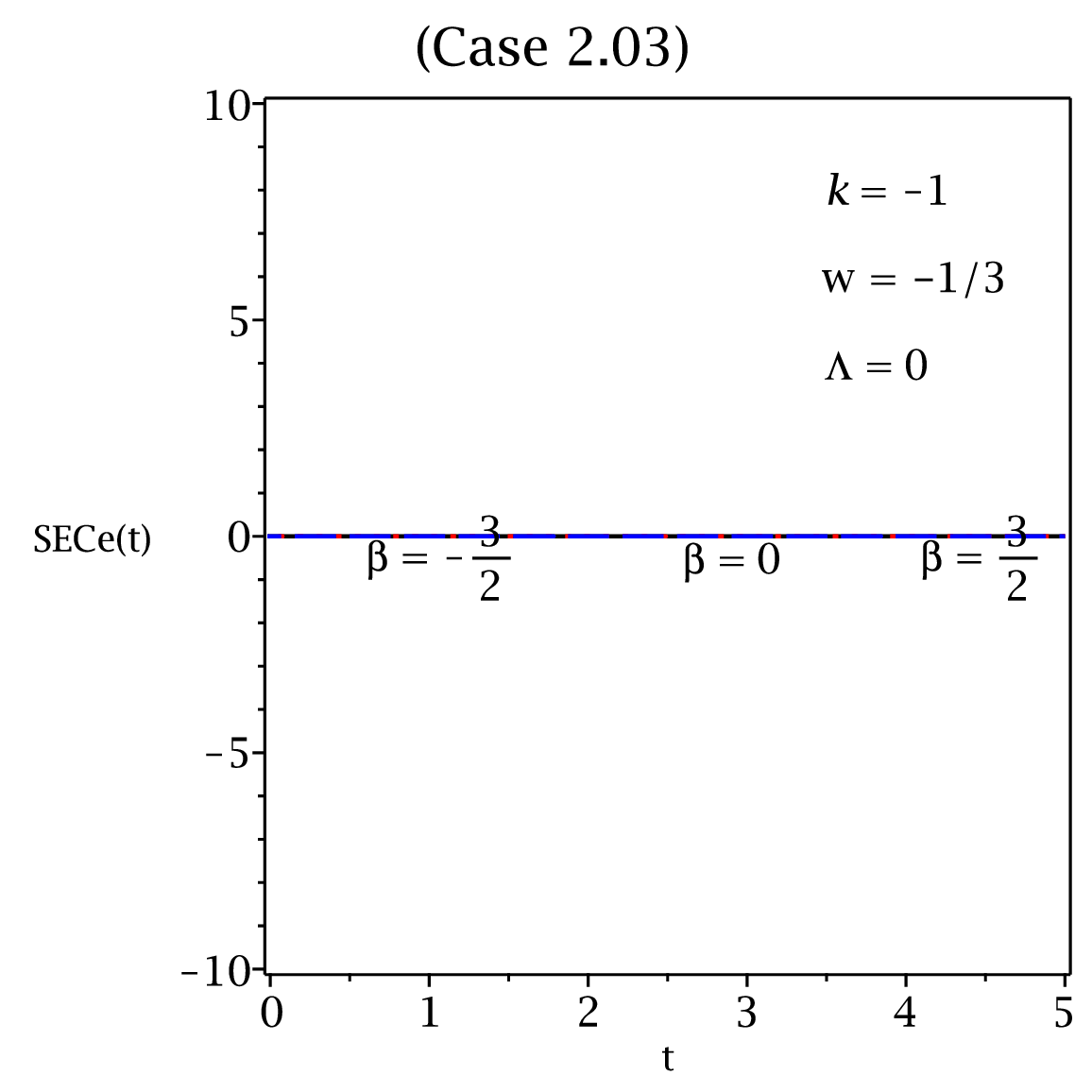}
	%
	%
	\includegraphics[width=3.4cm]{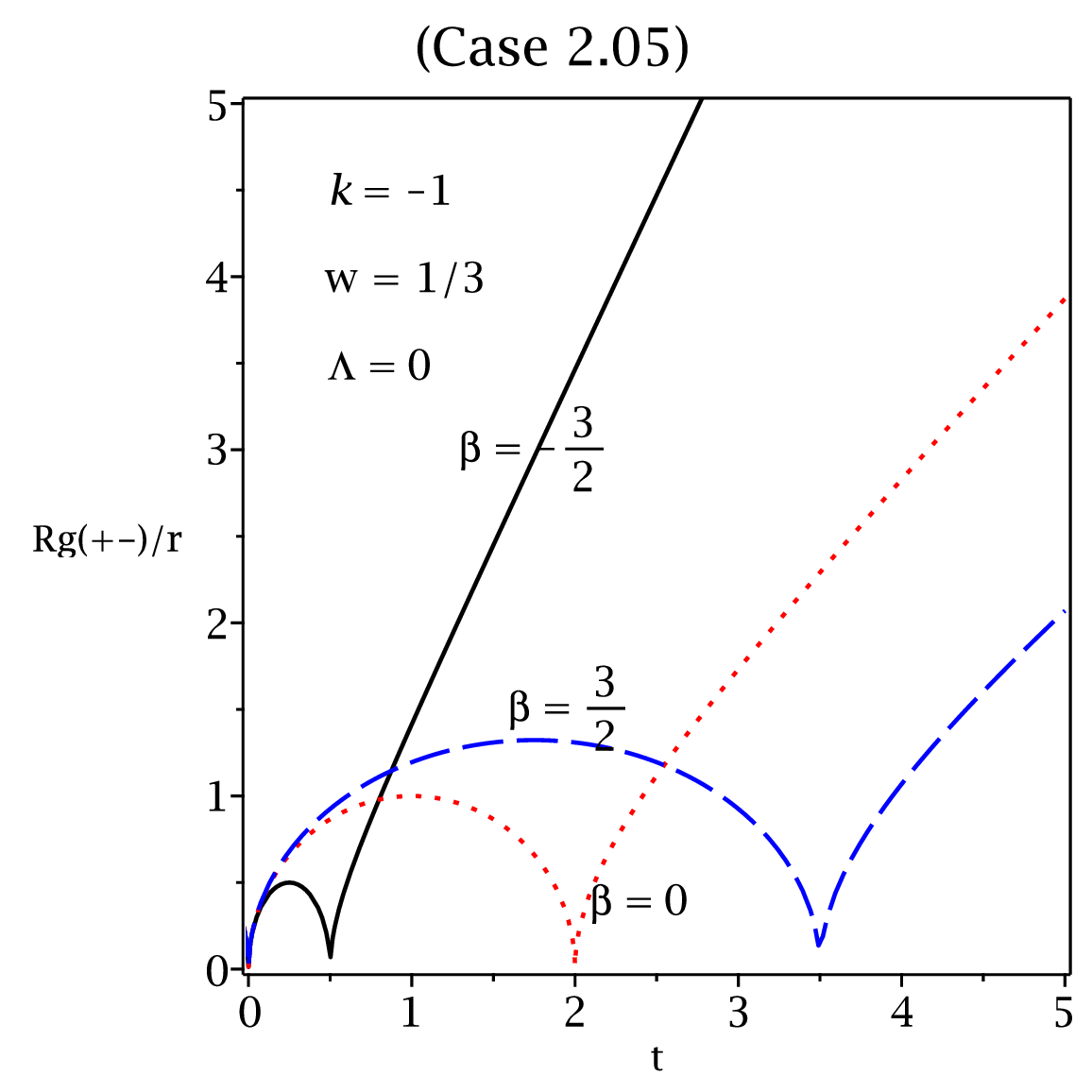}
	\includegraphics[width=3.4cm]{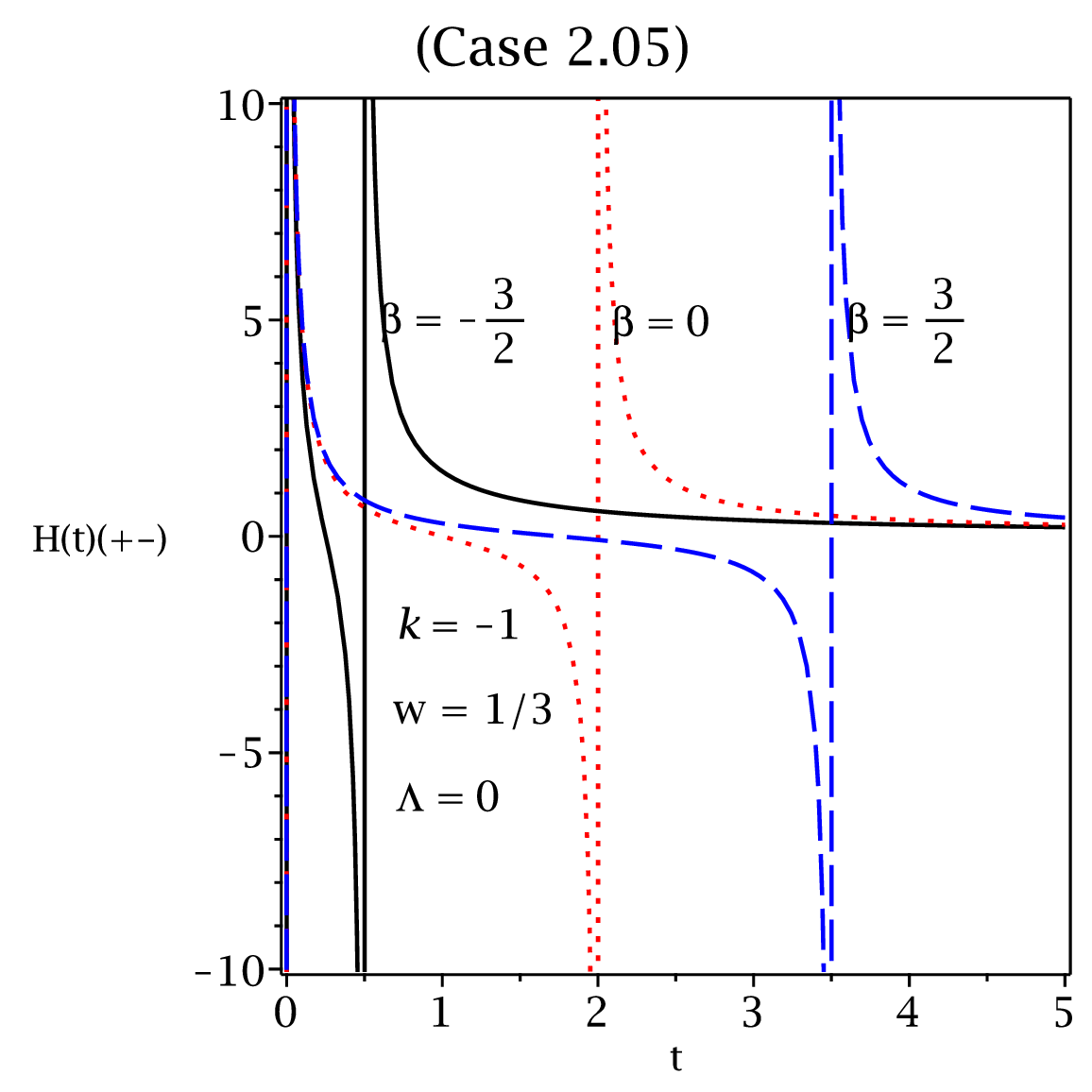}
	\includegraphics[width=3.4cm]{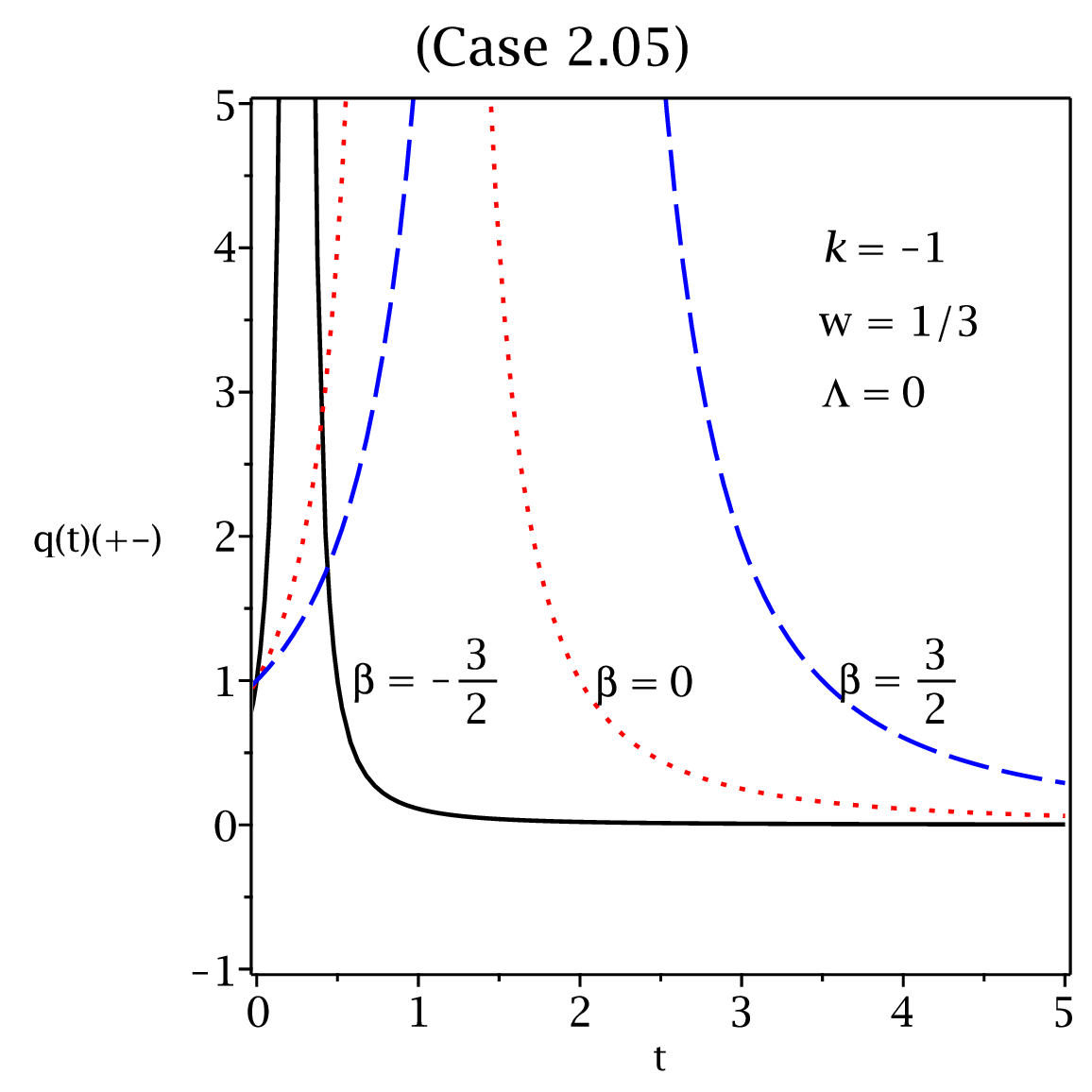}
	\includegraphics[width=3.4cm]{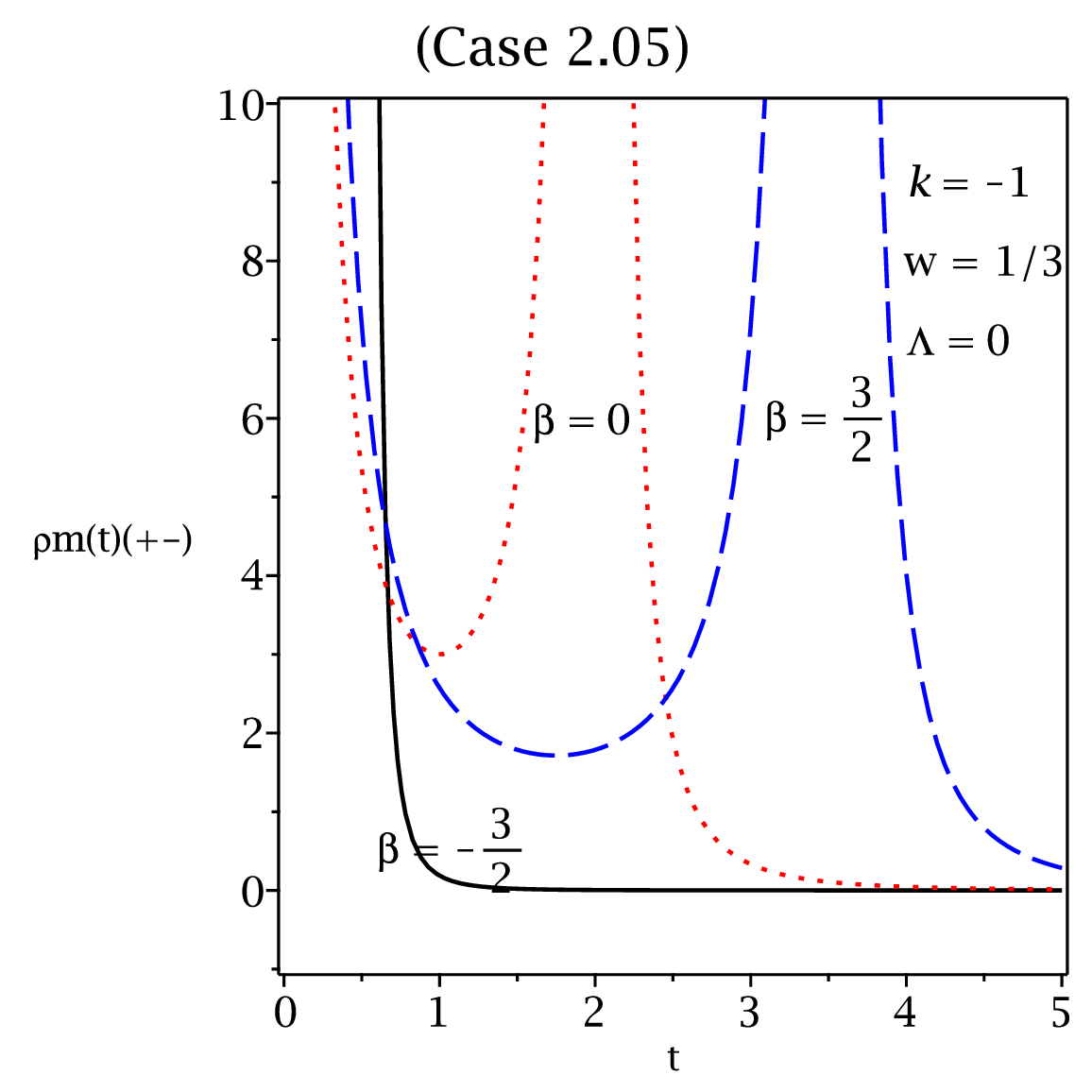}
	\includegraphics[width=3.4cm]{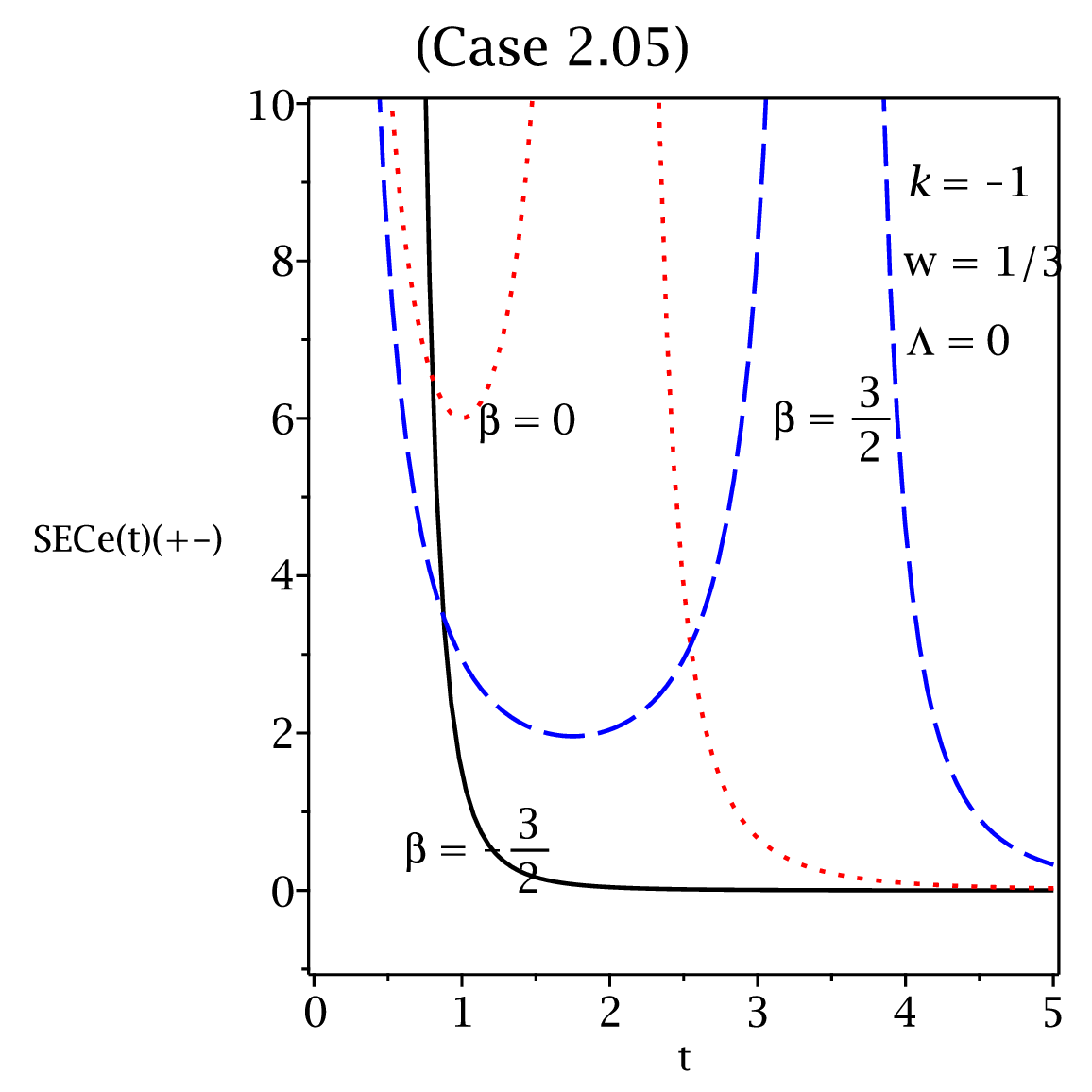}
	\caption{These figures are for $\Lambda=0$ and $k=-1$.
		These figures represent the quantities $R_g$ (geometrical radius), 
		$H(t)$ (Hubble parameter) and $q(t)$ (deceleration parameter) $\rho_m(t)$ 
		(energy density of the aether fluid) and $SEC_{e} \equiv SEC_{\rm eff}$ 
		(strong energy condition for the effective fluid) for the different
		values of $\beta=-3/2$ (black solid line), $\beta=0$ (red dotted line), 
		$\beta=3/2$ (blue dashed line). Assuming that $8 \pi G=1$ and
		$R_g(t=0)=0$. Assuming also that $C_1=1$, $C_2=0$ (Cases 2.01, 2.03 and 2.05); 
		$C_1=0$, $C_2=1$ (Case 2.02). The subscripts $(+)$ and $(-)$, denote the two different 
		solutions for $B(t)$.}
	\label{Figure-201-205}
\end{minipage}	
\end{figure}

\begin{figure}[!htp]
\begin{minipage}{175 mm}
	\centering	
	\includegraphics[width=3.4cm]{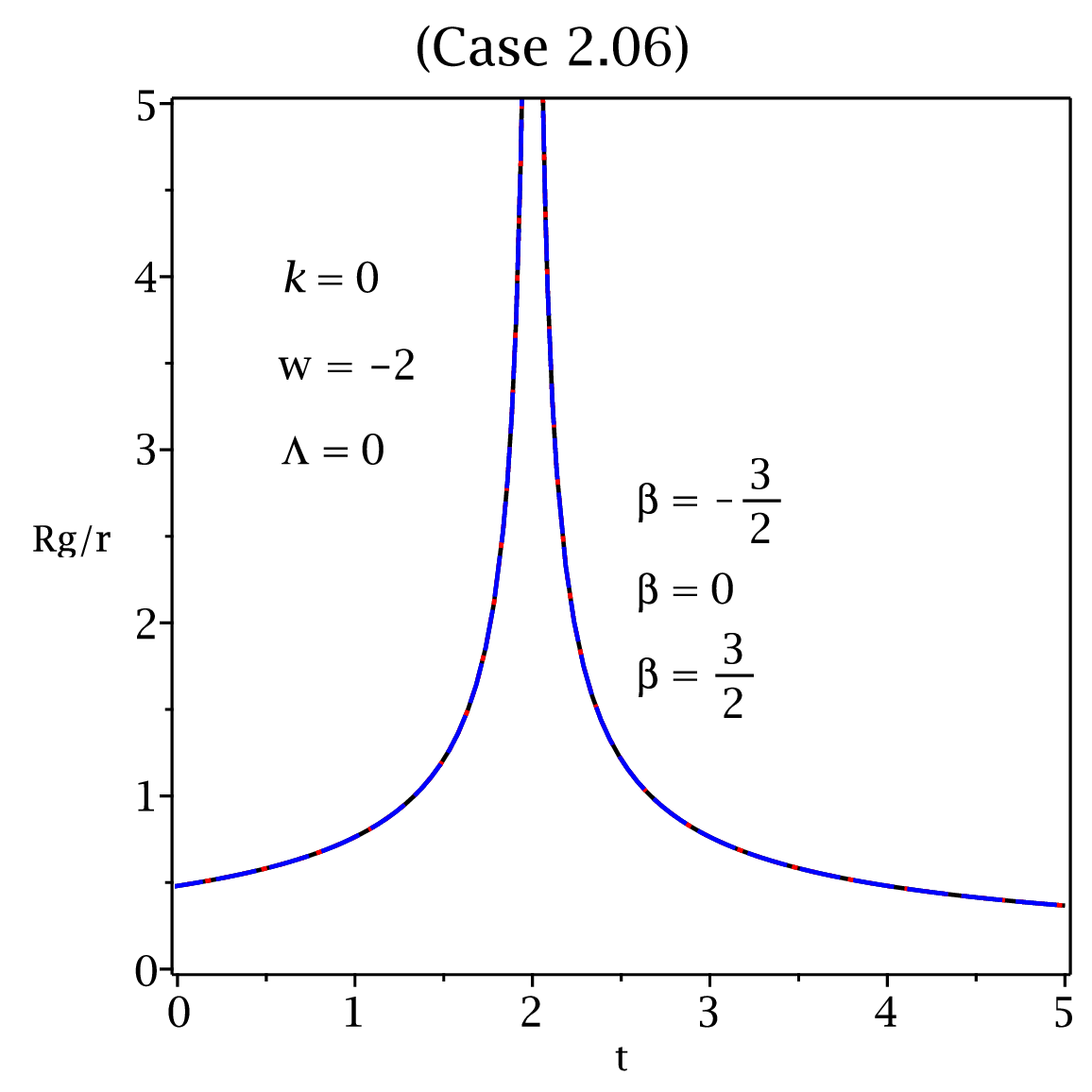}
	\includegraphics[width=3.4cm]{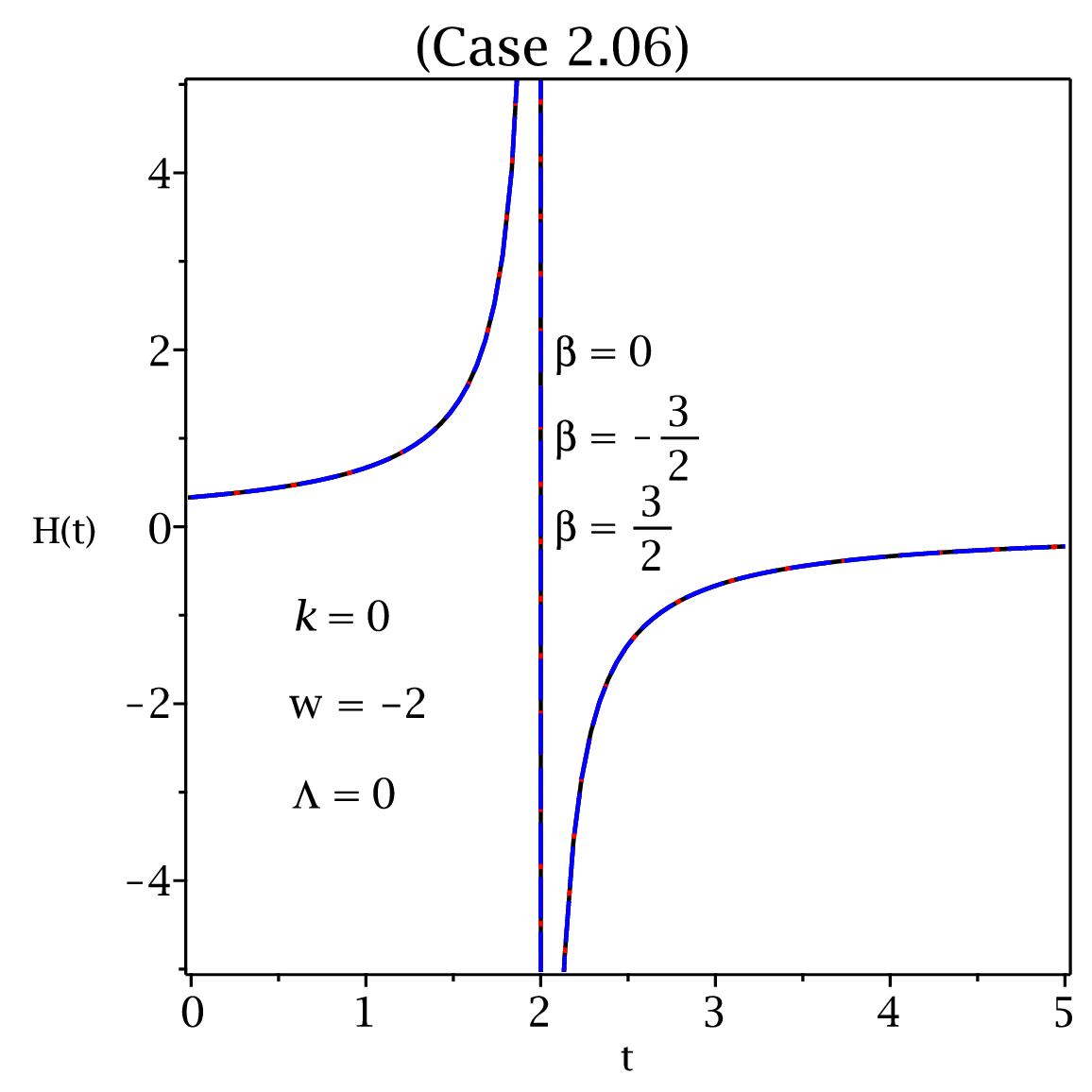}
	\includegraphics[width=3.4cm]{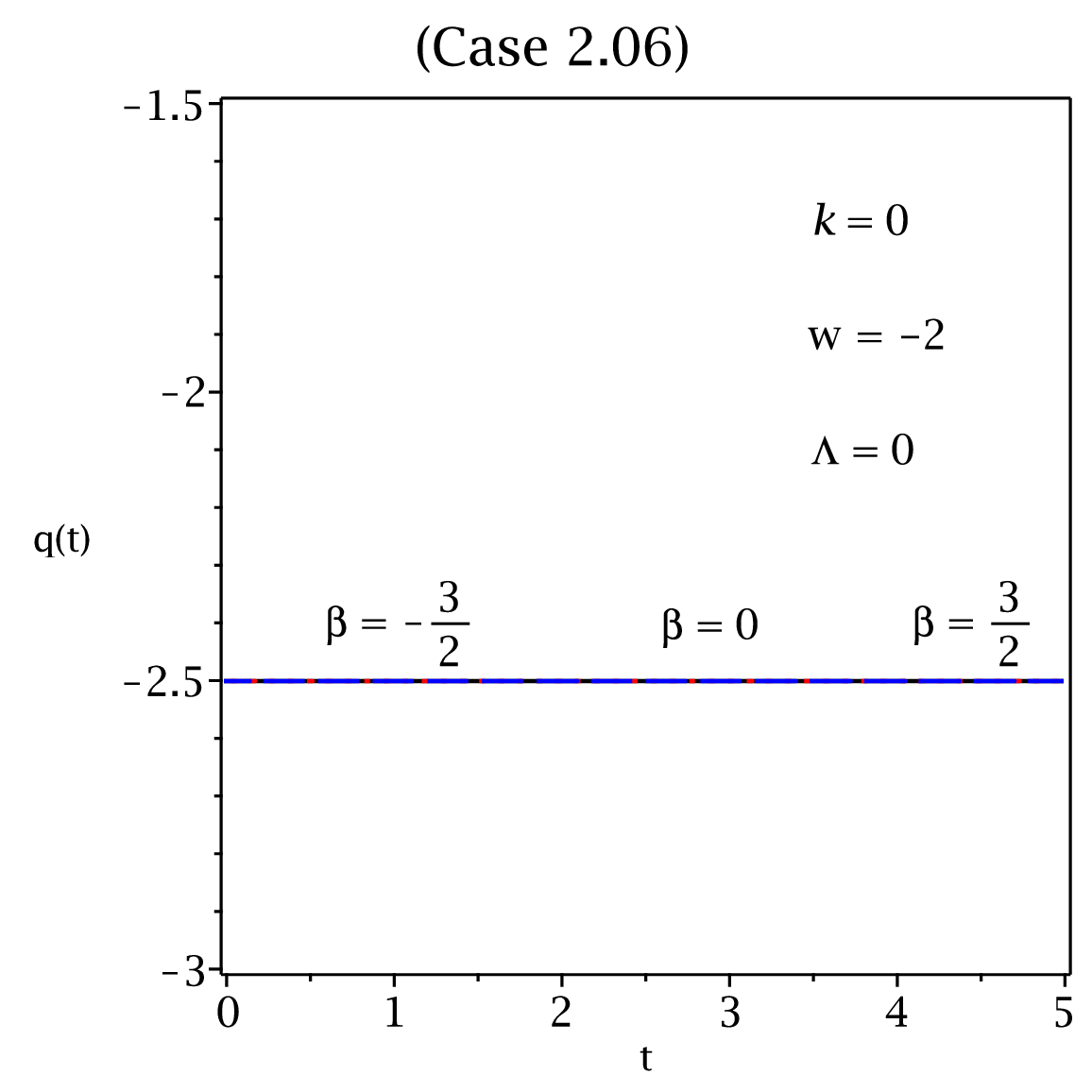}
	\includegraphics[width=3.4cm]{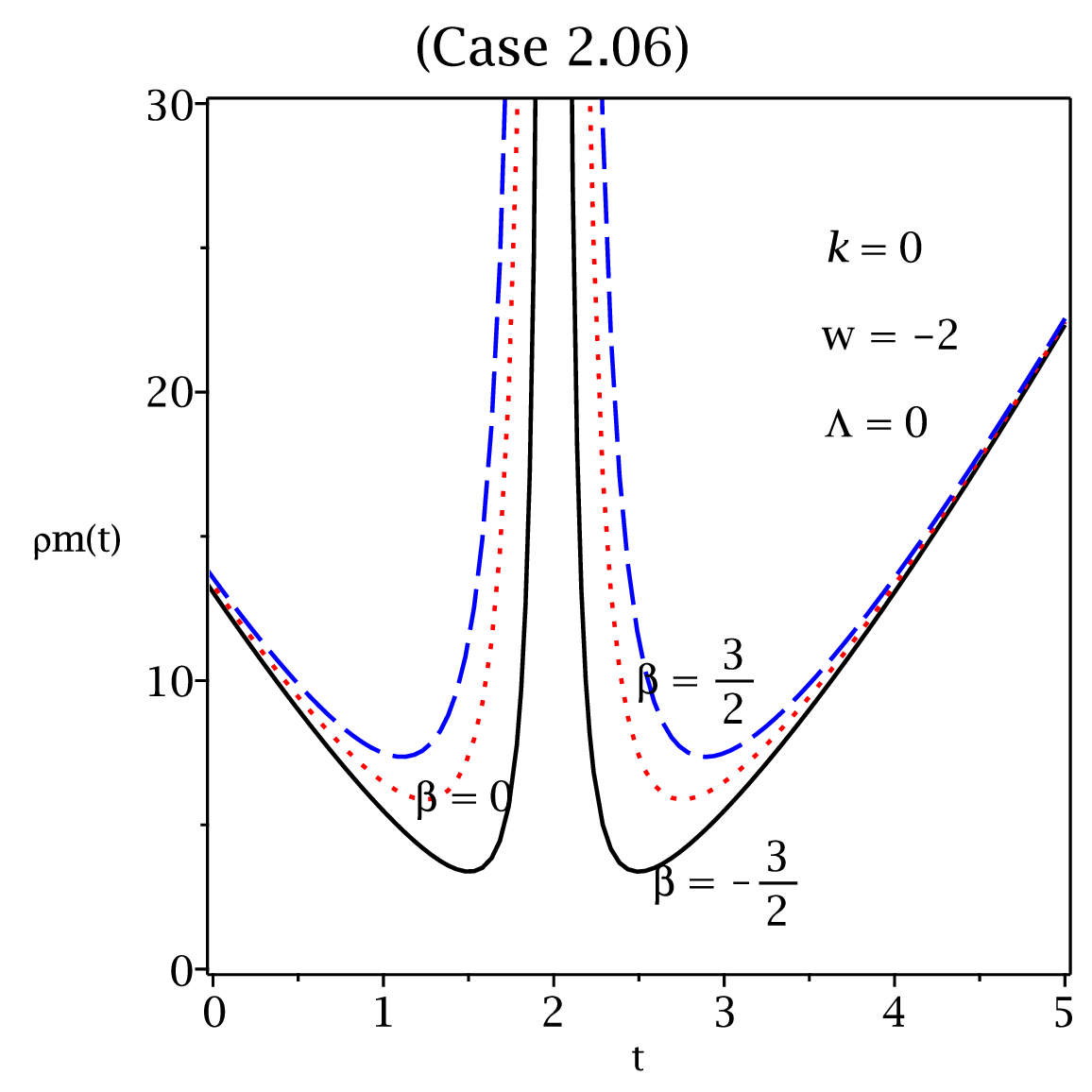}
	\includegraphics[width=3.4cm]{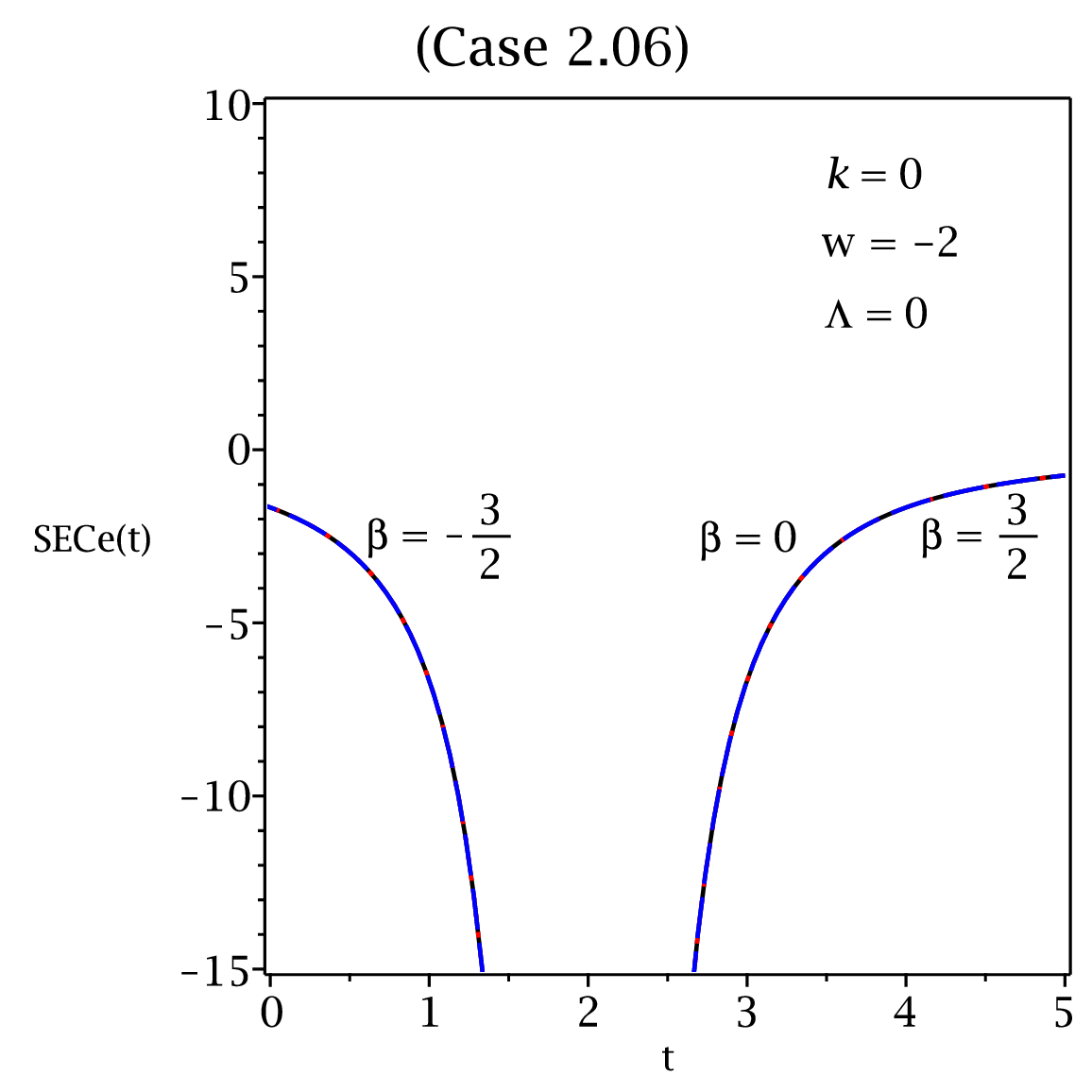}
	\includegraphics[width=3.4cm]{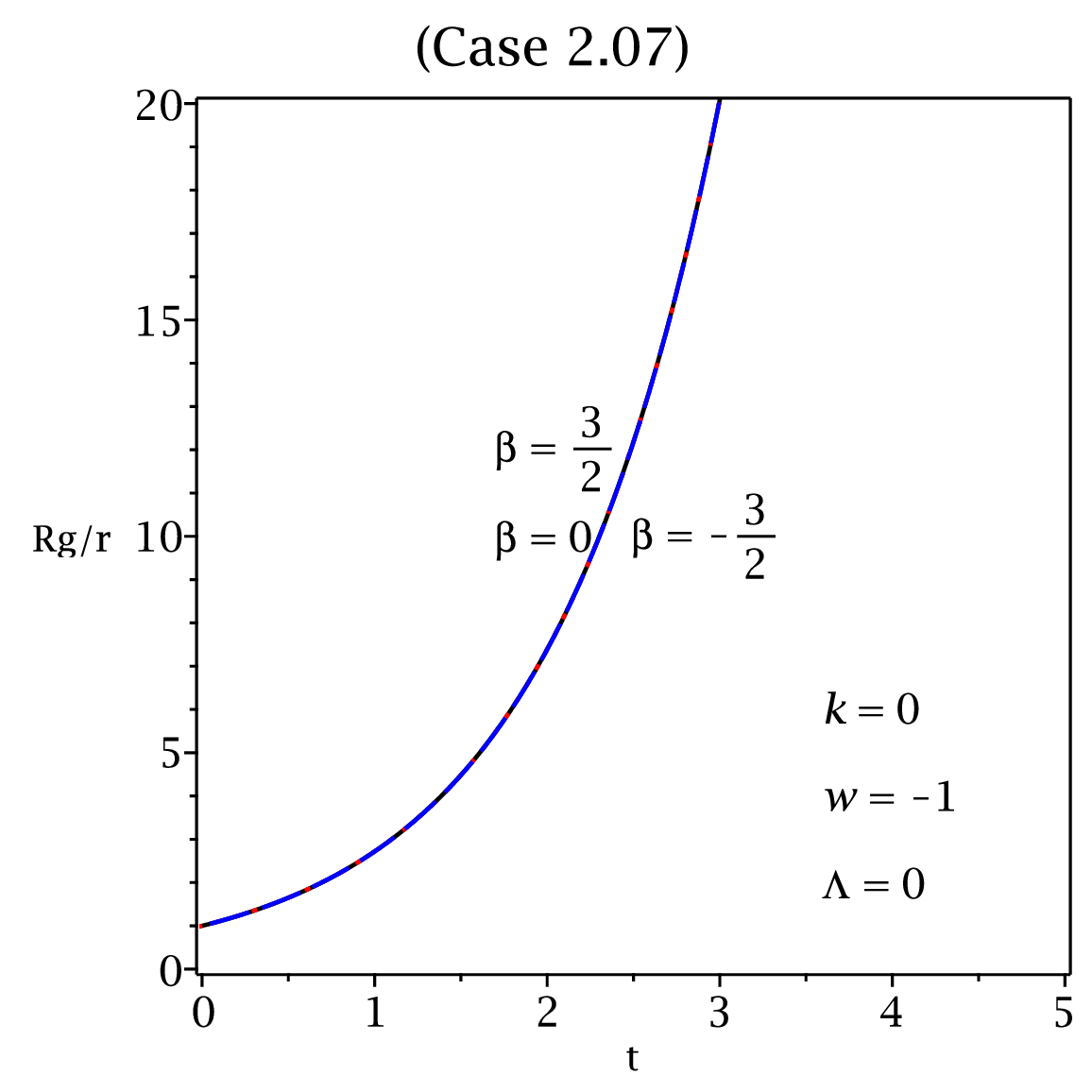}
	\includegraphics[width=3.4cm]{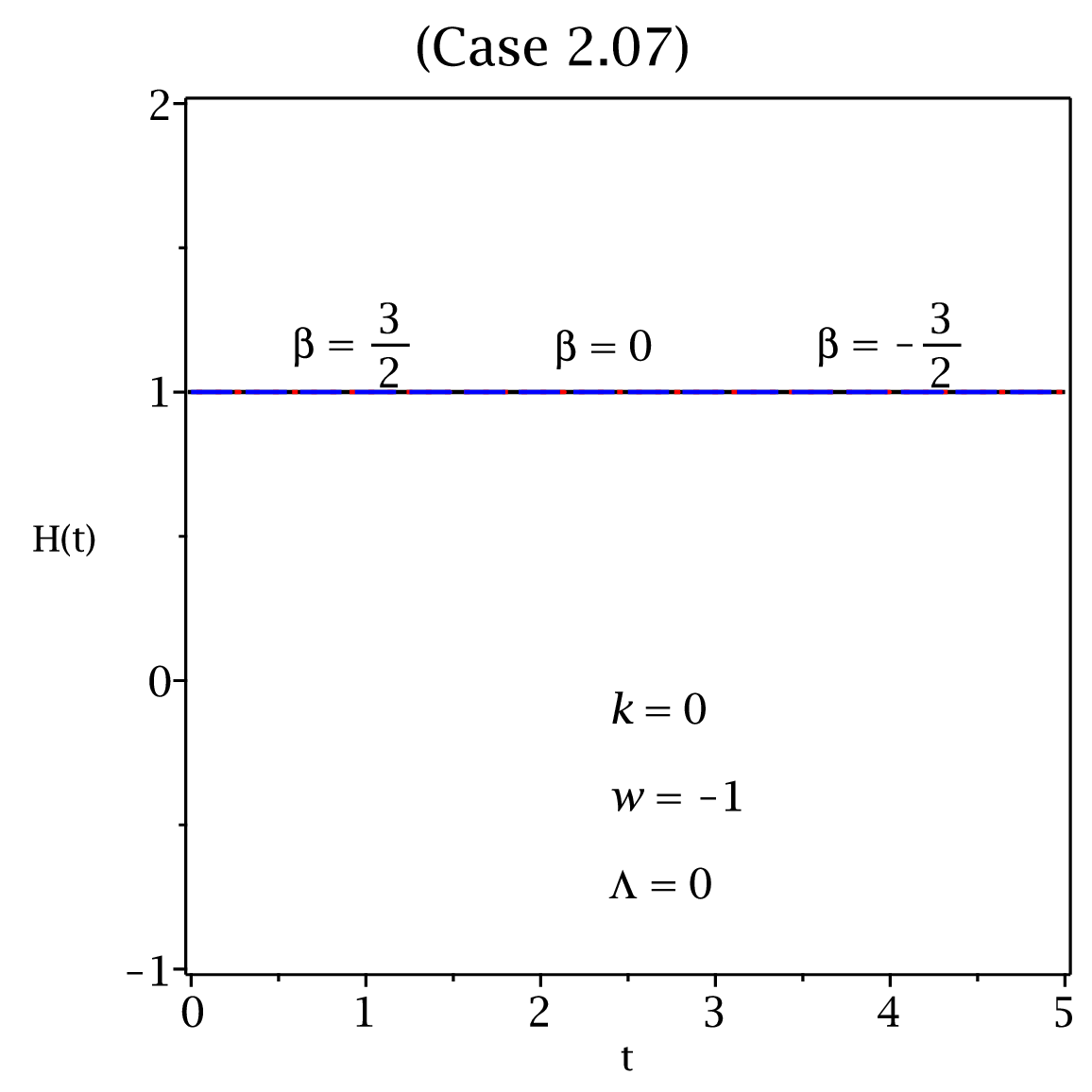}
	\includegraphics[width=3.4cm]{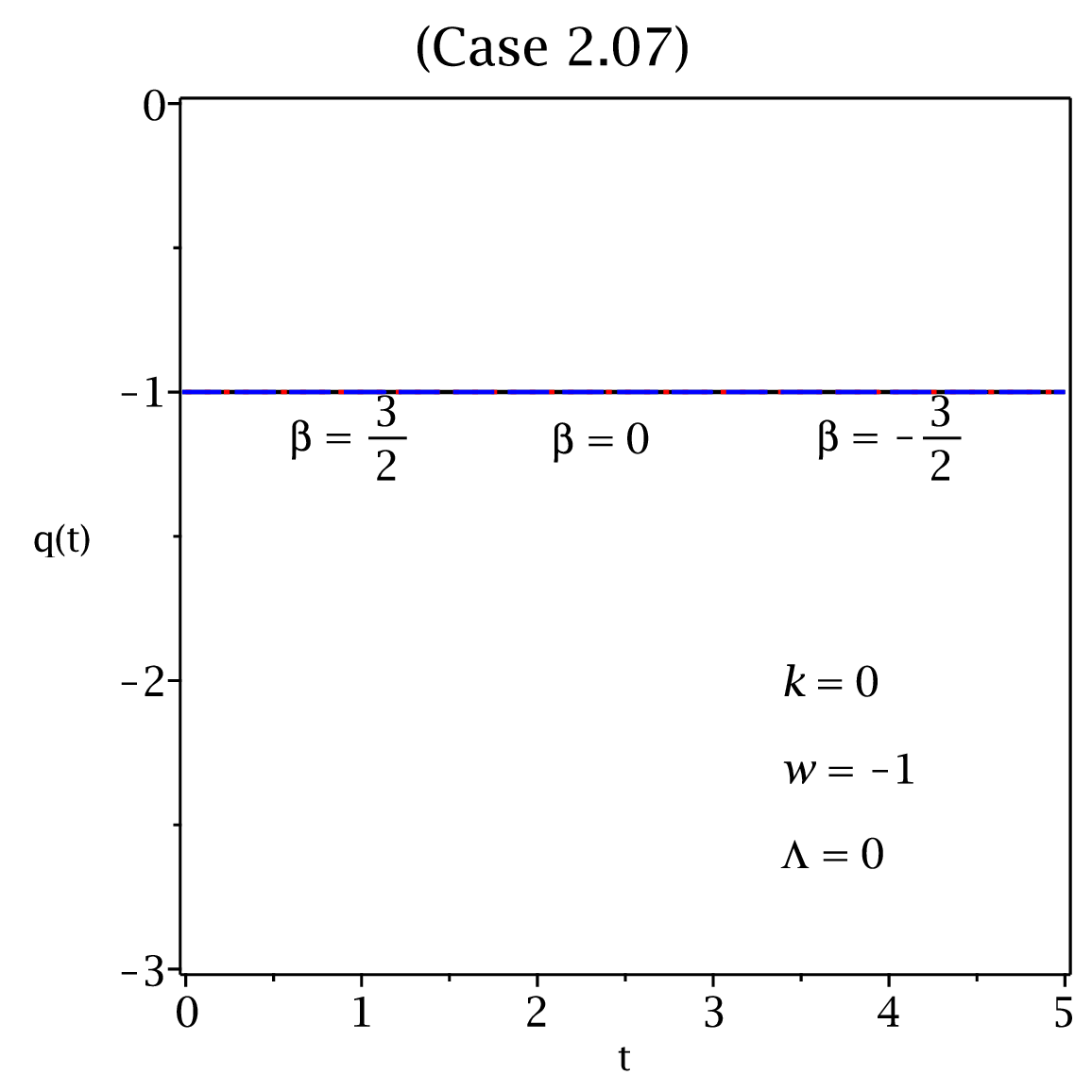}
	\includegraphics[width=3.4cm]{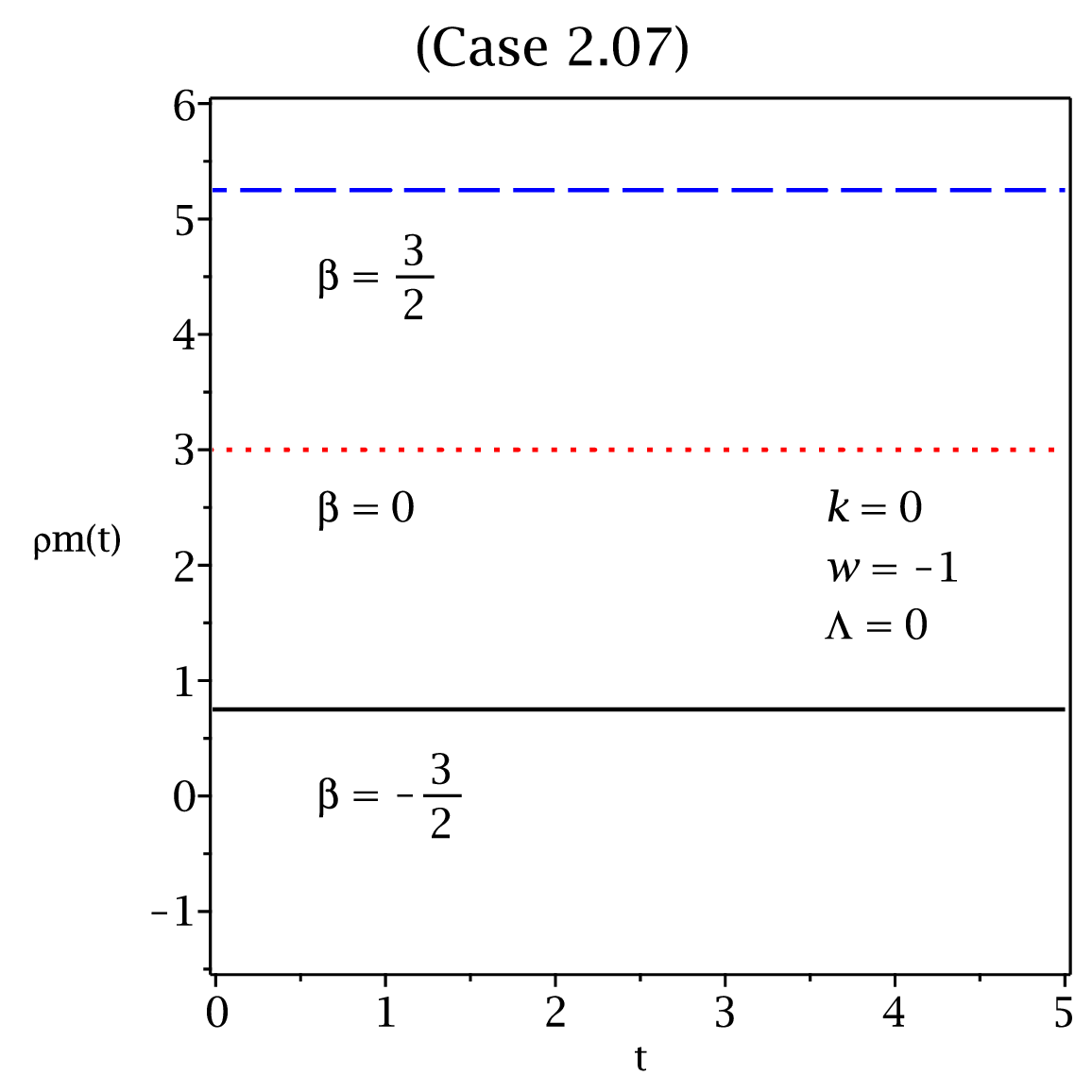}
	\includegraphics[width=3.4cm]{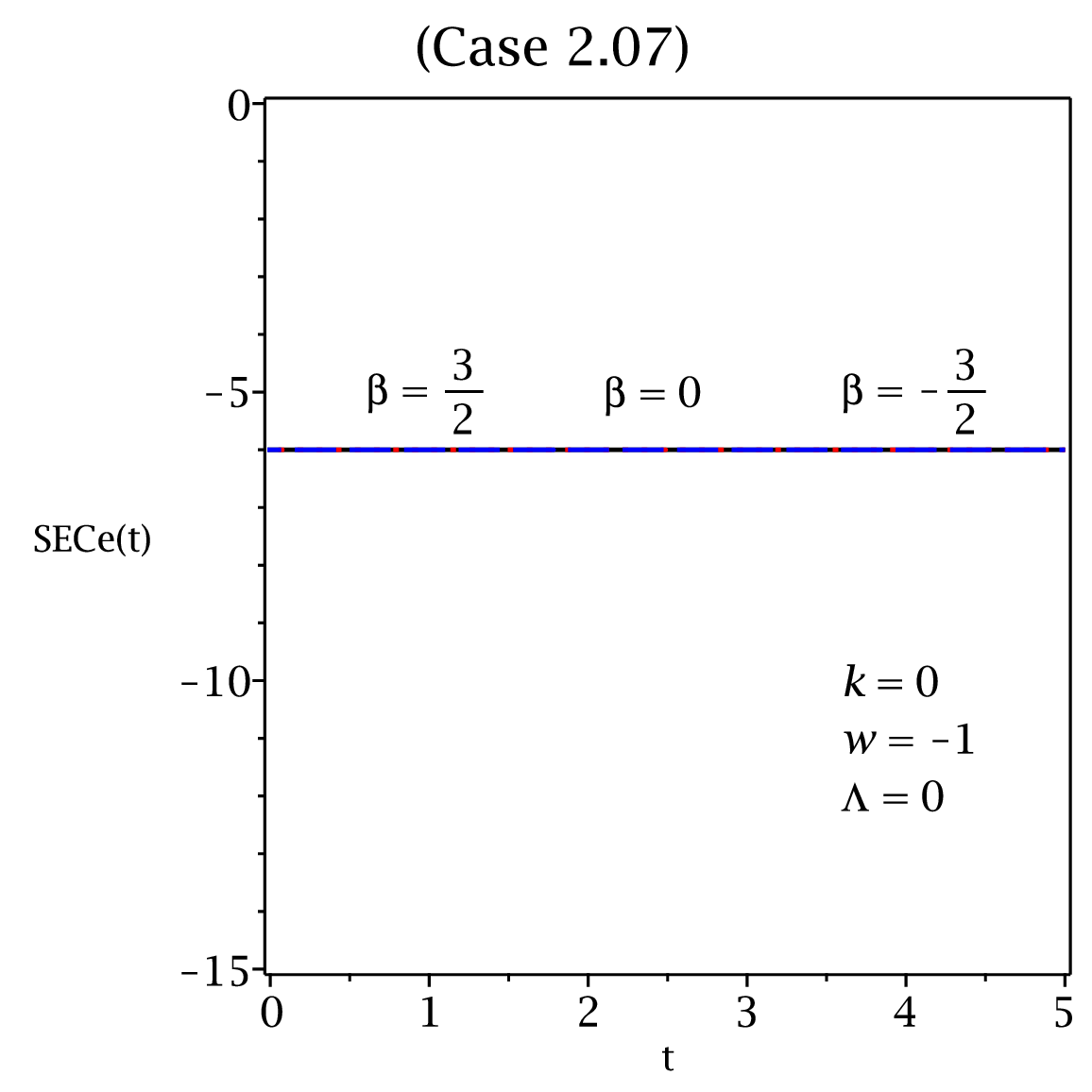}
	\includegraphics[width=3.4cm]{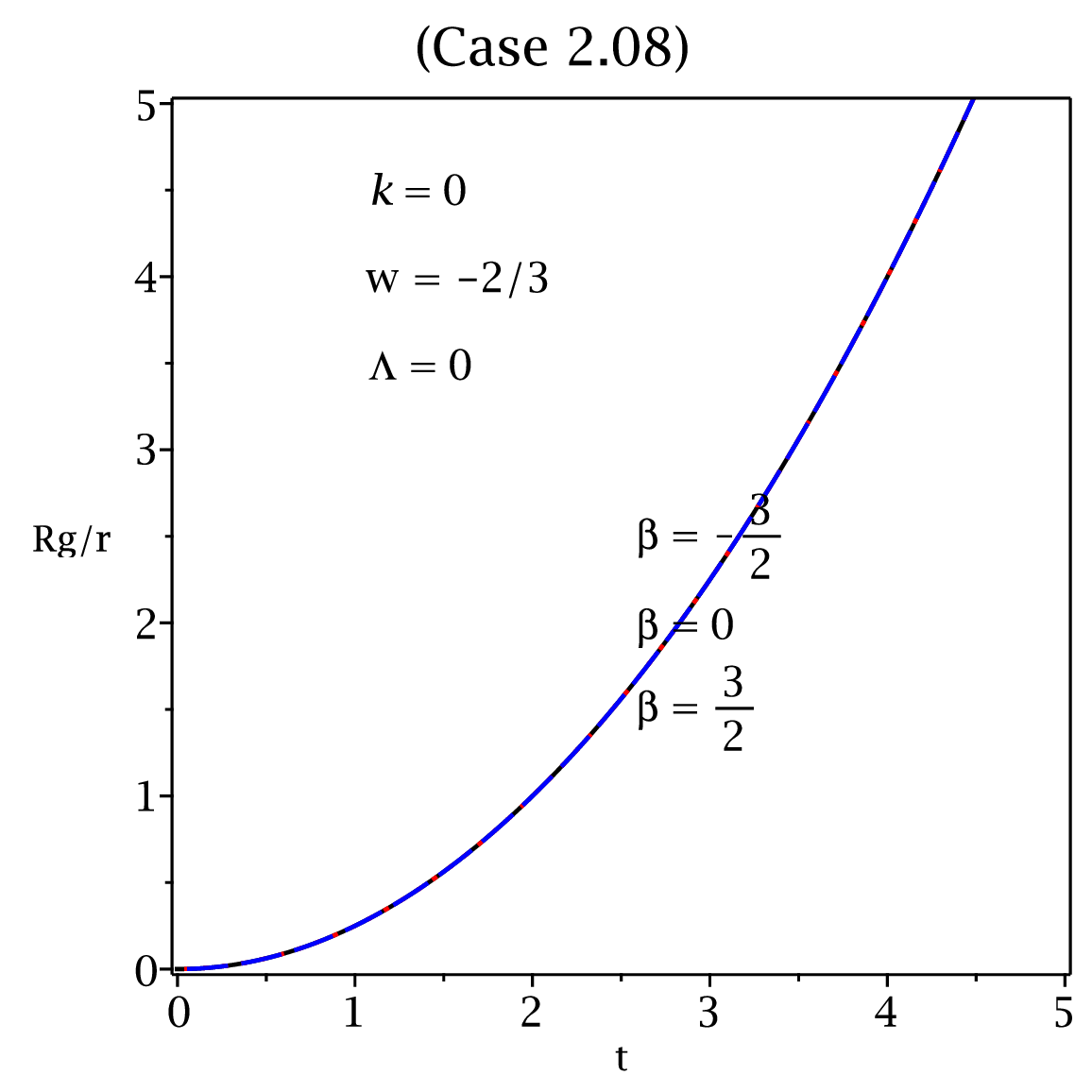}
	\includegraphics[width=3.4cm]{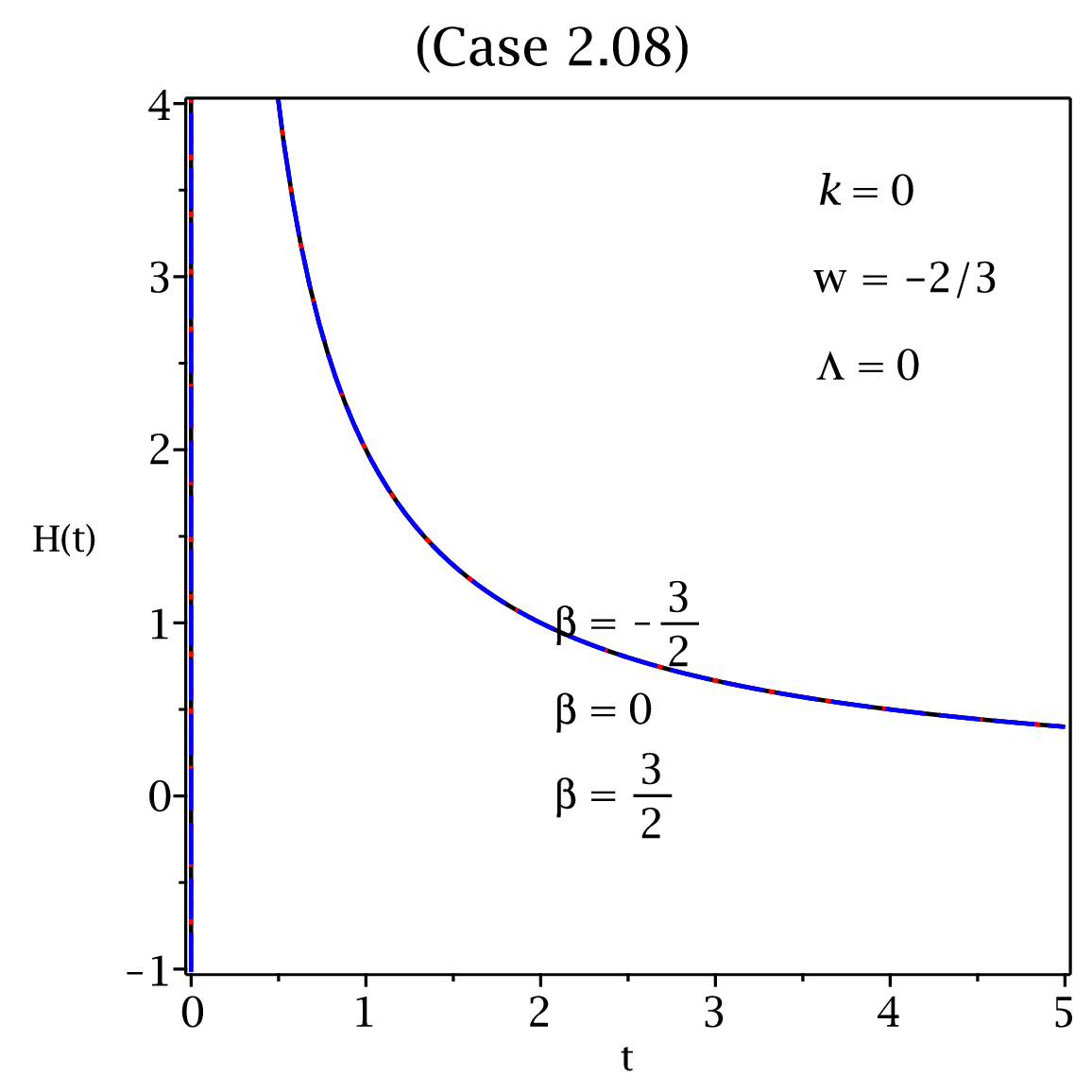}
	\includegraphics[width=3.4cm]{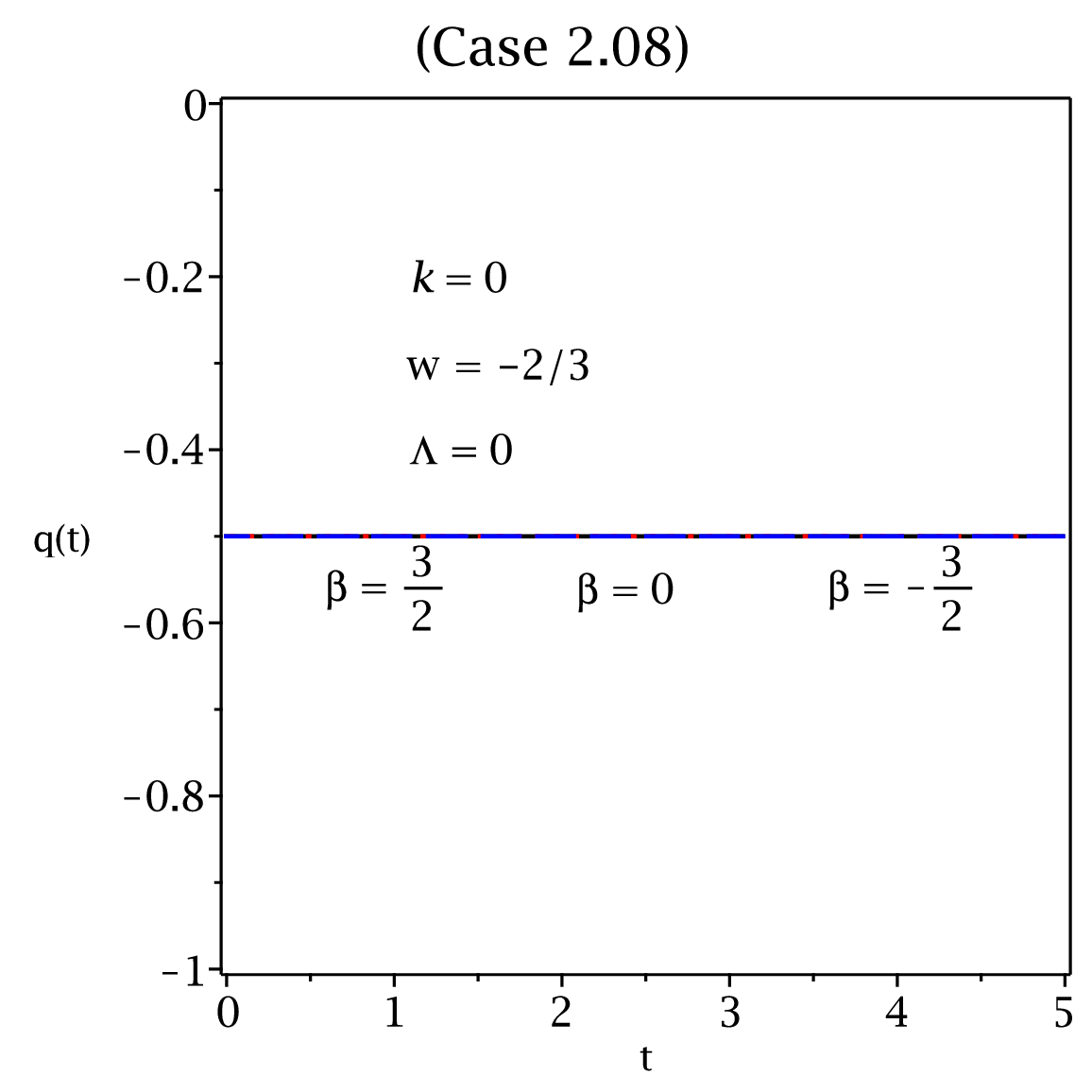}
	\includegraphics[width=3.4cm]{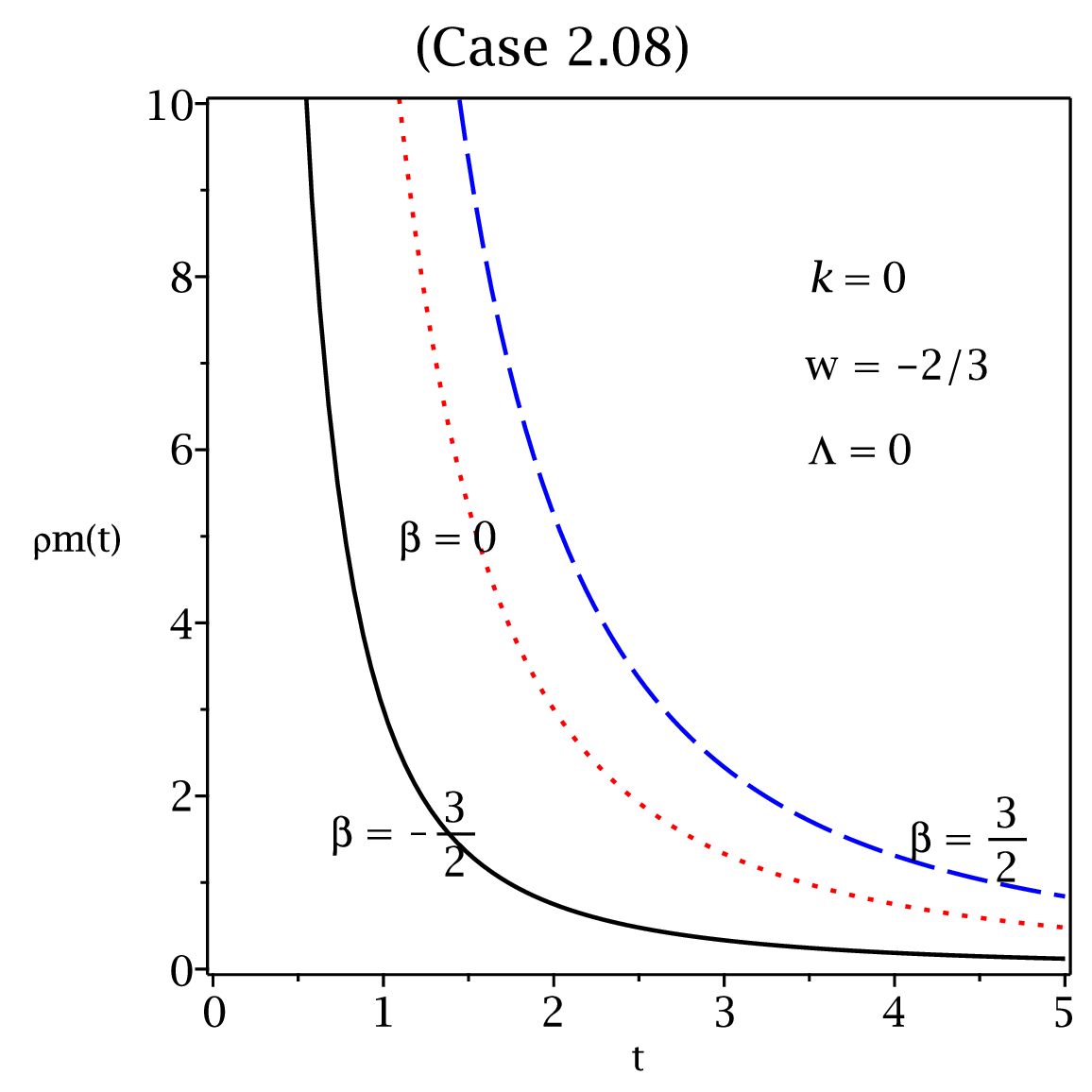}
	\includegraphics[width=3.4cm]{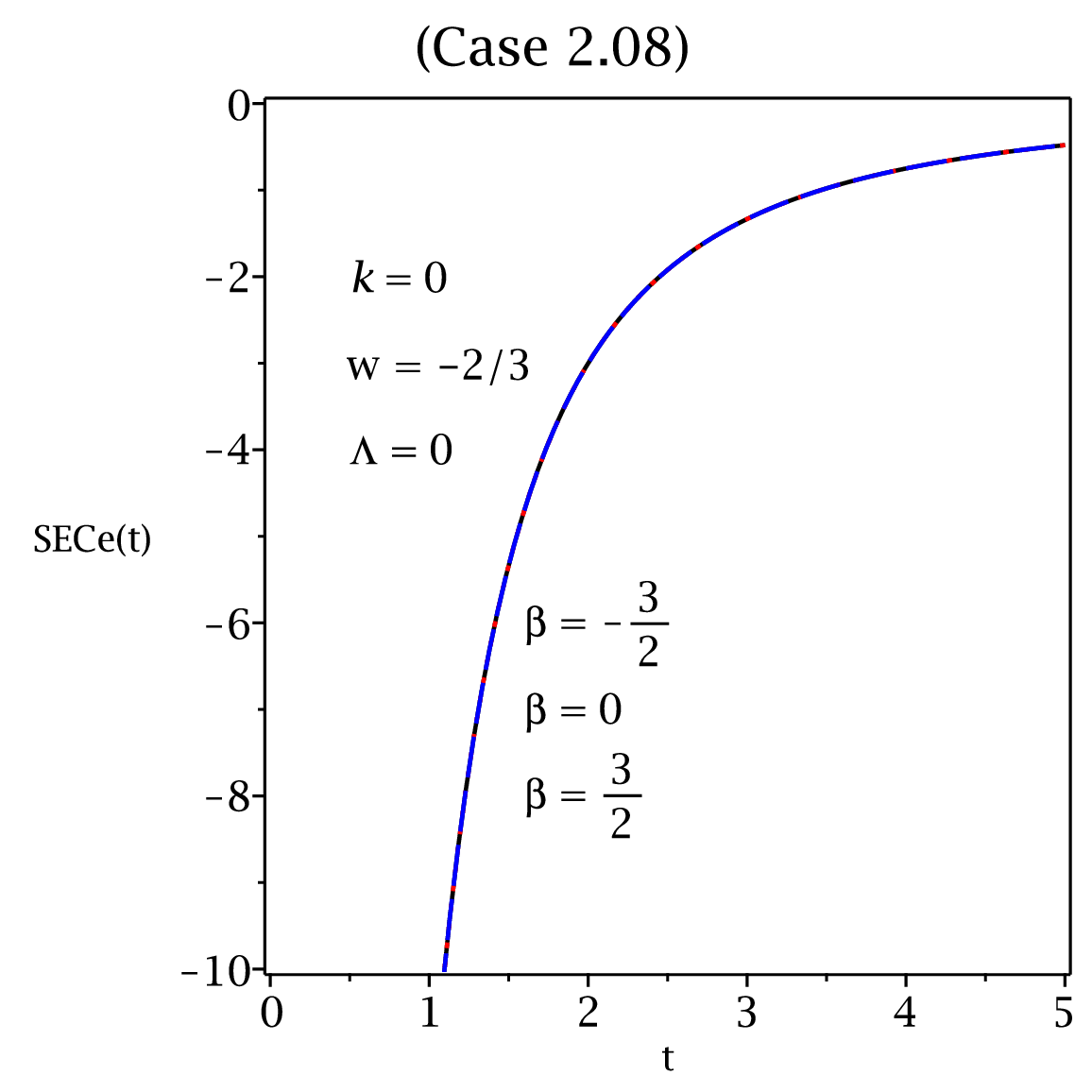}
	\includegraphics[width=3.4cm]{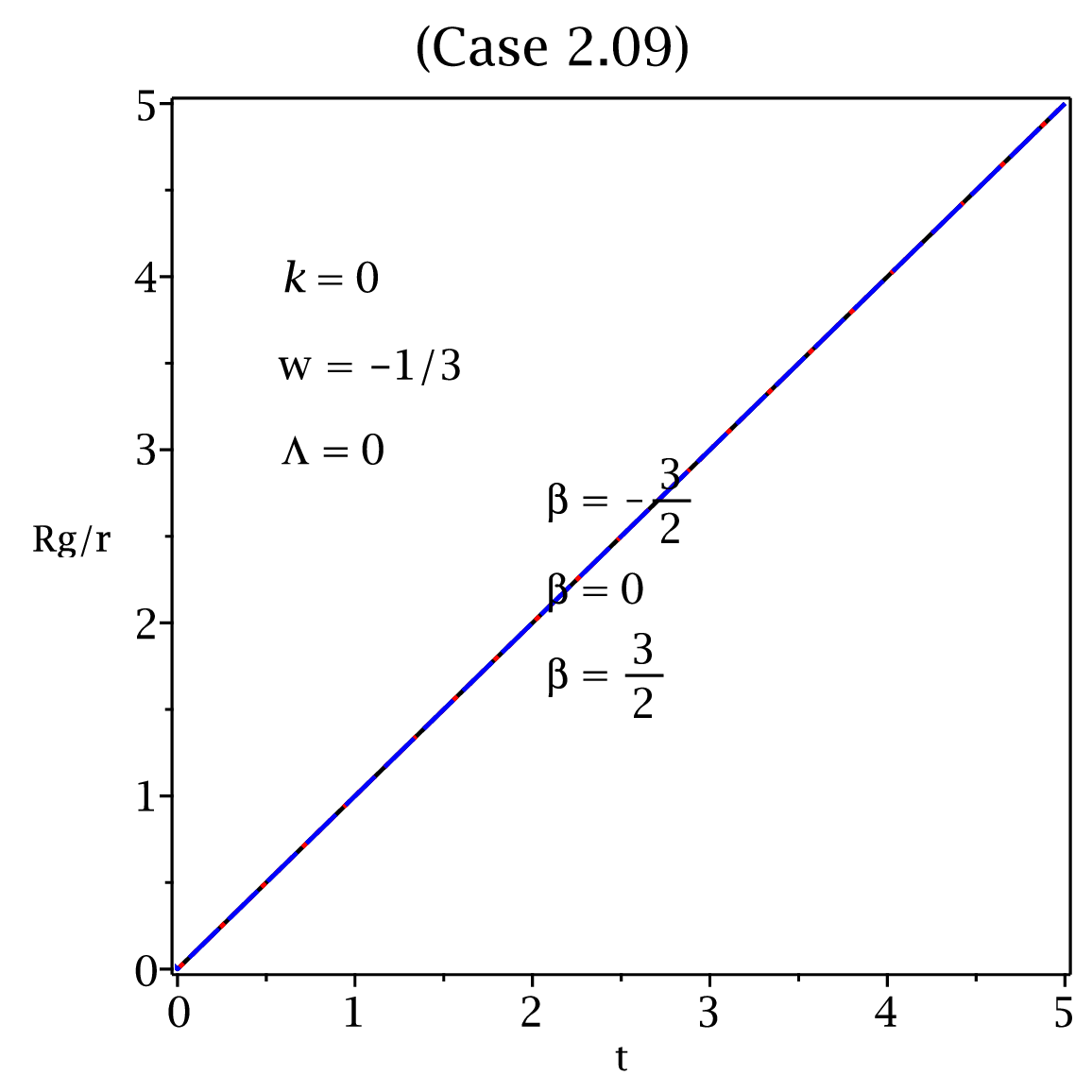}
	\includegraphics[width=3.4cm]{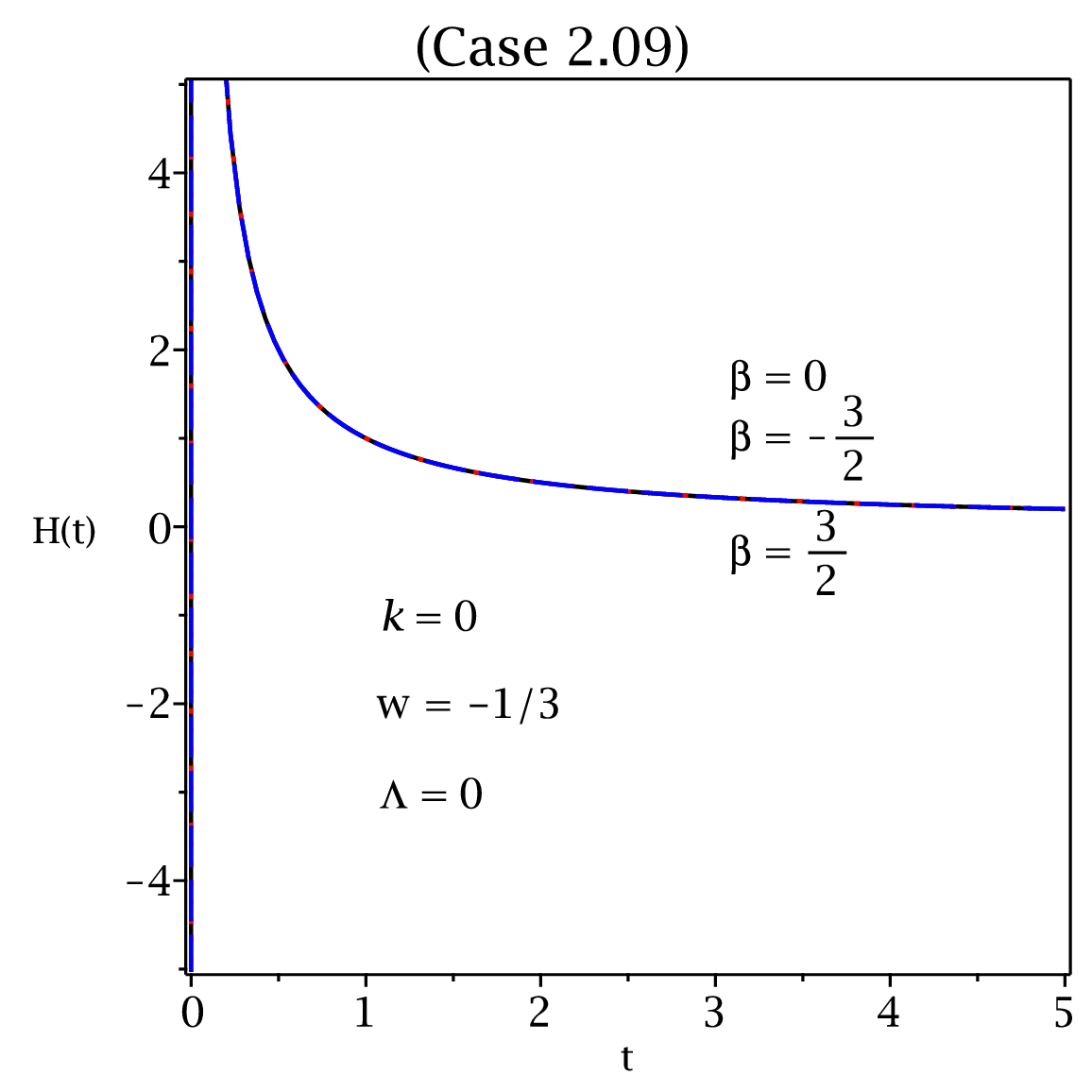}
	\includegraphics[width=3.4cm]{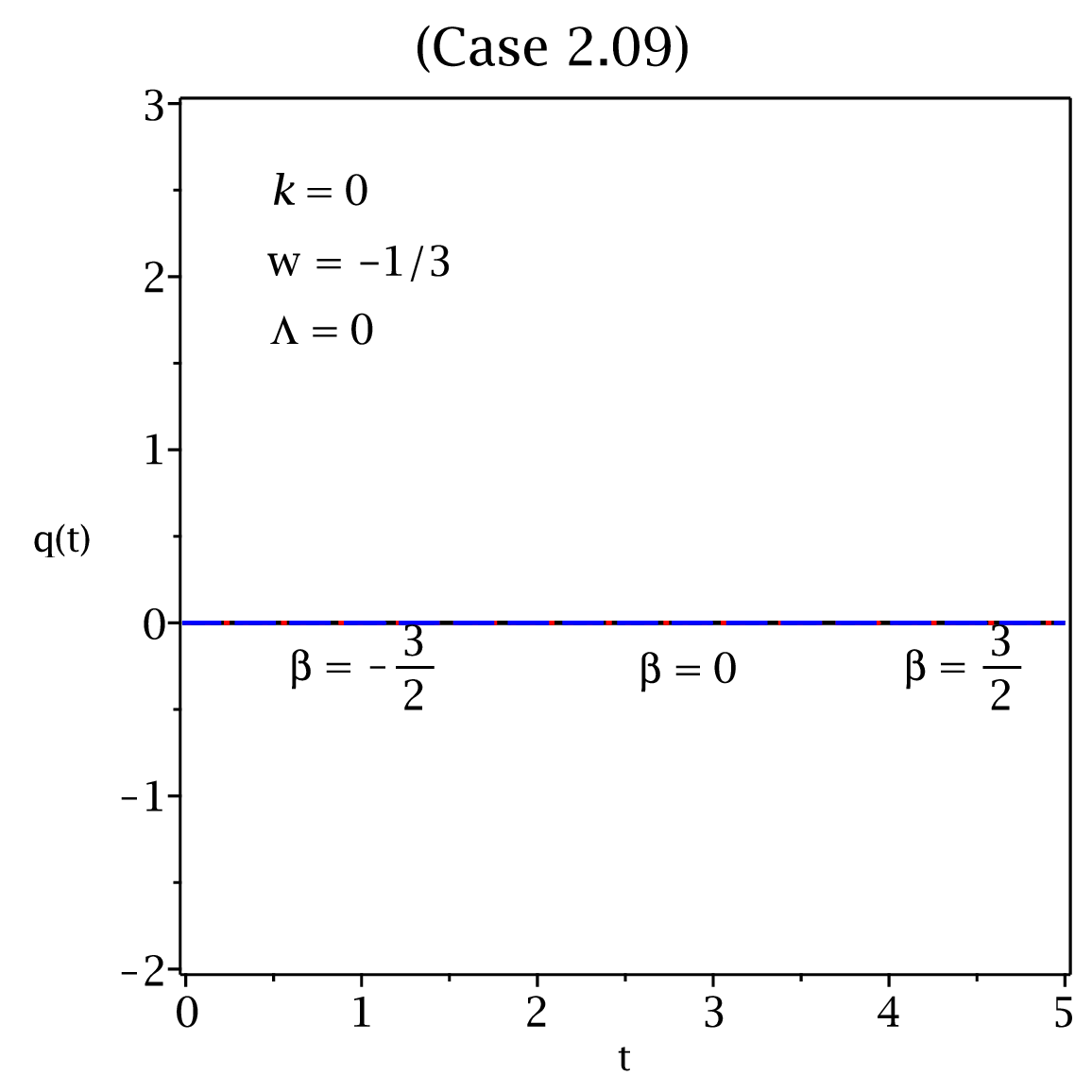}
	\includegraphics[width=3.4cm]{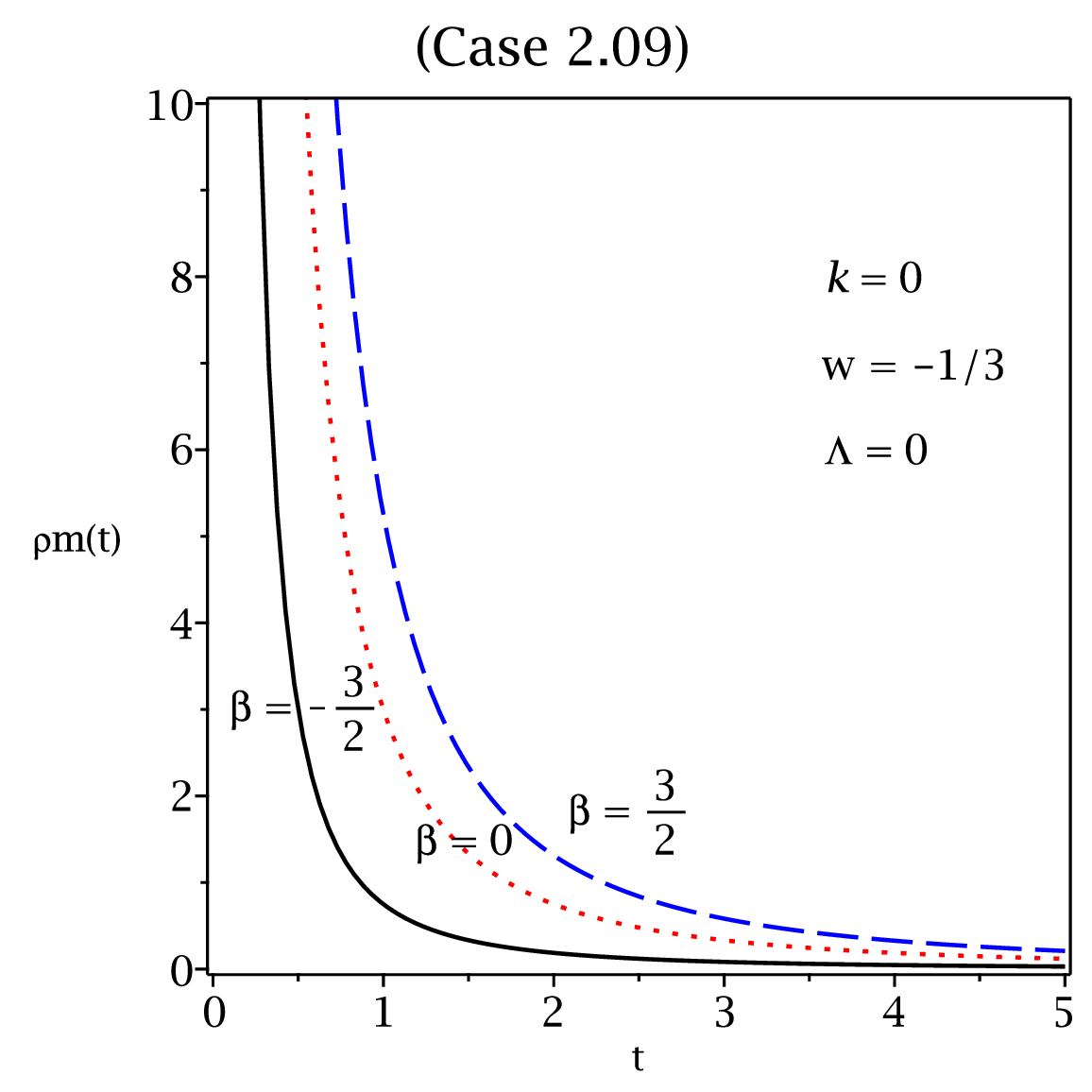}
	\includegraphics[width=3.4cm]{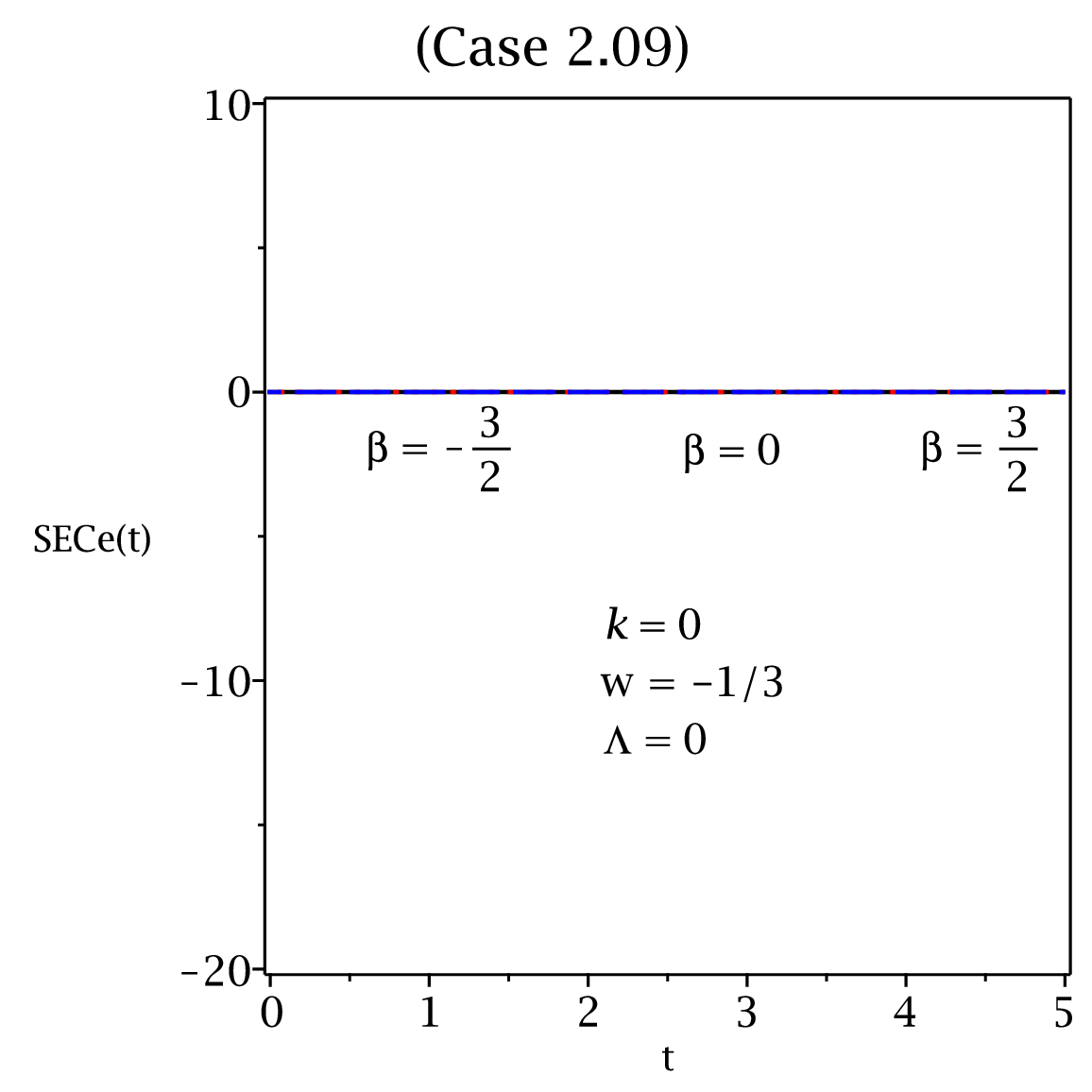}
	\includegraphics[width=3.4cm]{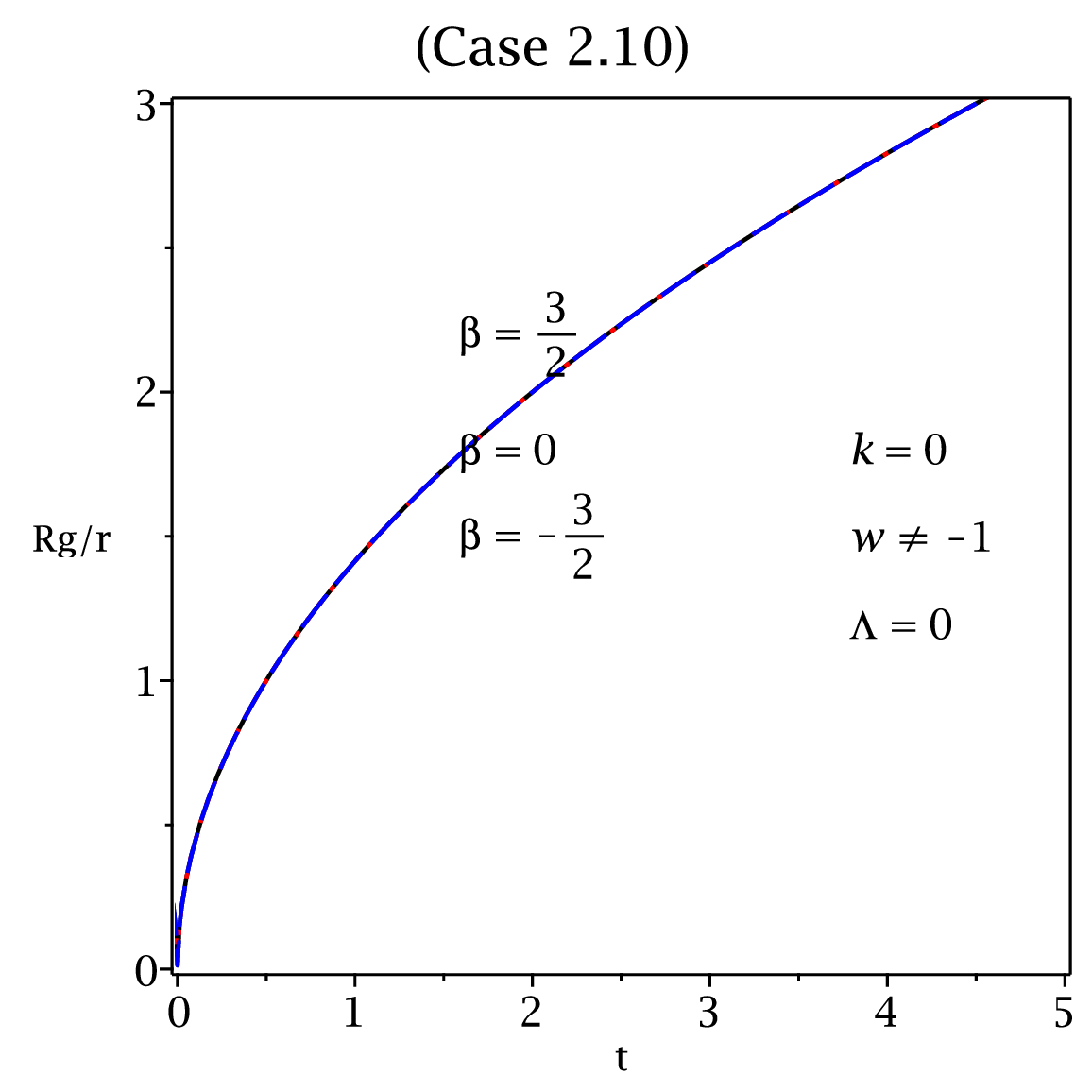}
	\includegraphics[width=3.4cm]{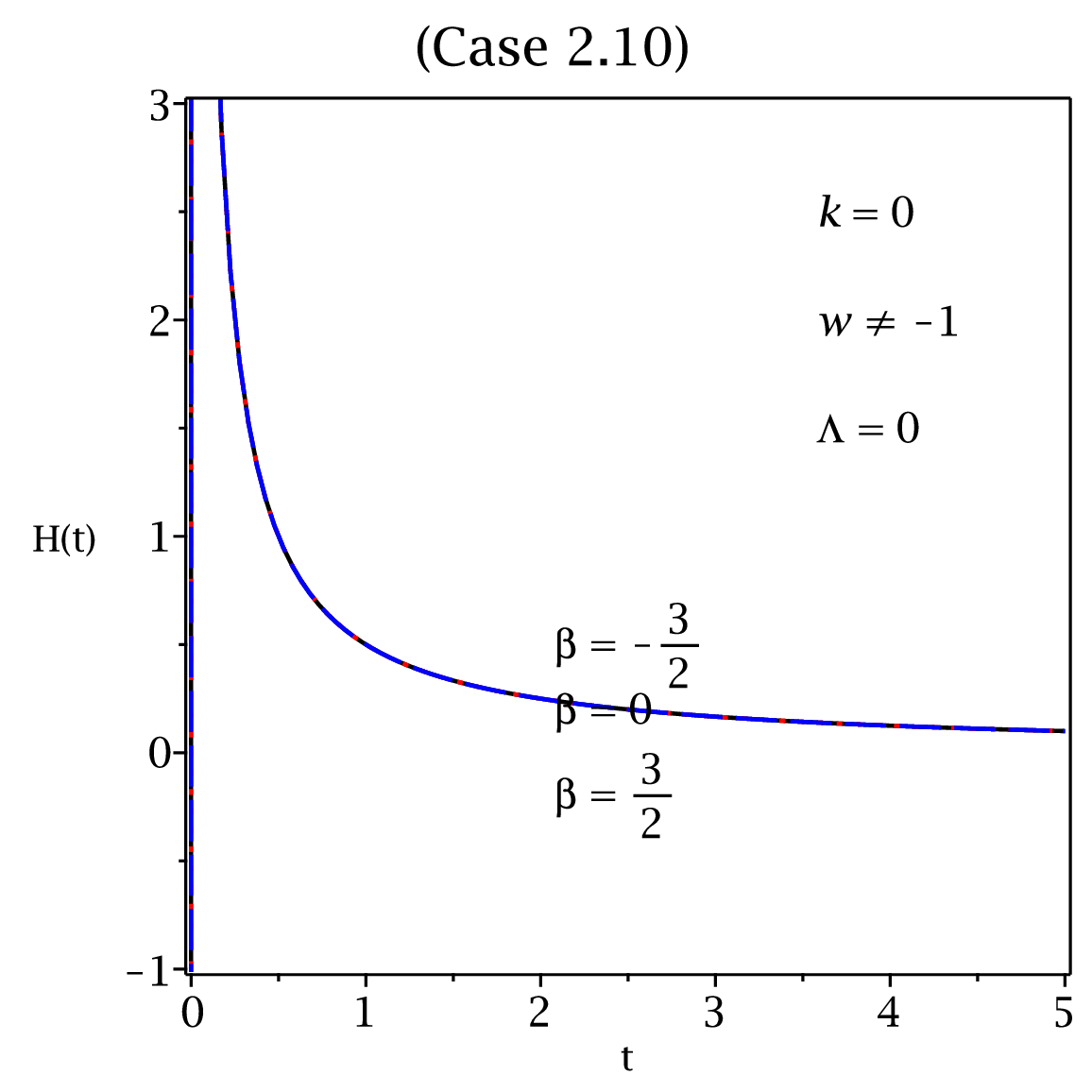}
	\includegraphics[width=3.4cm]{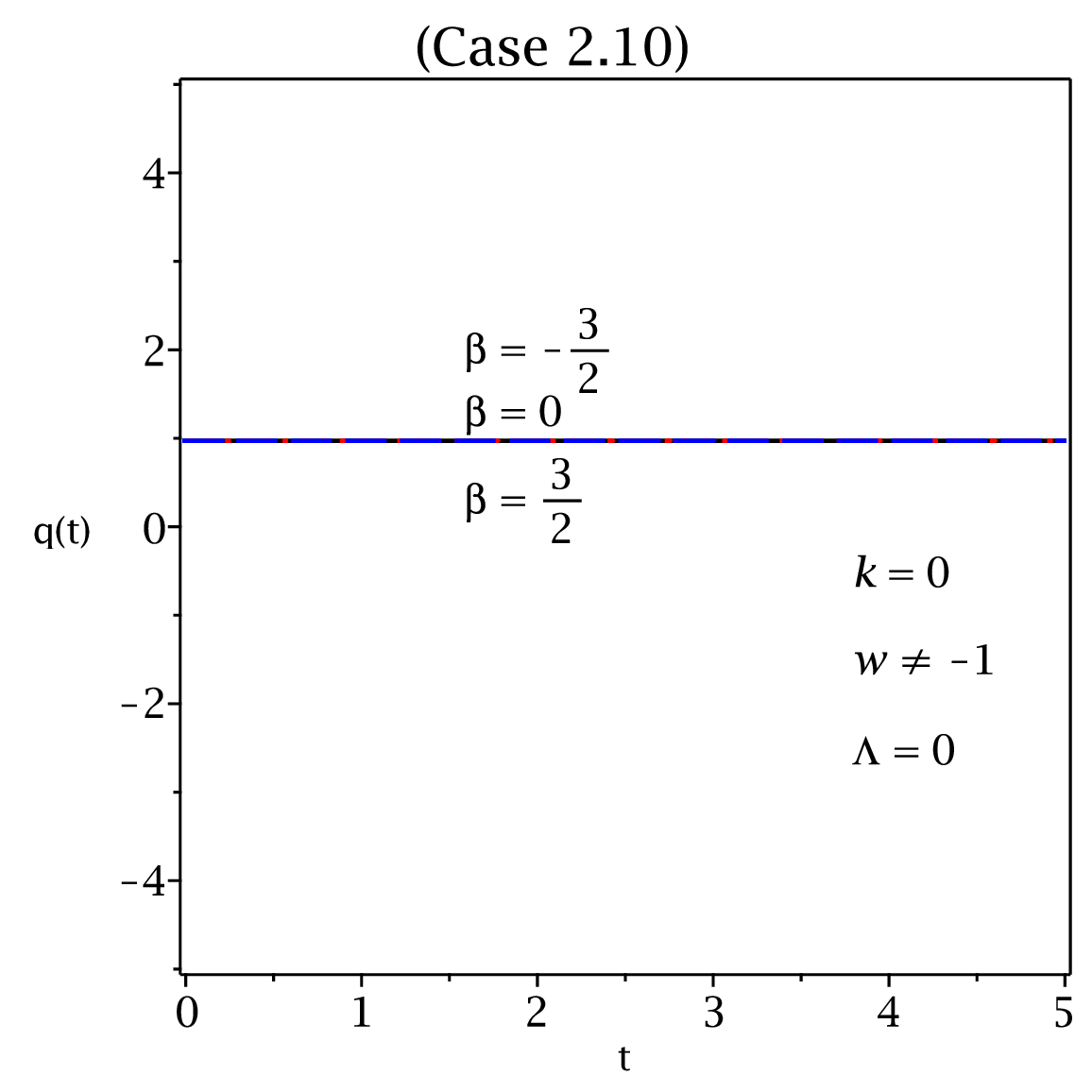}
	\includegraphics[width=3.4cm]{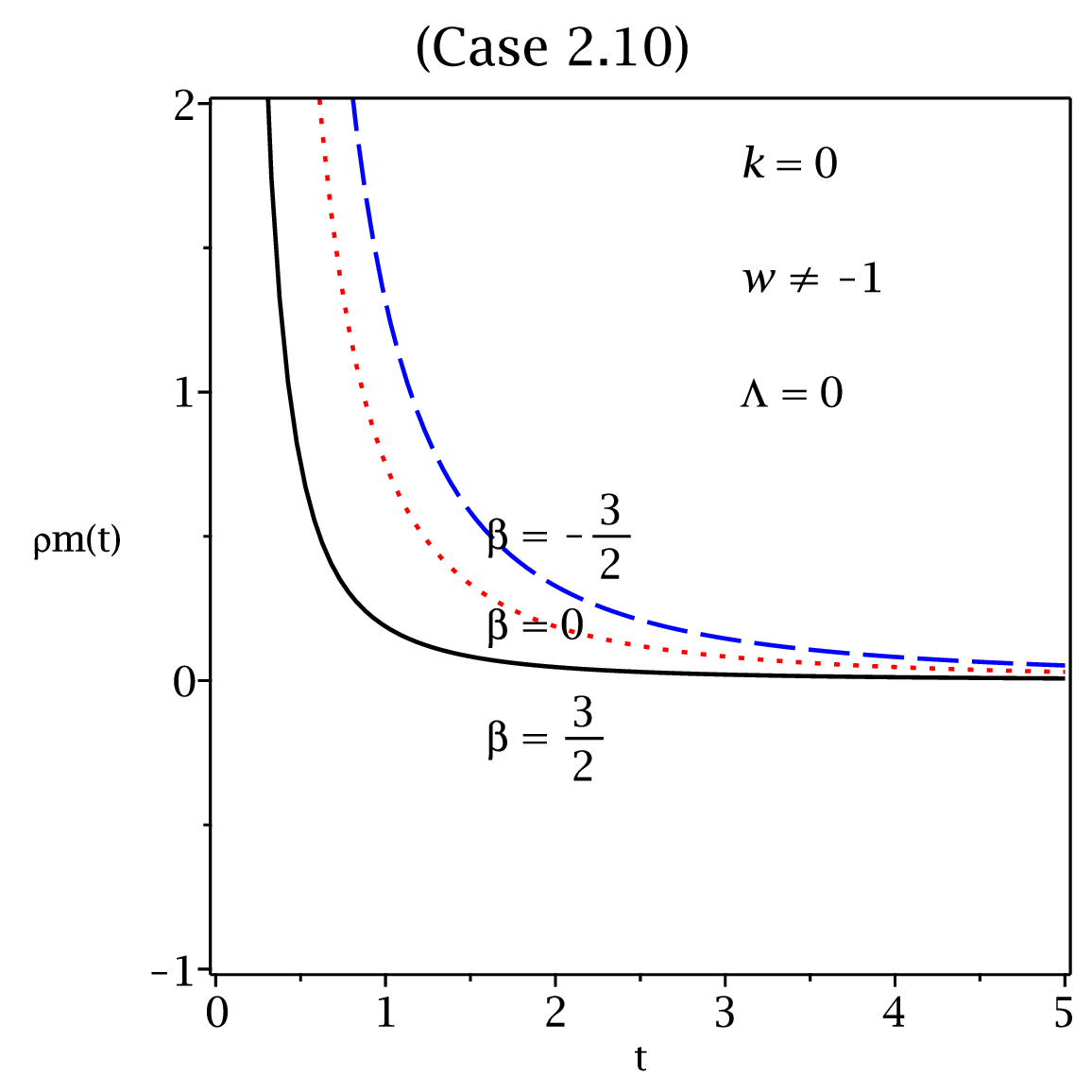}
	\includegraphics[width=3.4cm]{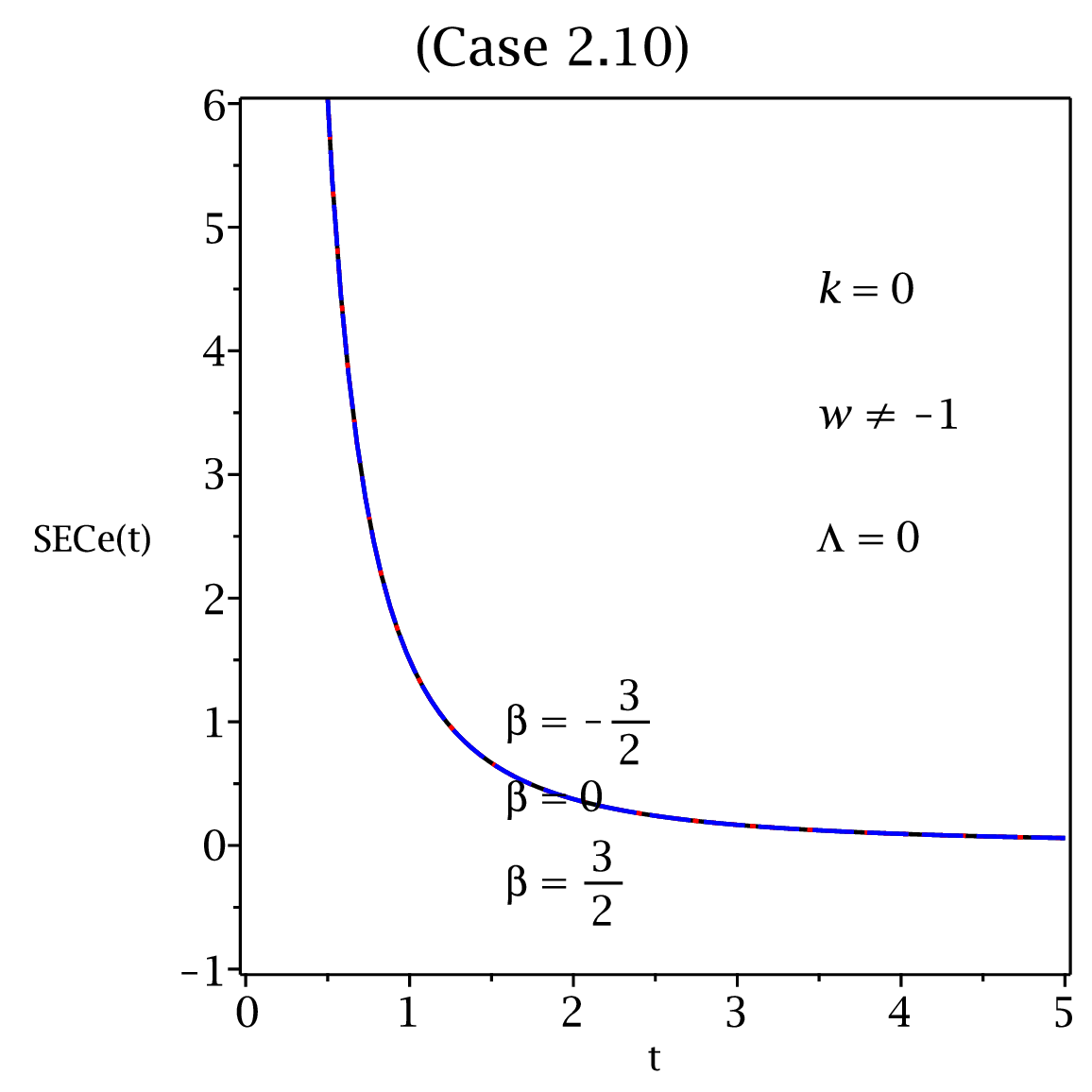}
	\includegraphics[width=3.4cm]{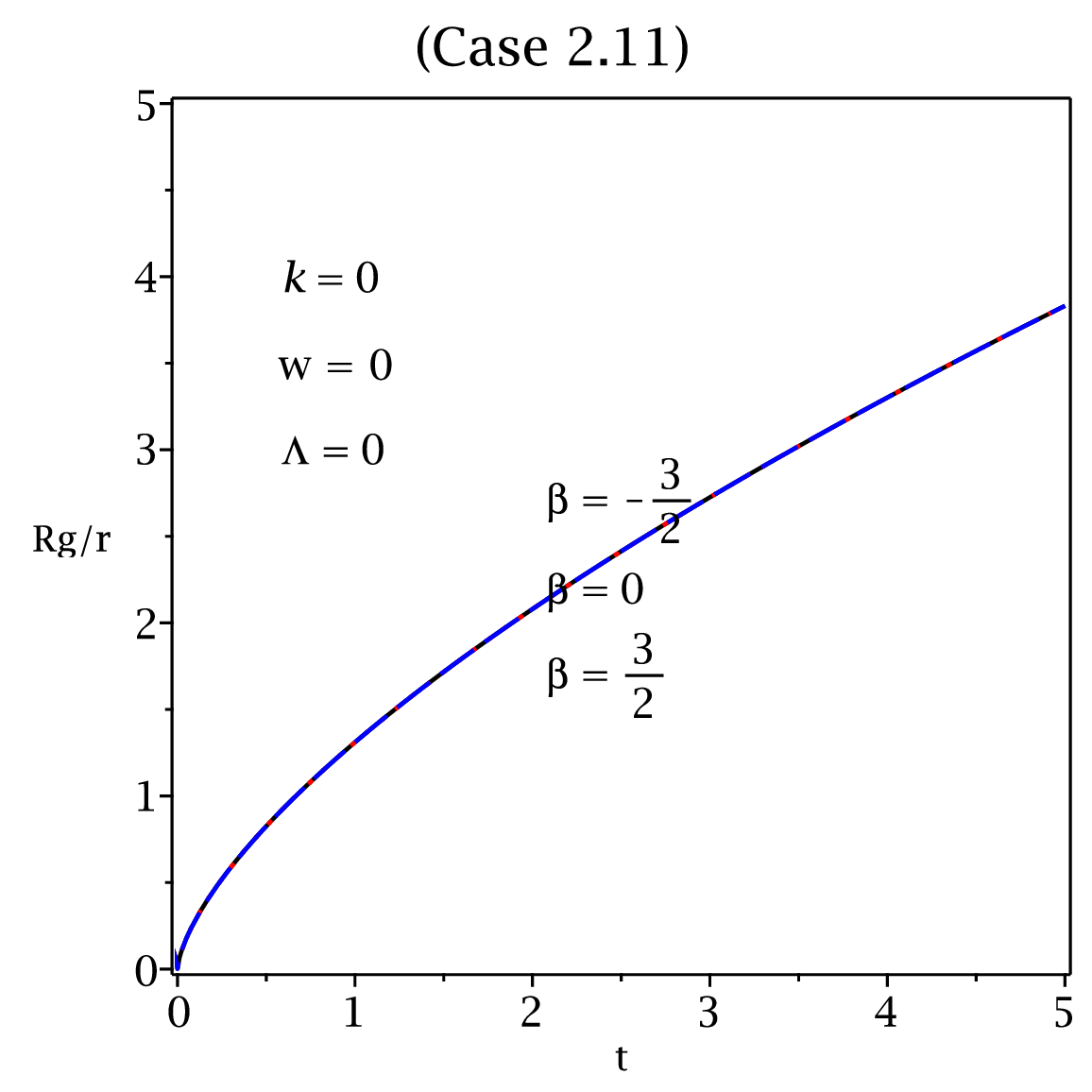}
	\includegraphics[width=3.4cm]{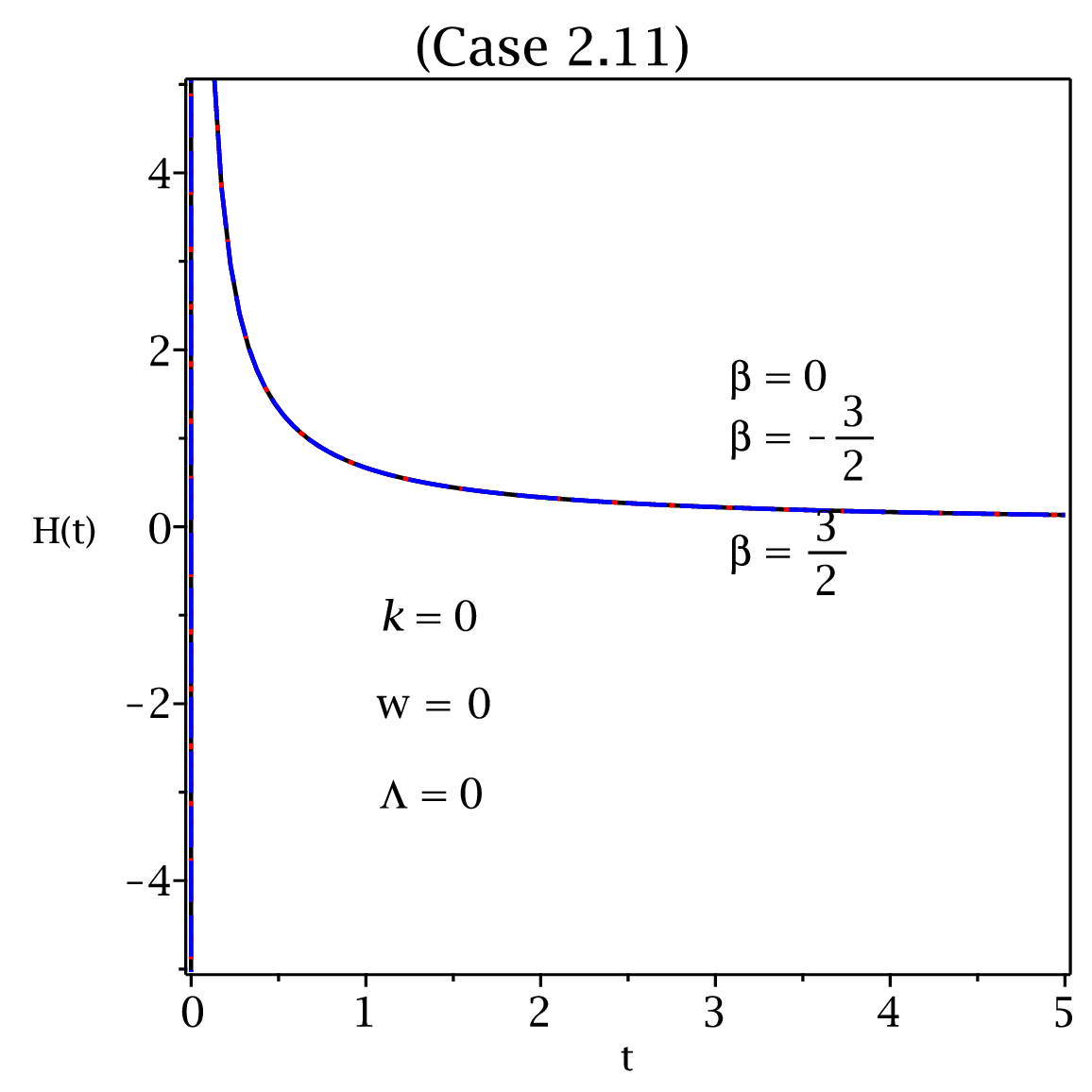}
	\includegraphics[width=3.4cm]{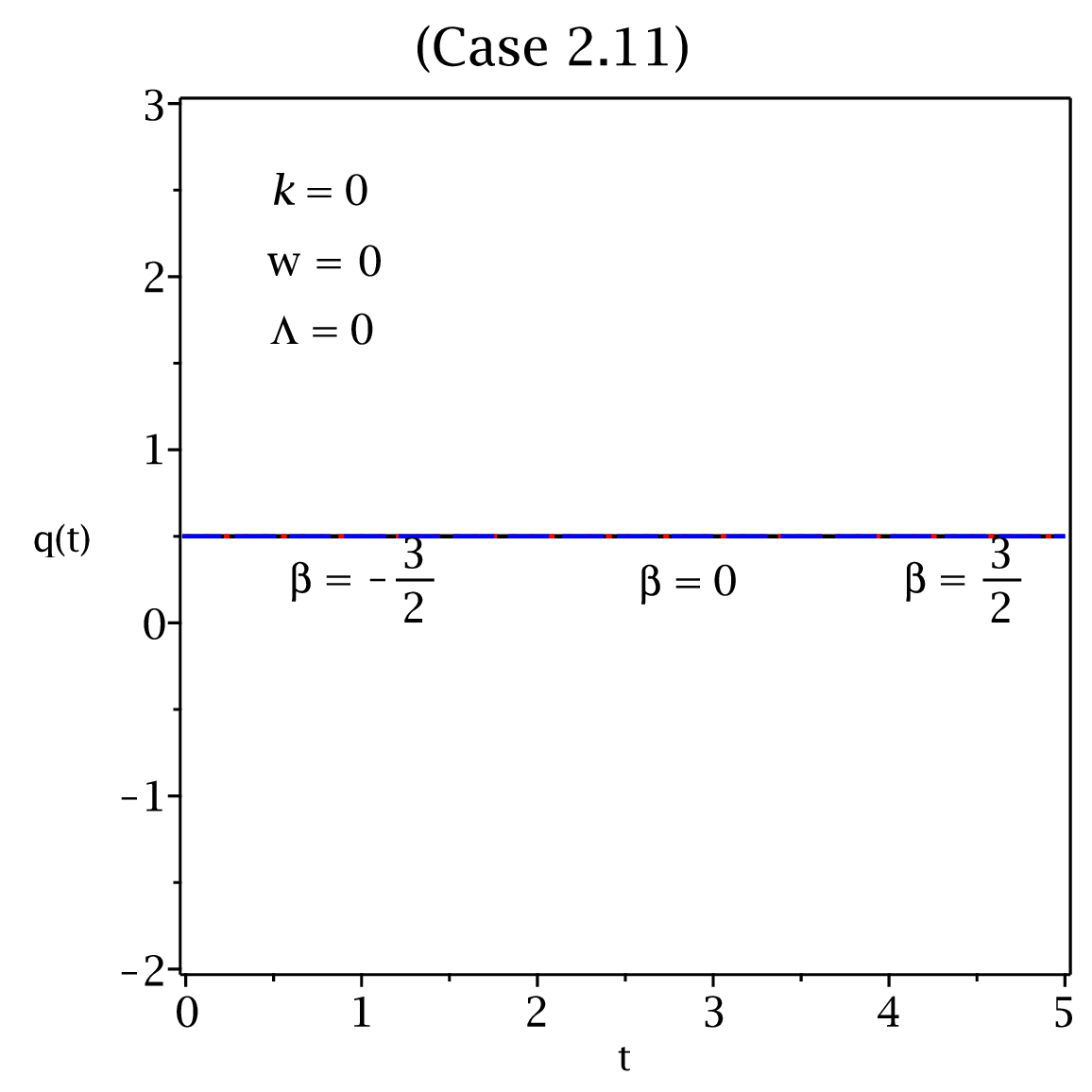}
	\includegraphics[width=3.4cm]{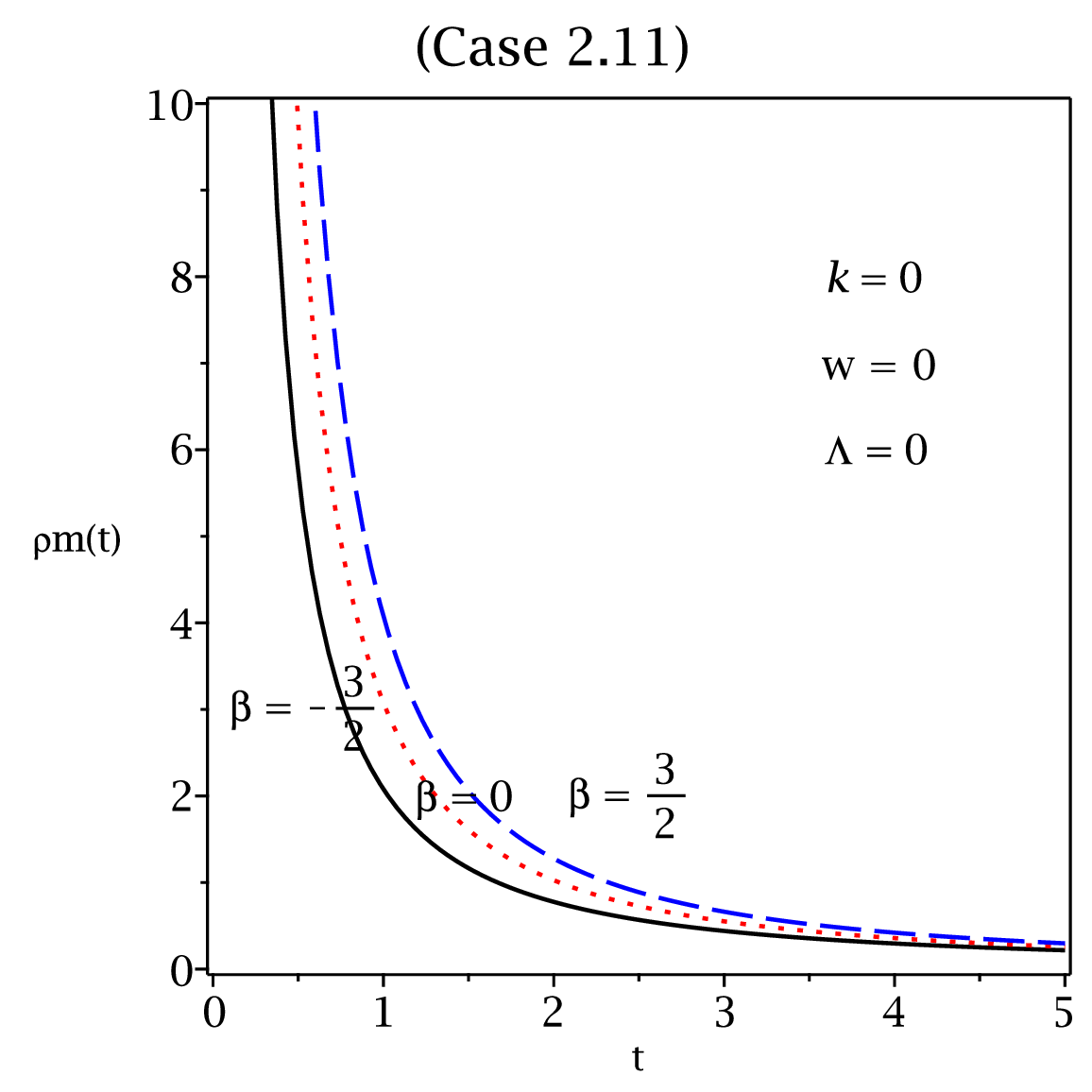}
	\includegraphics[width=3.4cm]{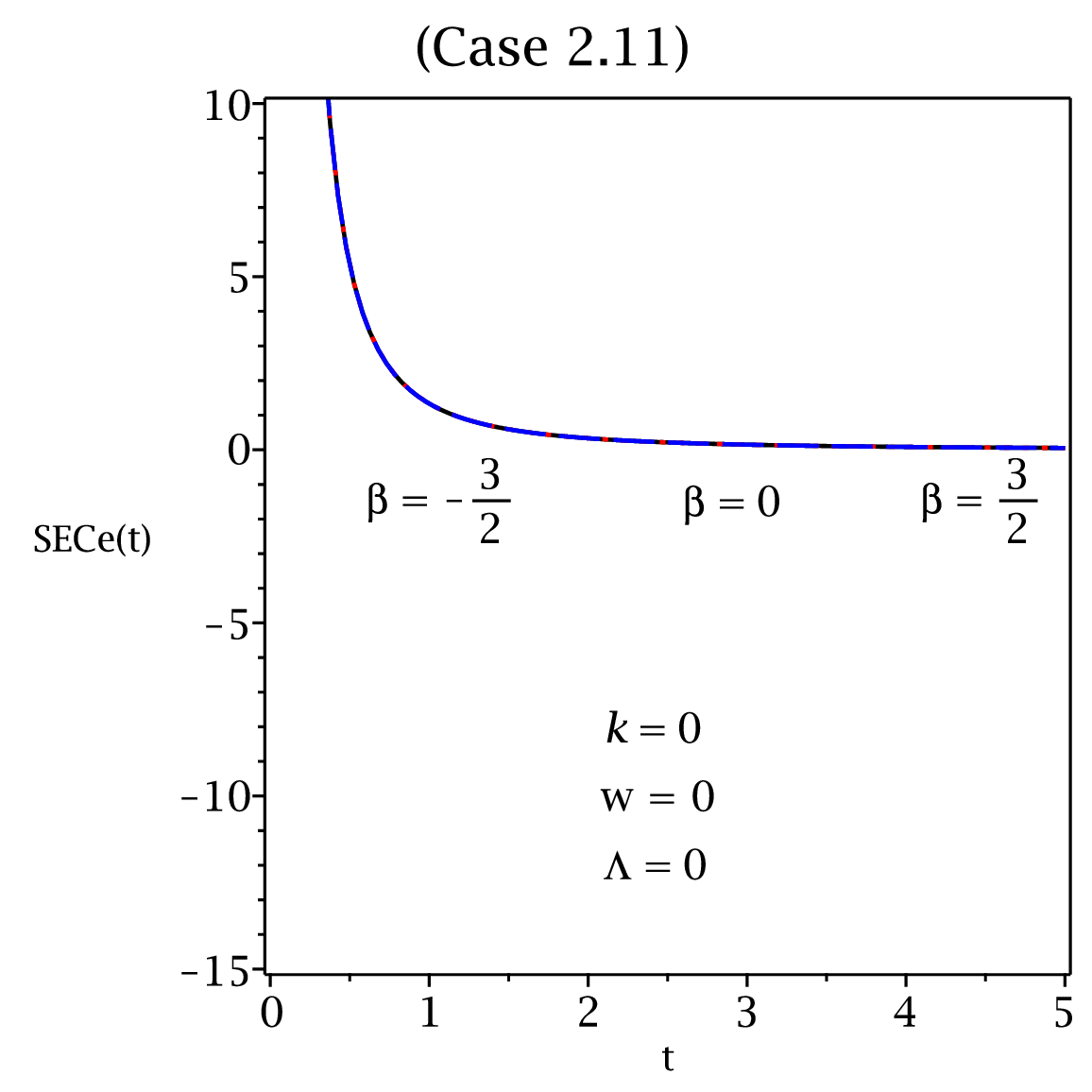}
	\caption{These figures are for $\Lambda=0$ and $k=0$.
		These figures represent the quantities $R_g$ (geometrical radius), 
		$H(t)$ (Hubble parameter) and $q(t)$ (deceleration parameter) $\rho_m(t)$ 
		(energy density of the aether fluid) and $SEC_{e} \equiv SEC_{\rm eff}$ 
		(strong energy condition for the effective fluid) for the different
		values of $\beta=-3/2$ (black solid line), $\beta=0$ (red dotted line), 
		$\beta=3/2$ (blue dashed line). Assuming that $8 \pi G=1$ and
		$R_g(t=0)=0$. Assuming also that $C_1=1$, $C_2=0$ (Case 2.10); 
		$C_1=1$, $C_2=1$ (Case 2.07); $C_1=1$, $C_2=0$ (Cases 2.08, 2.09 and 2.11);
		$C_1=1$, $C_2=-2$ (Case 2.06).}
	\label{Figure-206-211}
\end{minipage}	
\end{figure}

\begin{figure}[!htp]
\begin{minipage}{175 mm}
	\centering	
	\includegraphics[width=3.4cm]{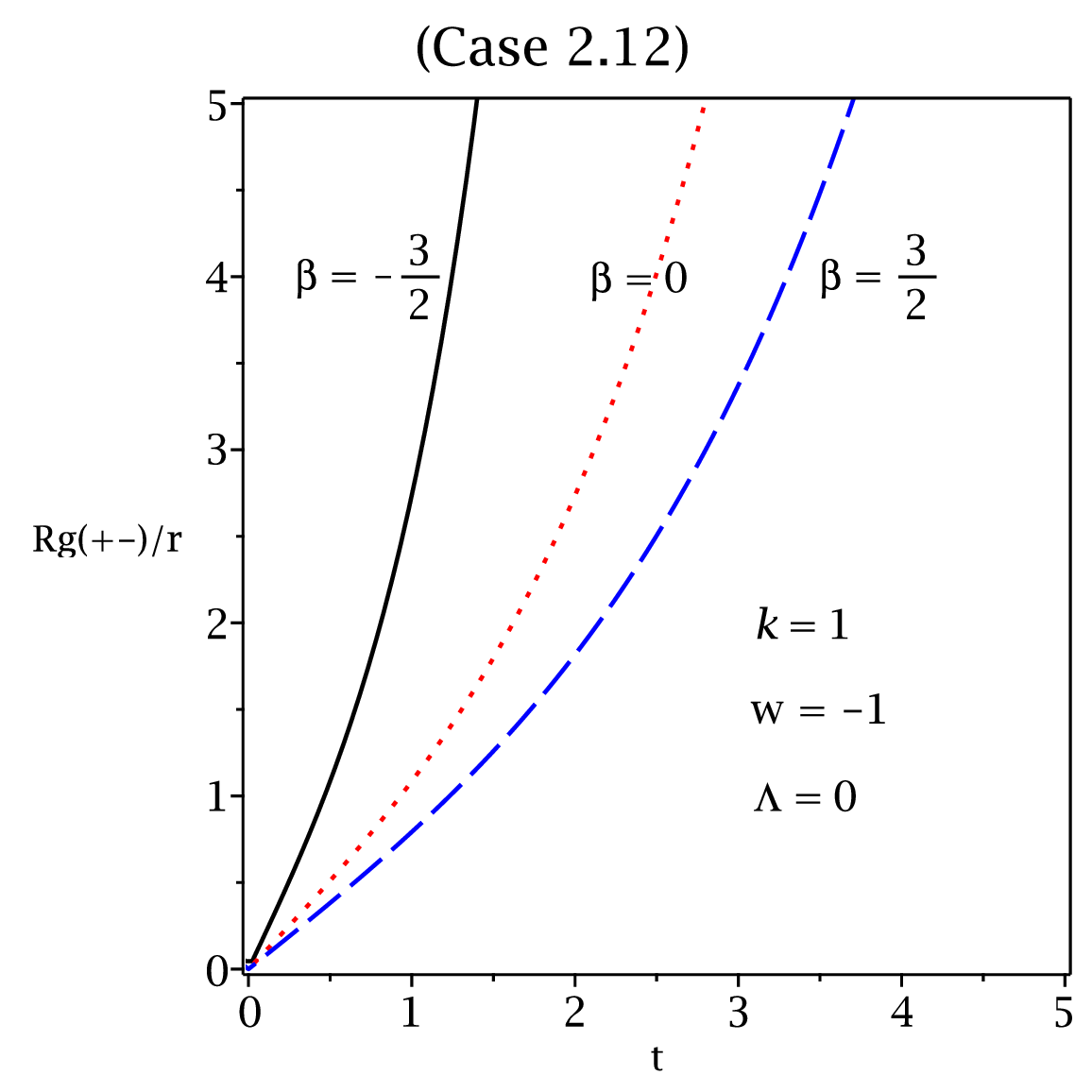}
	\includegraphics[width=3.4cm]{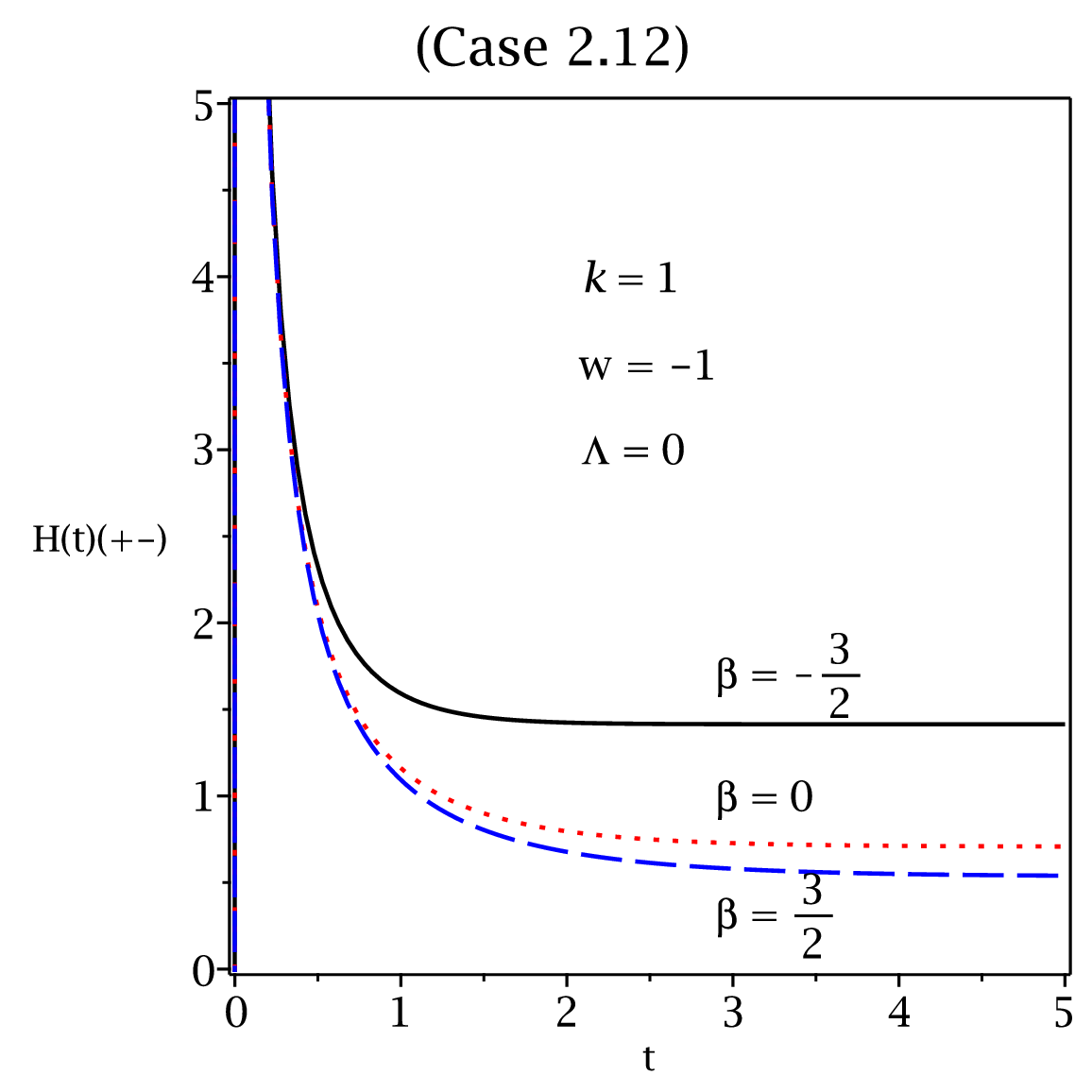}
	\includegraphics[width=3.4cm]{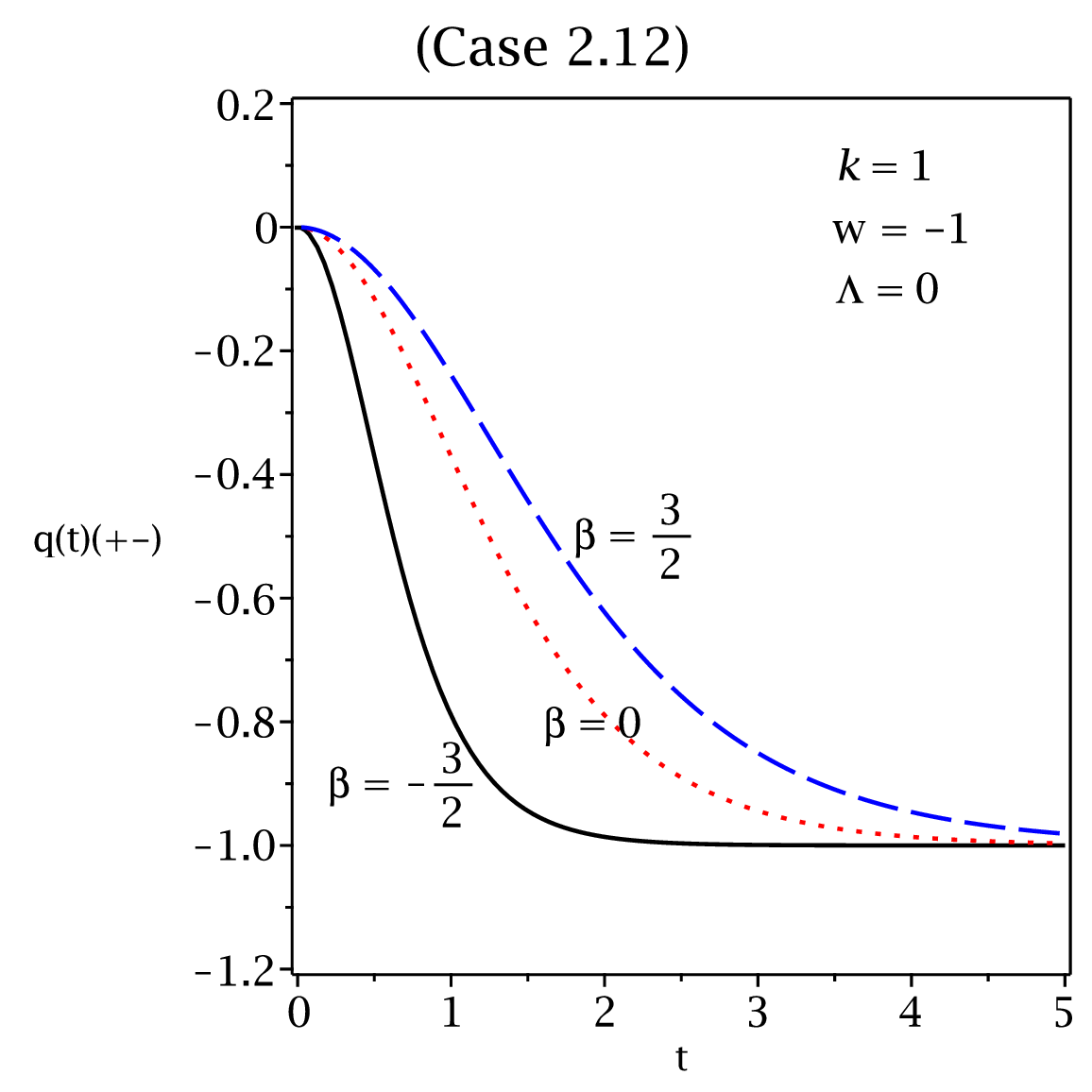}
	\includegraphics[width=3.4cm]{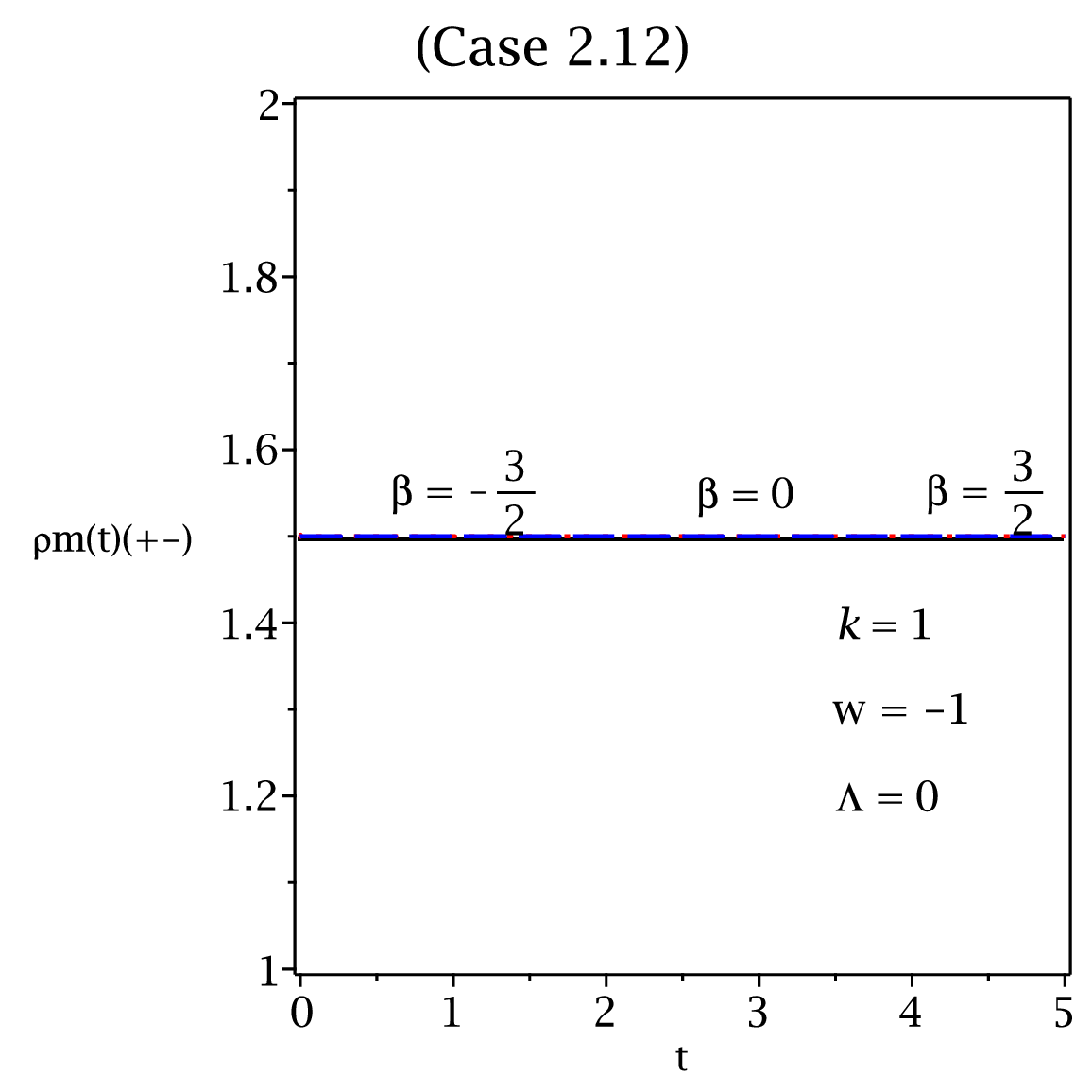}
	\includegraphics[width=3.4cm]{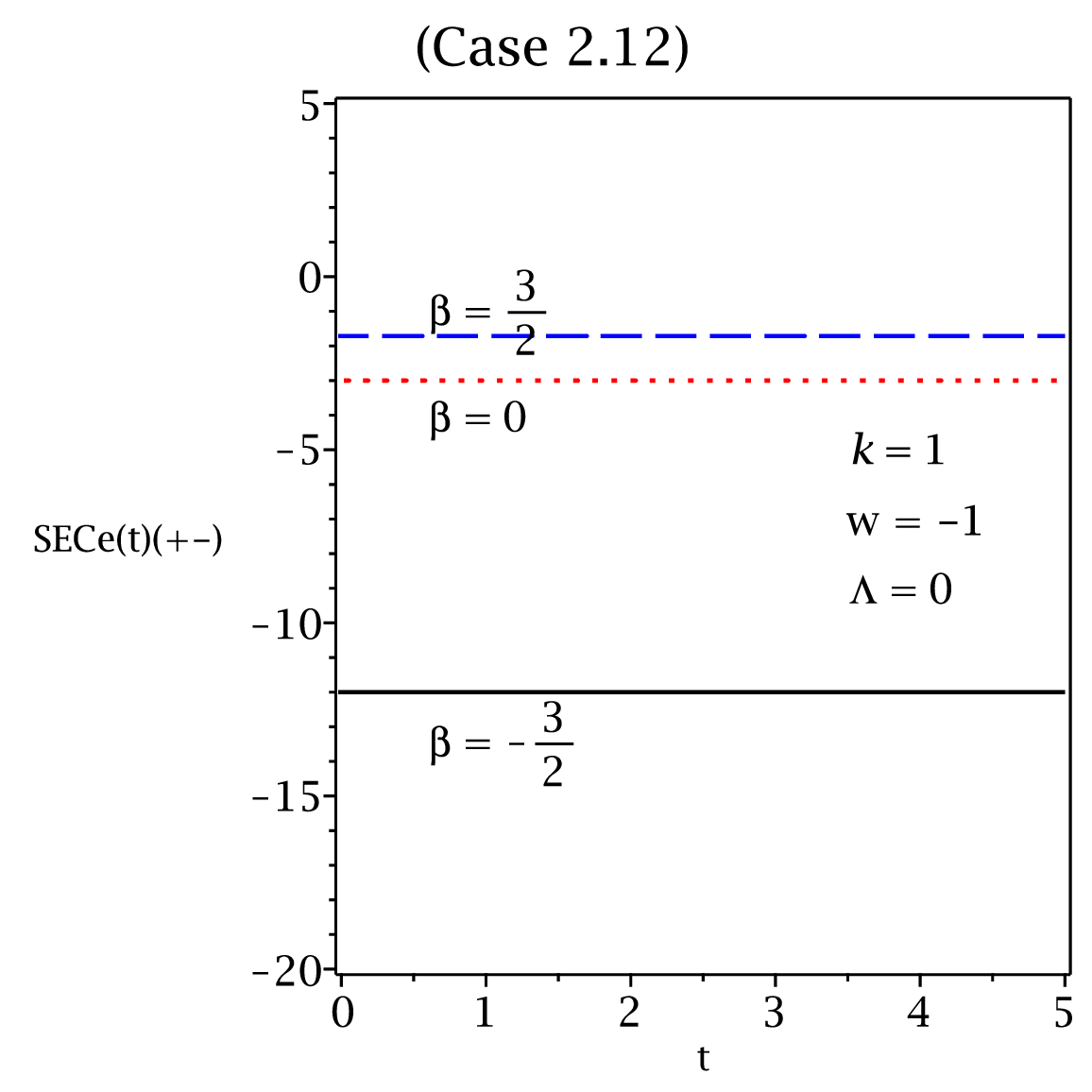}
	\includegraphics[width=3.4cm]{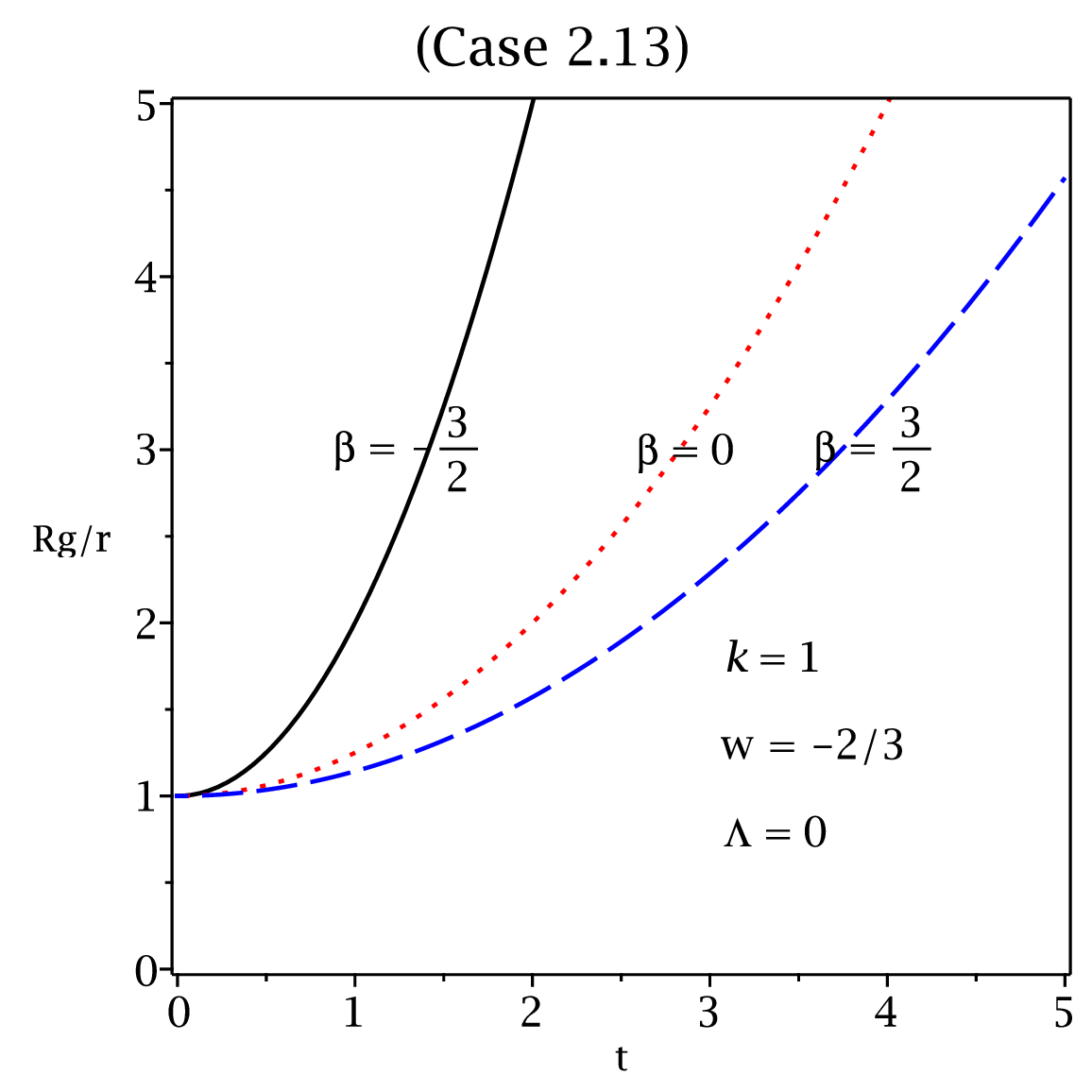}
	\includegraphics[width=3.4cm]{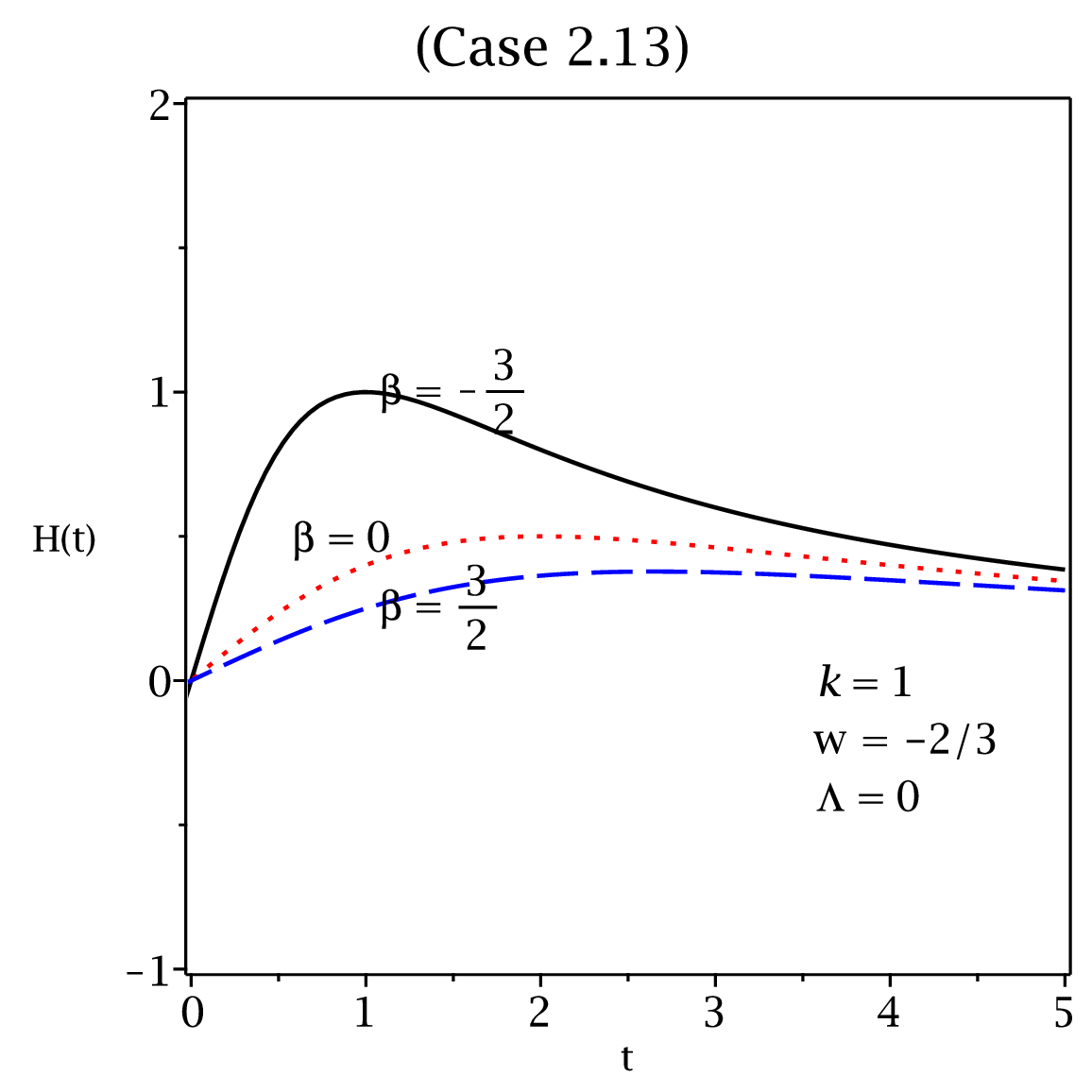}
	\includegraphics[width=3.4cm]{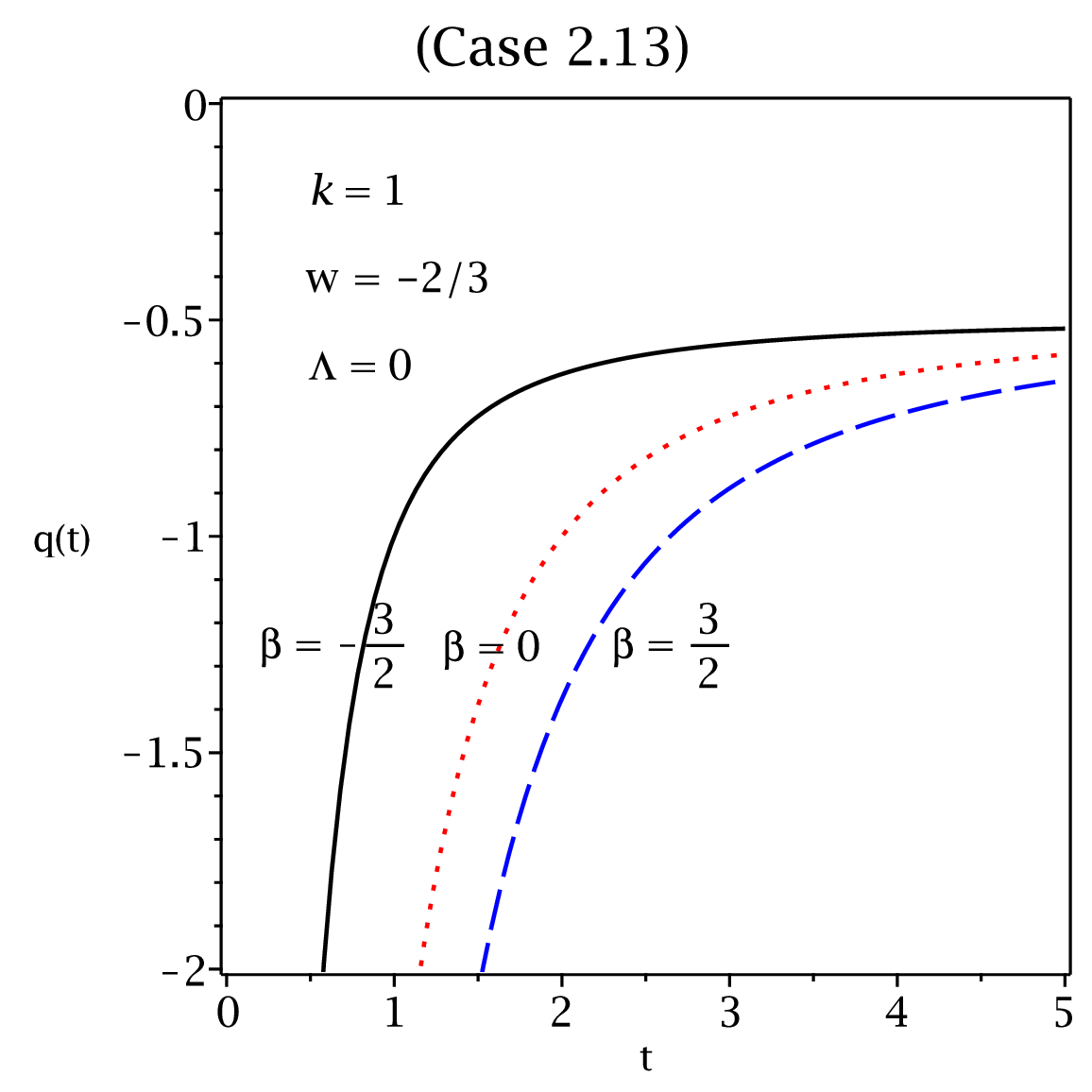}
	\includegraphics[width=3.4cm]{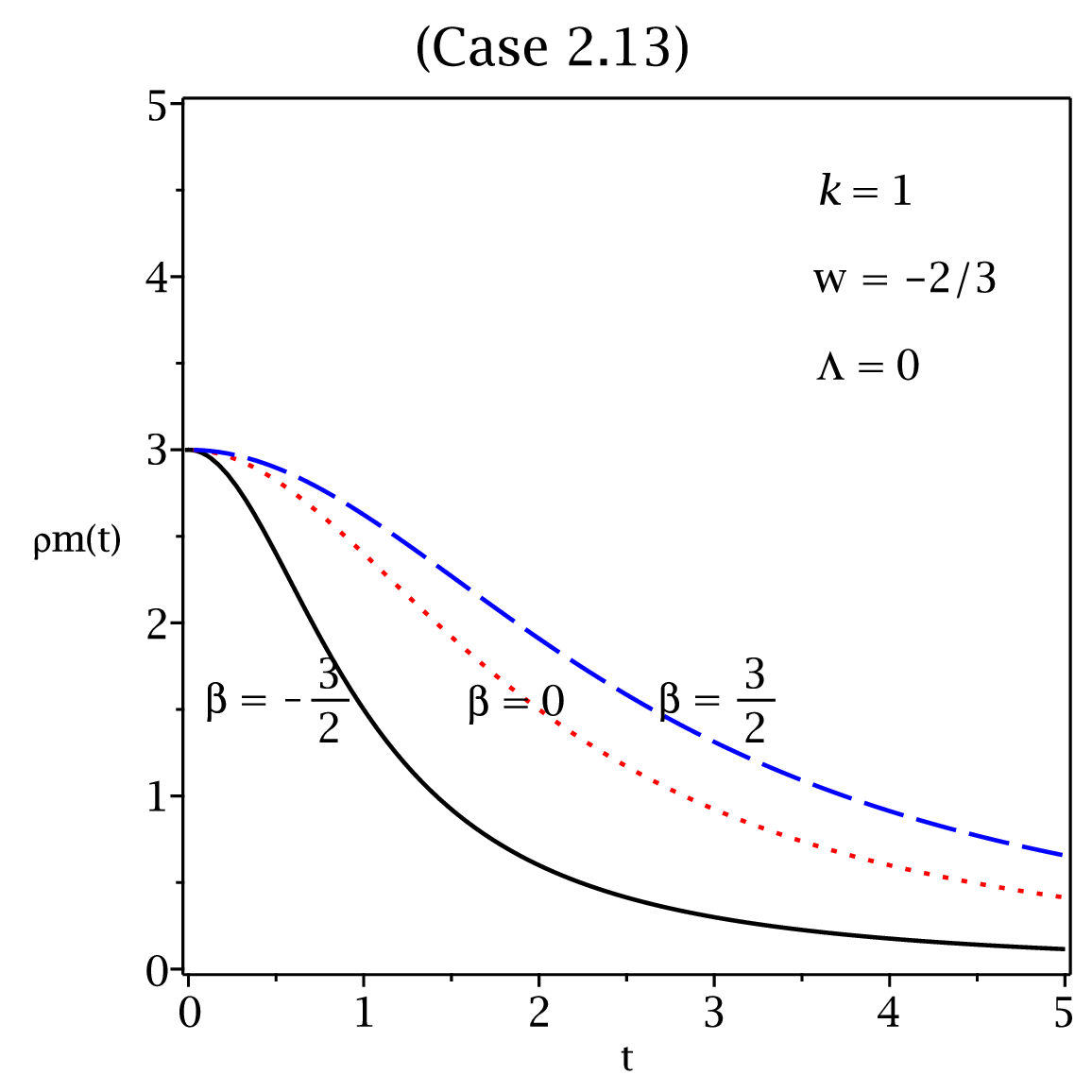}
	\includegraphics[width=3.4cm]{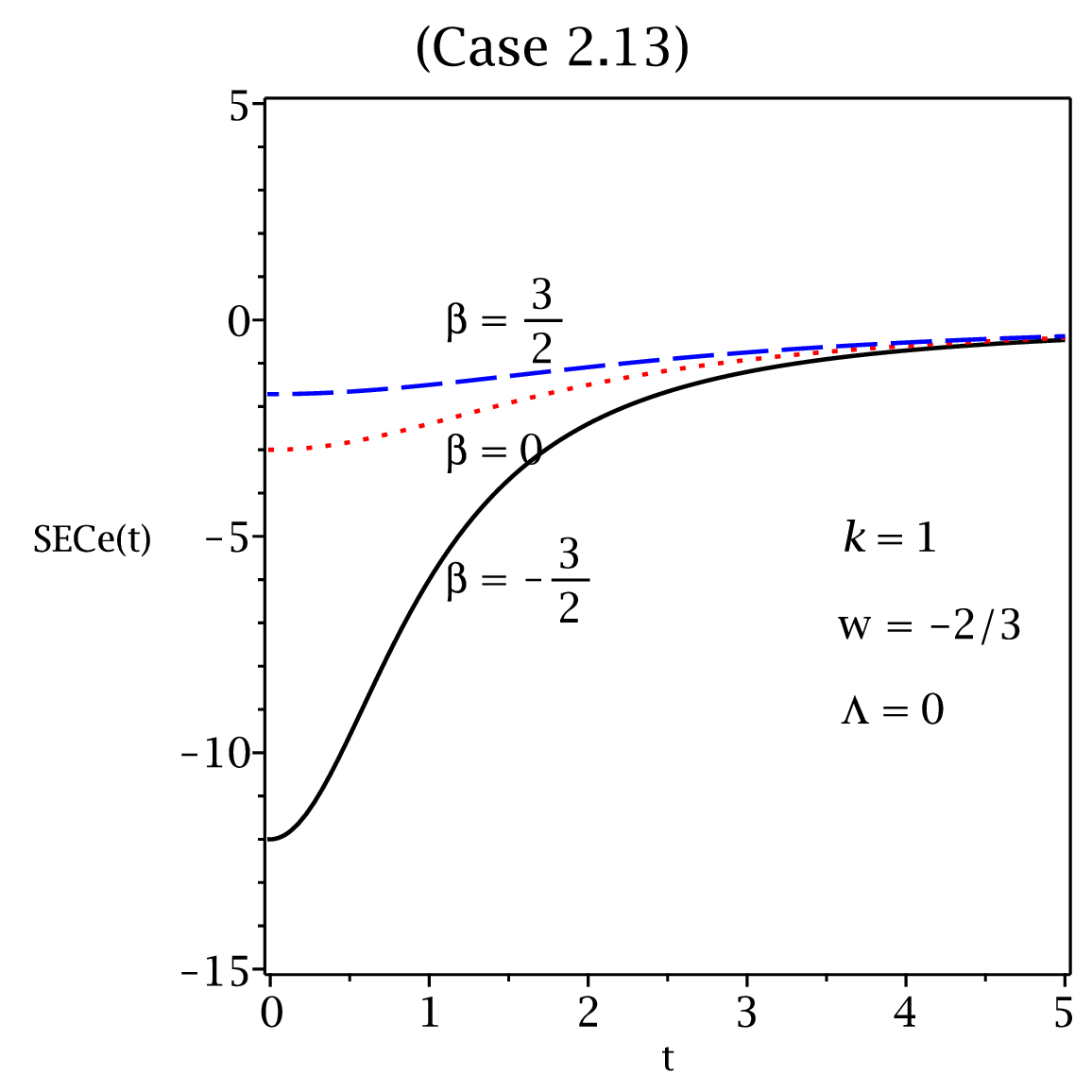}
	\includegraphics[width=3.4cm]{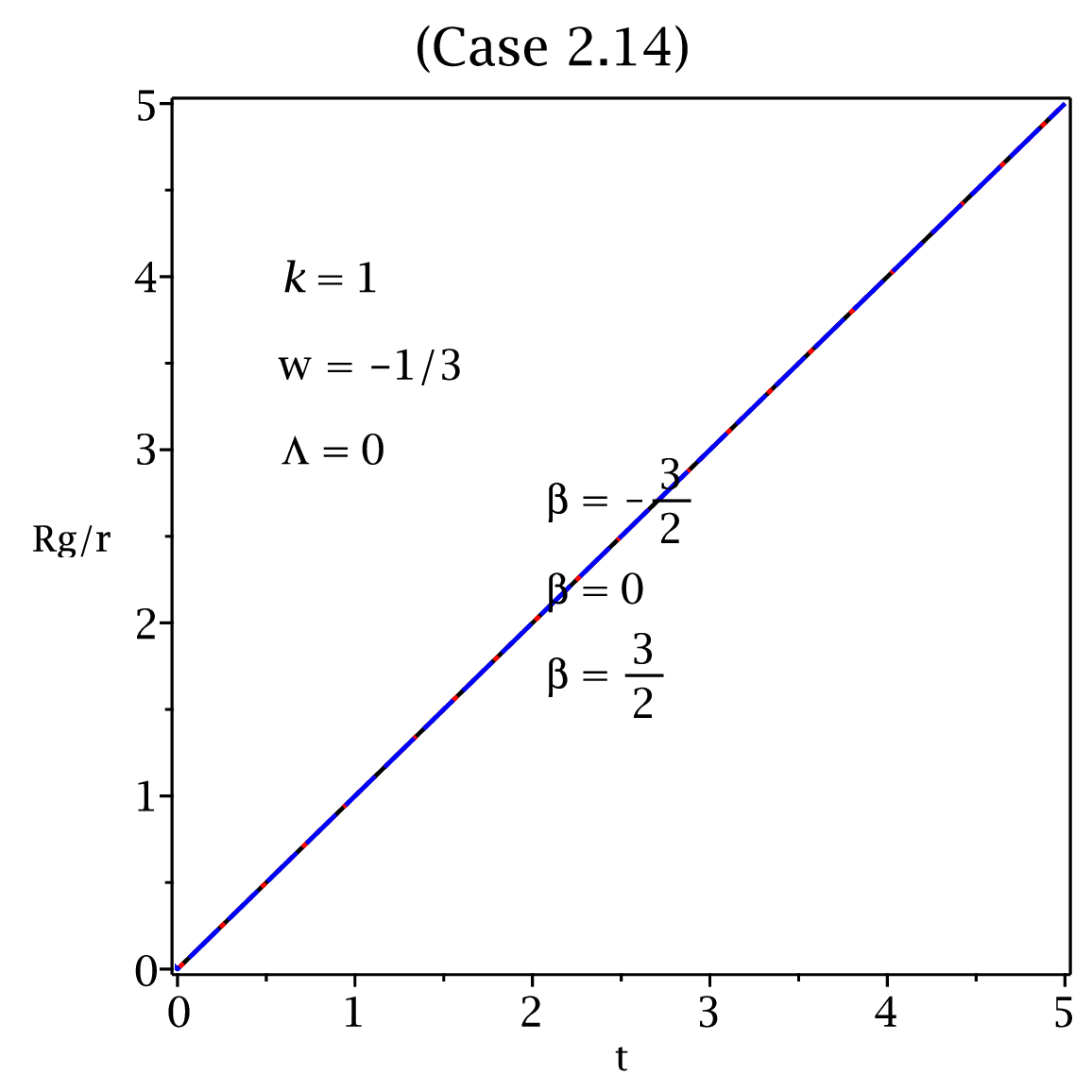}
	\includegraphics[width=3.4cm]{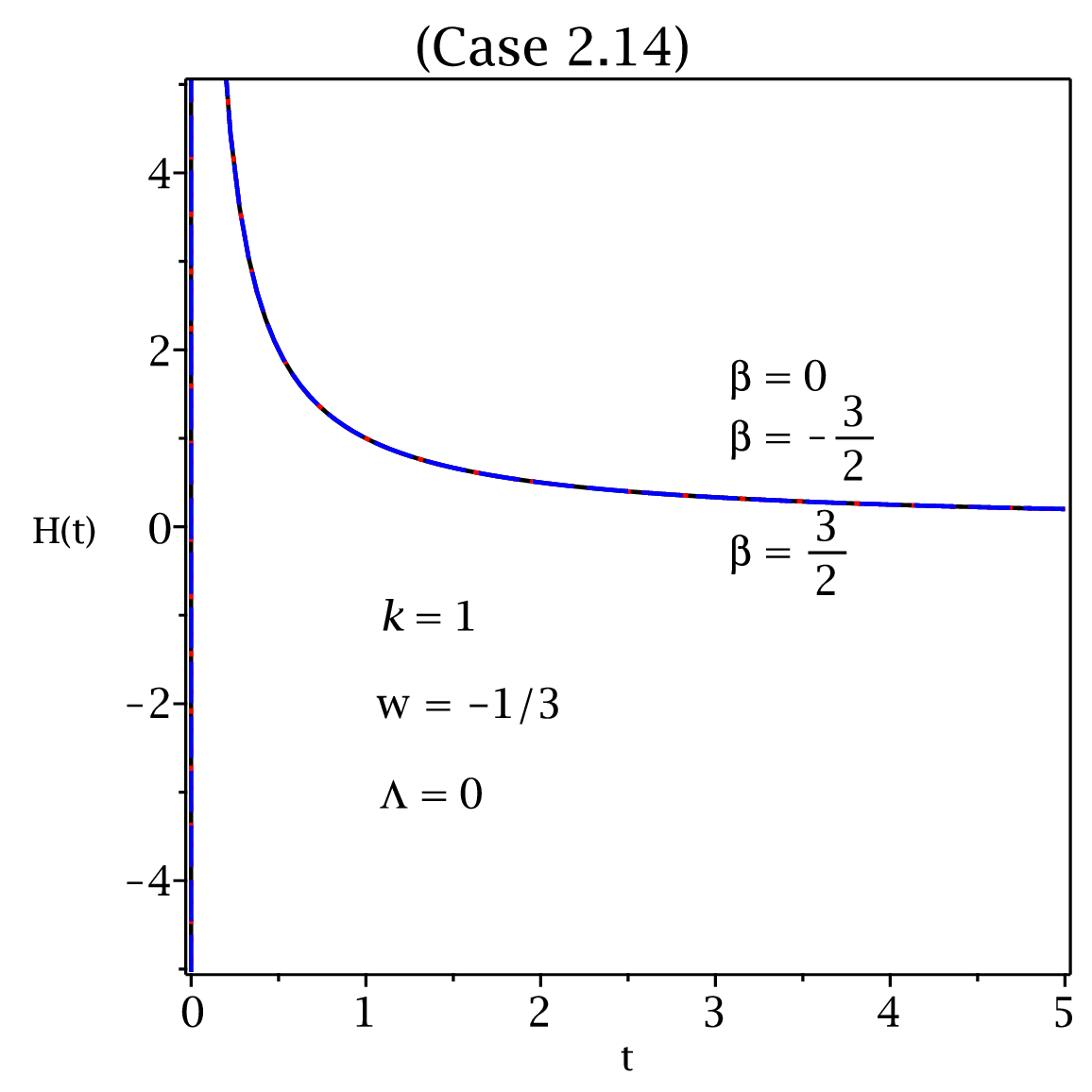}
	\includegraphics[width=3.4cm]{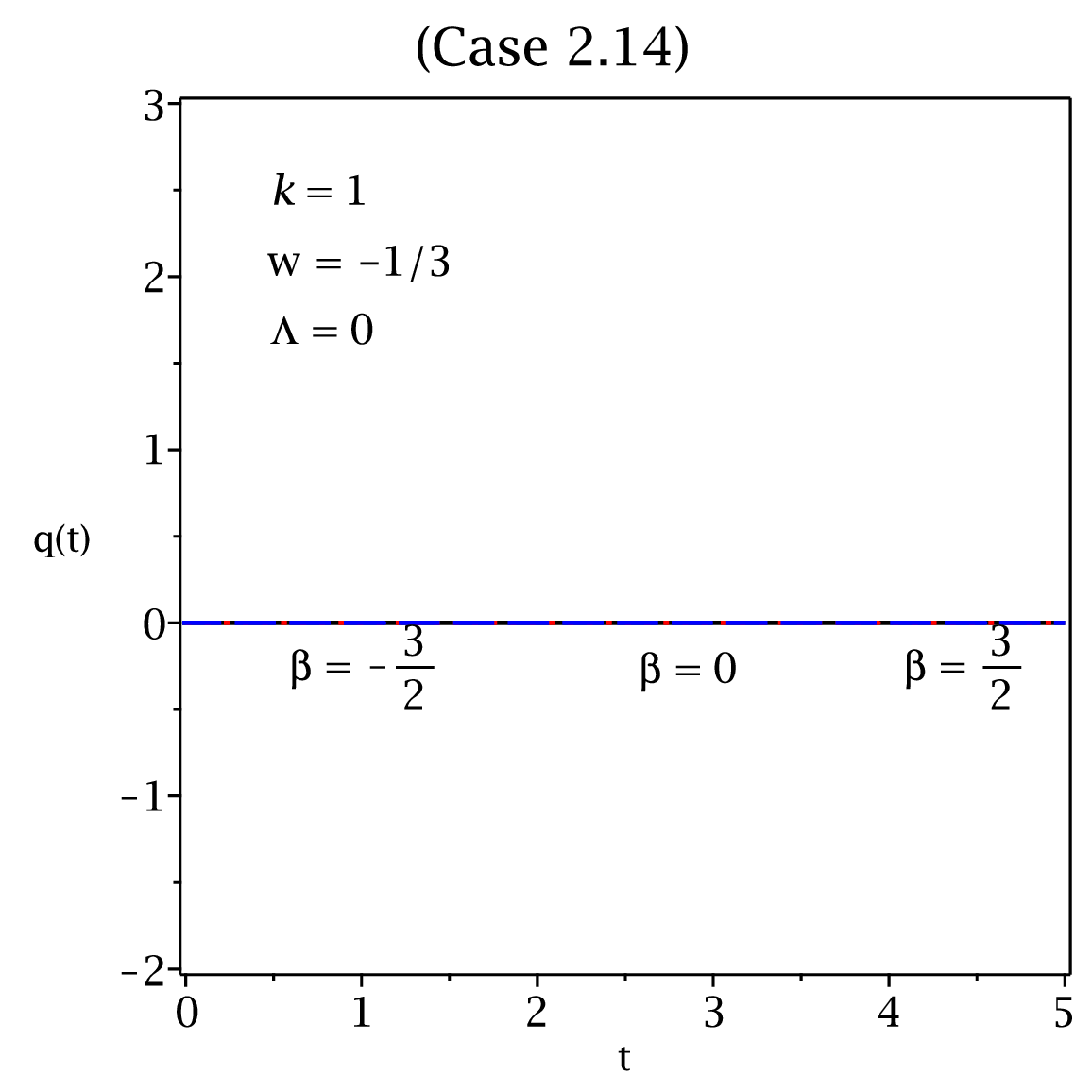}
	\includegraphics[width=3.4cm]{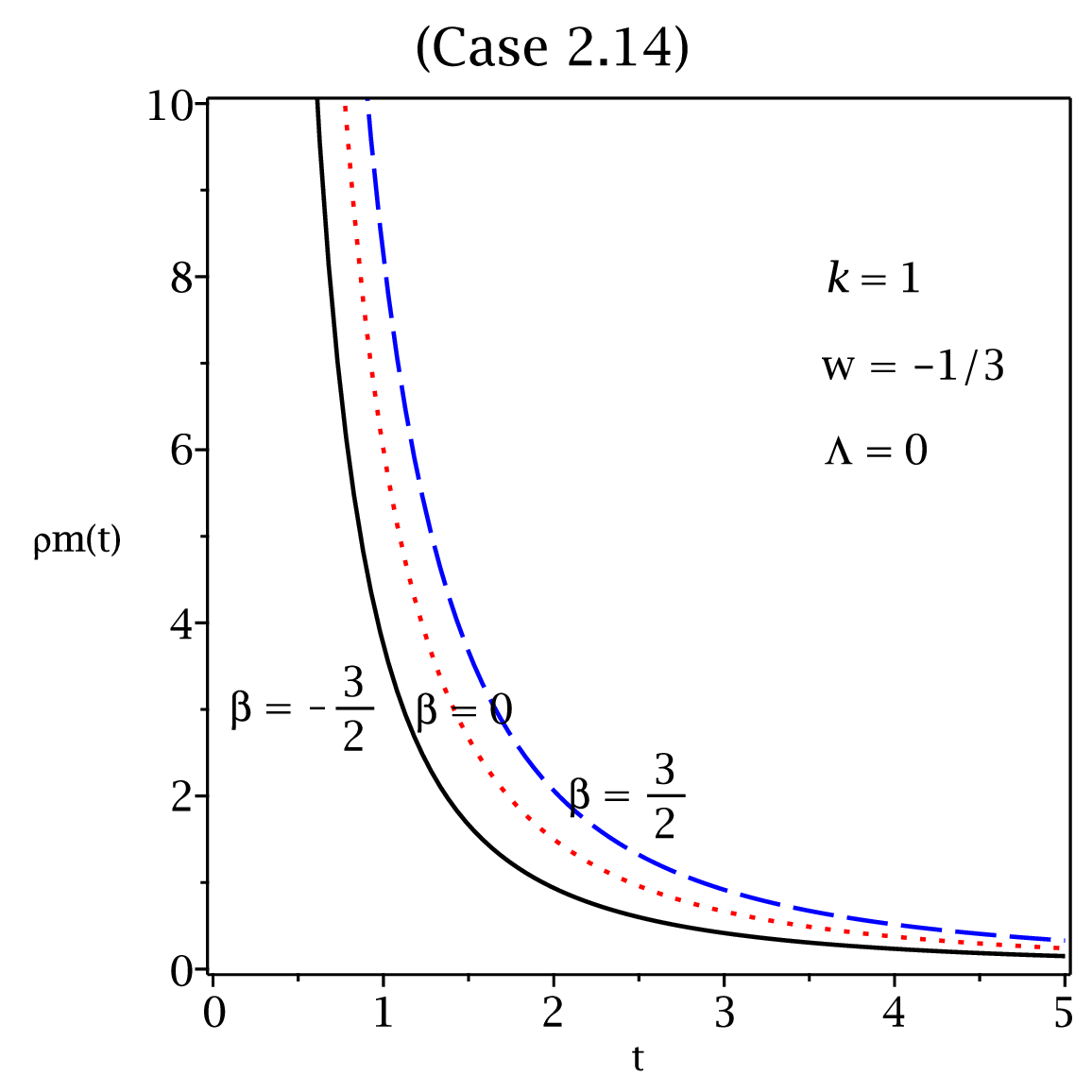}
	\includegraphics[width=3.4cm]{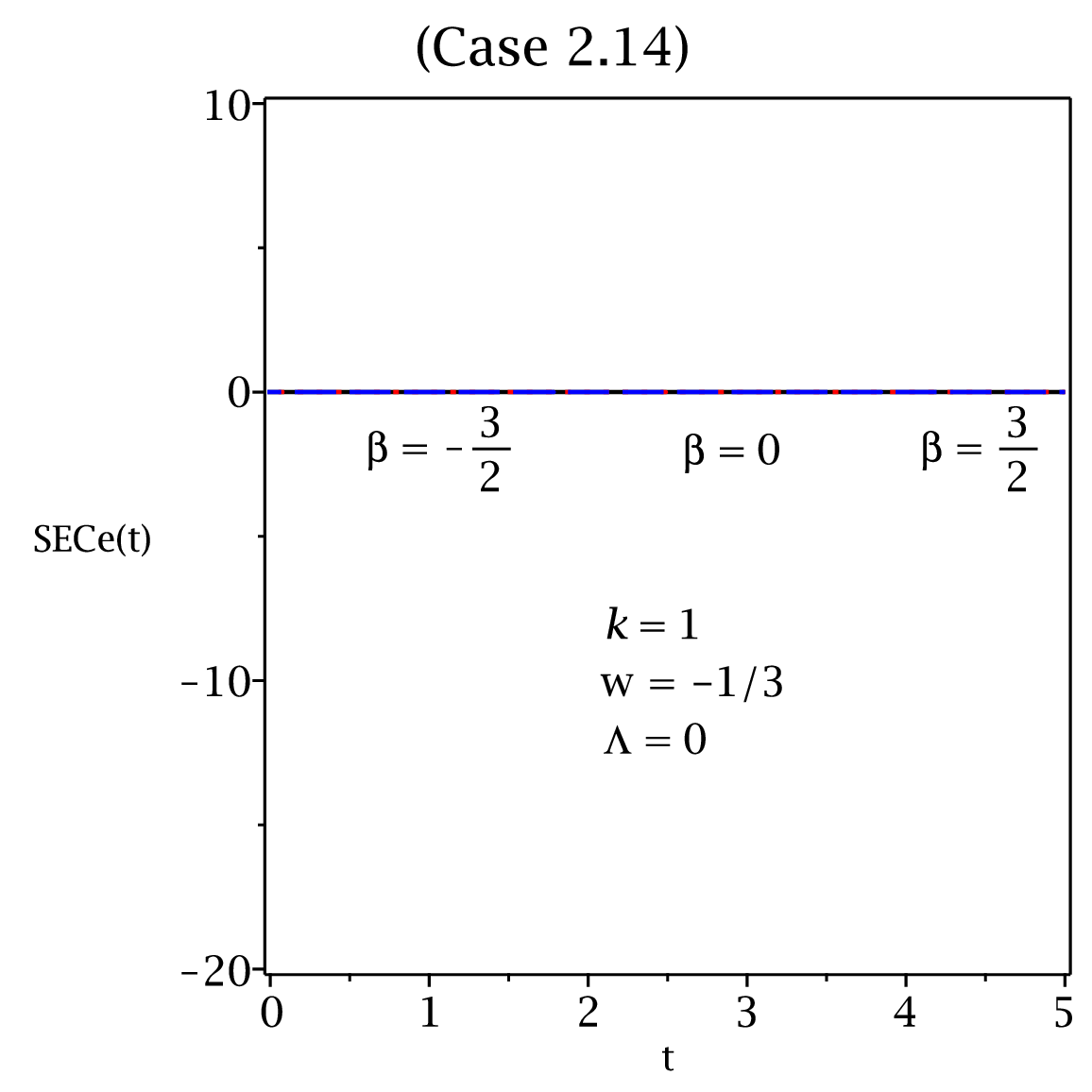}
	%
	%
	\includegraphics[width=3.4cm]{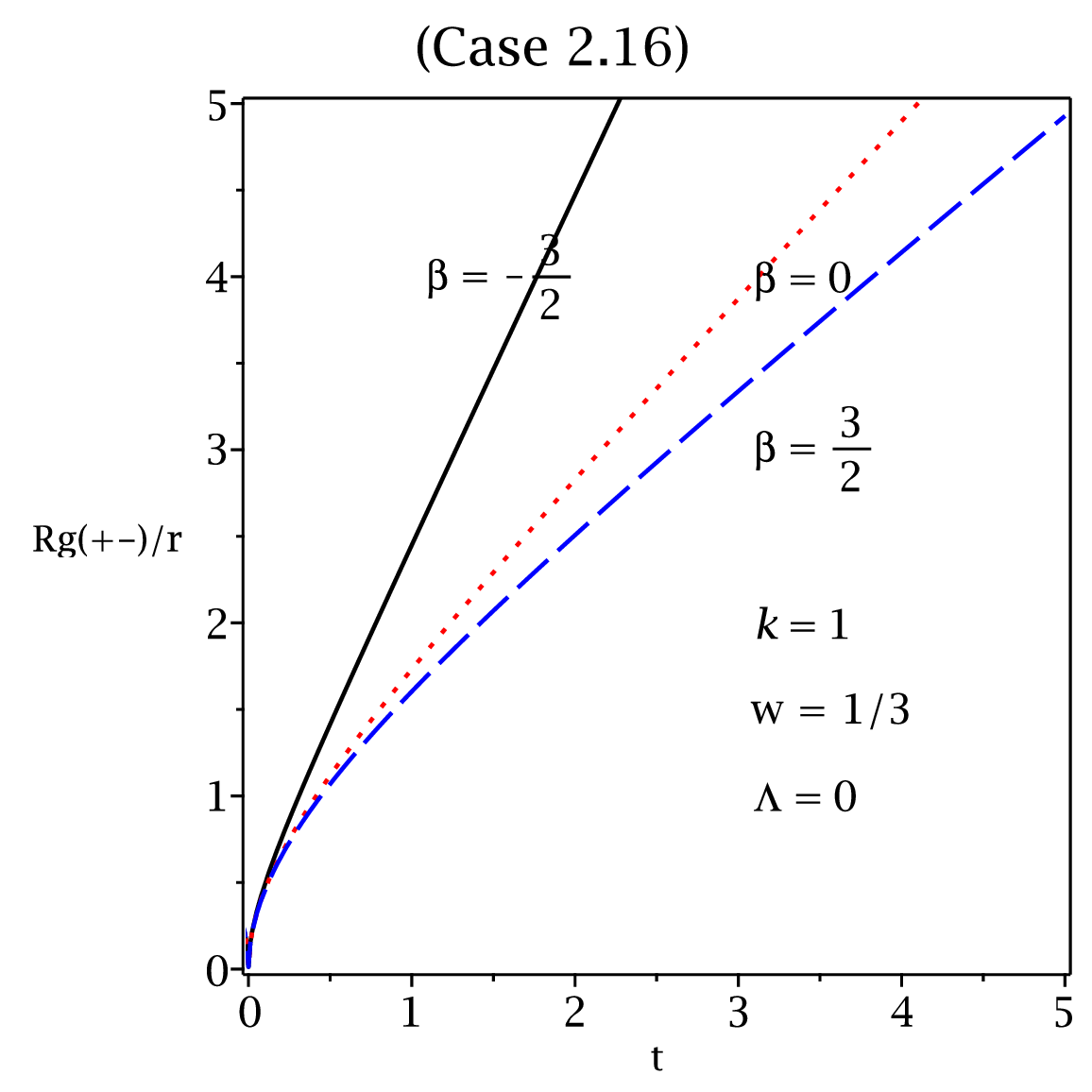}
	\includegraphics[width=3.4cm]{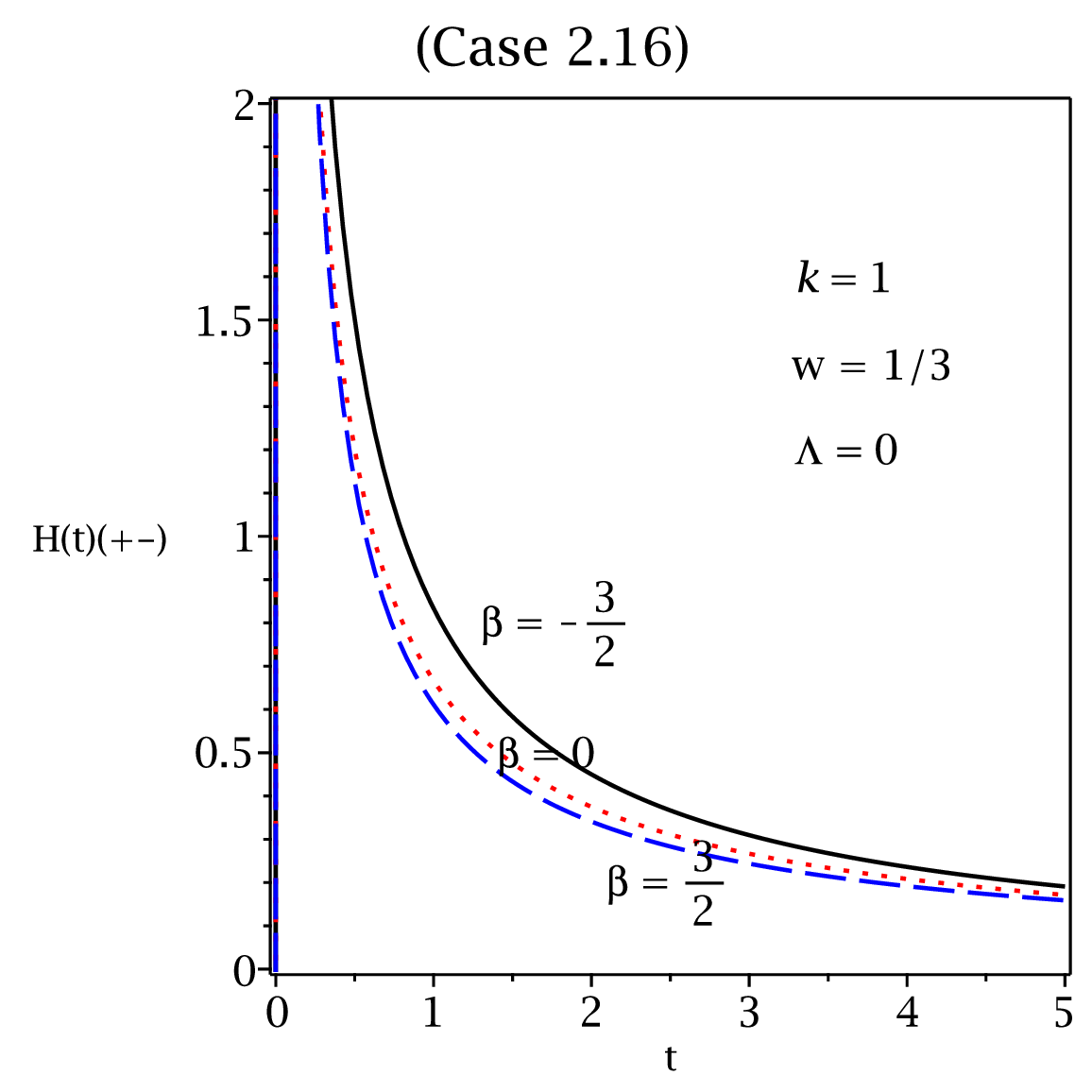}
	\includegraphics[width=3.4cm]{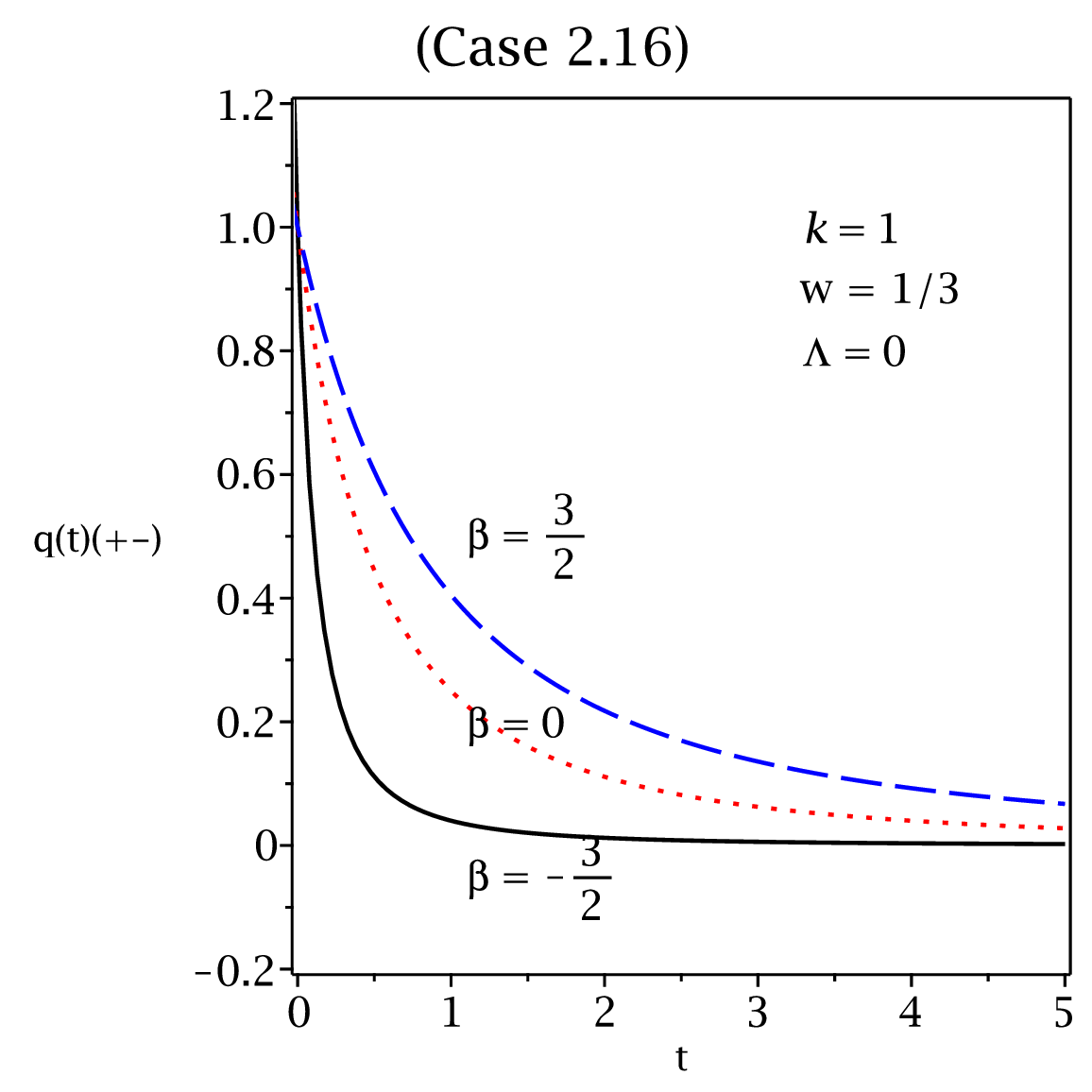}
	\includegraphics[width=3.4cm]{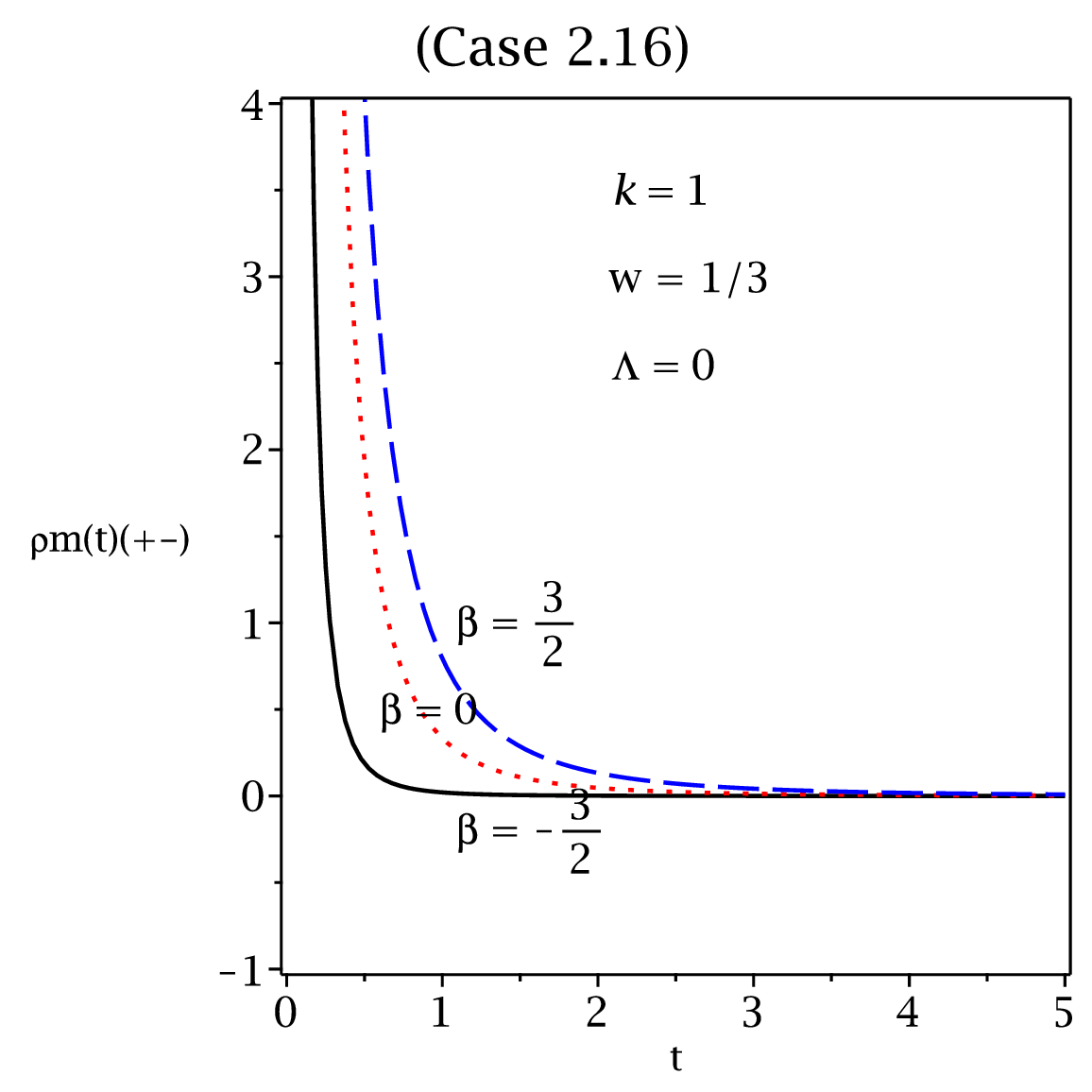}
	\includegraphics[width=3.4cm]{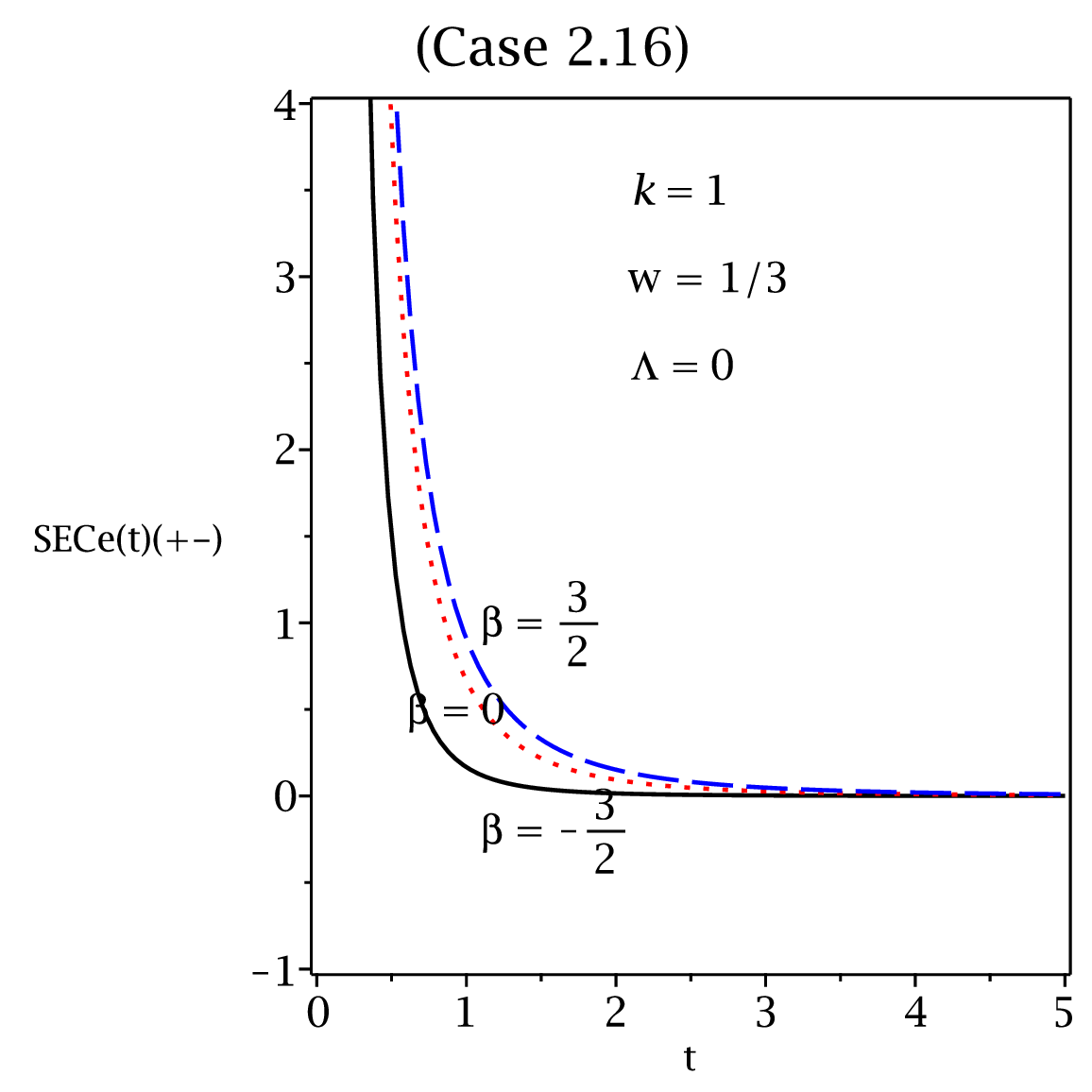}
	\caption{These figures are for $\Lambda=0$ and $k=1$.
		These figures represent the quantities $R_g$ (geometrical radius), 
		$H(t)$ (Hubble parameter) and $q(t)$ (deceleration parameter) $\rho_m(t)$ 
		(energy density of the aether fluid) and $SEC_{e} \equiv SEC_{\rm eff}$ 
		(strong energy condition for the effective fluid) for the different
		values of $\beta=-3/2$ (black solid line), $\beta=0$ (red dotted line), 
		$\beta=3/2$ (blue dashed line). Assuming that $8 \pi G=1$ and
		$R_g(t=0)=0$. Assuming also that $C_1=1$, $C_2=0$ (Cases 2.12, 2.14 and 2.16); 
		$C_1=0$, $C_2=1$ (Case 2.13).The subscripts $(+)$ and $(-)$, denote the two different 
		solutions for $B(t)$.}
	\label{Figure-212-216}
\end{minipage}	
\end{figure}

\begin{figure}[!htp]
\begin{minipage}{175 mm}
	\centering	
	\includegraphics[width=3.4cm]{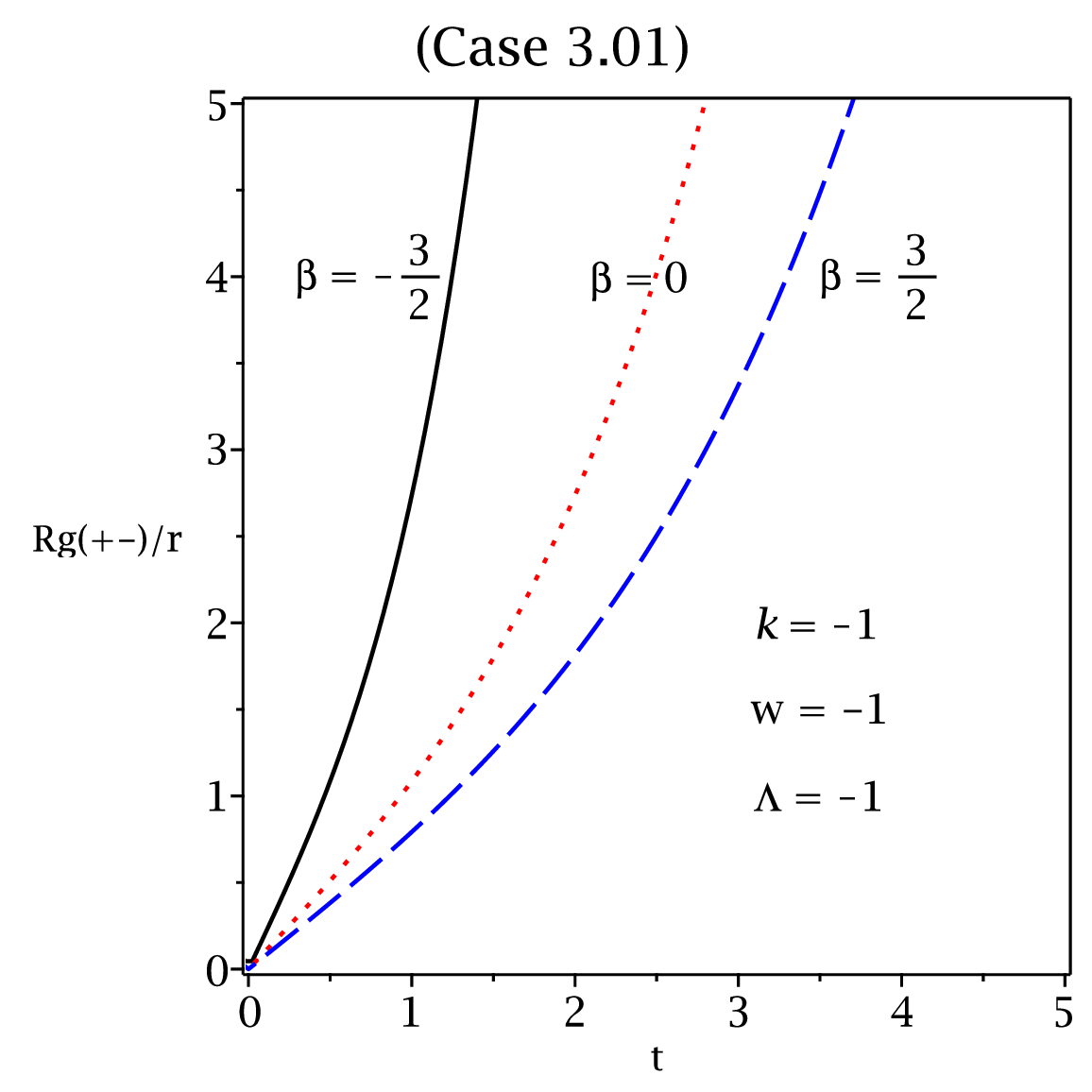}
	\includegraphics[width=3.4cm]{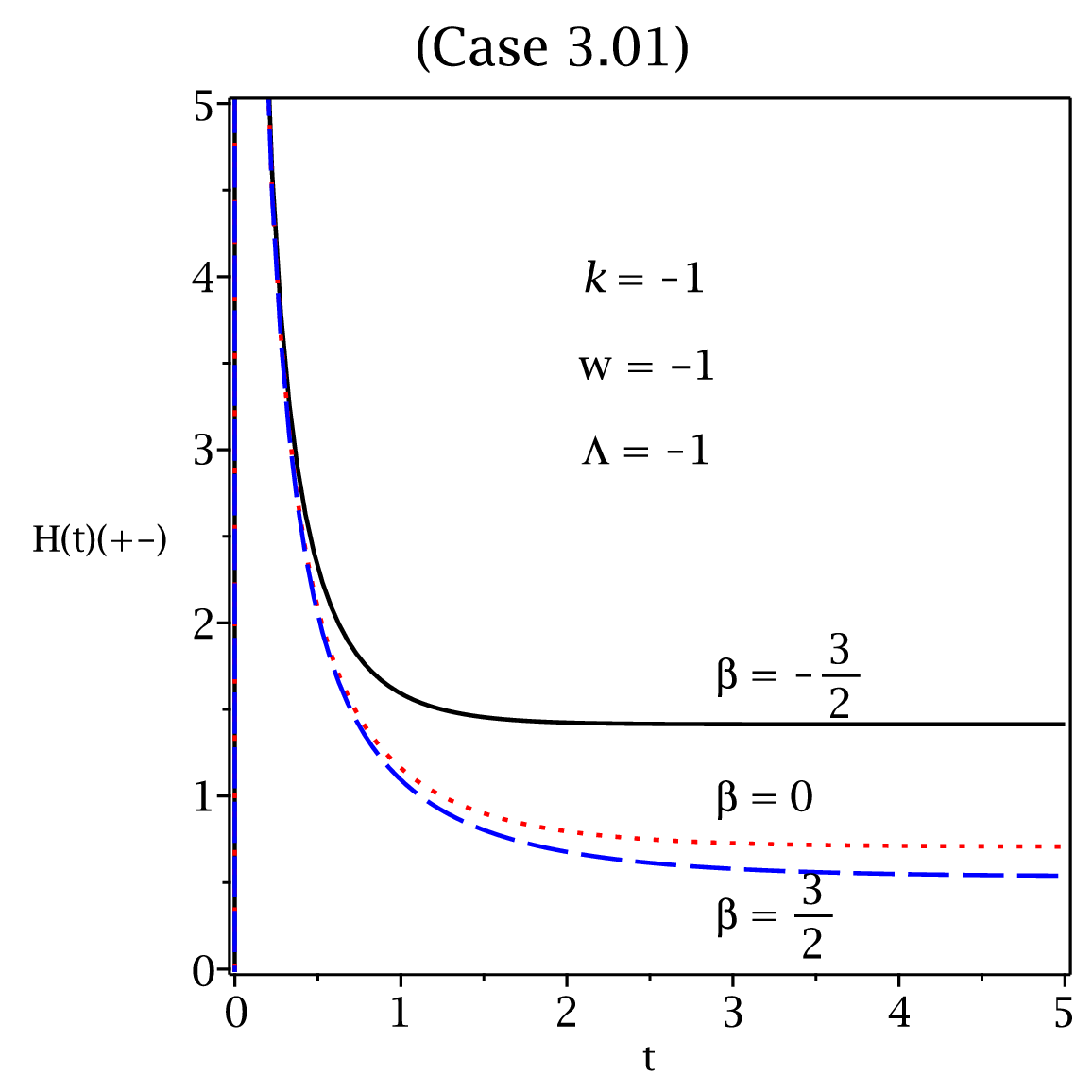}
	\includegraphics[width=3.4cm]{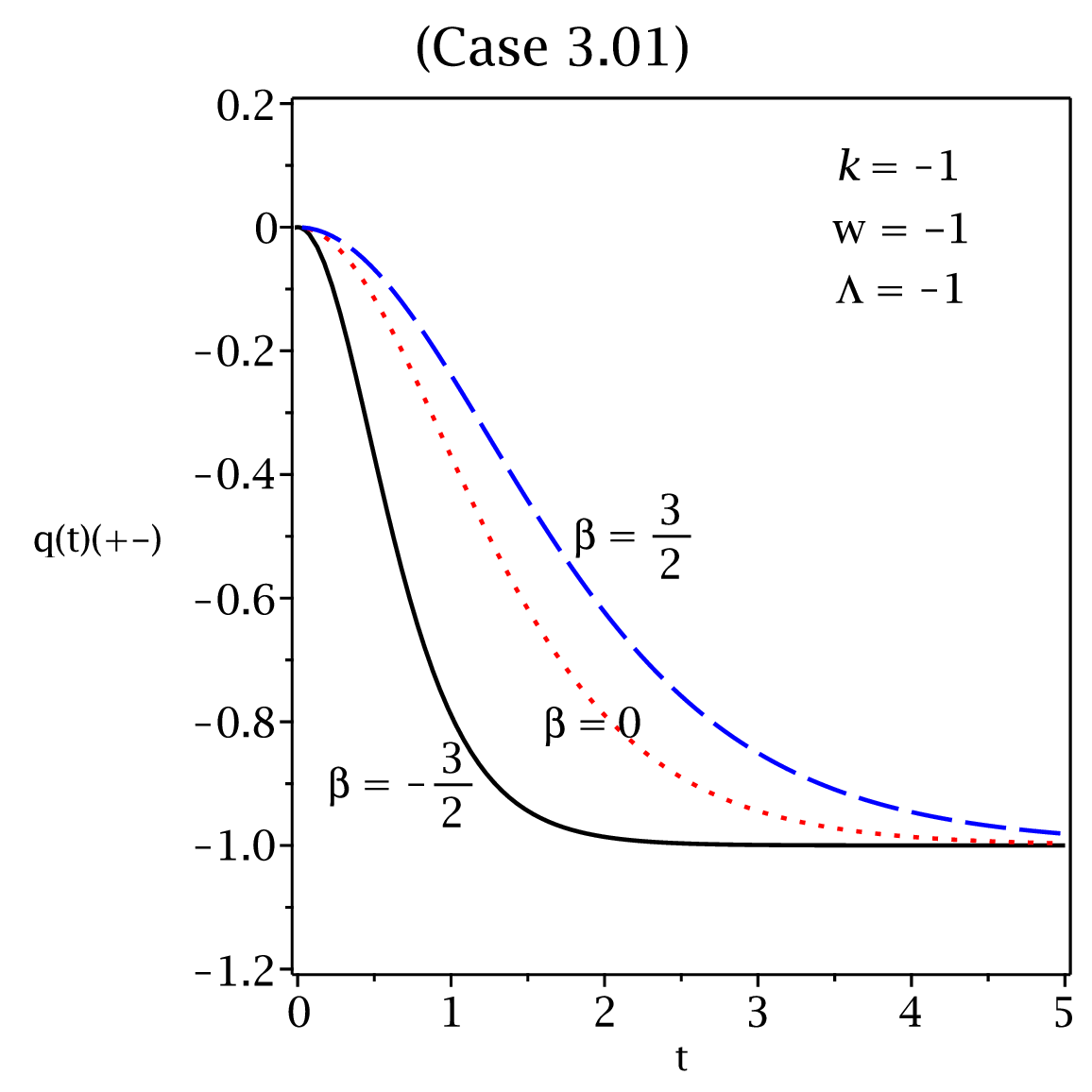}
	\includegraphics[width=3.4cm]{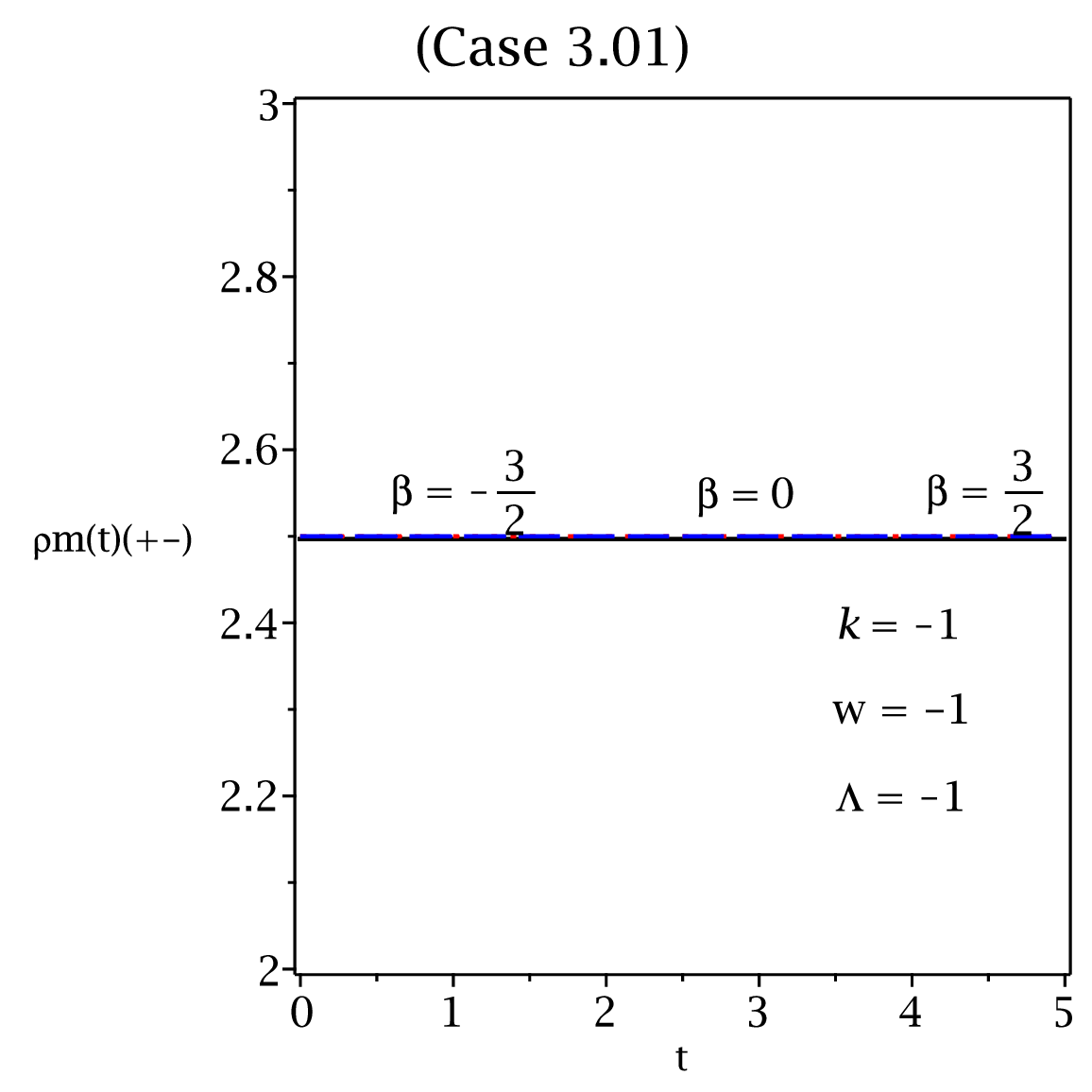}
	\includegraphics[width=3.4cm]{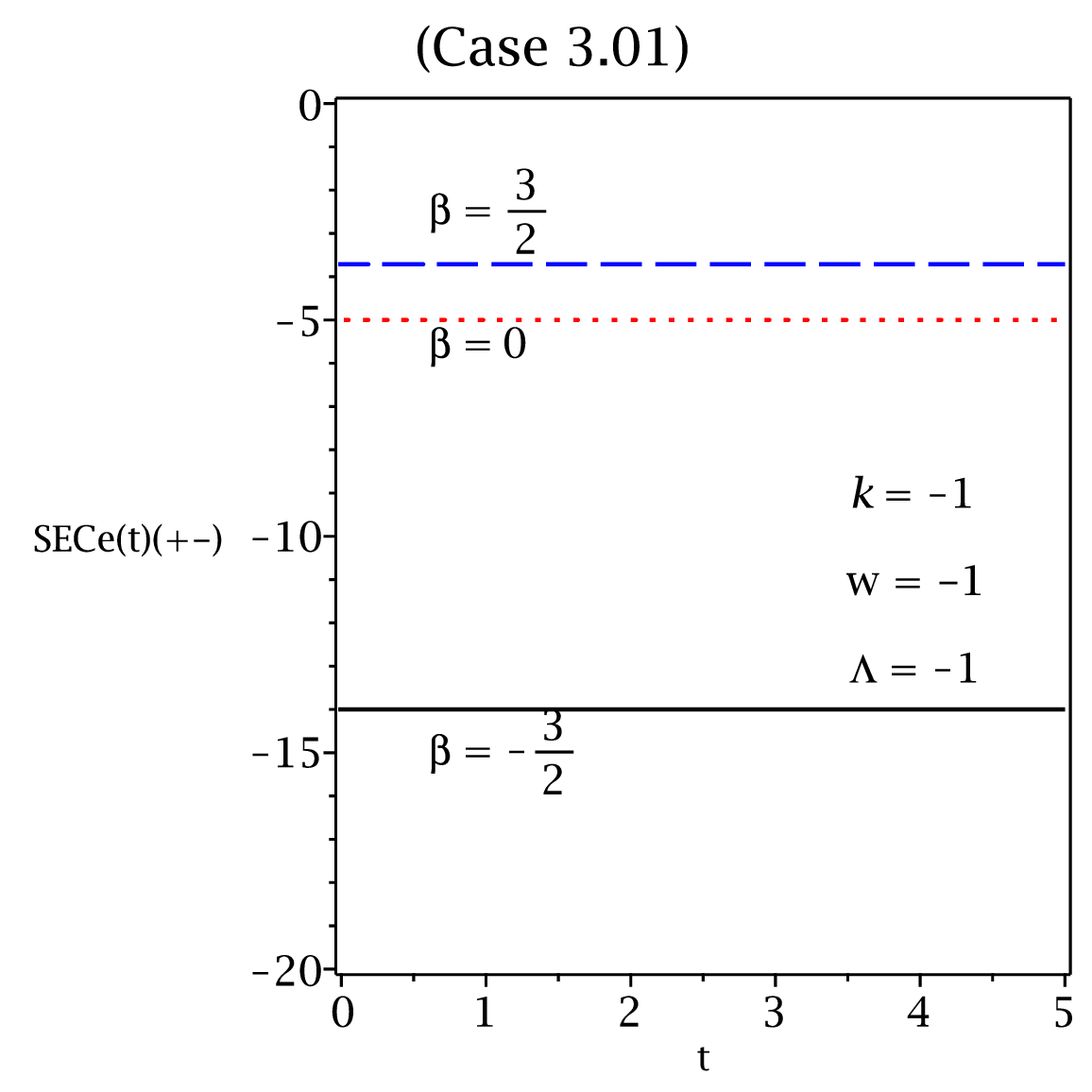}
	\includegraphics[width=3.4cm]{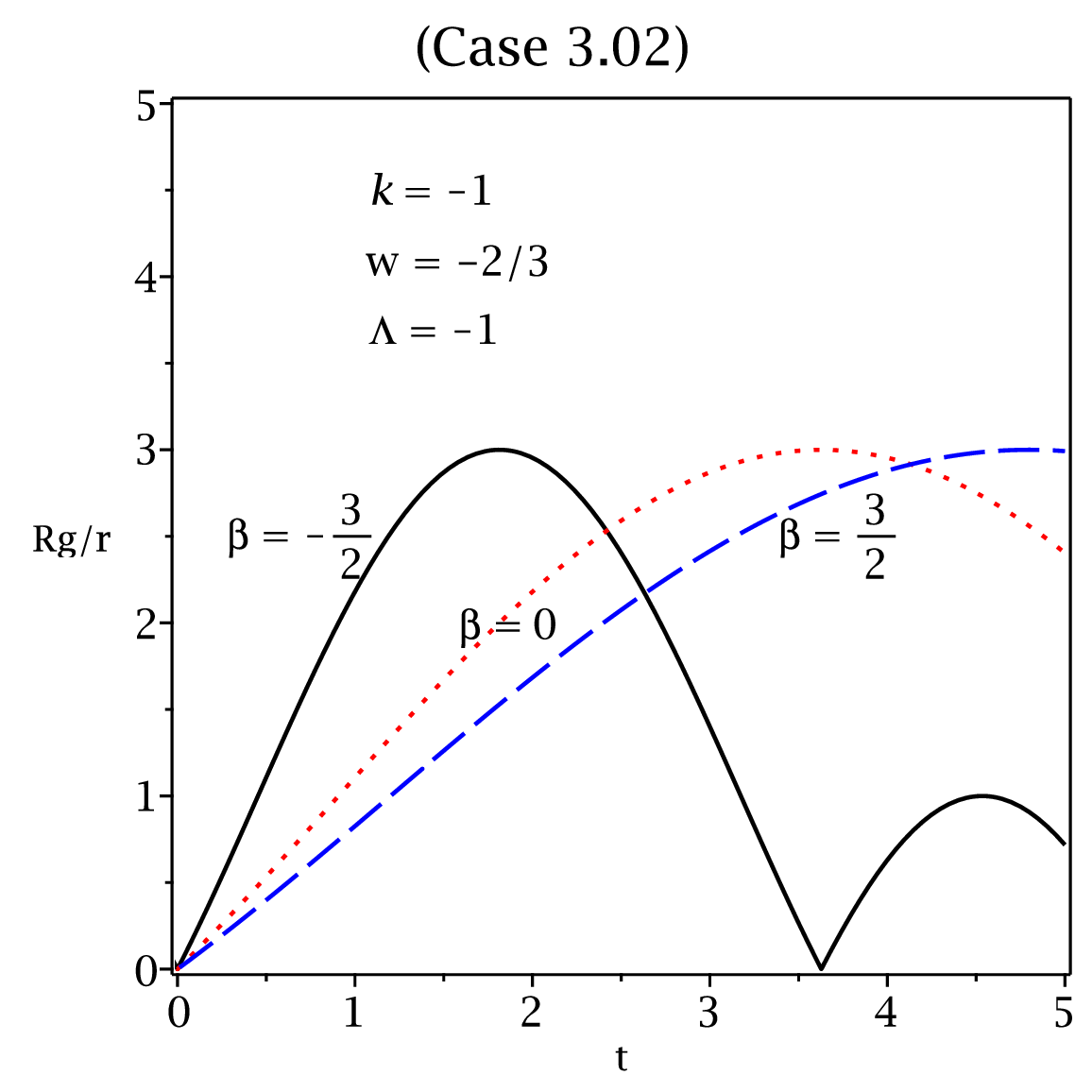}
	\includegraphics[width=3.4cm]{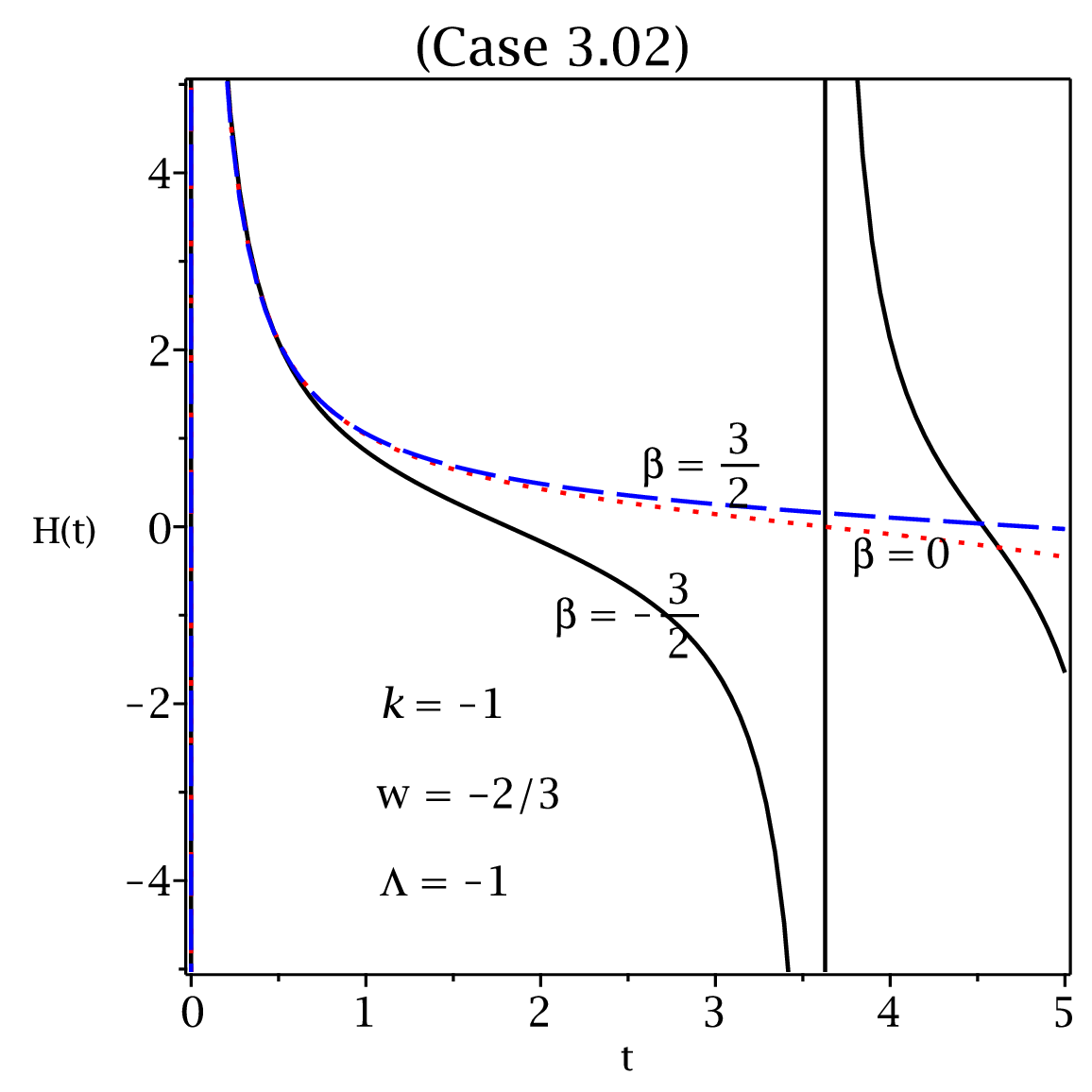}
	\includegraphics[width=3.4cm]{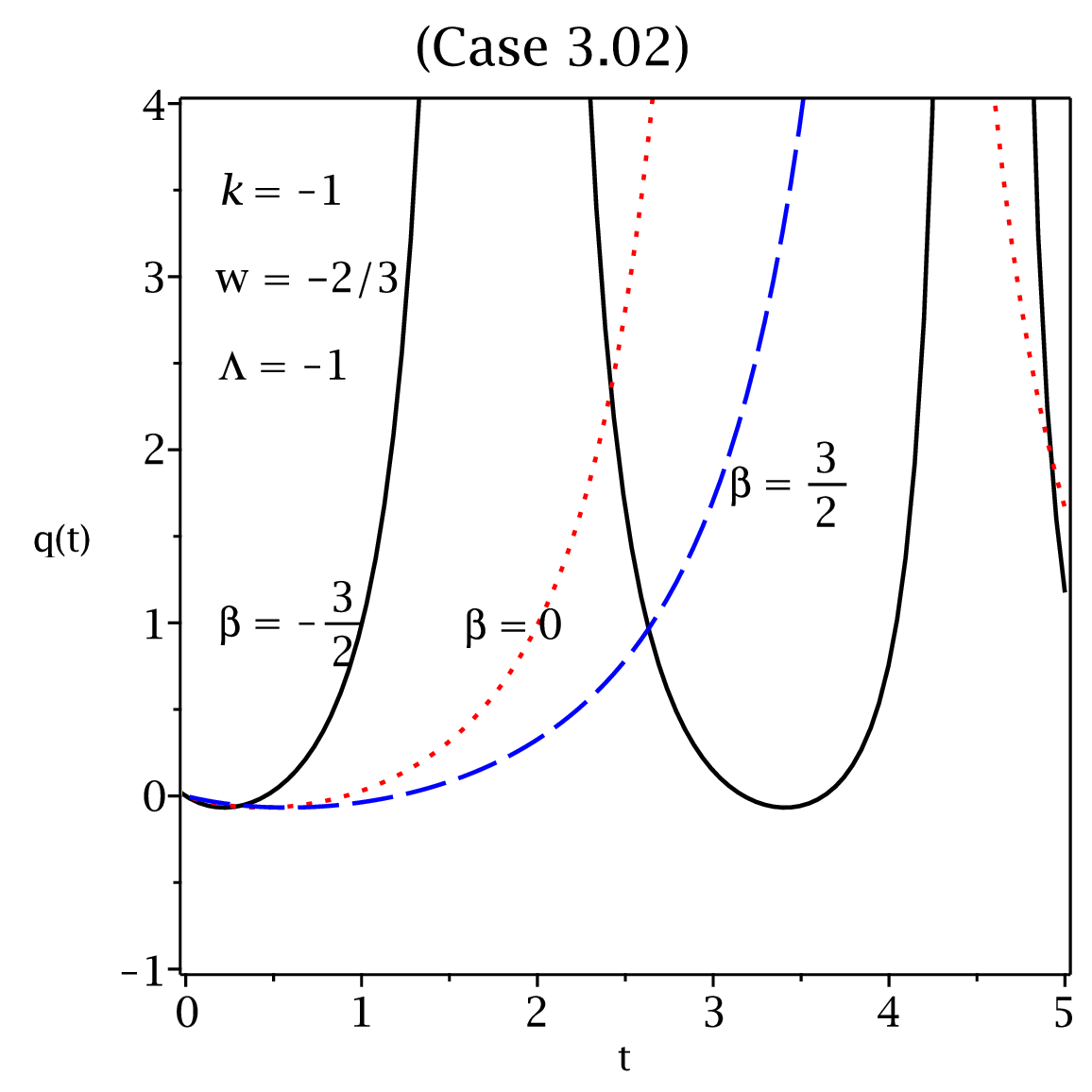}
	\includegraphics[width=3.4cm]{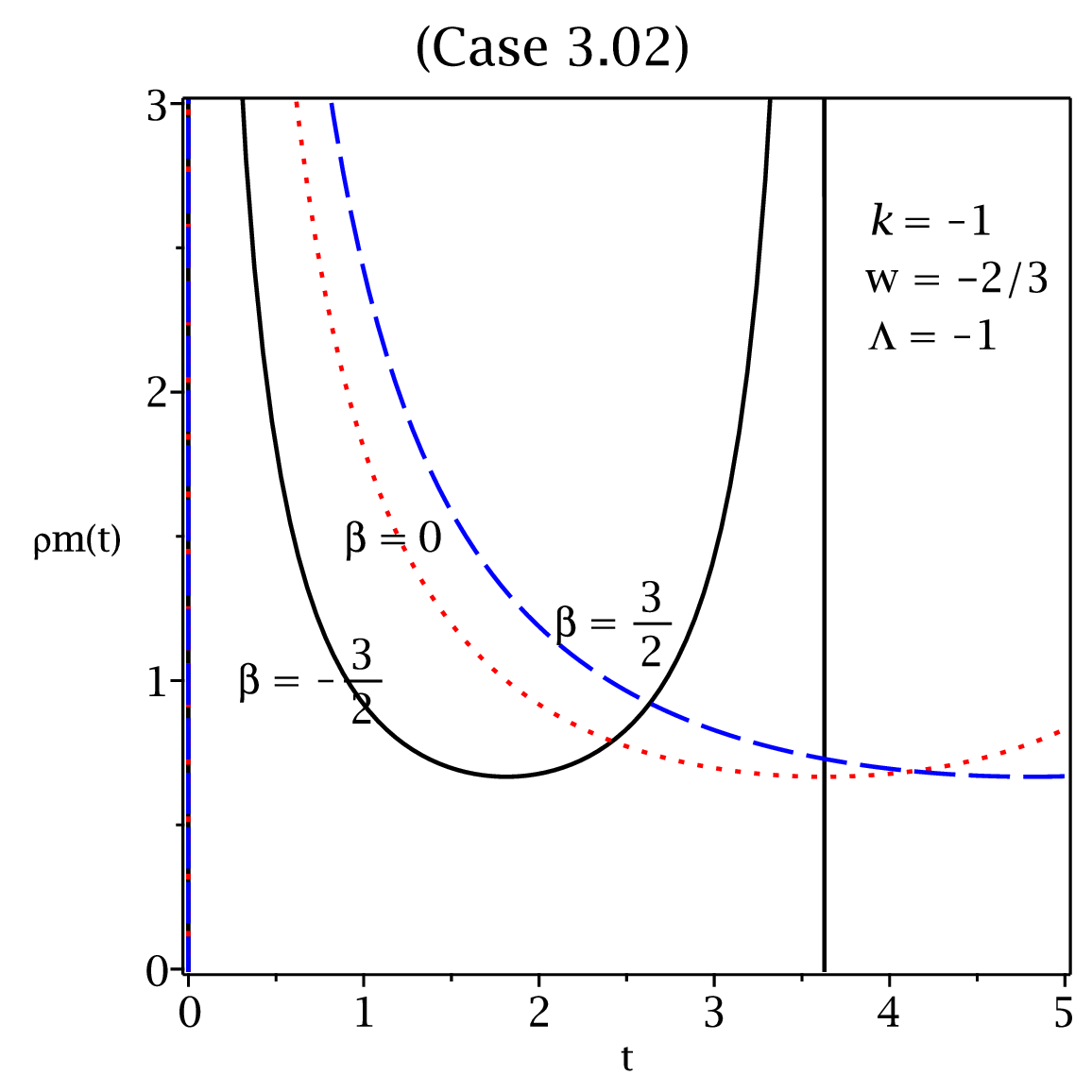}
	\includegraphics[width=3.4cm]{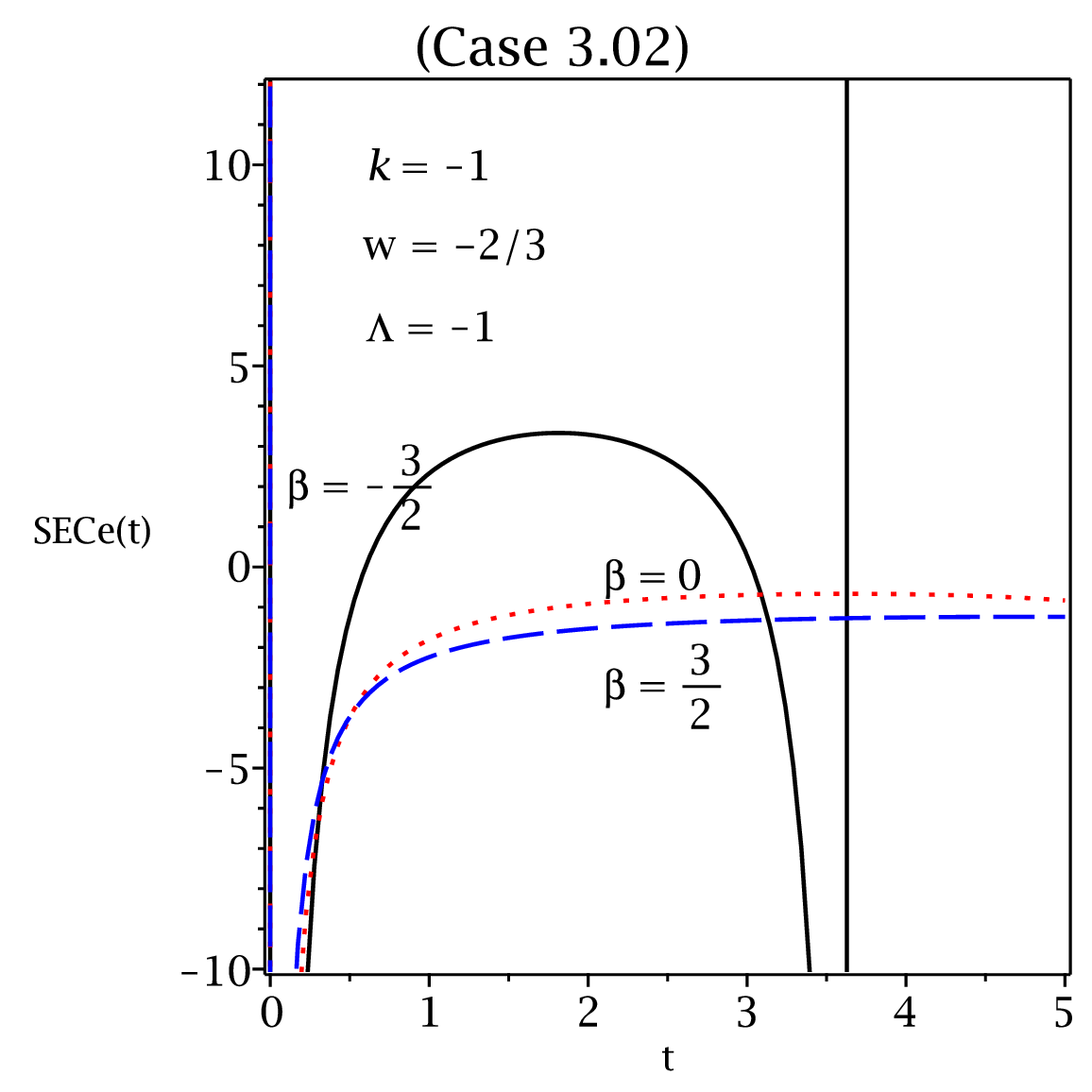}
	\includegraphics[width=3.4cm]{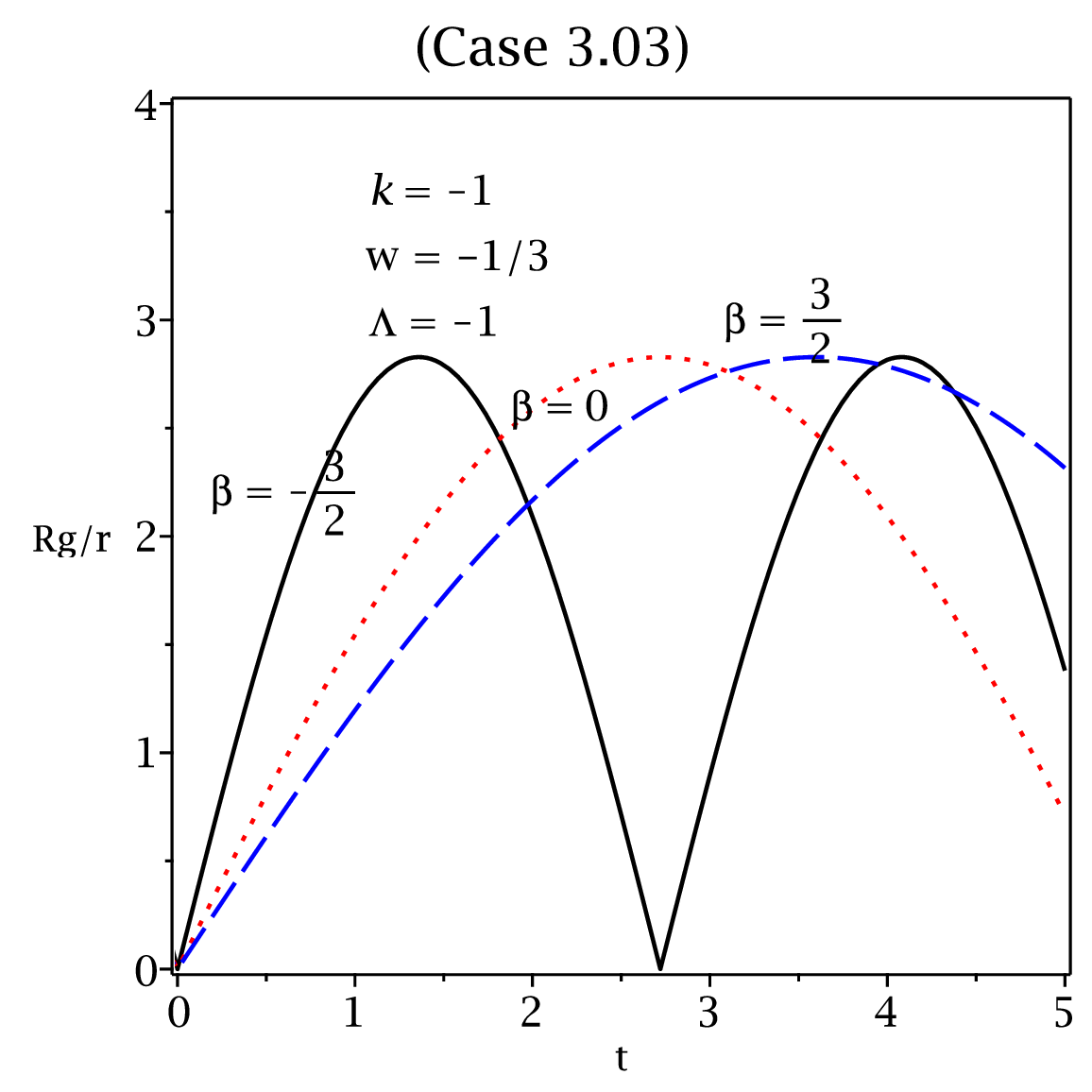}
	\includegraphics[width=3.4cm]{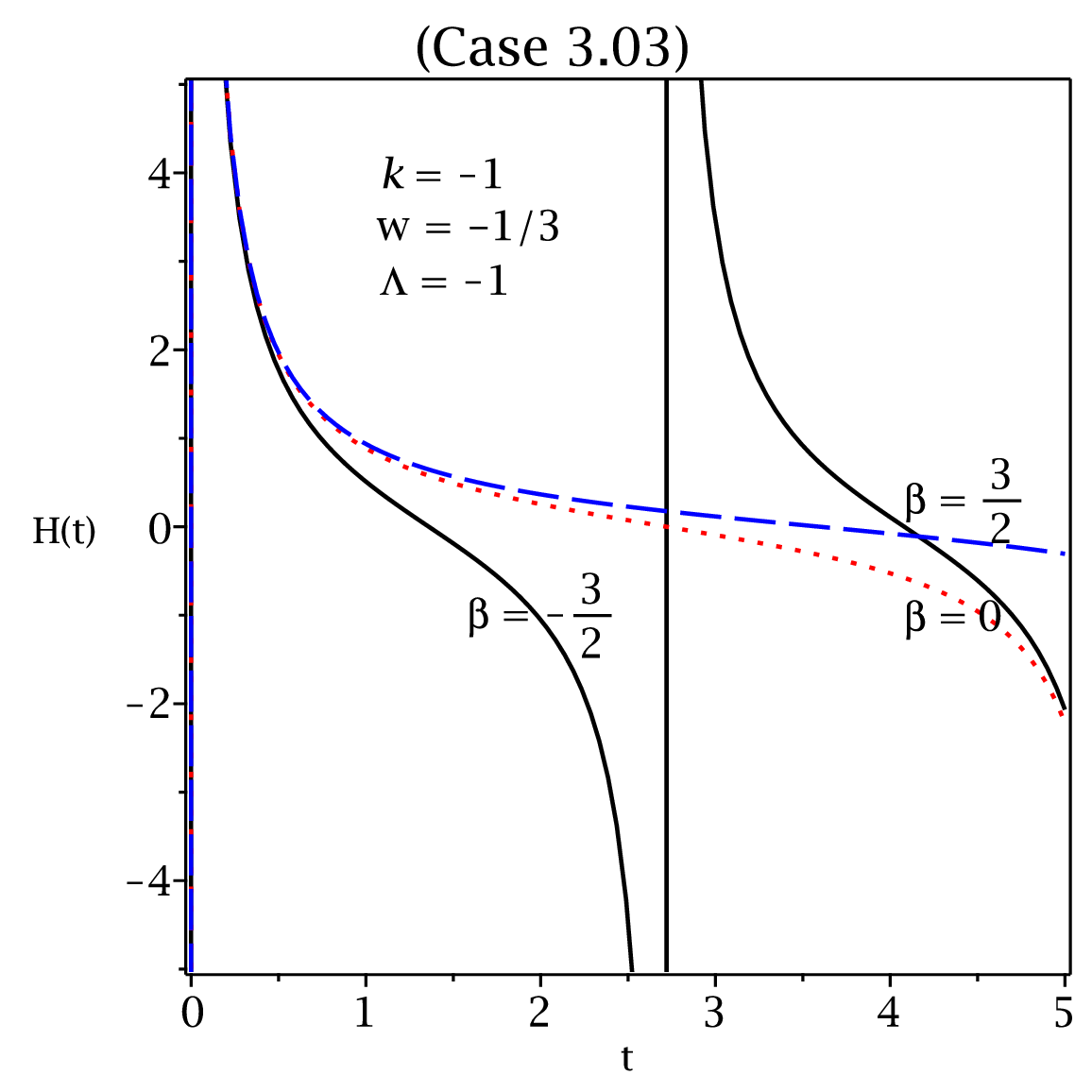}
	\includegraphics[width=3.4cm]{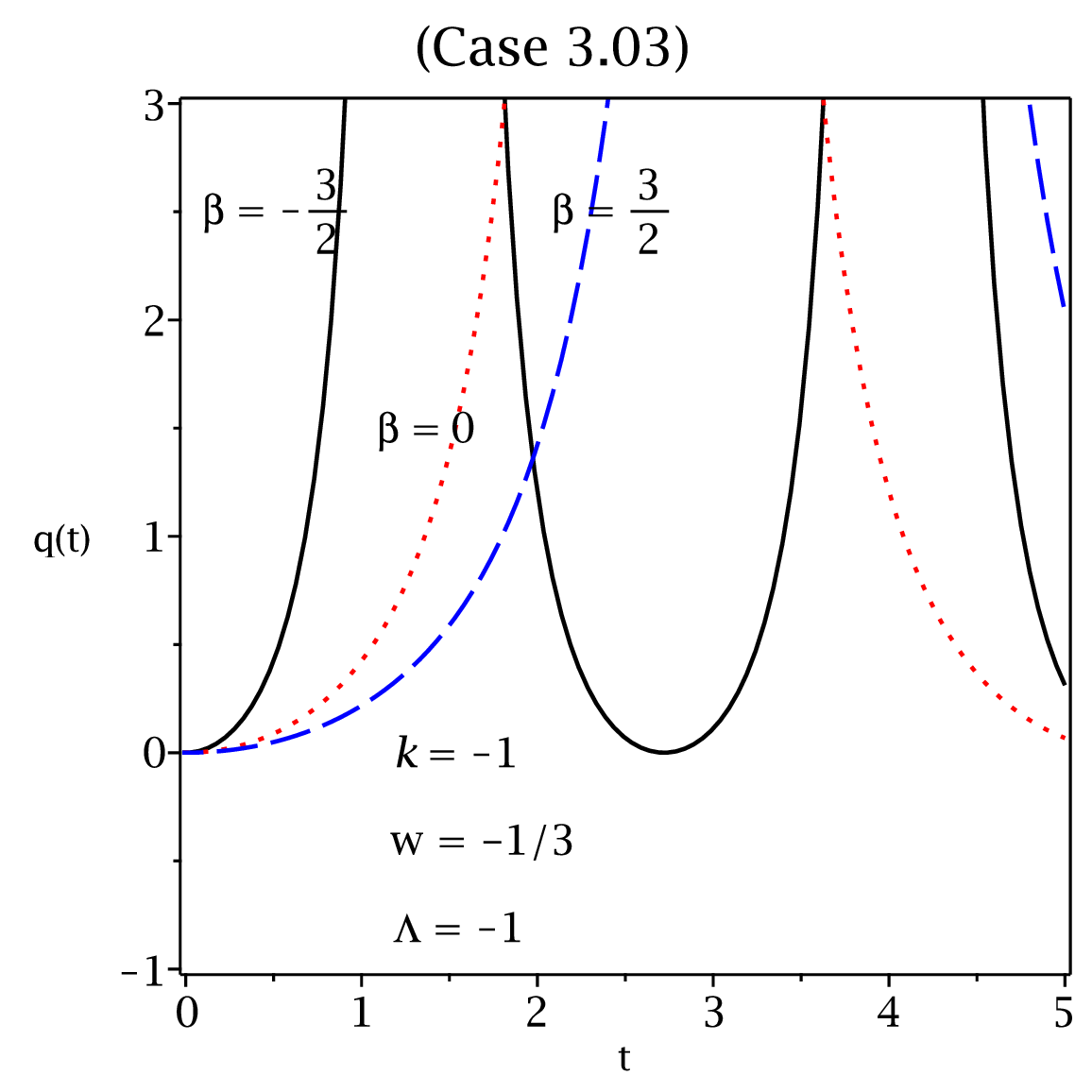}
	\includegraphics[width=3.4cm]{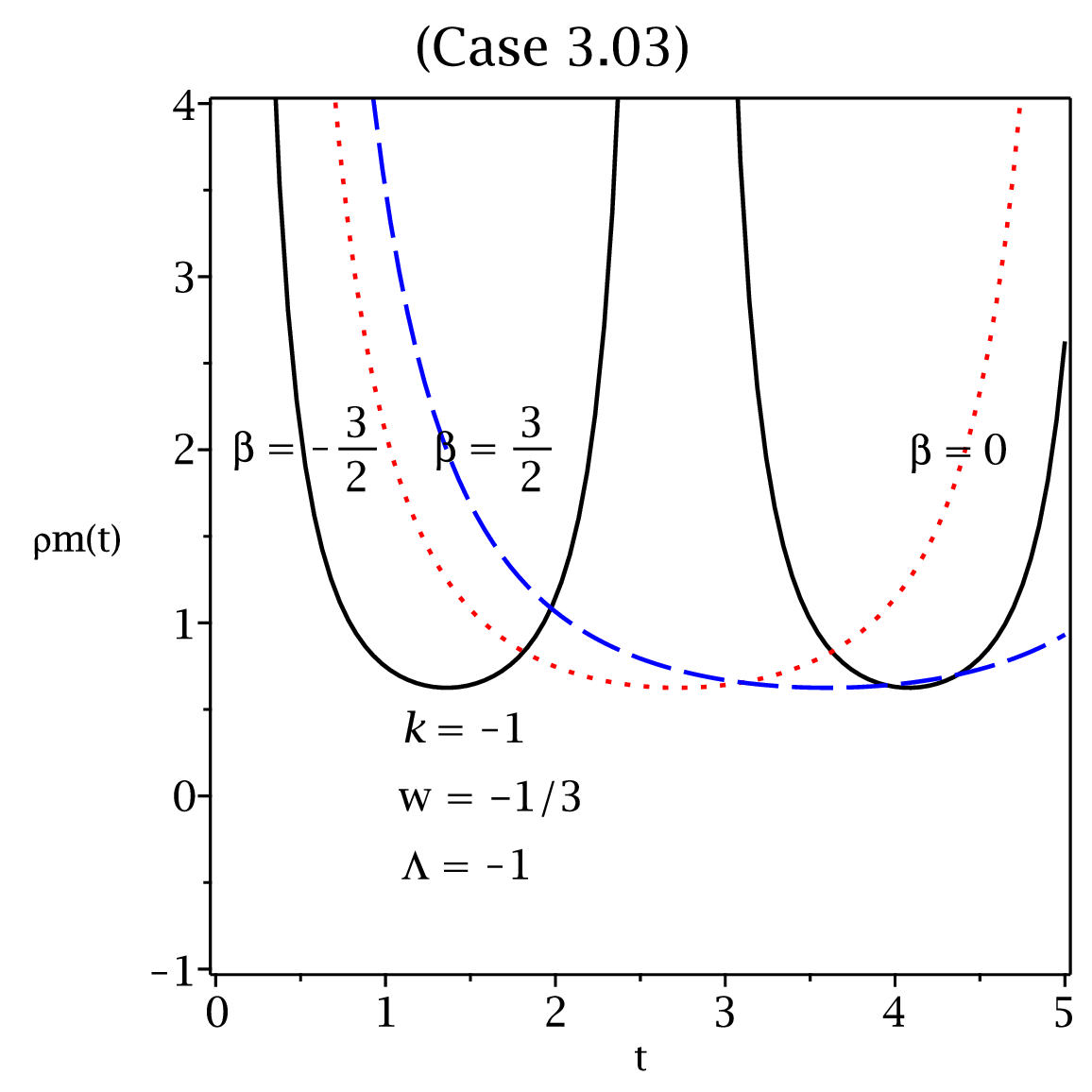}
	\includegraphics[width=3.4cm]{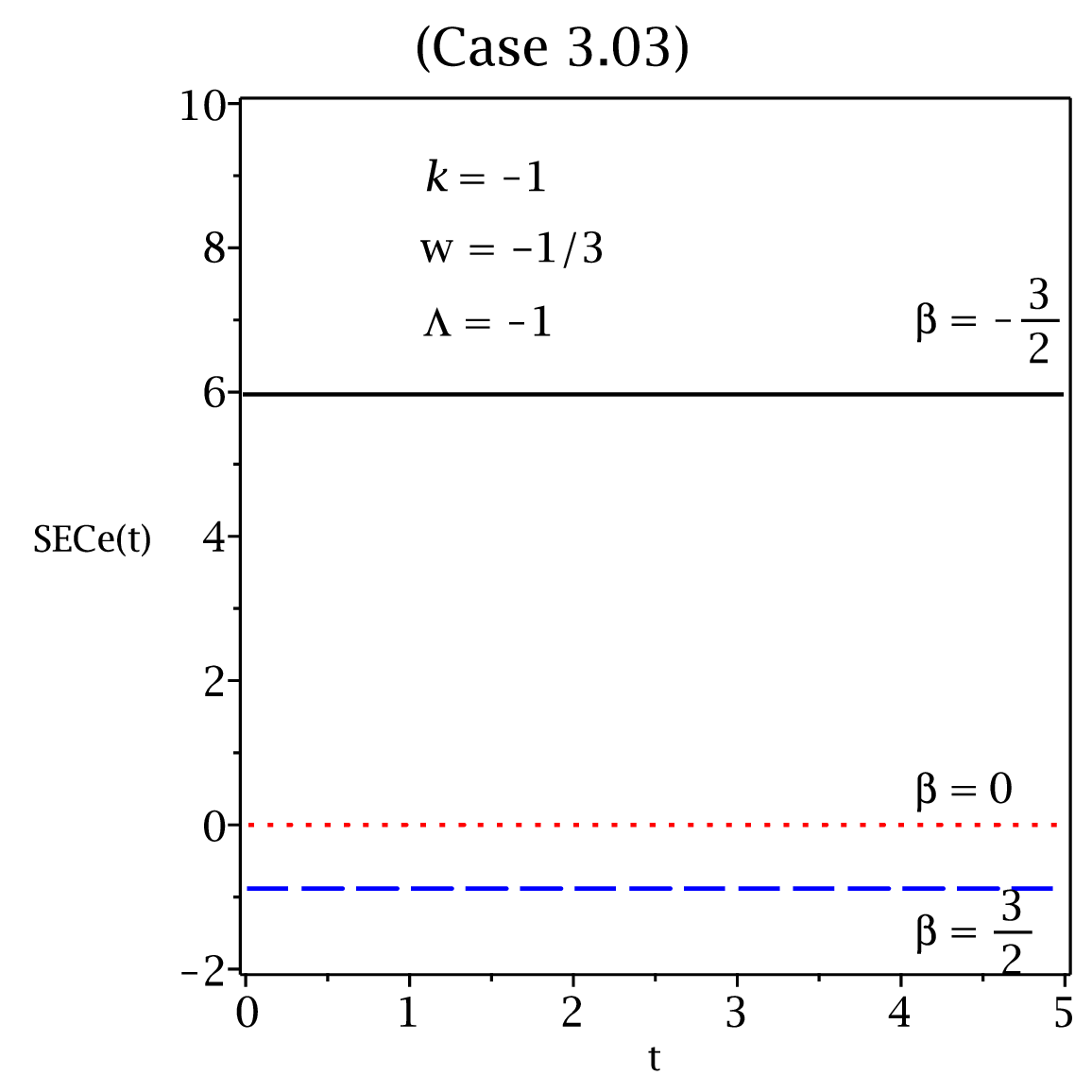}
	\includegraphics[width=3.4cm]{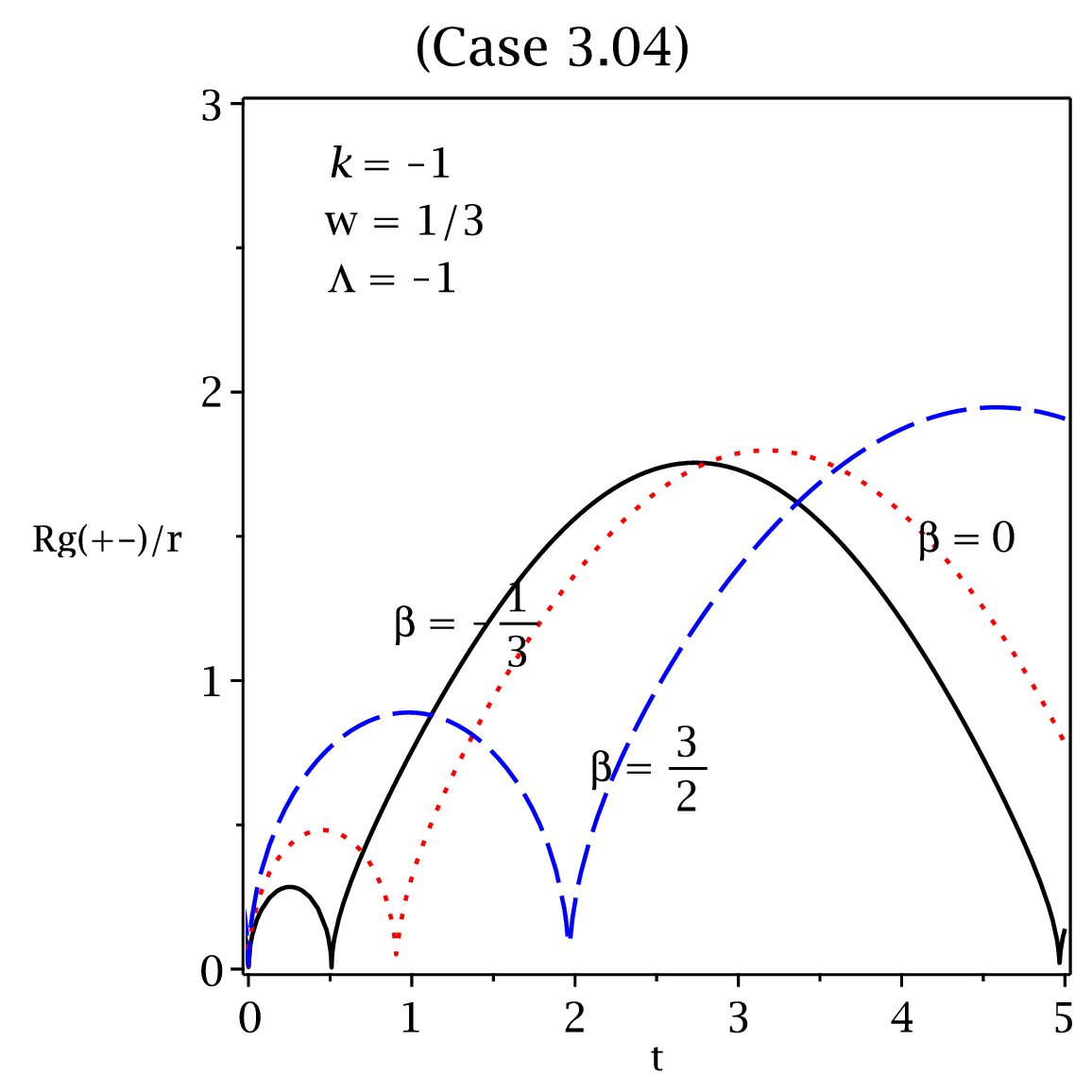}
	\includegraphics[width=3.4cm]{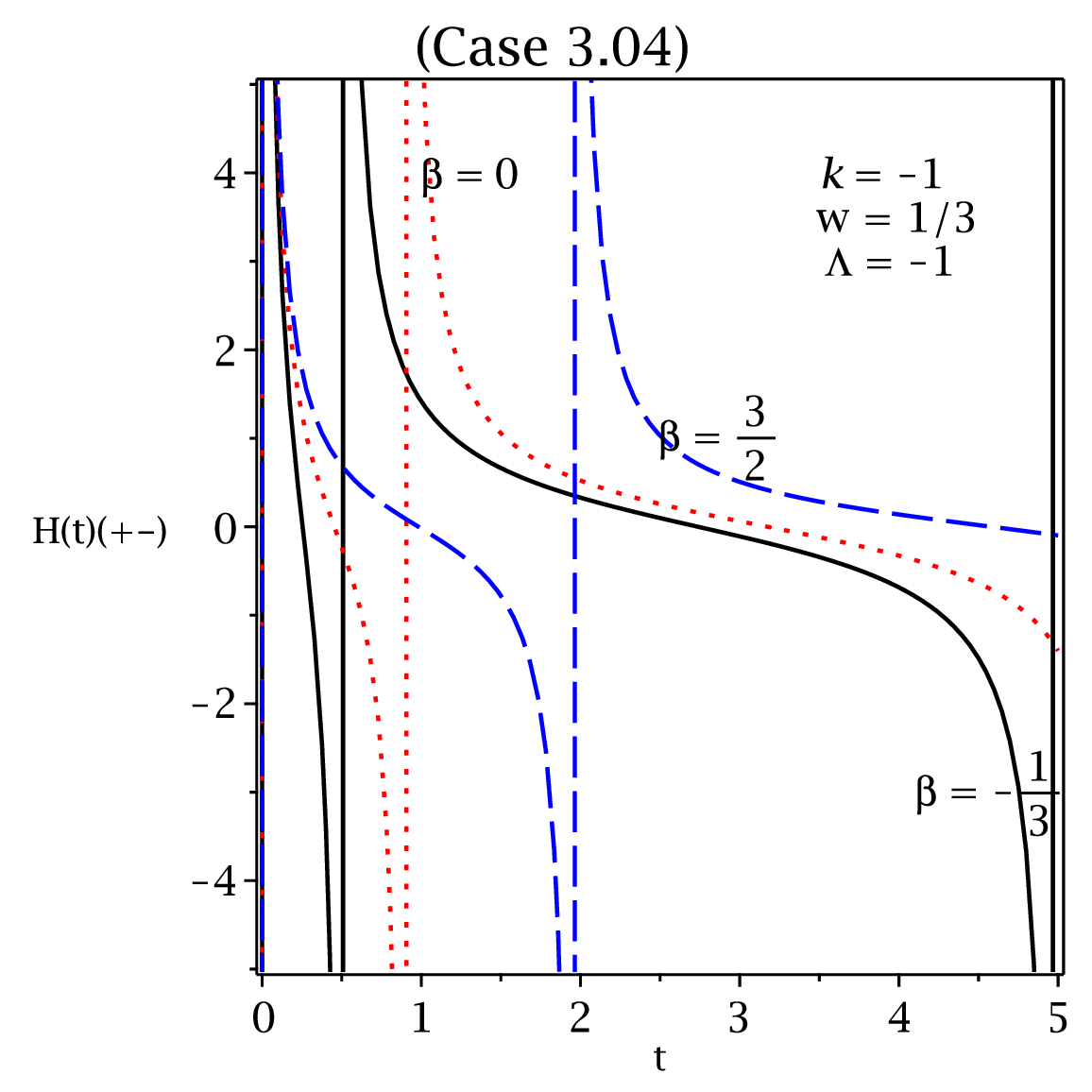}
	\includegraphics[width=3.4cm]{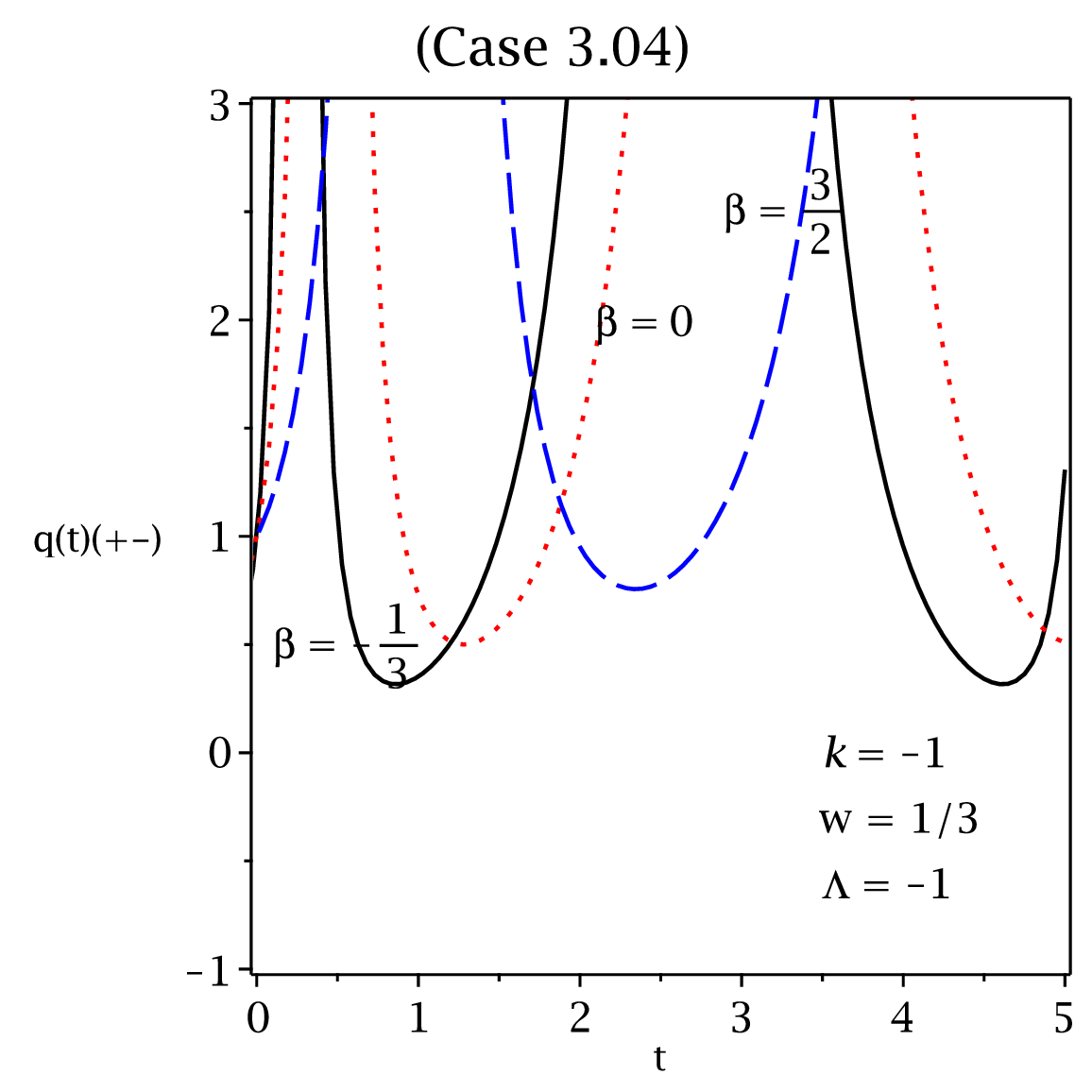}
	\includegraphics[width=3.4cm]{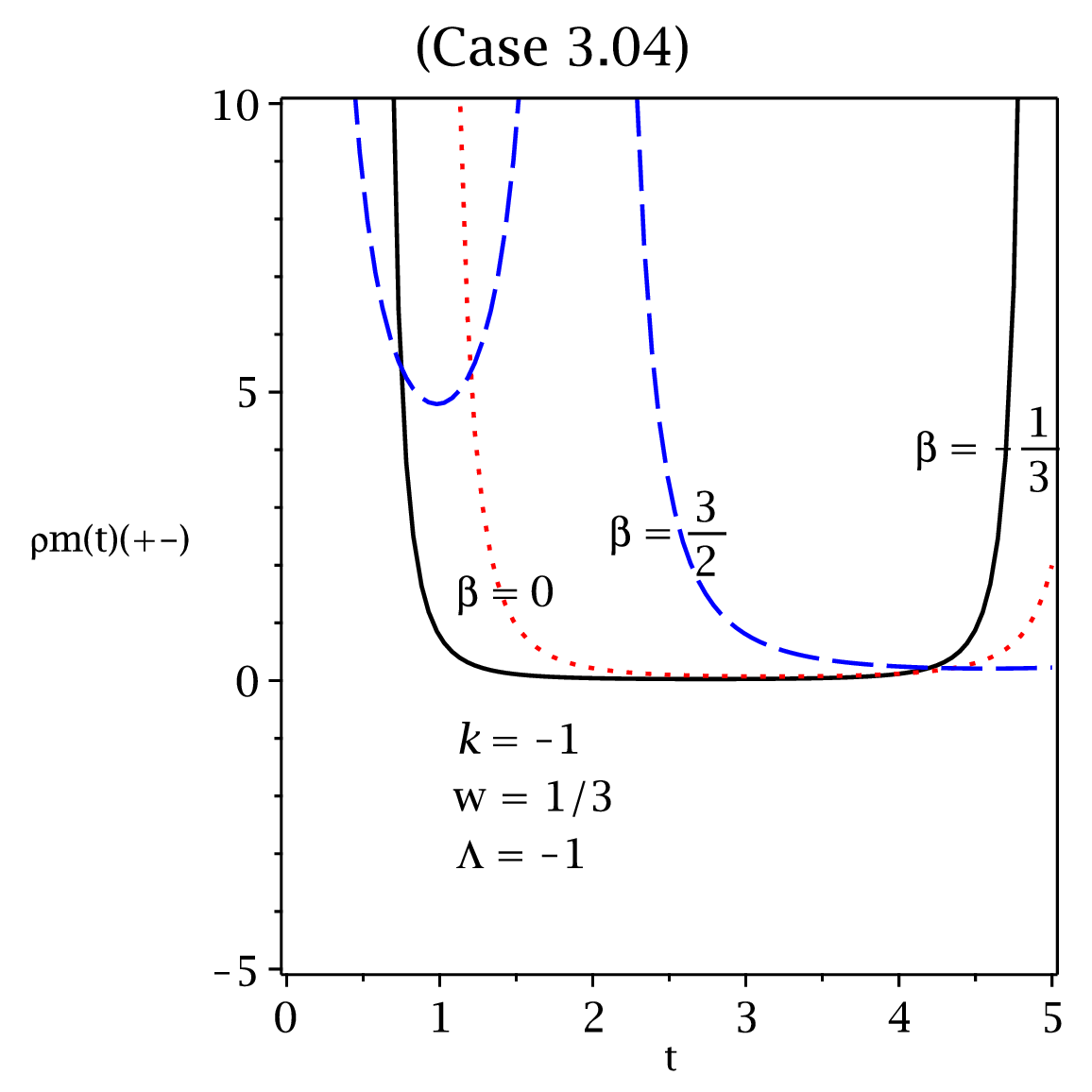}
	\includegraphics[width=3.4cm]{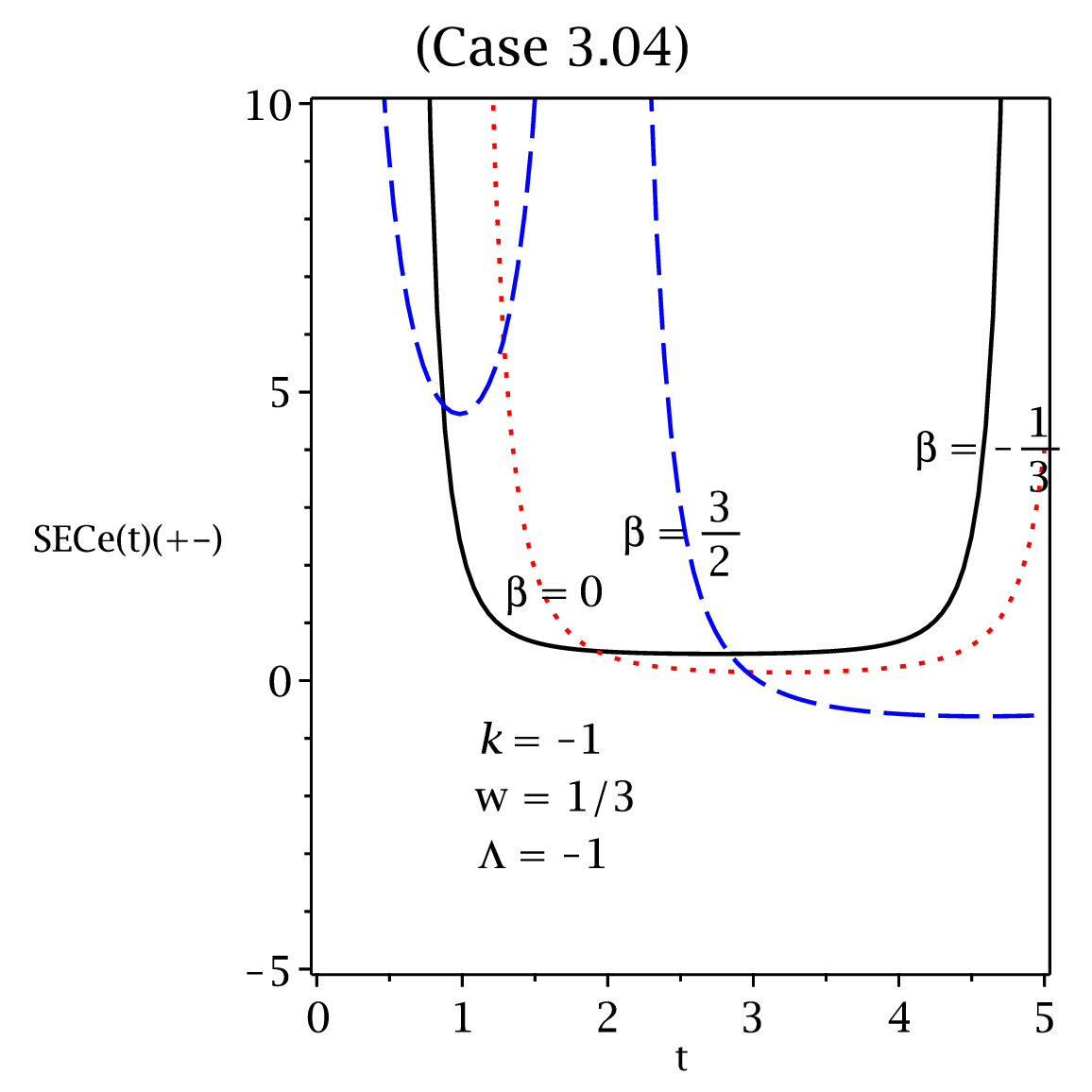}
	\caption{These figures are for $\Lambda<0$ and $k=-1$.
		These figures represent the quantities $R_g$ (geometrical radius), 
		$H(t)$ (Hubble parameter) and $q(t)$ (deceleration parameter) $\rho_m(t)$ 
		(energy density of the aether fluid) and $SEC_{e} \equiv SEC_{\rm eff}$ 
		(strong energy condition for the effective fluid) for the different
		values of $\beta=-3/2$ (black solid line), $\beta=0$ (red dotted line), 
		$\beta=3/2$ (blue dashed line). Assuming that $8 \pi G=1$ and
		$R_g(t=0)=0$. Assuming also that $C_1=1$, $C_2=0$ (Cases 3.01 and 3.04); 
		$C_1=2$, $C_2=0$ (Case 3.02); $C_1=2$, $C_2=2$ (Case 3.03).}
	\label{Figure-301-304}
\end{minipage}	
\end{figure}

\begin{figure}[!htp]
\begin{minipage}{175 mm}
	\centering	
	\includegraphics[width=3.4cm]{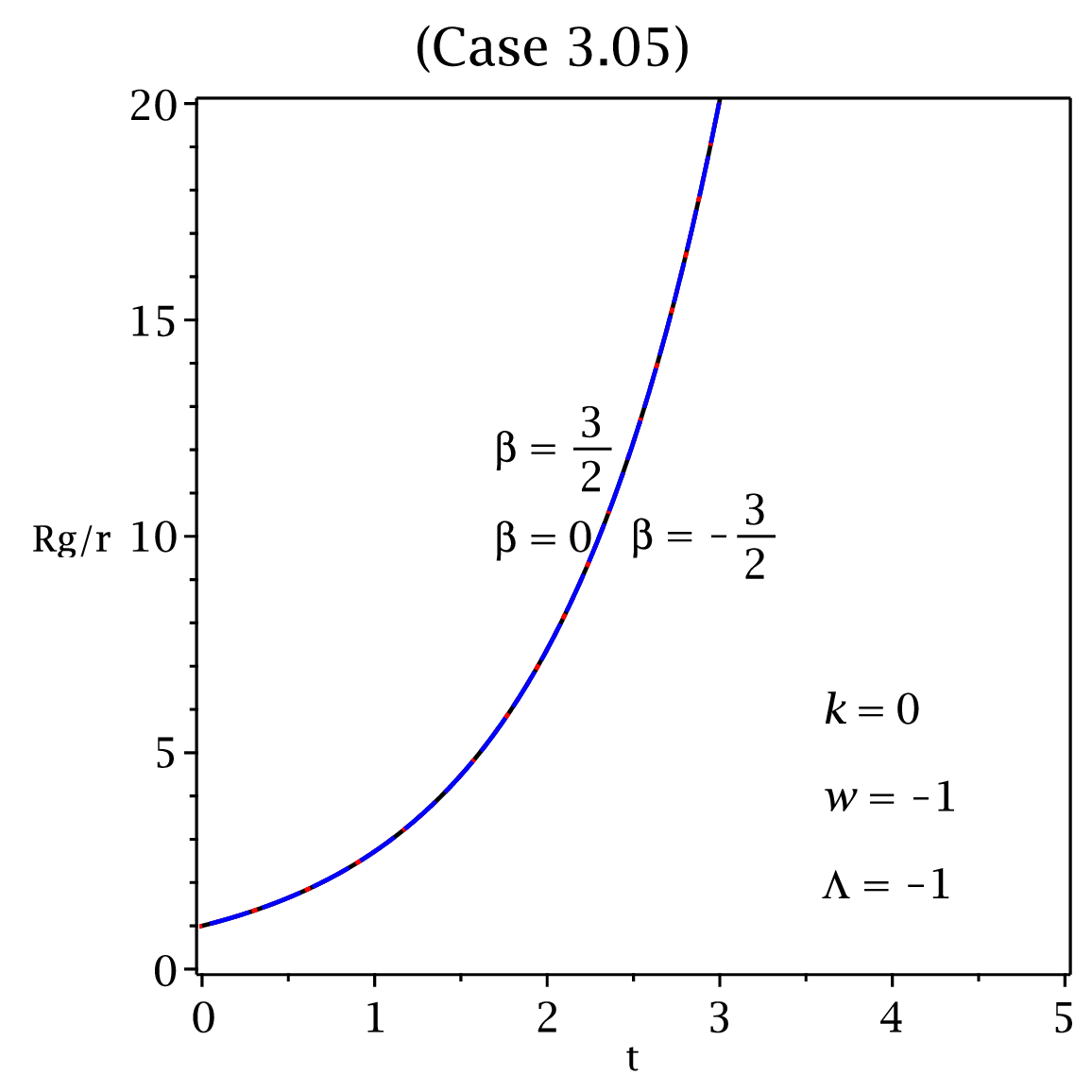}
	\includegraphics[width=3.4cm]{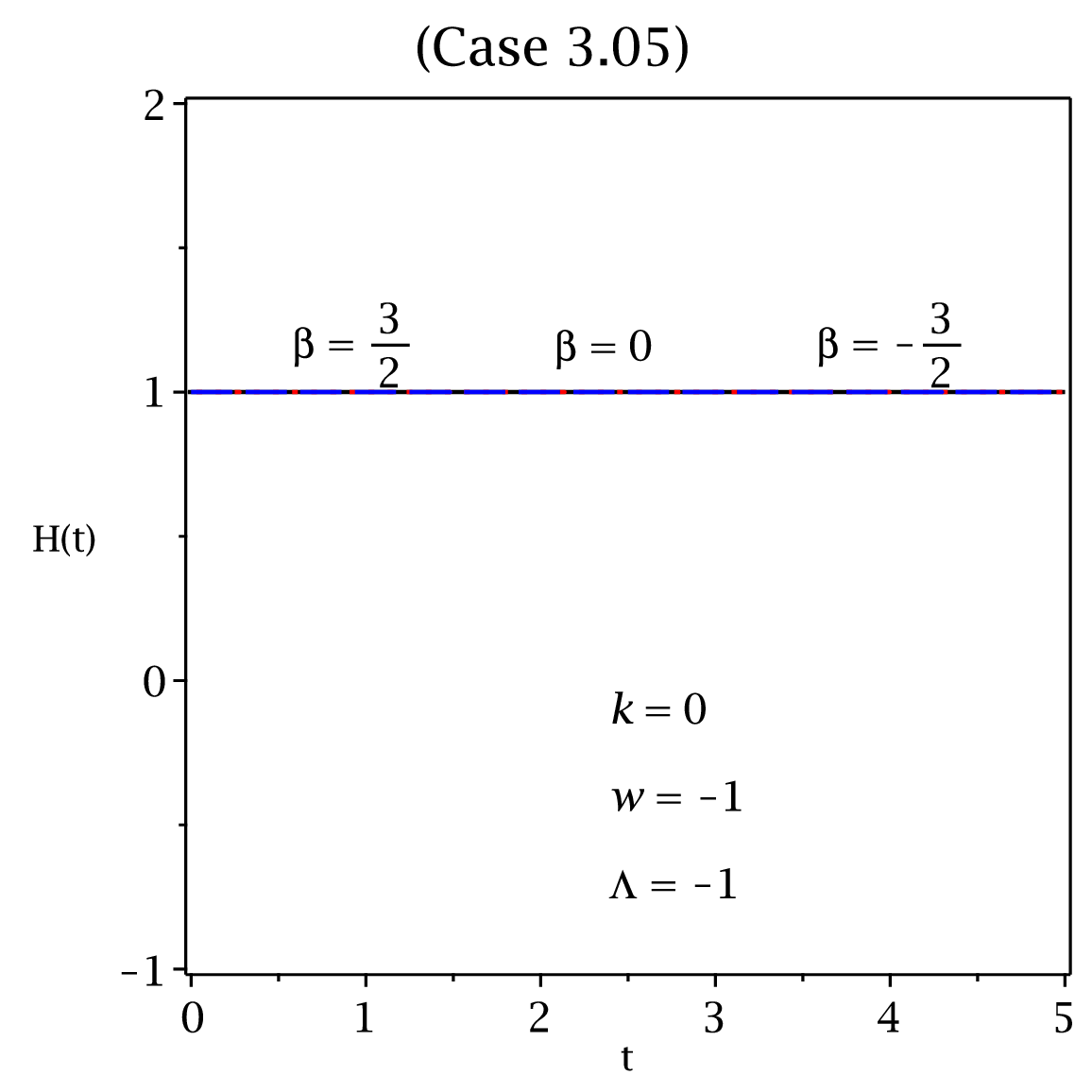}
	\includegraphics[width=3.4cm]{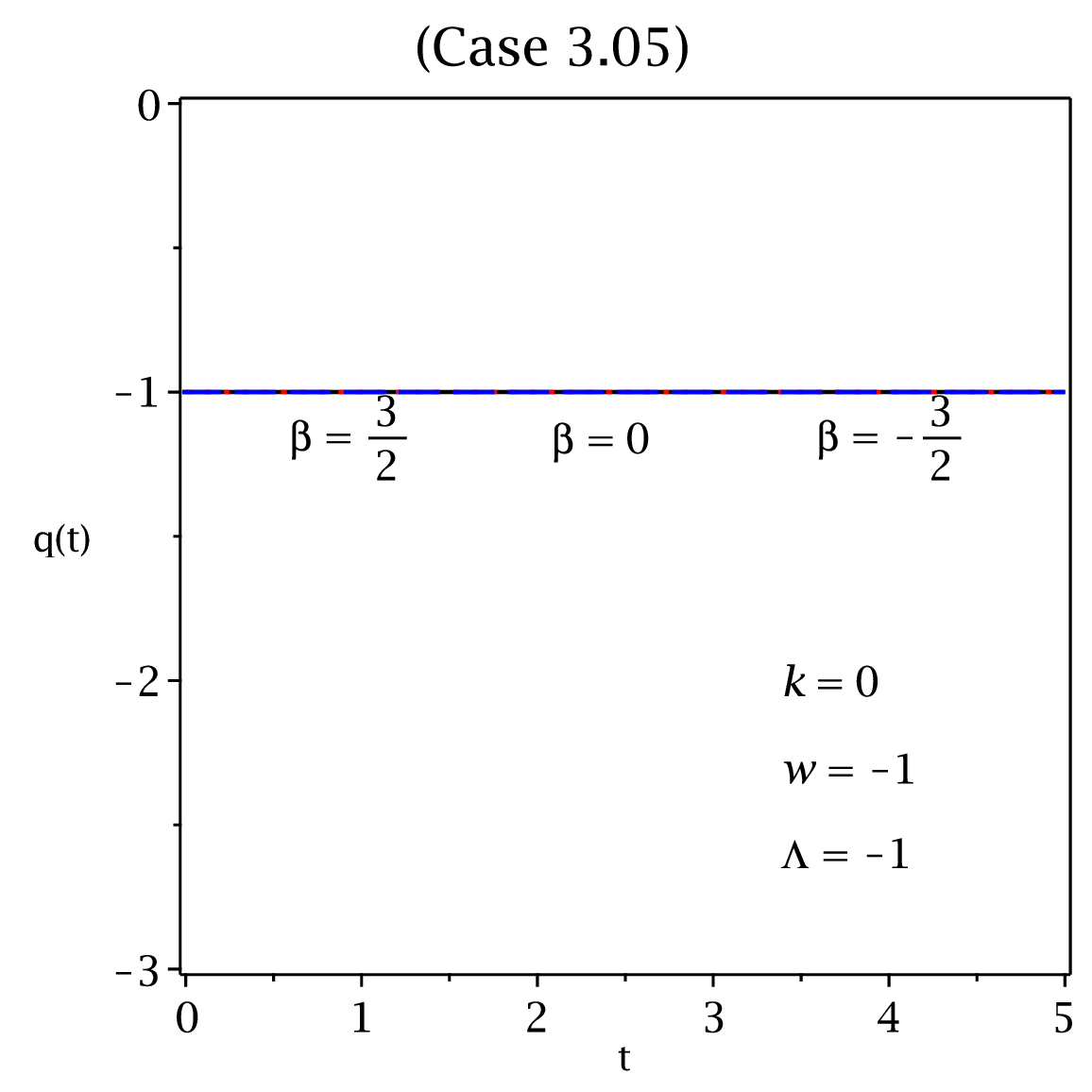}
	\includegraphics[width=3.4cm]{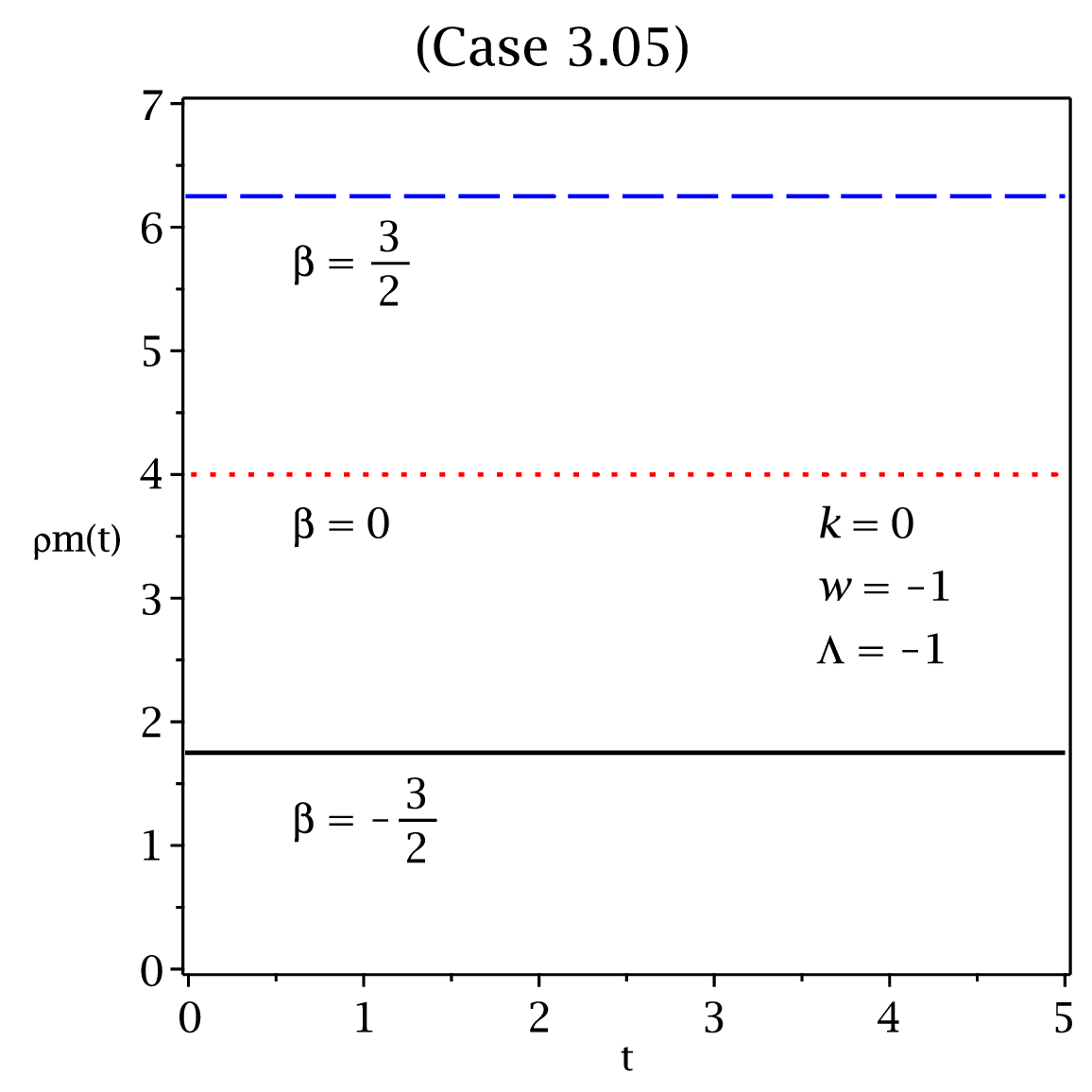}
	\includegraphics[width=3.4cm]{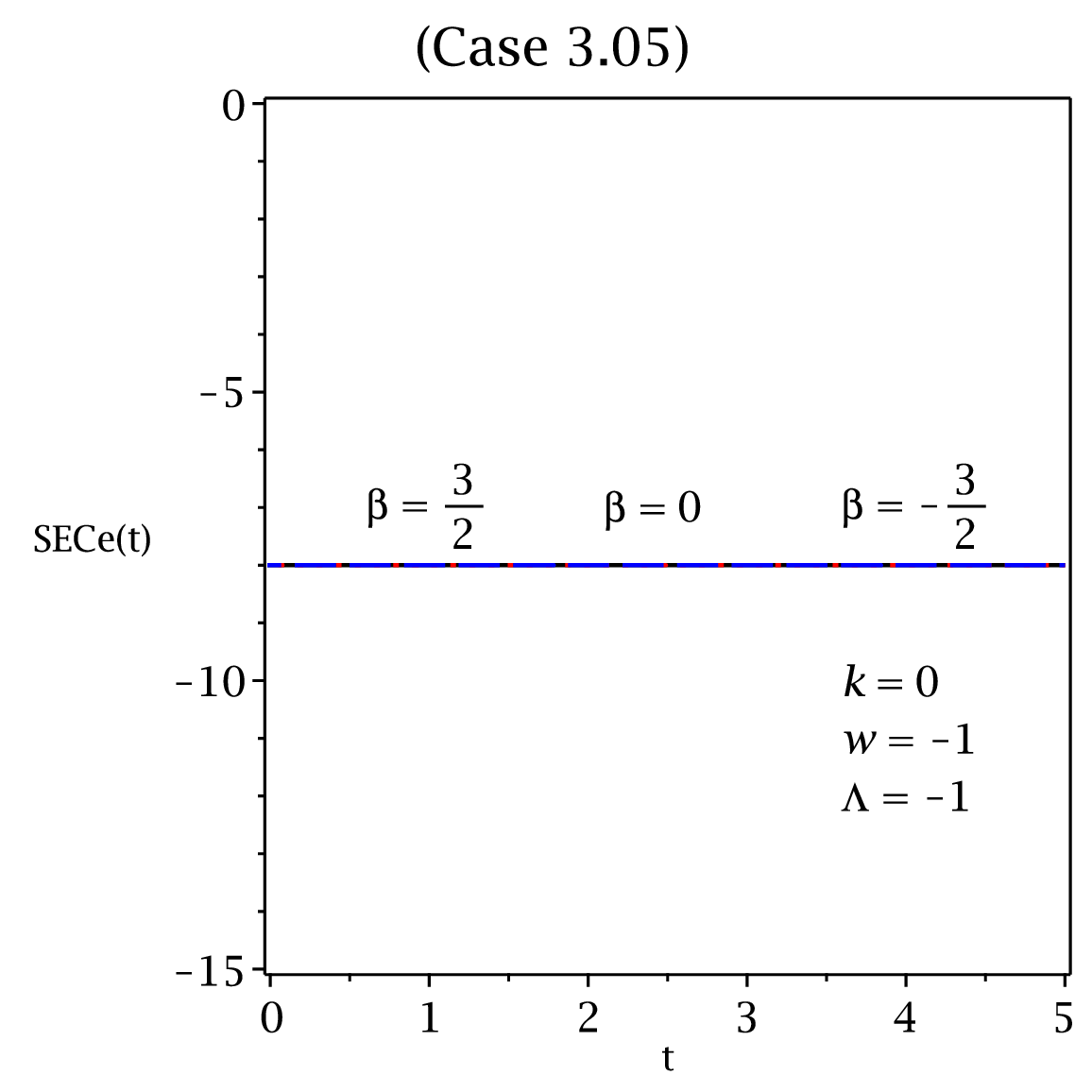}
	\includegraphics[width=3.4cm]{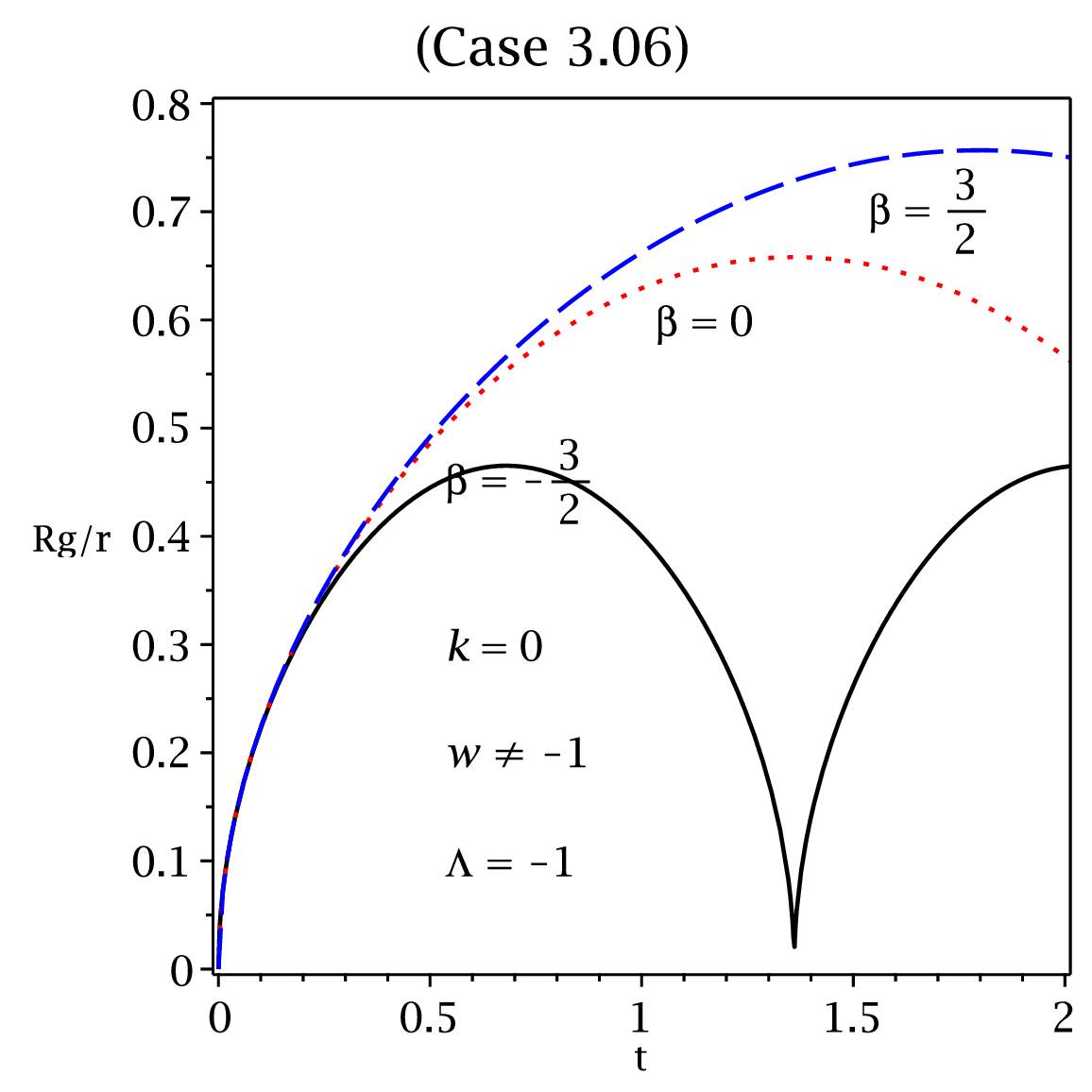}
	\includegraphics[width=3.4cm]{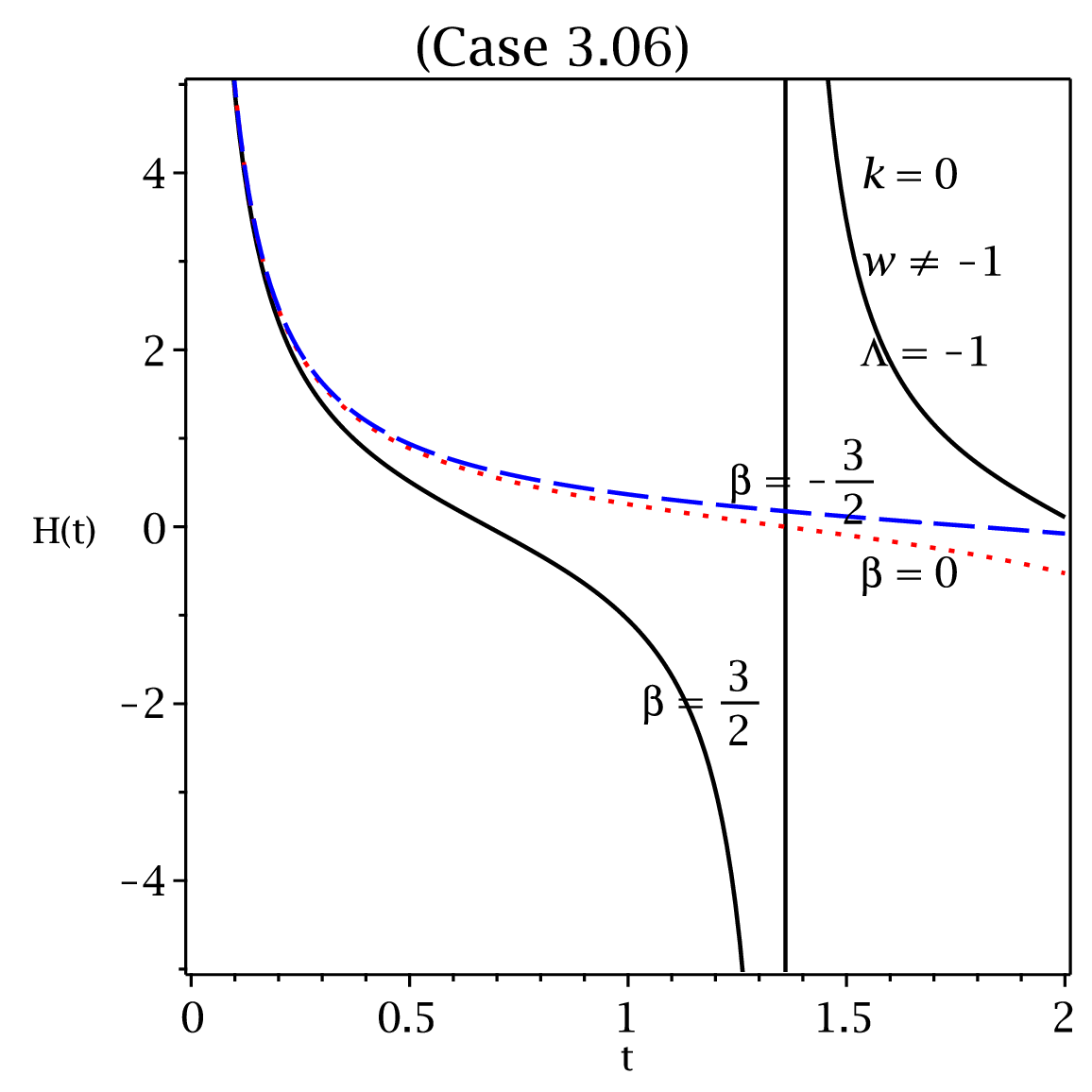}
	\includegraphics[width=3.4cm]{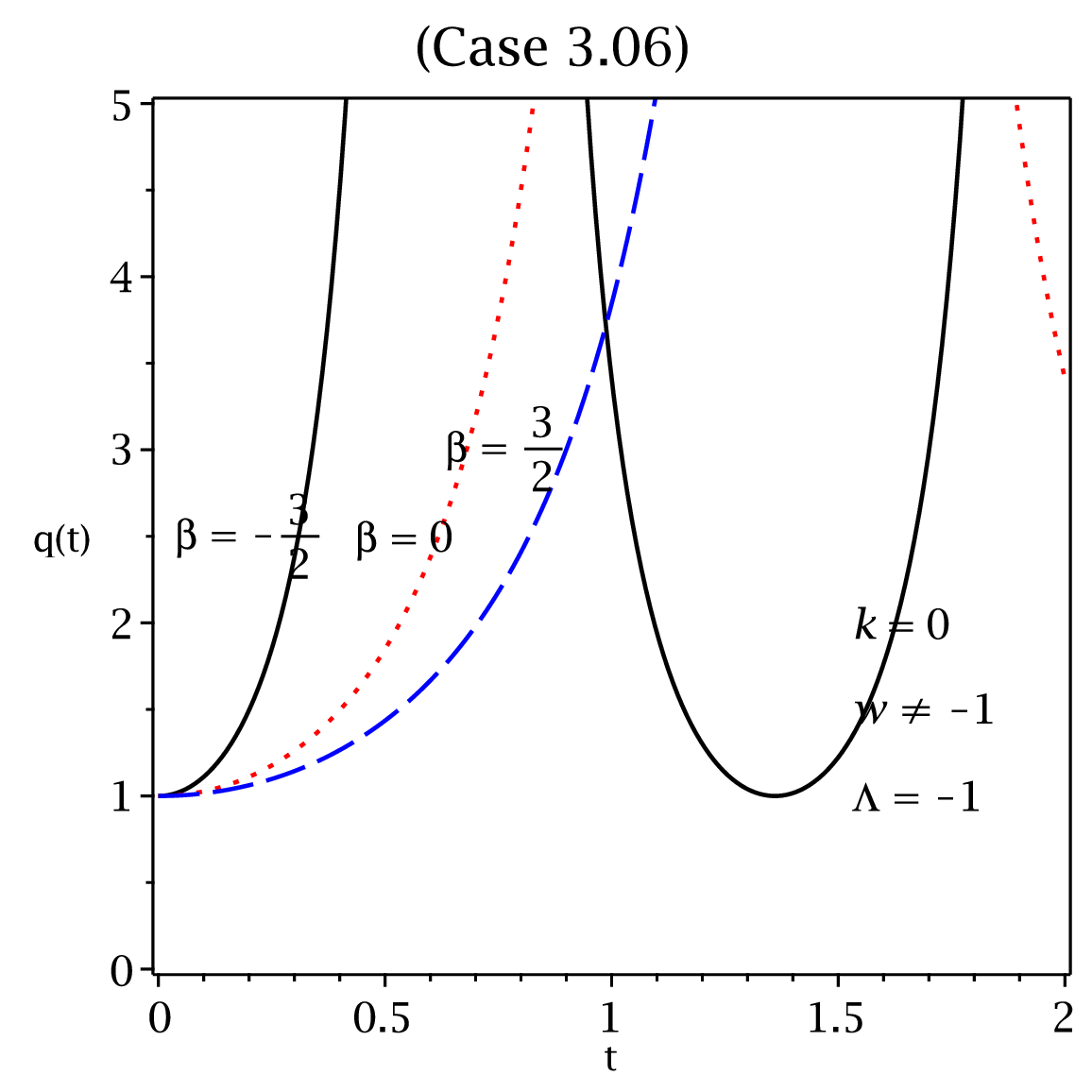}
	\includegraphics[width=3.4cm]{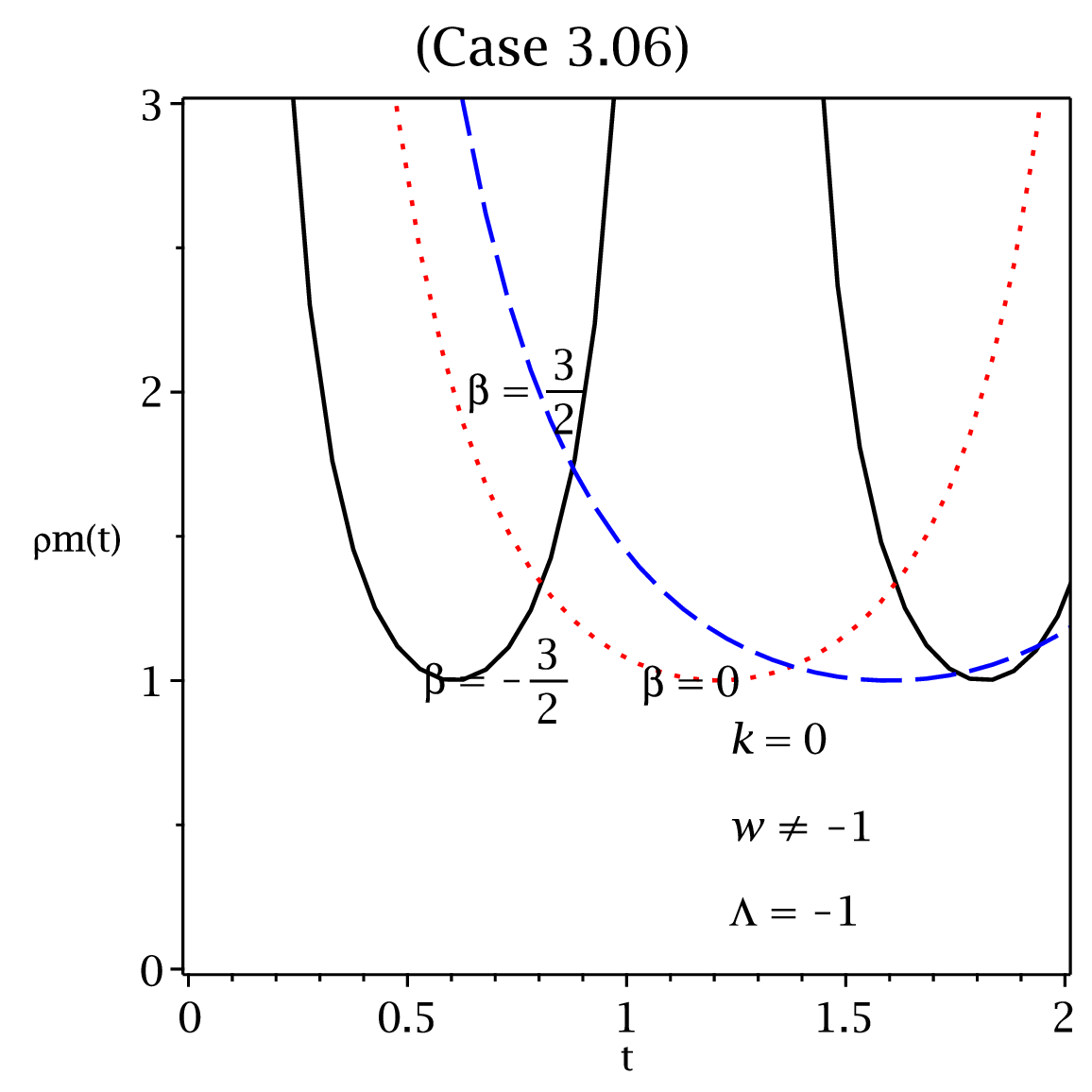}
	\includegraphics[width=3.4cm]{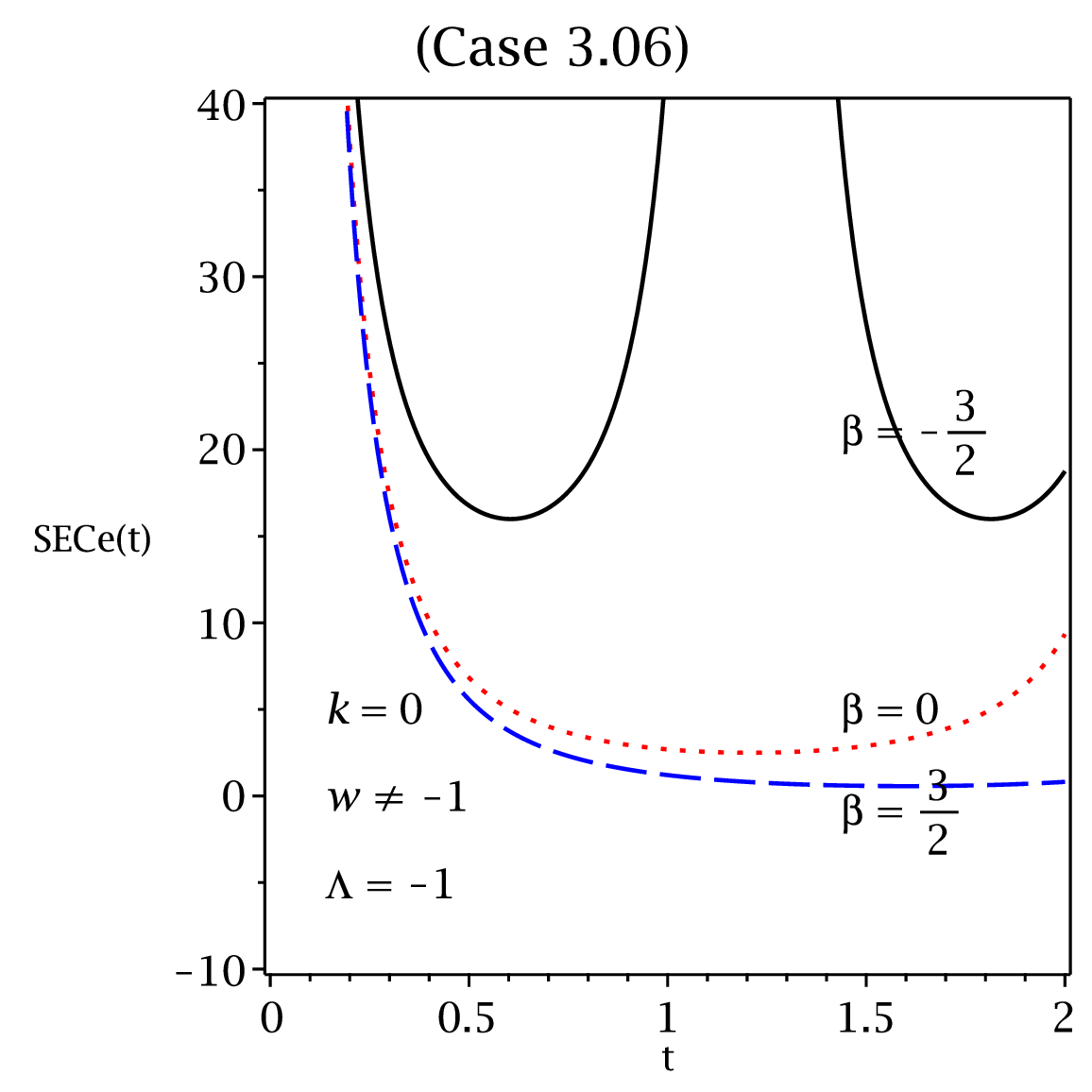}
	\includegraphics[width=3.4cm]{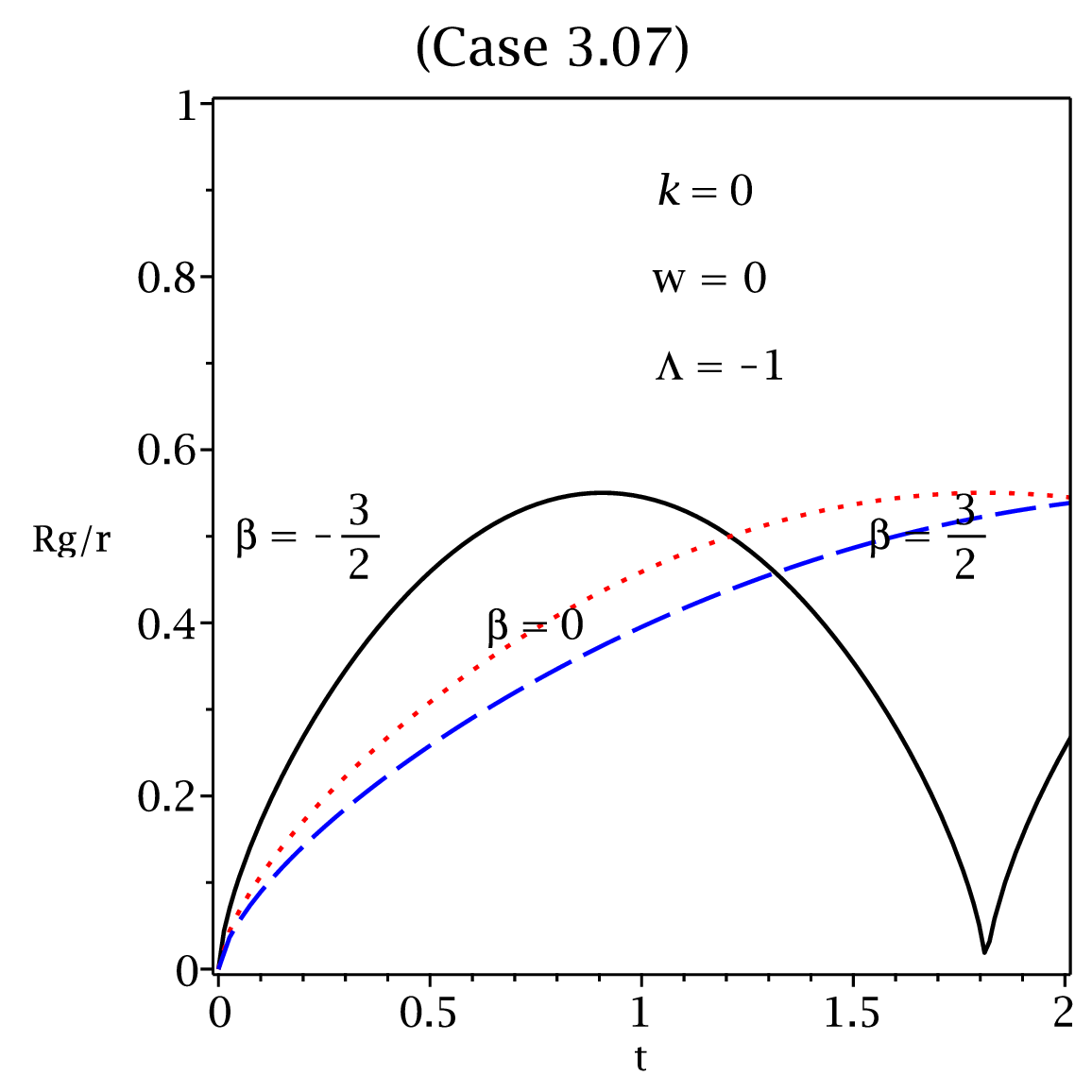}
	\includegraphics[width=3.4cm]{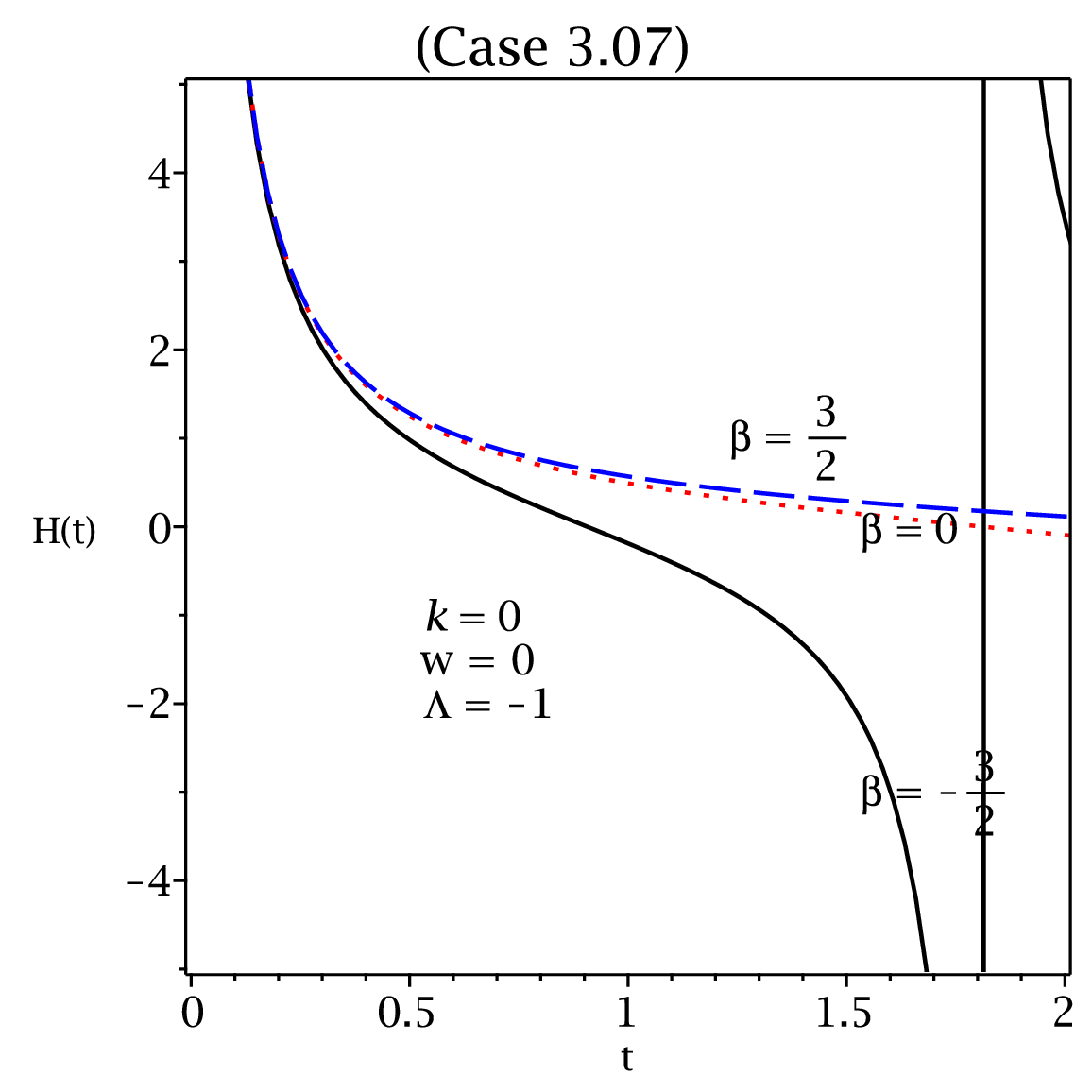}
	\includegraphics[width=3.4cm]{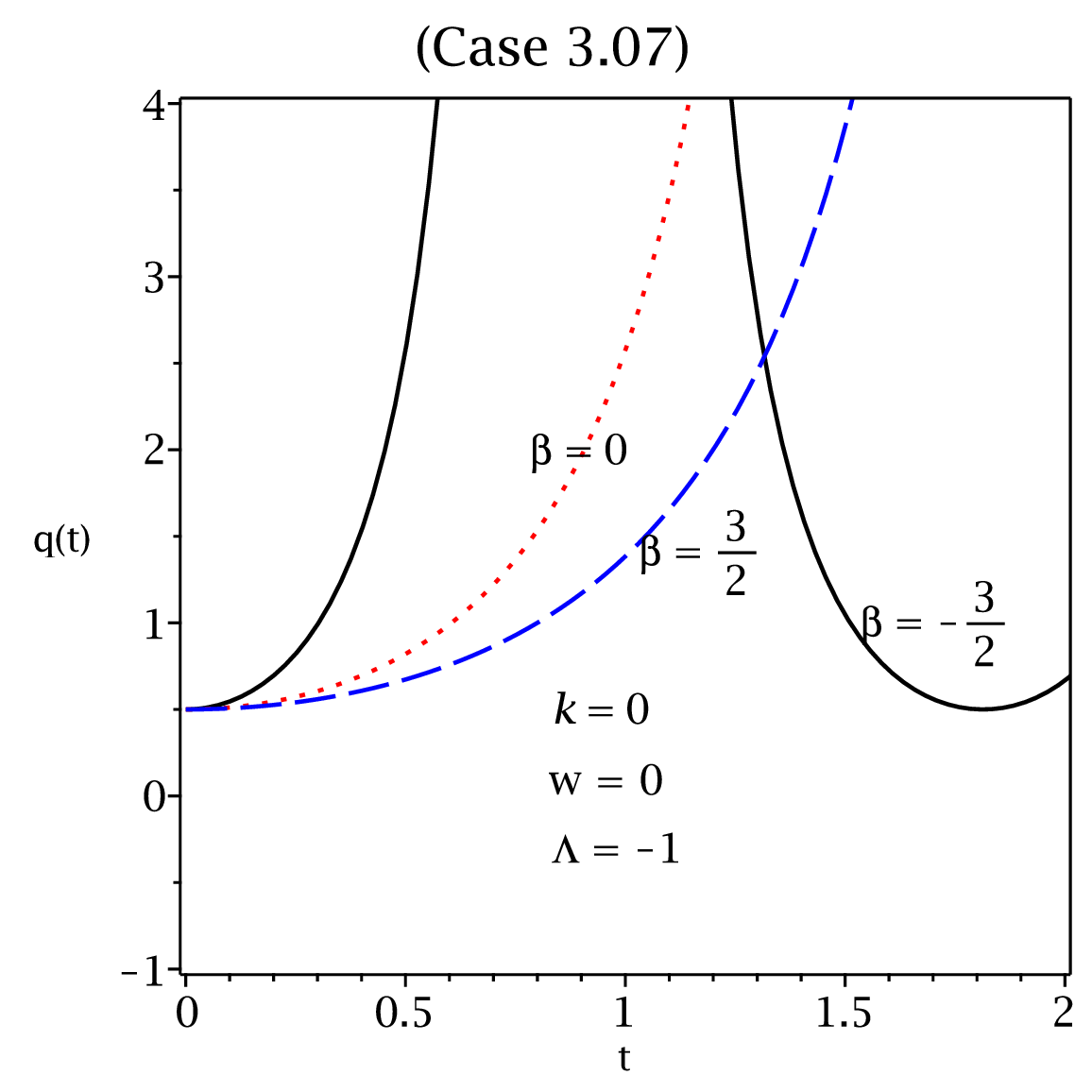}
	\includegraphics[width=3.4cm]{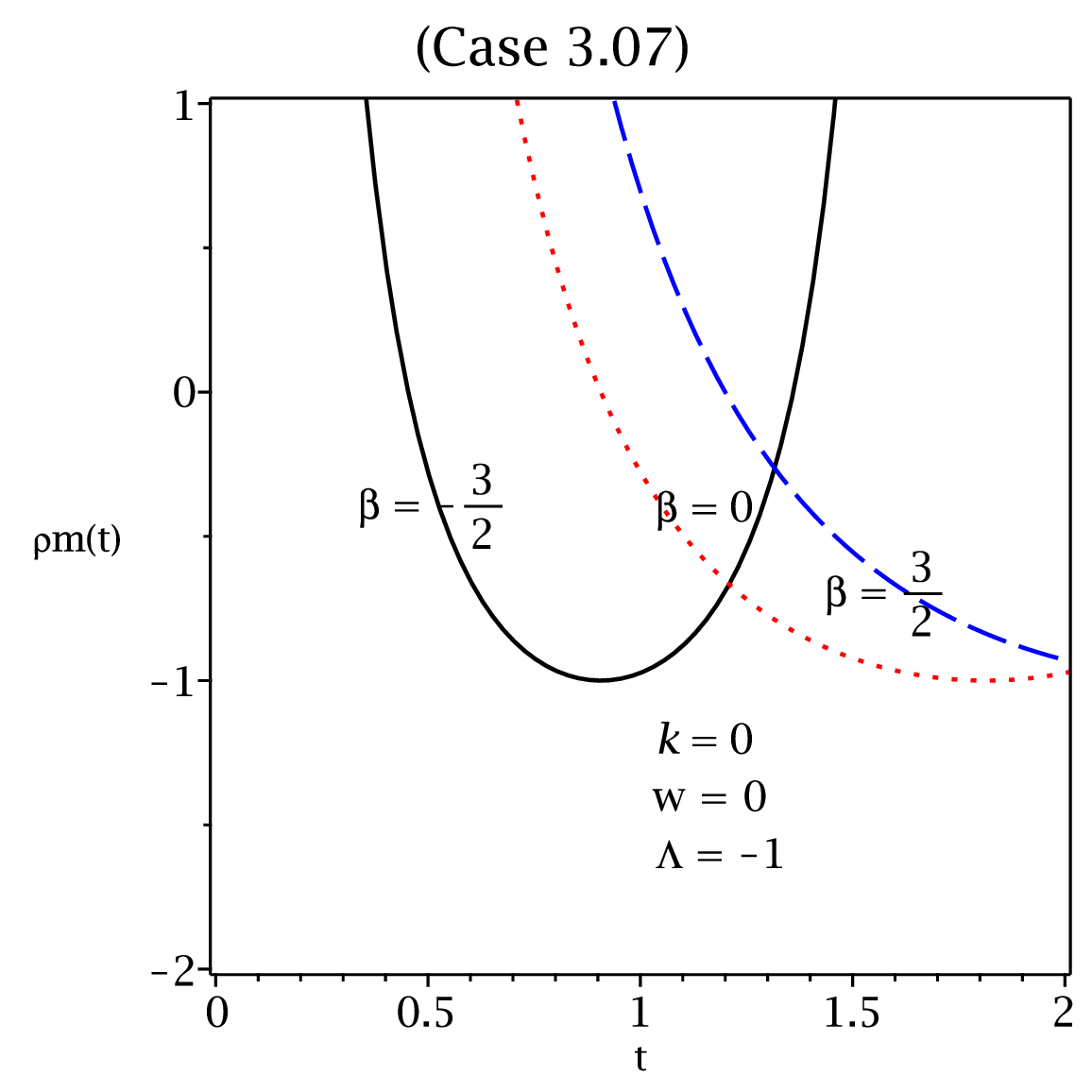}
	\includegraphics[width=3.4cm]{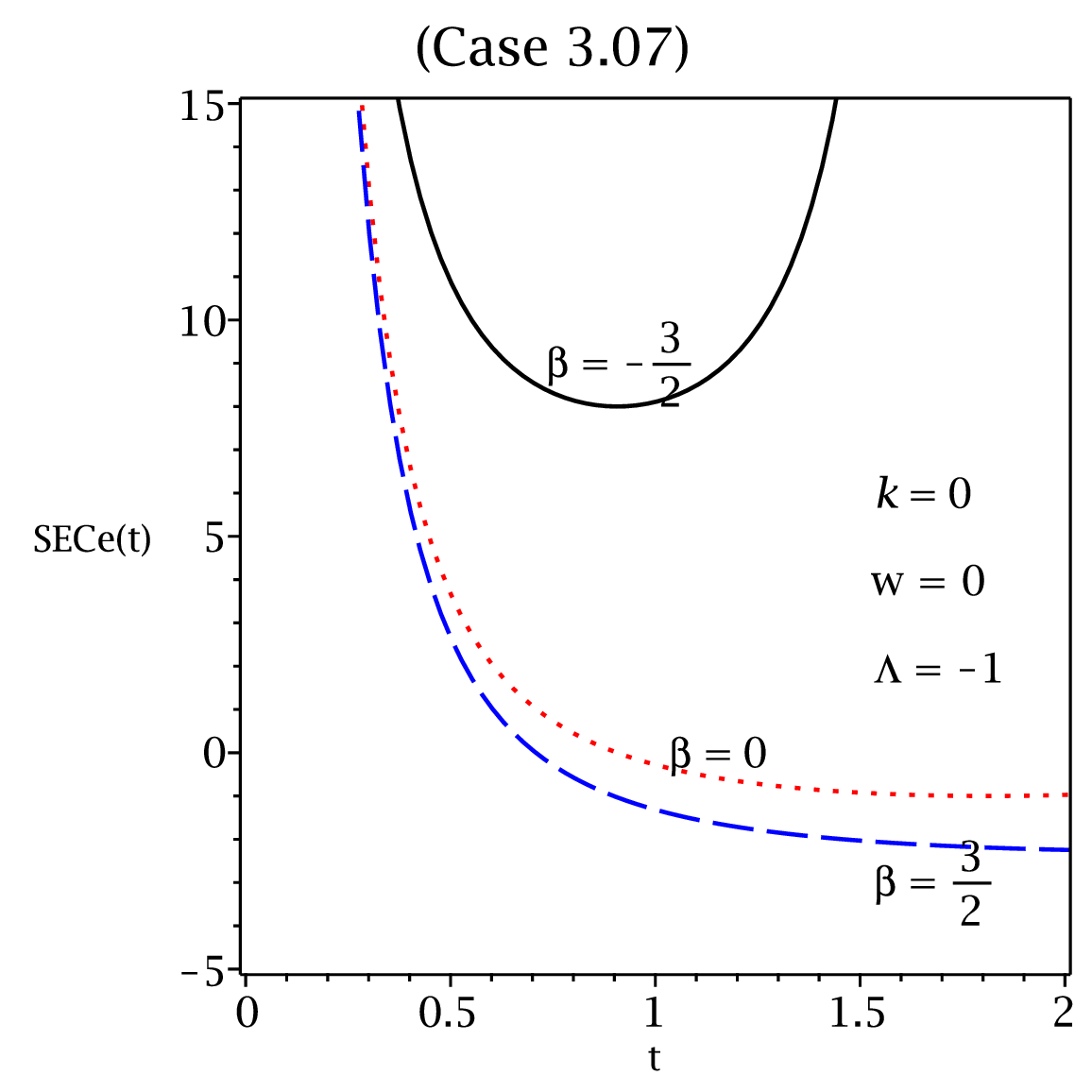}
	\includegraphics[width=3.4cm]{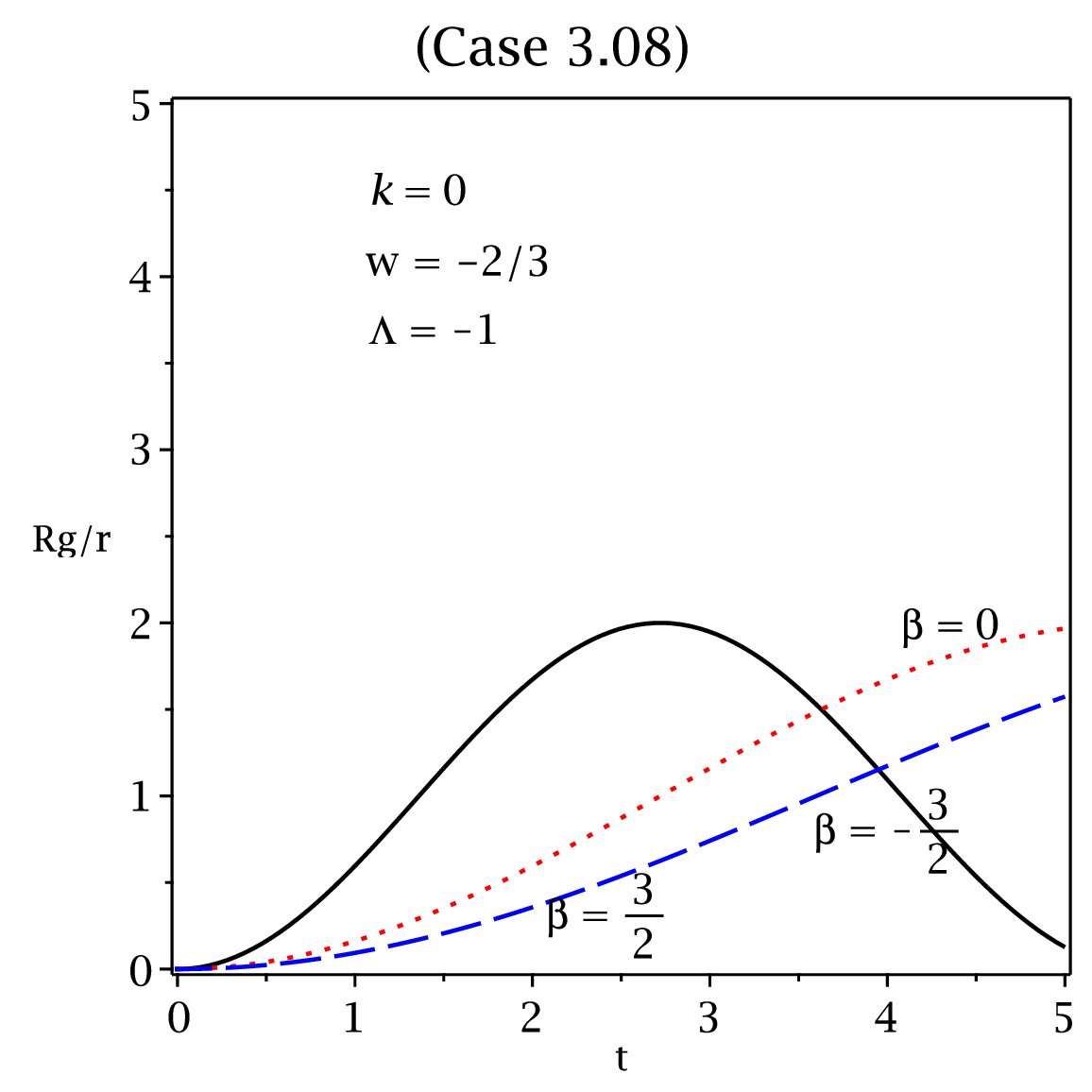}
	\includegraphics[width=3.4cm]{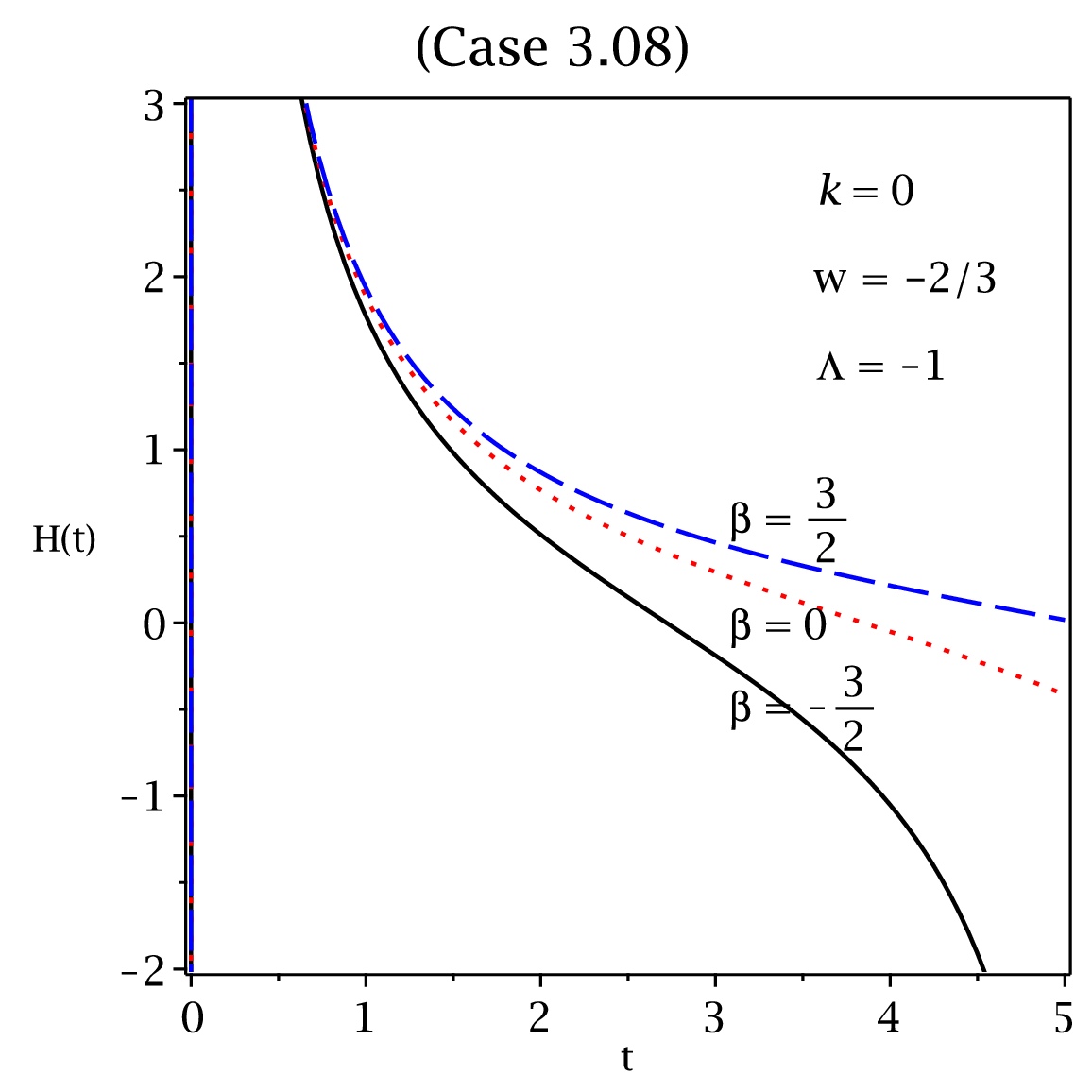}
	\includegraphics[width=3.4cm]{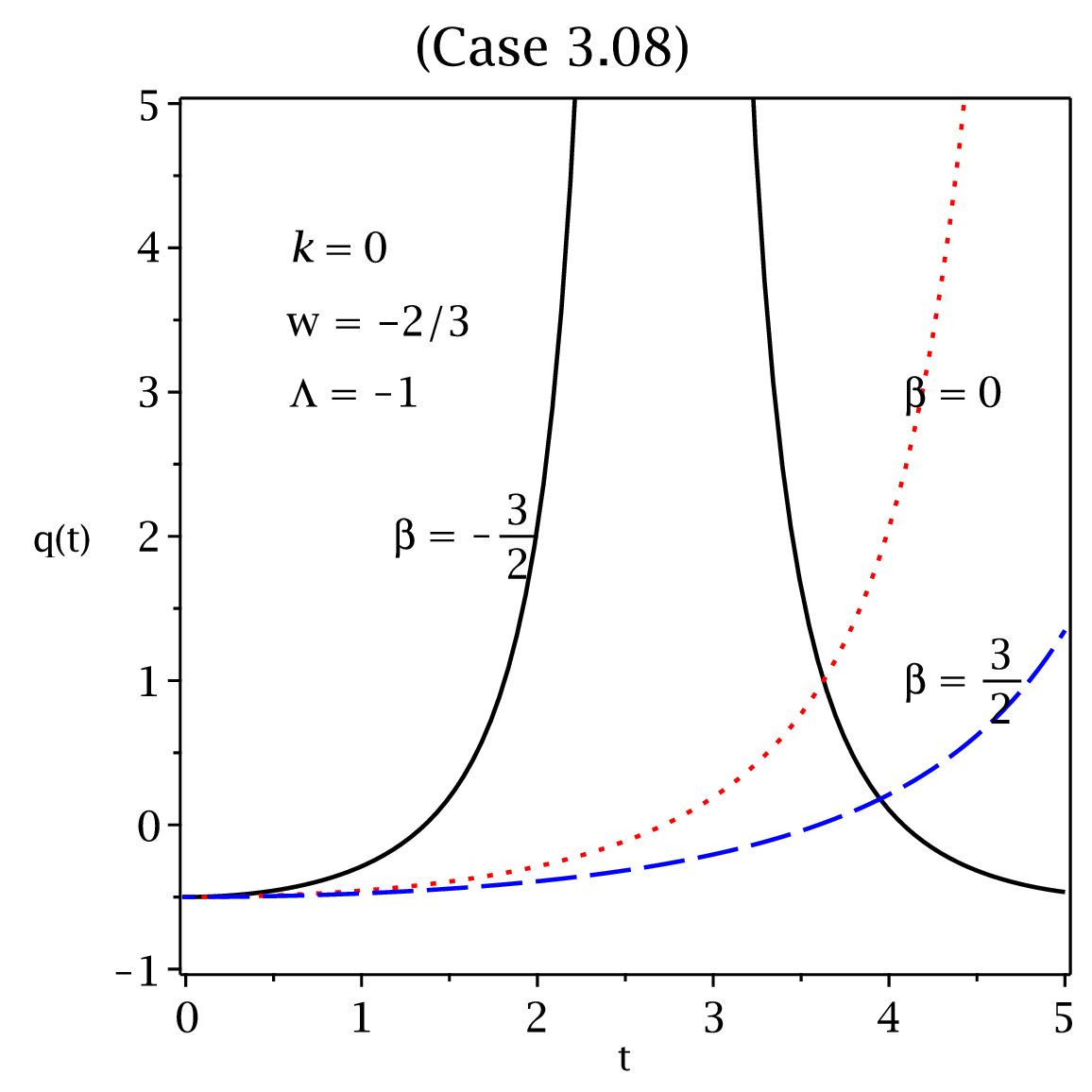}
	\includegraphics[width=3.4cm]{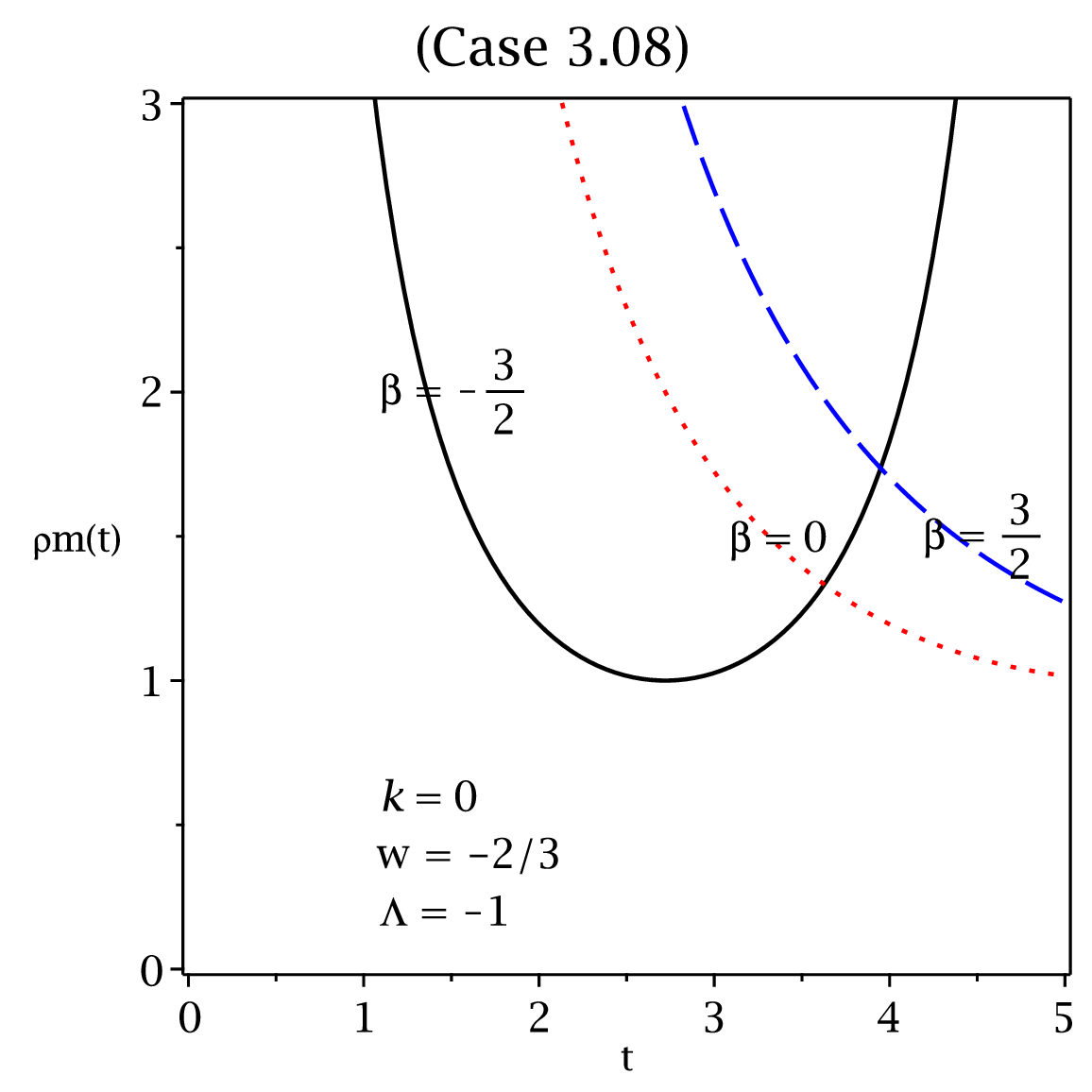}
	\includegraphics[width=3.4cm]{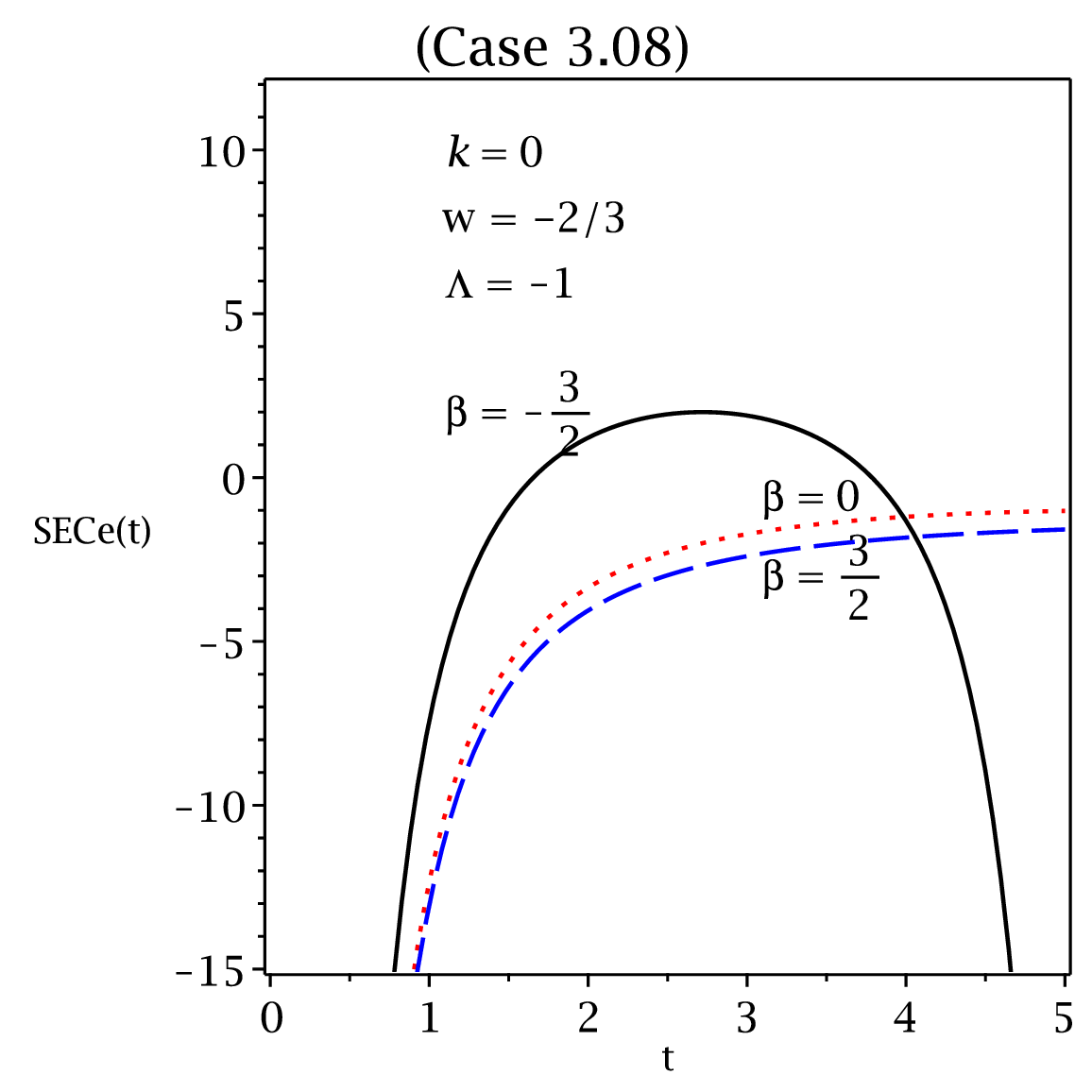}
	\includegraphics[width=3.4cm]{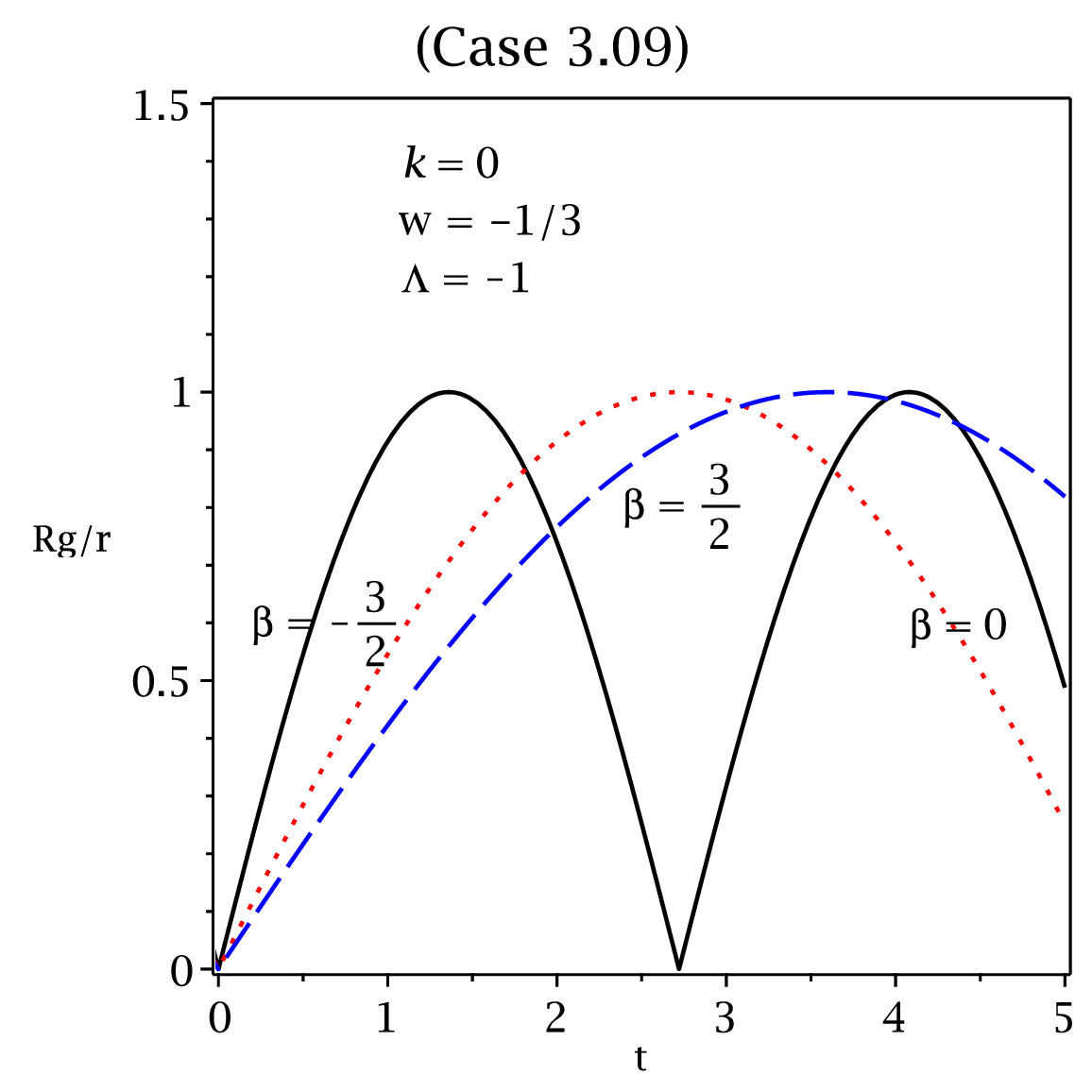}
	\includegraphics[width=3.4cm]{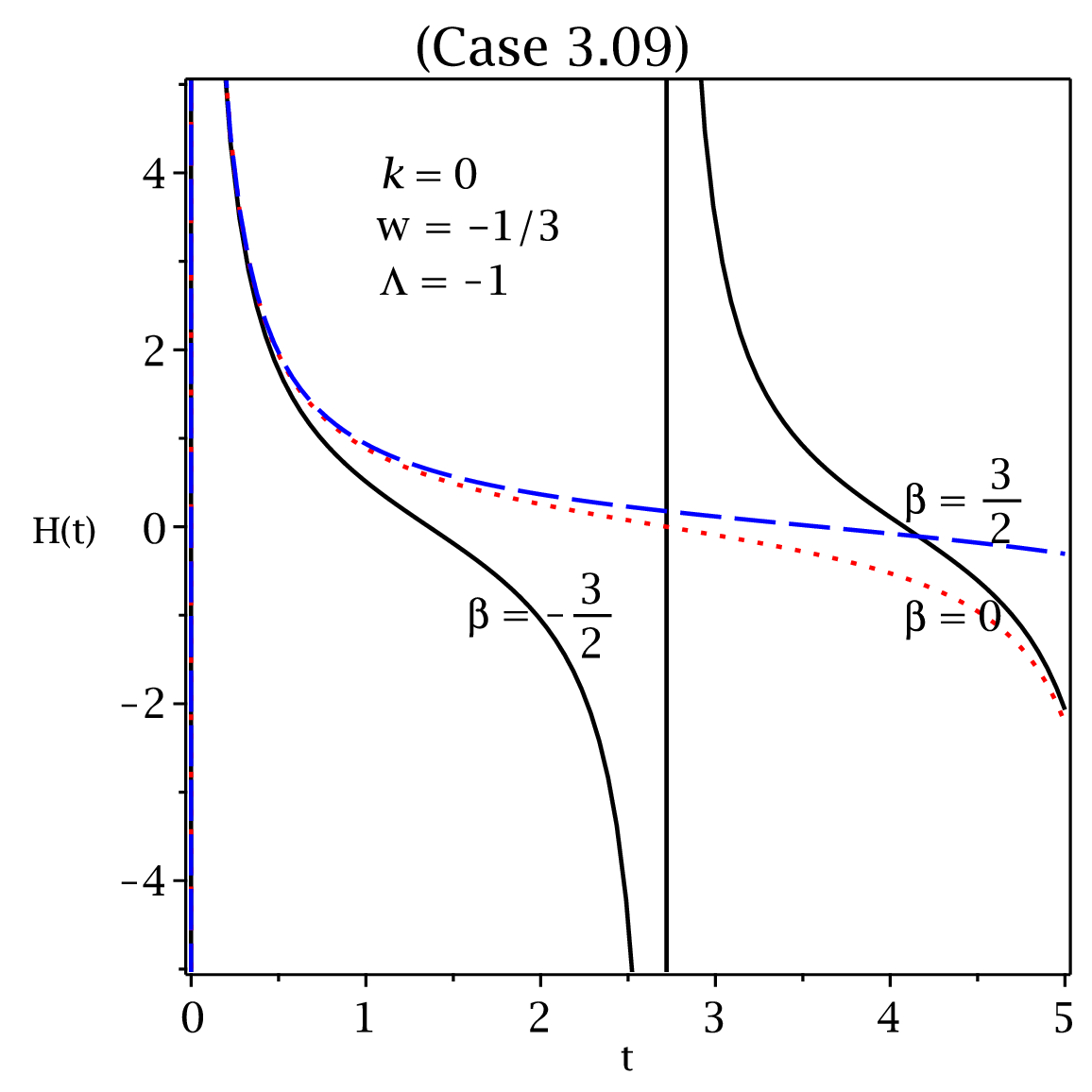}
	\includegraphics[width=3.4cm]{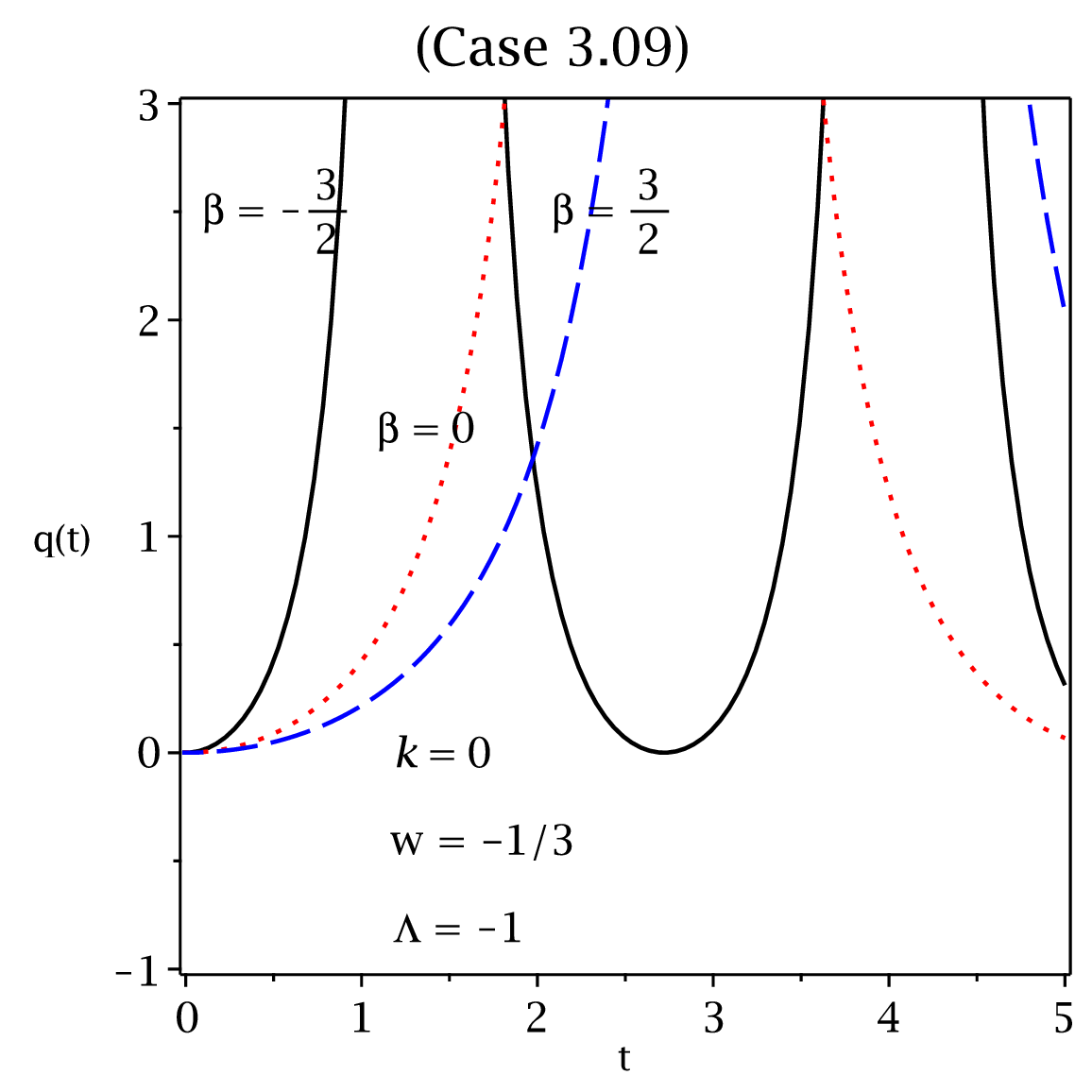}
	\includegraphics[width=3.4cm]{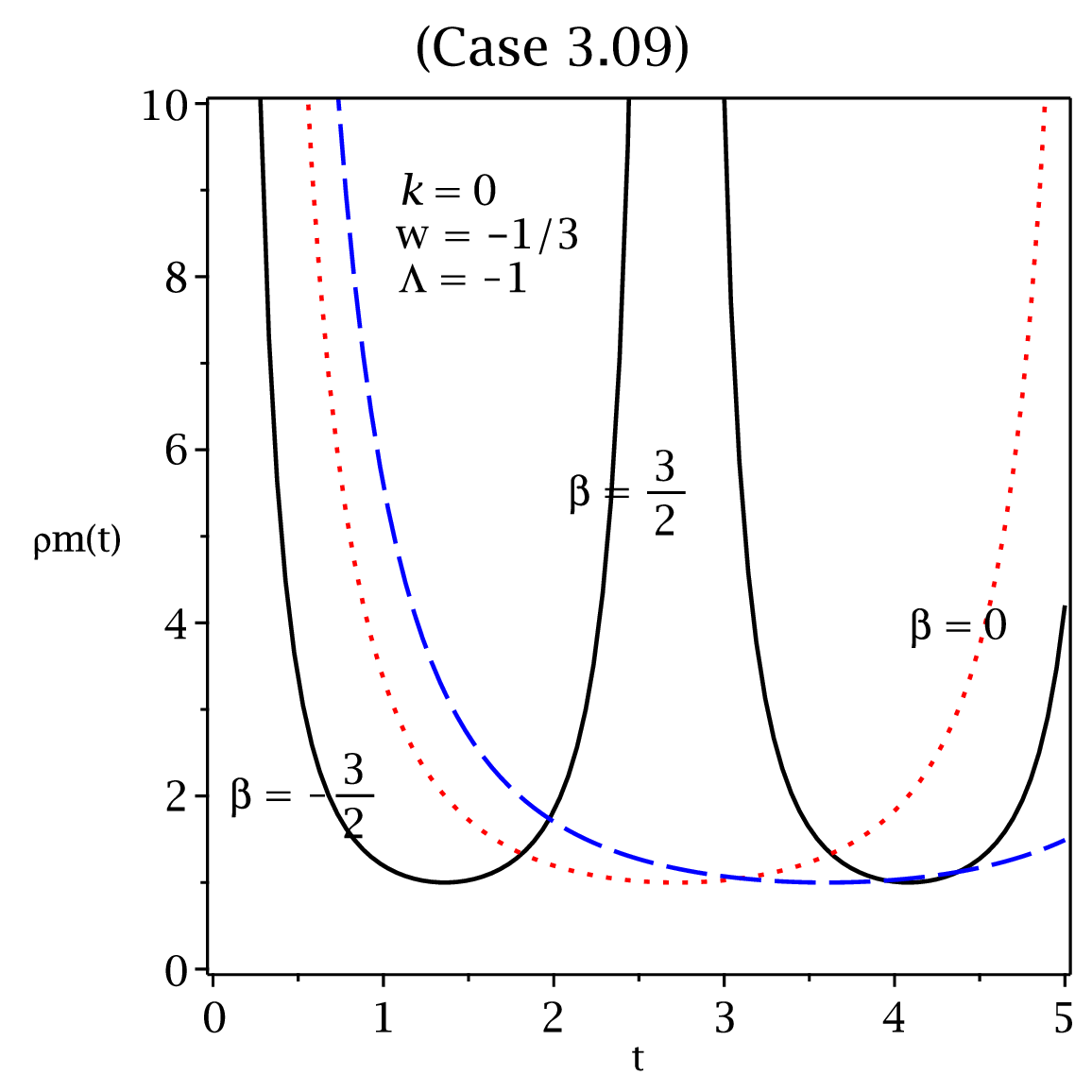}
	\includegraphics[width=3.4cm]{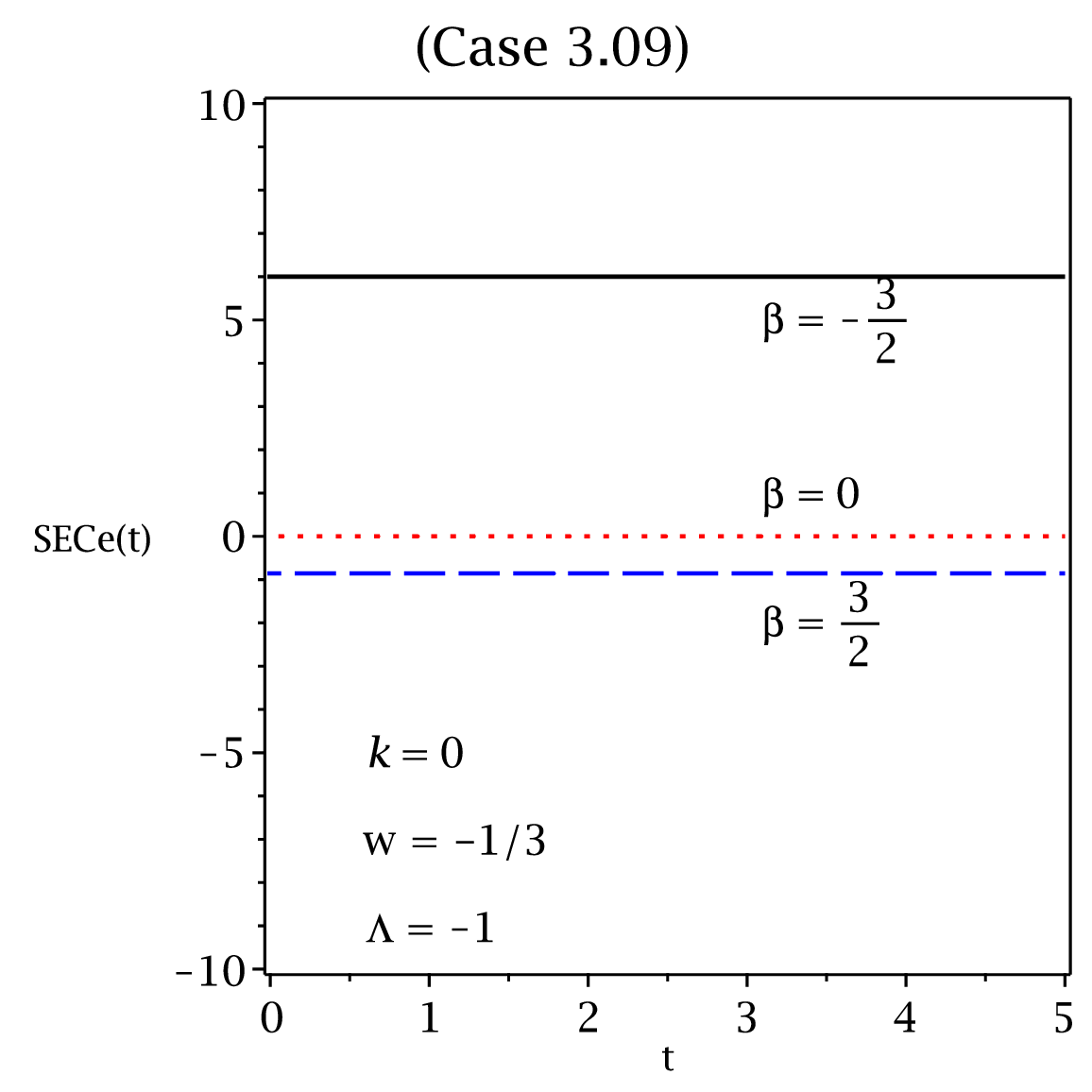}
	\includegraphics[width=3.4cm]{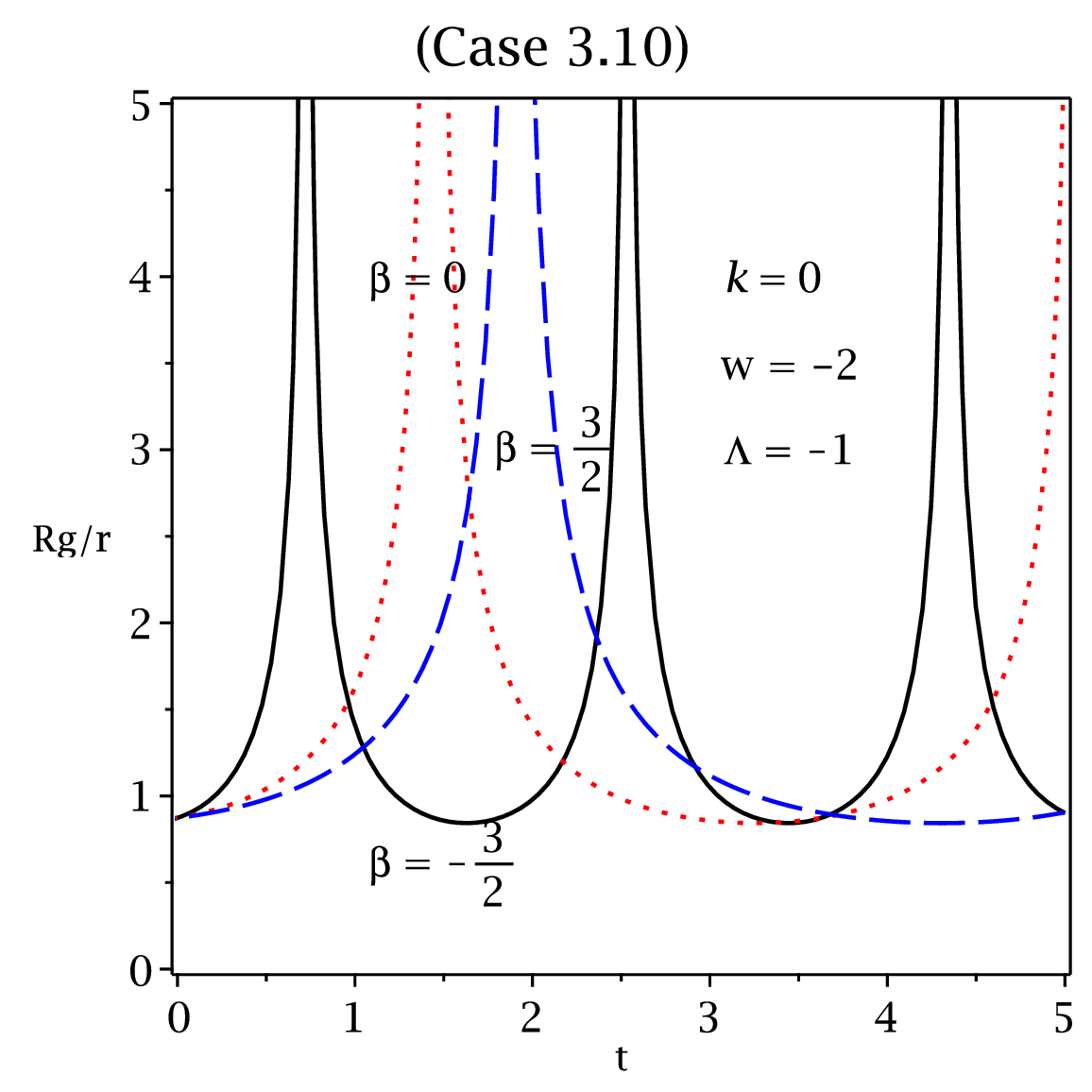}
	\includegraphics[width=3.4cm]{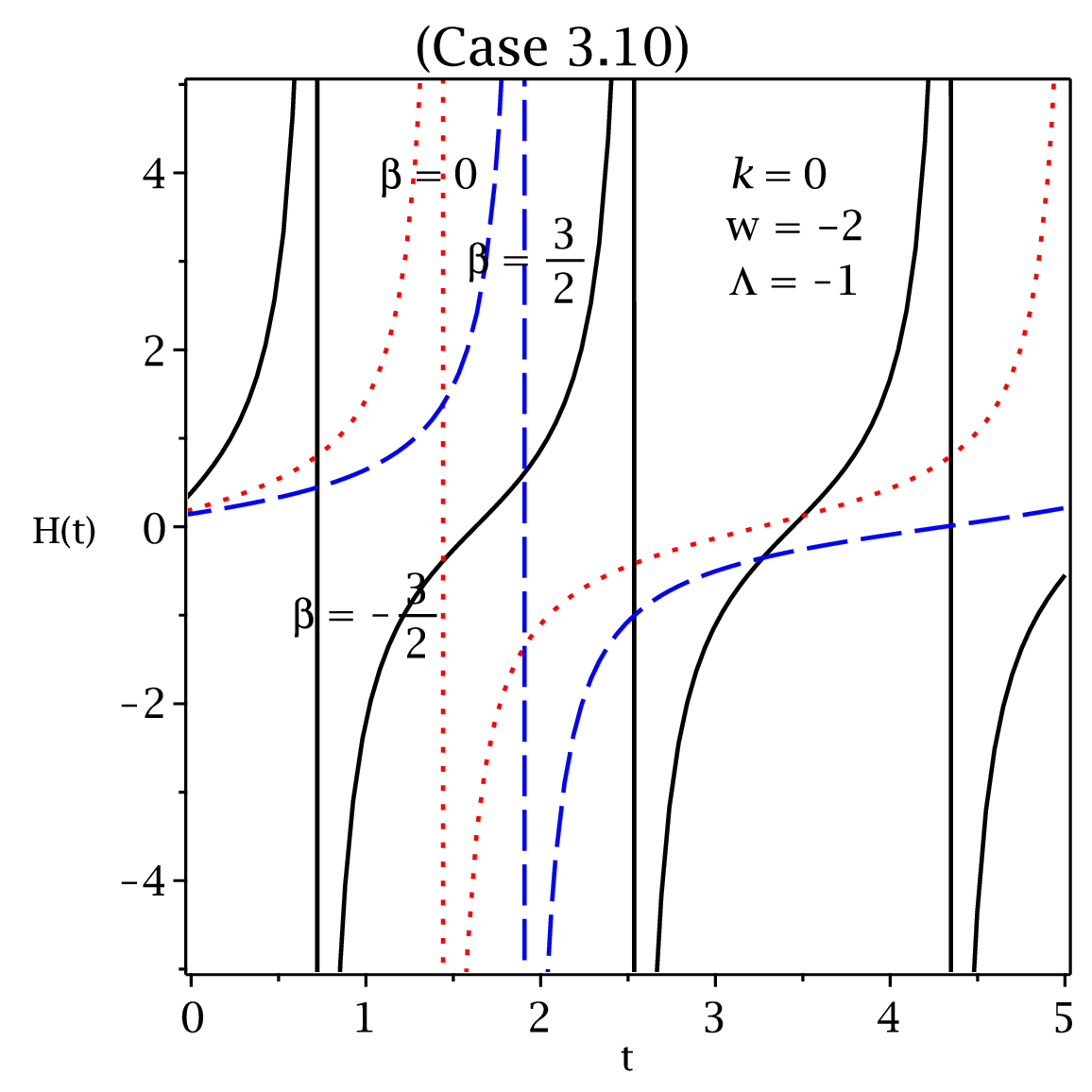}
	\includegraphics[width=3.4cm]{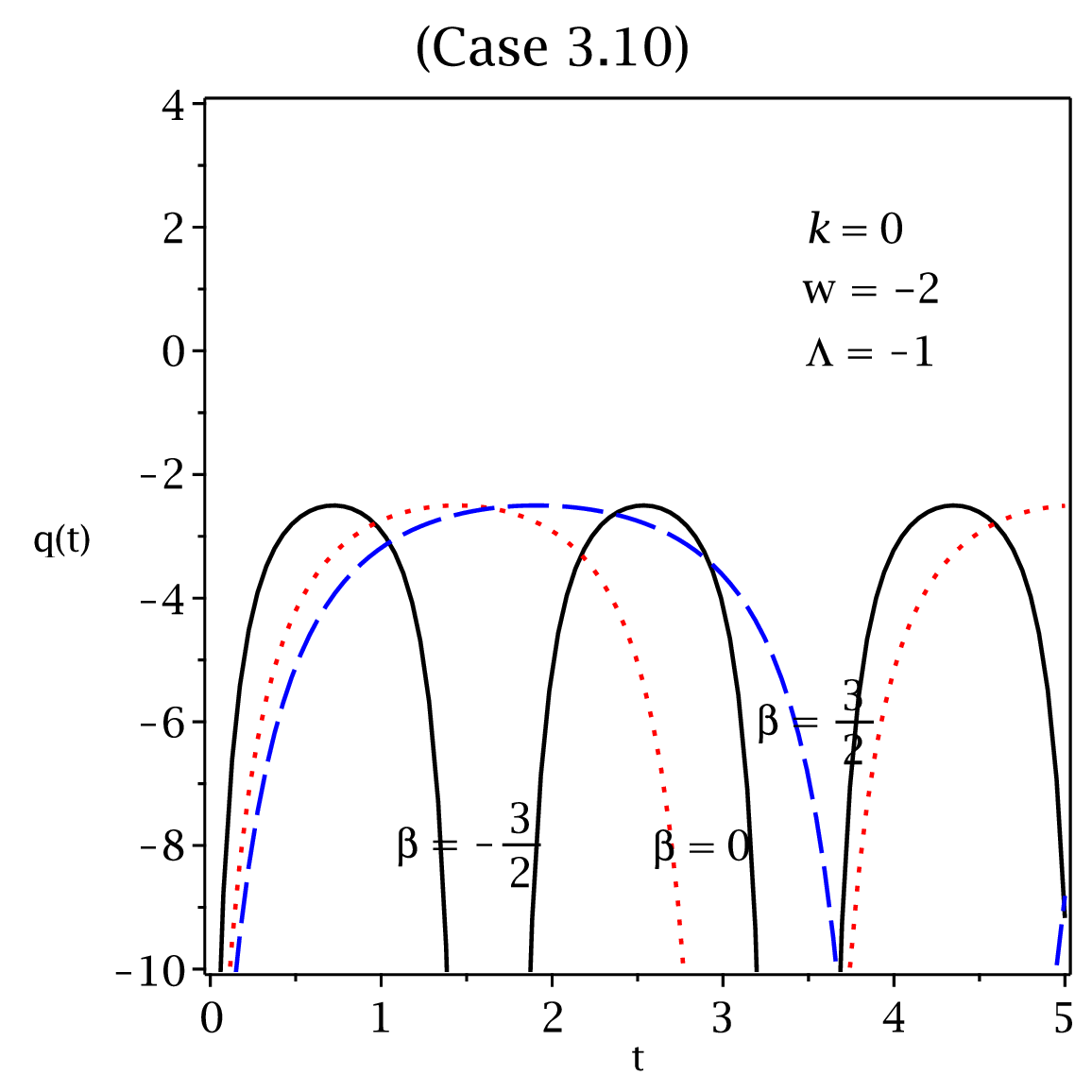}
	\includegraphics[width=3.4cm]{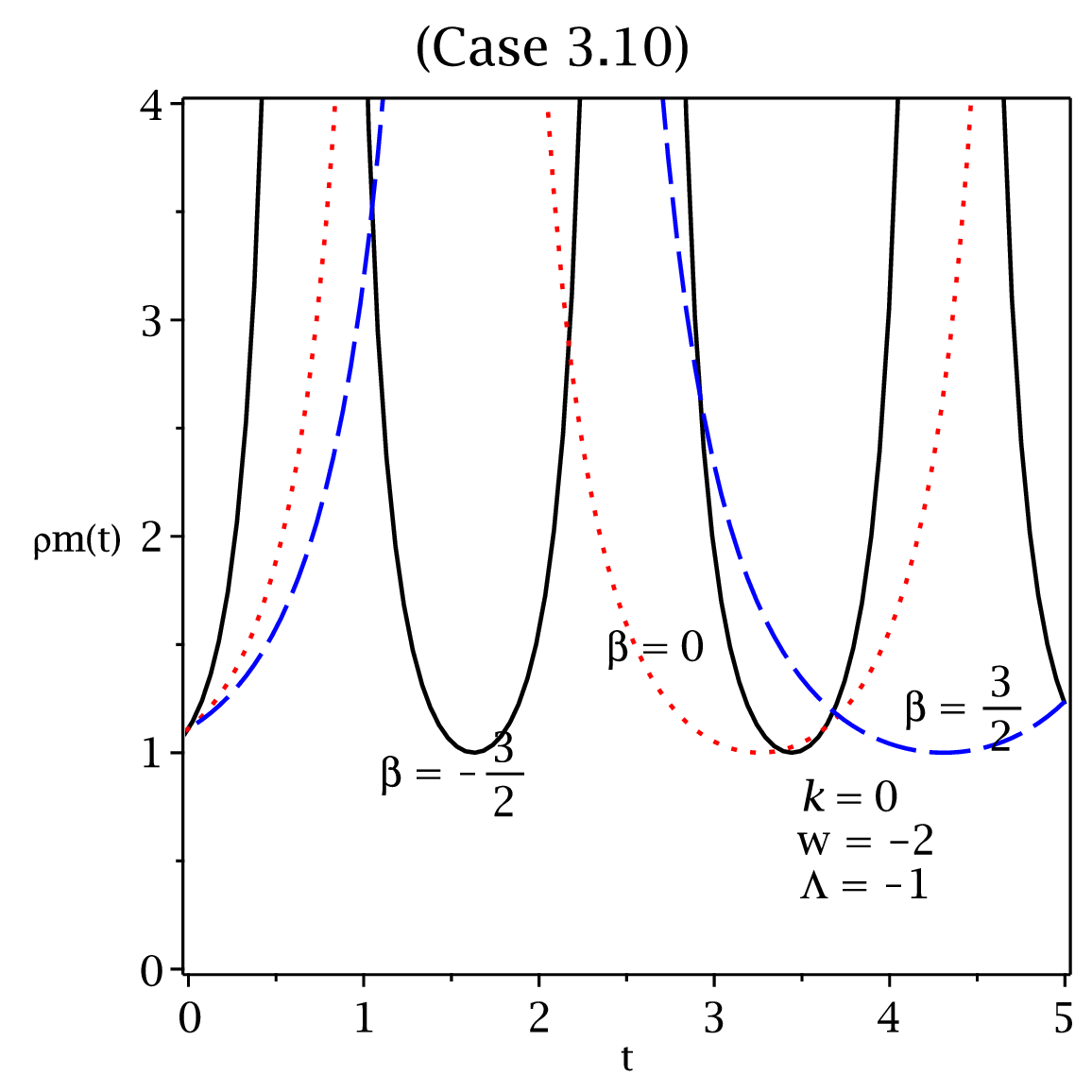}
	\includegraphics[width=3.4cm]{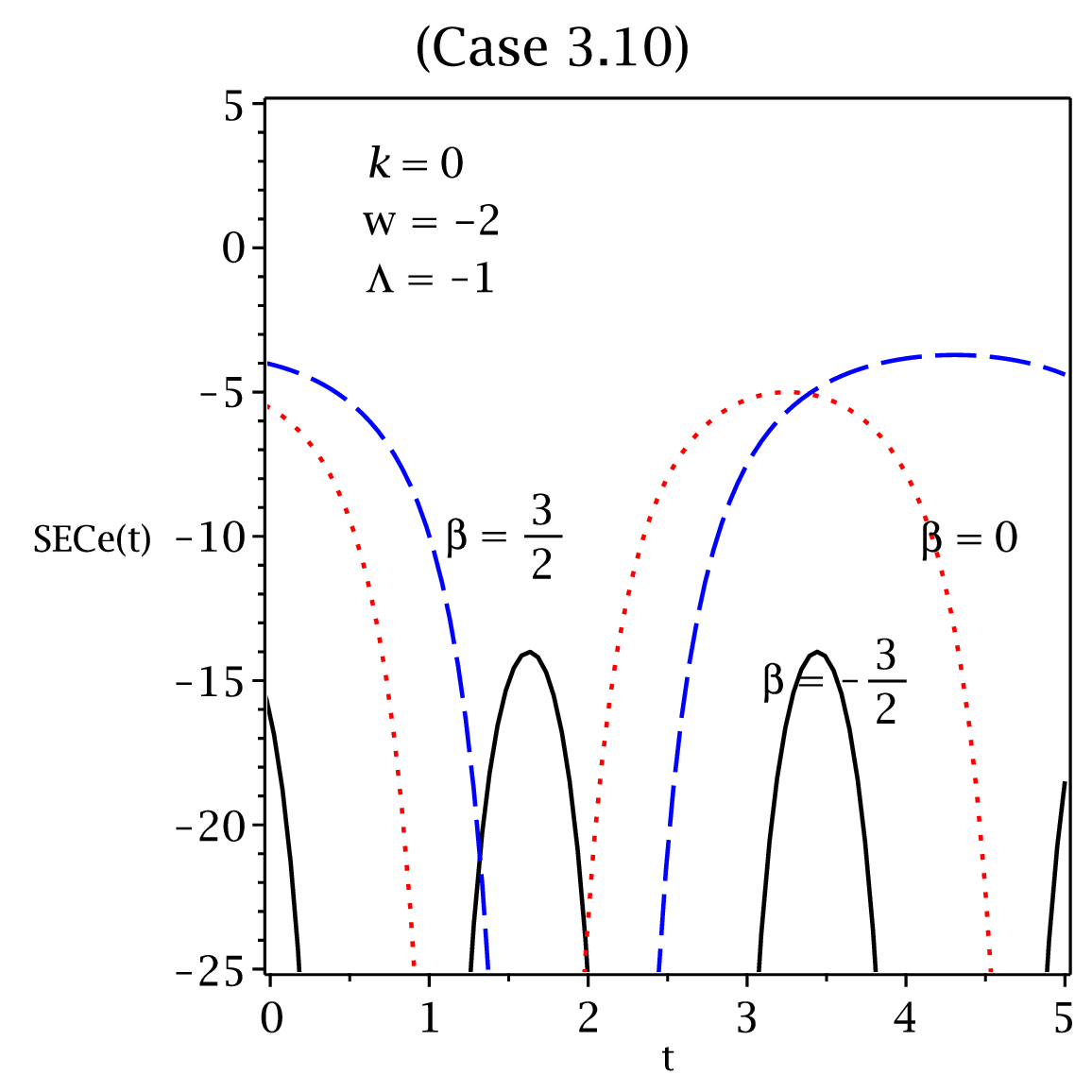}
	\caption{These figures are for $\Lambda<0$ and $k=0$.
		These figures represent the quantities $R_g$ (geometrical radius), 
		$H(t)$ (Hubble parameter) and $q(t)$ (deceleration parameter) $\rho_m(t)$ 
		(energy density of the aether fluid) and $SEC_{e} \equiv SEC_{\rm eff}$ 
		(strong energy condition for the effective fluid) for the different
		values of $\beta=-3/2$ (black solid line), $\beta=0$ (red dotted line), 
		$\beta=3/2$ (blue dashed line). Assuming that $8 \pi G=1$ and
		$R_g(t=0)=0$. Assuming also that $C_1=1$, $C_2=0$ (Cases 3.07 and 3.06); 
		$C_1=1$, $C_2=1$ (Case 3.05); $C_1=1$, $C_2=-3$ (Case 3.10).}
	\label{Figure-305-310}
\end{minipage}	
\end{figure}

\begin{figure}[!htp]
\begin{minipage}{175 mm}
	\centering	
	
	\includegraphics[width=3.4cm]{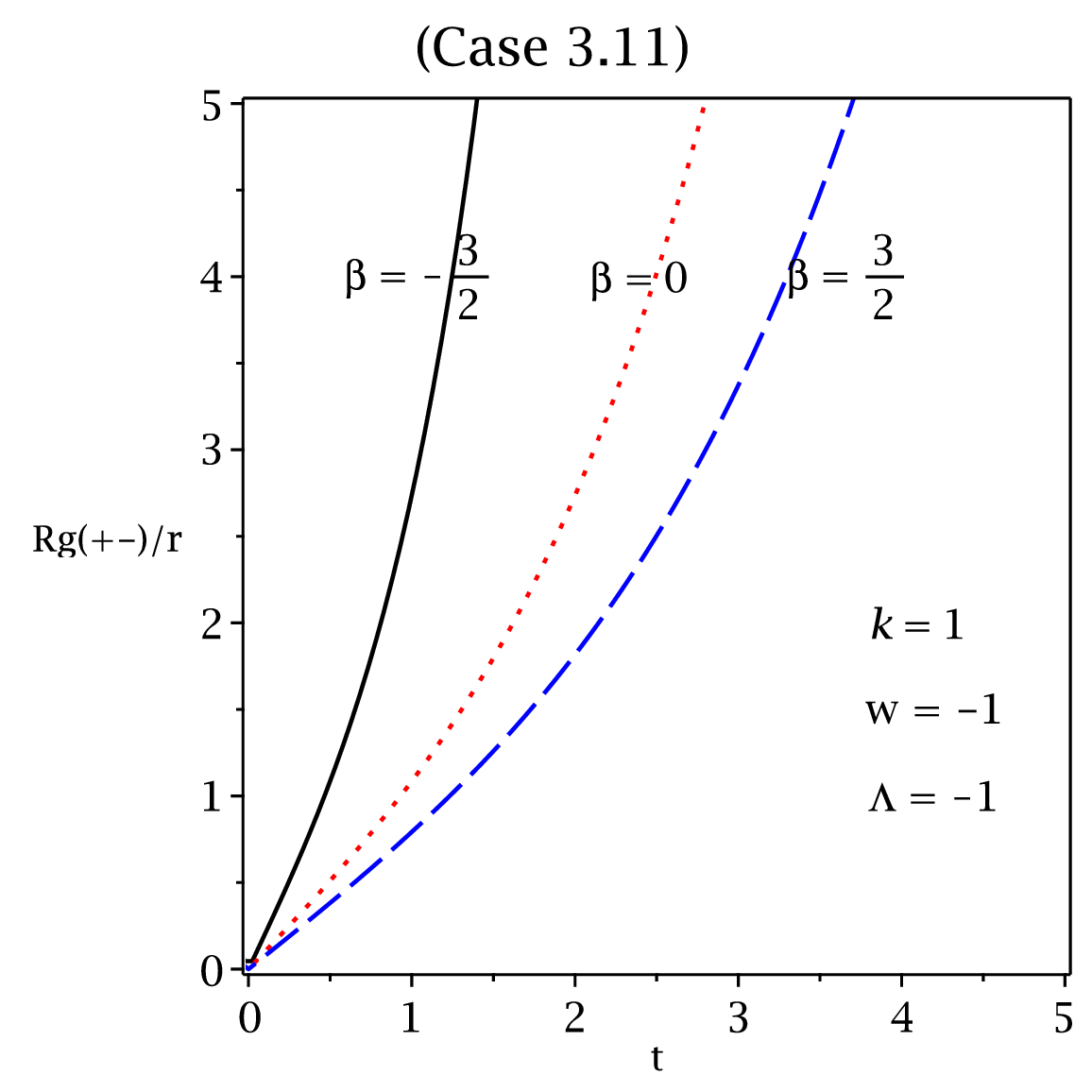}
	\includegraphics[width=3.4cm]{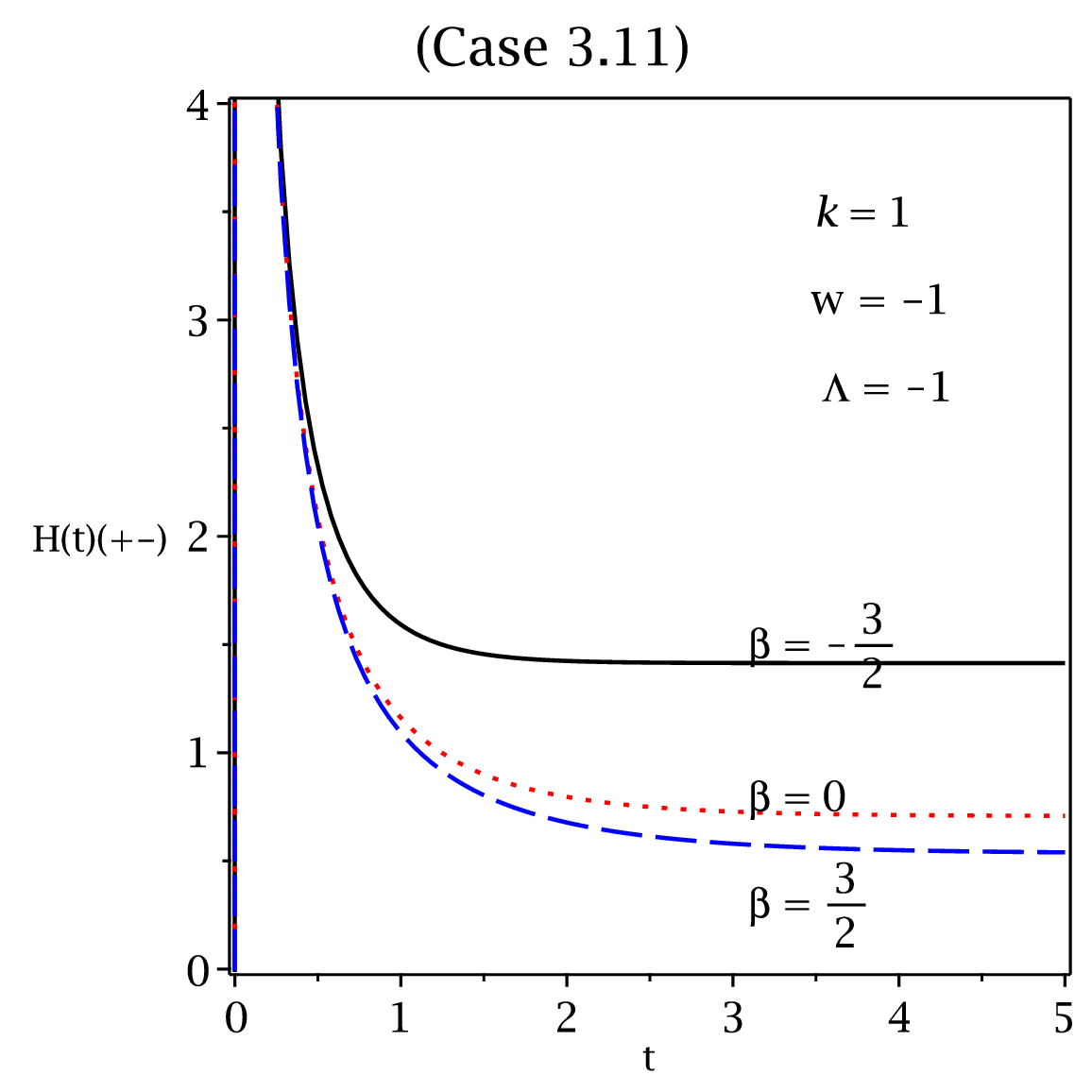}
	\includegraphics[width=3.4cm]{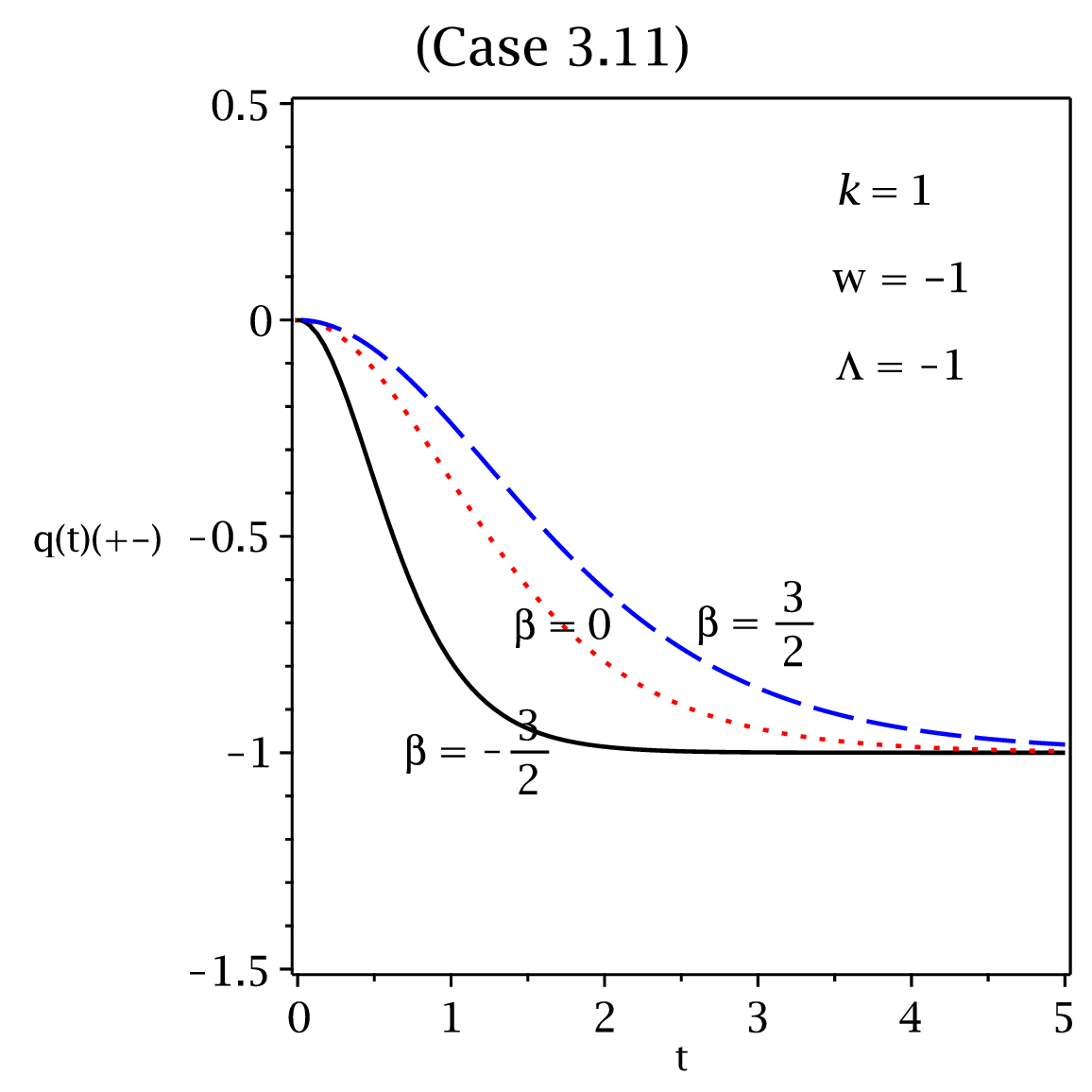}
	\includegraphics[width=3.4cm]{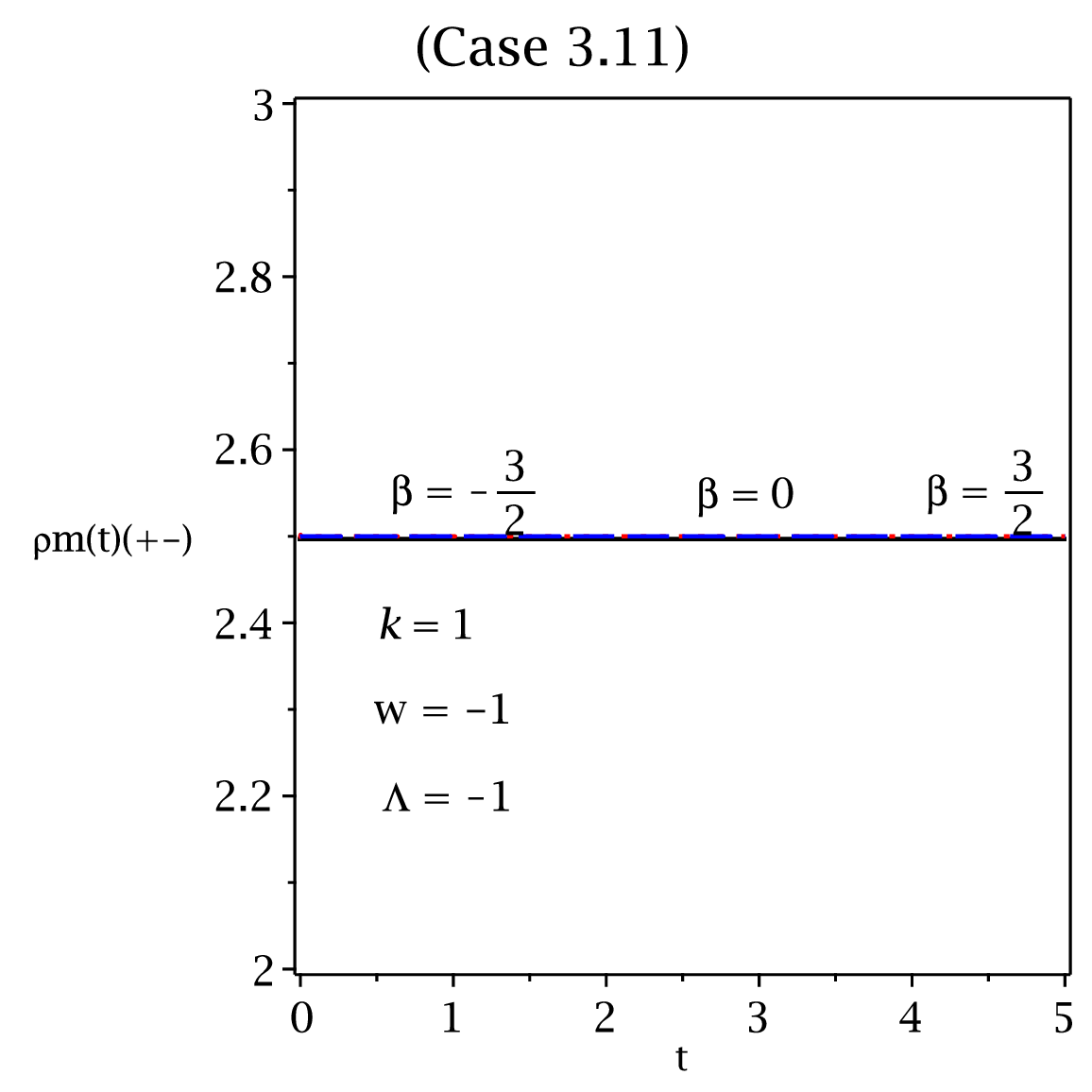}
	\includegraphics[width=3.4cm]{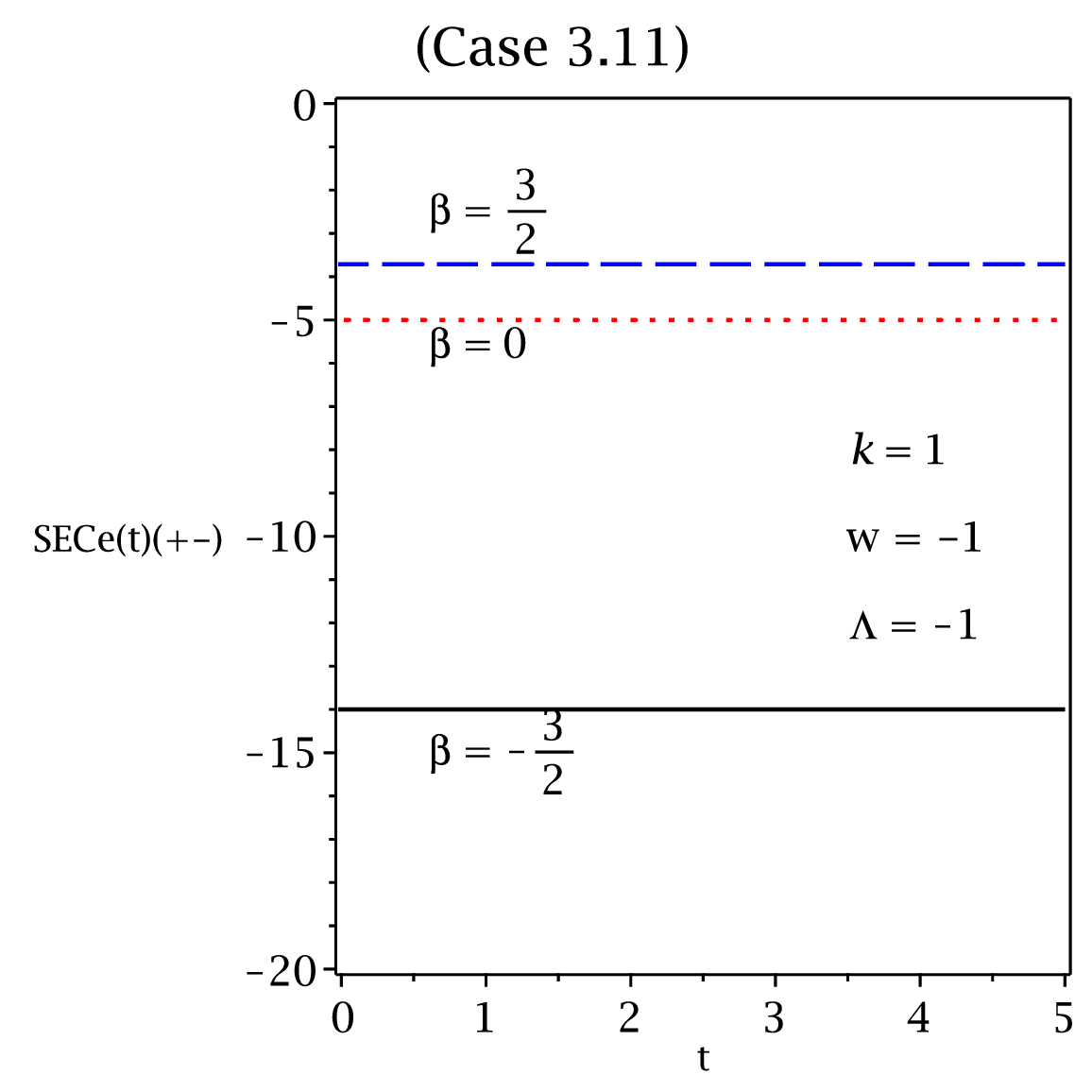}
	\includegraphics[width=3.4cm]{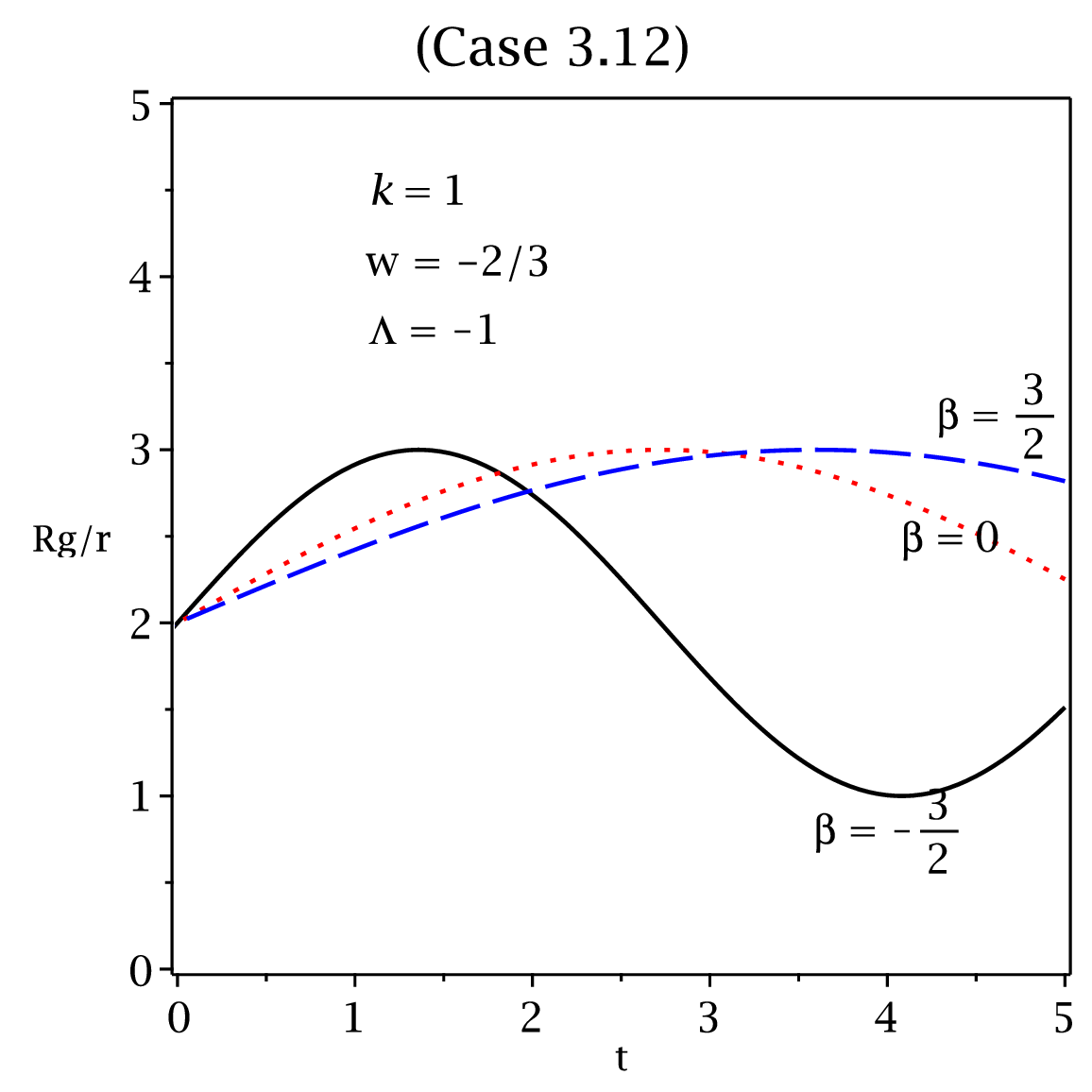}
	\includegraphics[width=3.4cm]{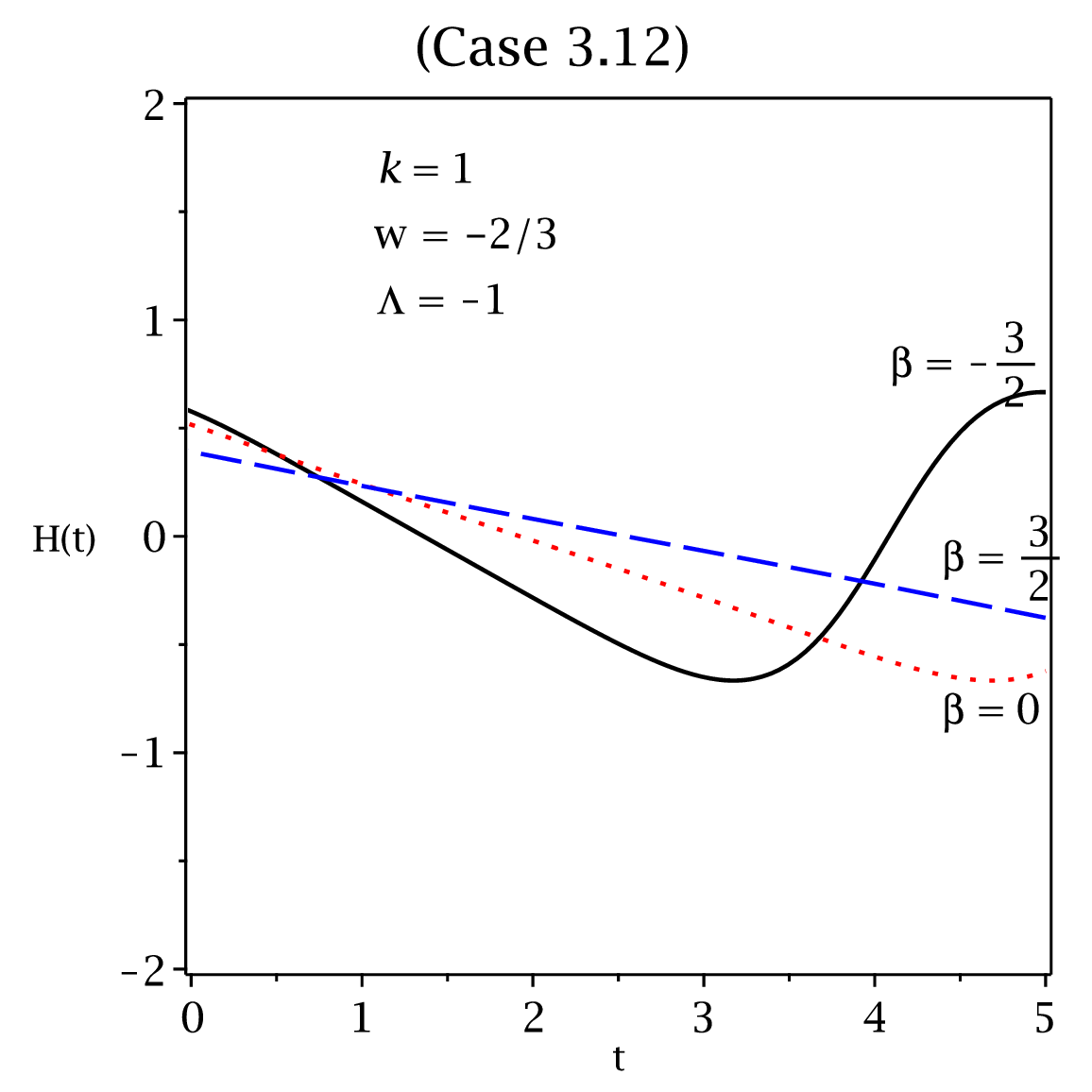}
	\includegraphics[width=3.4cm]{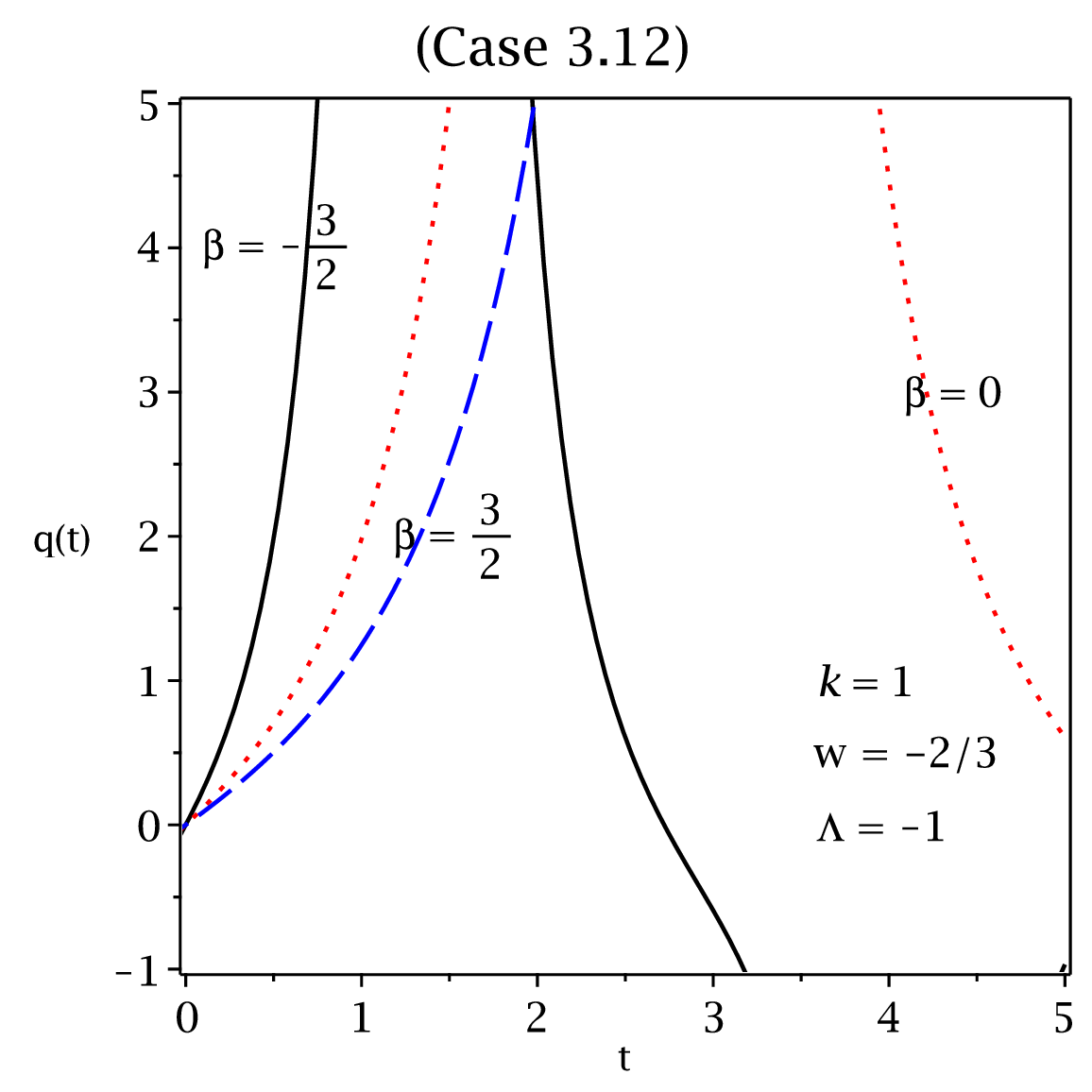}
	\includegraphics[width=3.4cm]{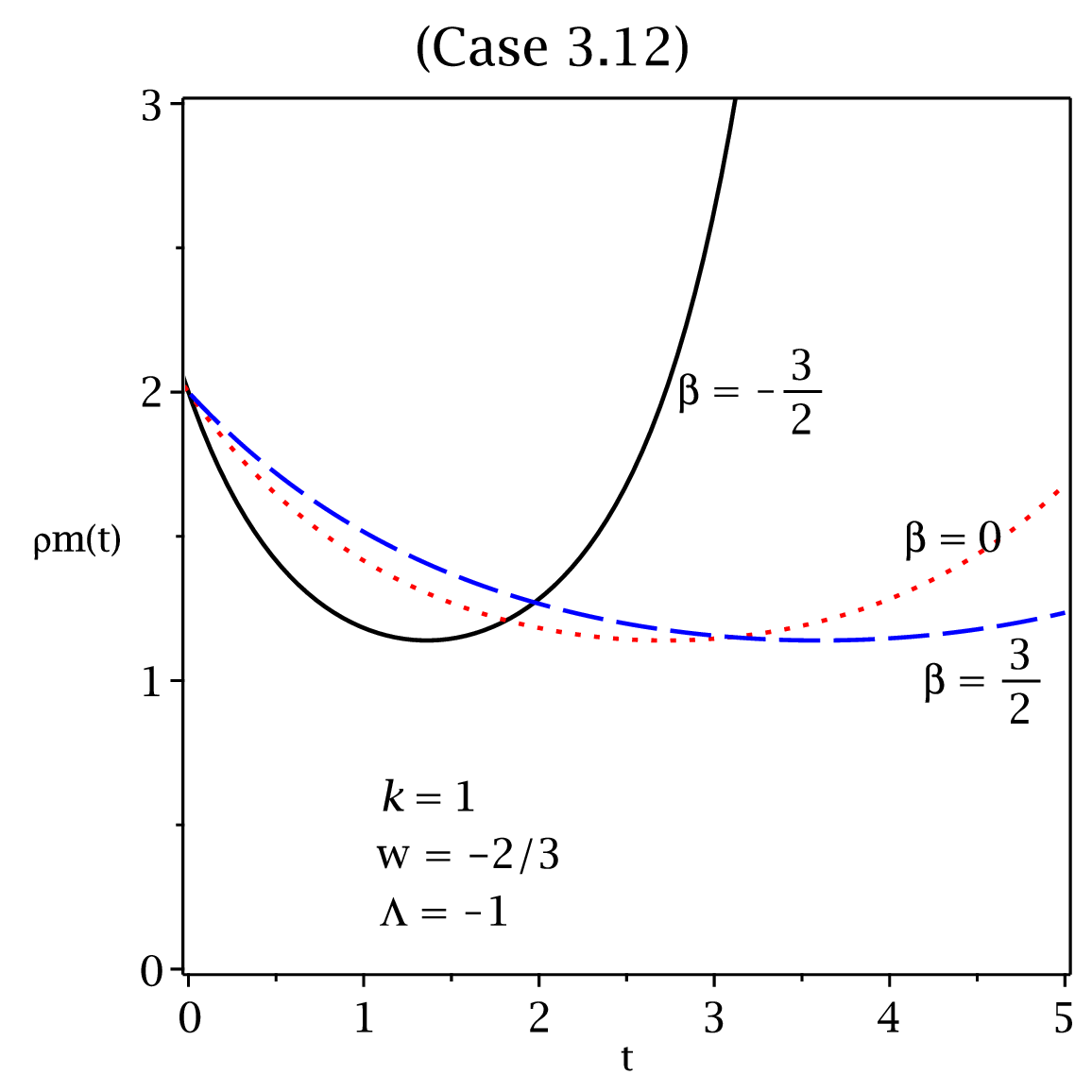}
	\includegraphics[width=3.4cm]{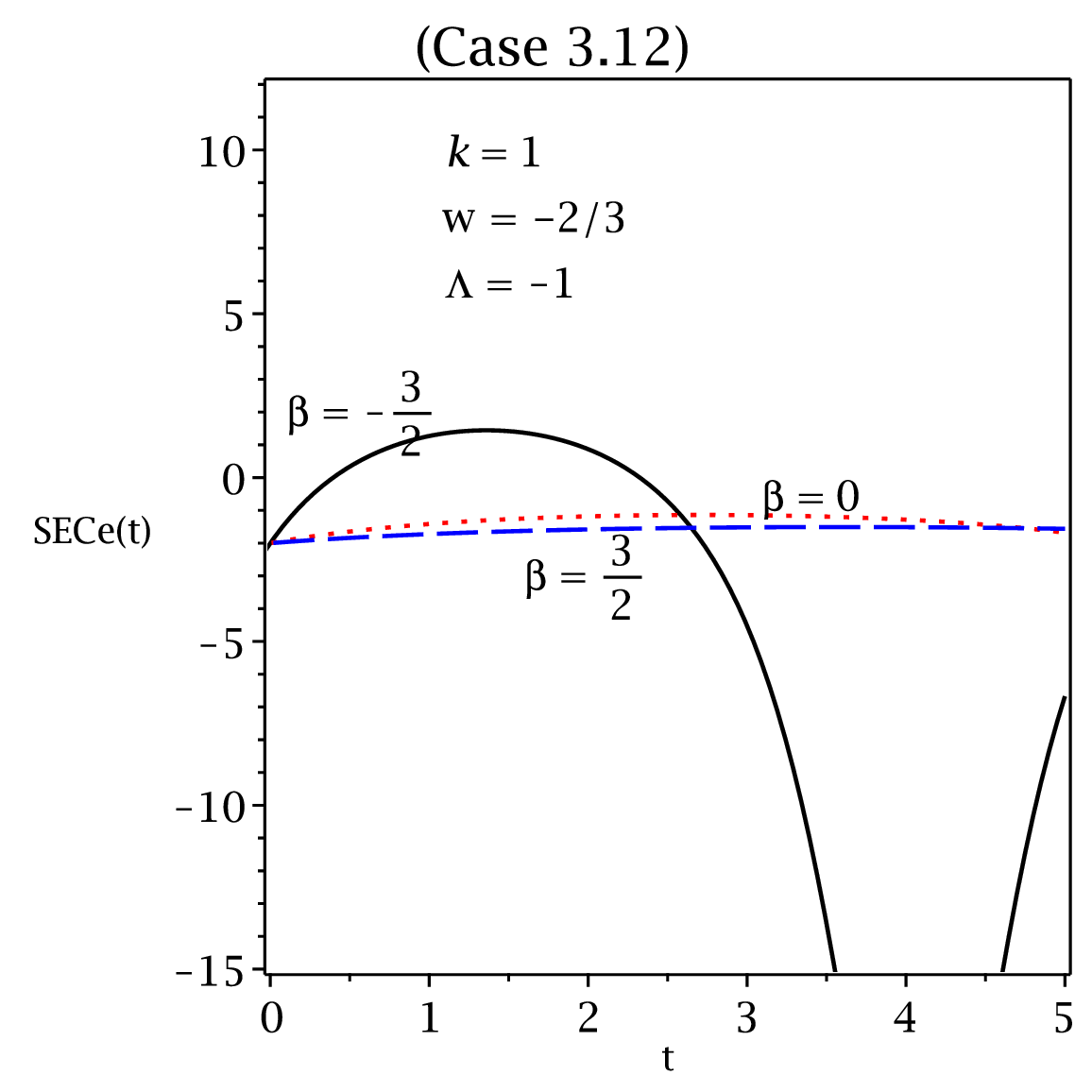}
	\includegraphics[width=3.4cm]{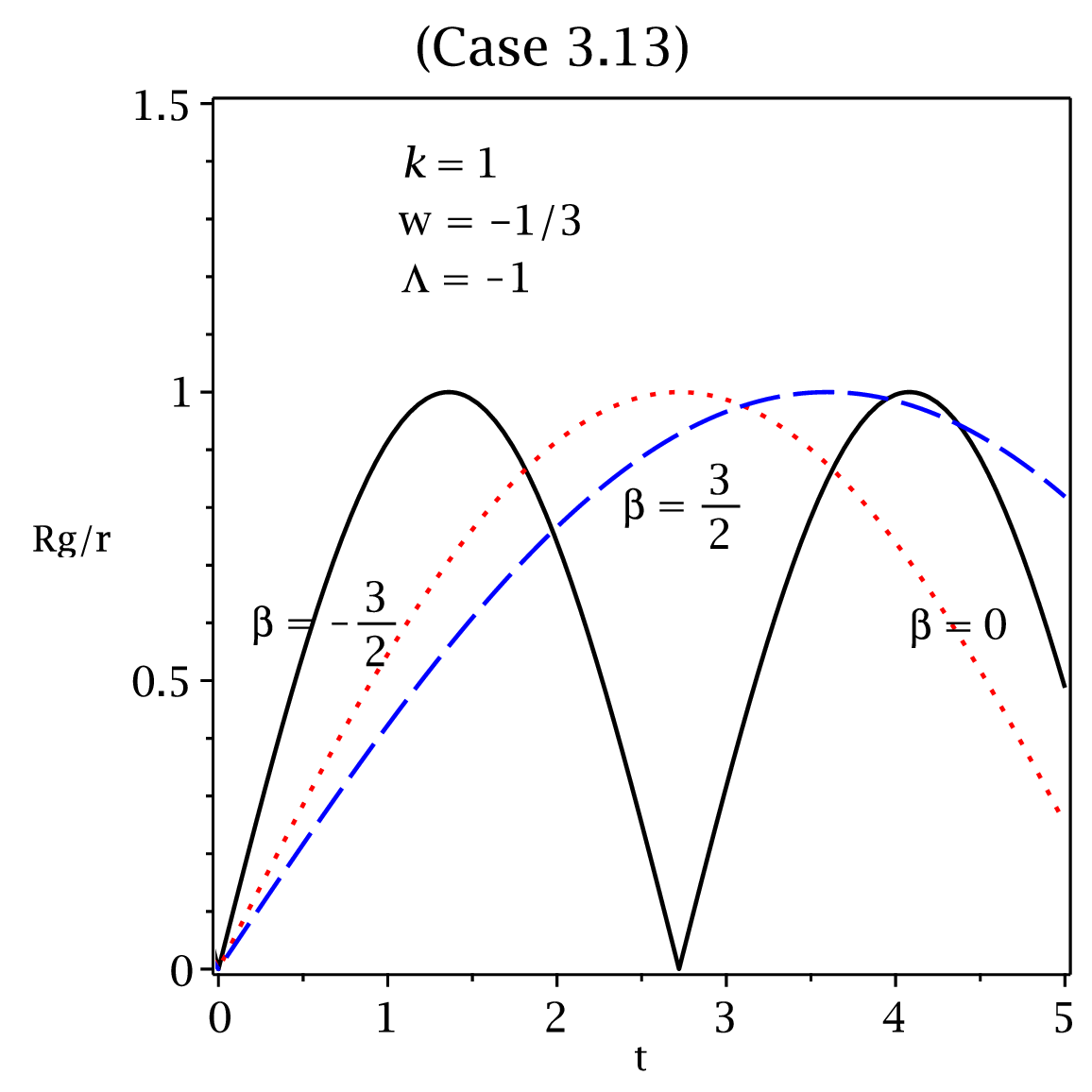}
	\includegraphics[width=3.4cm]{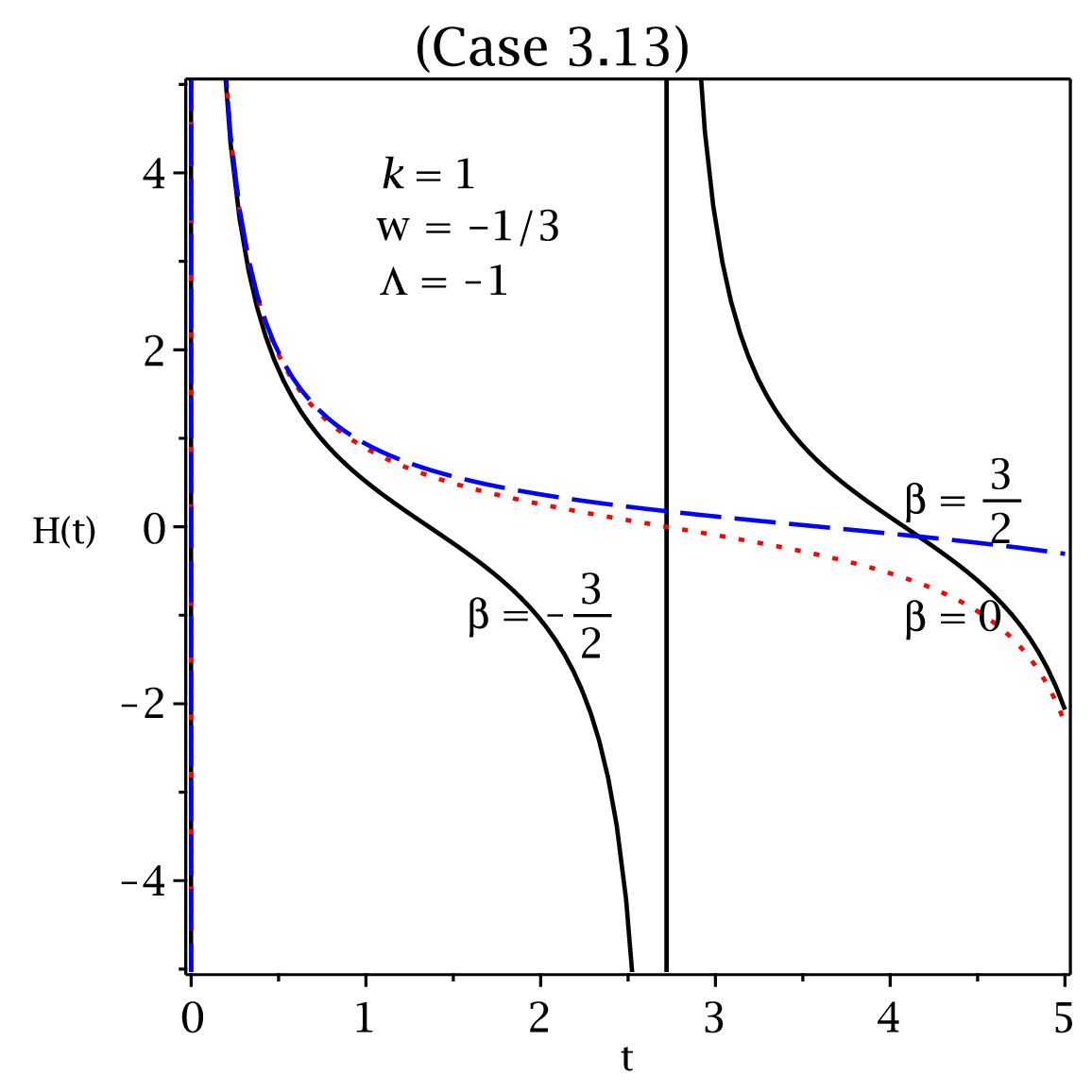}
	\includegraphics[width=3.4cm]{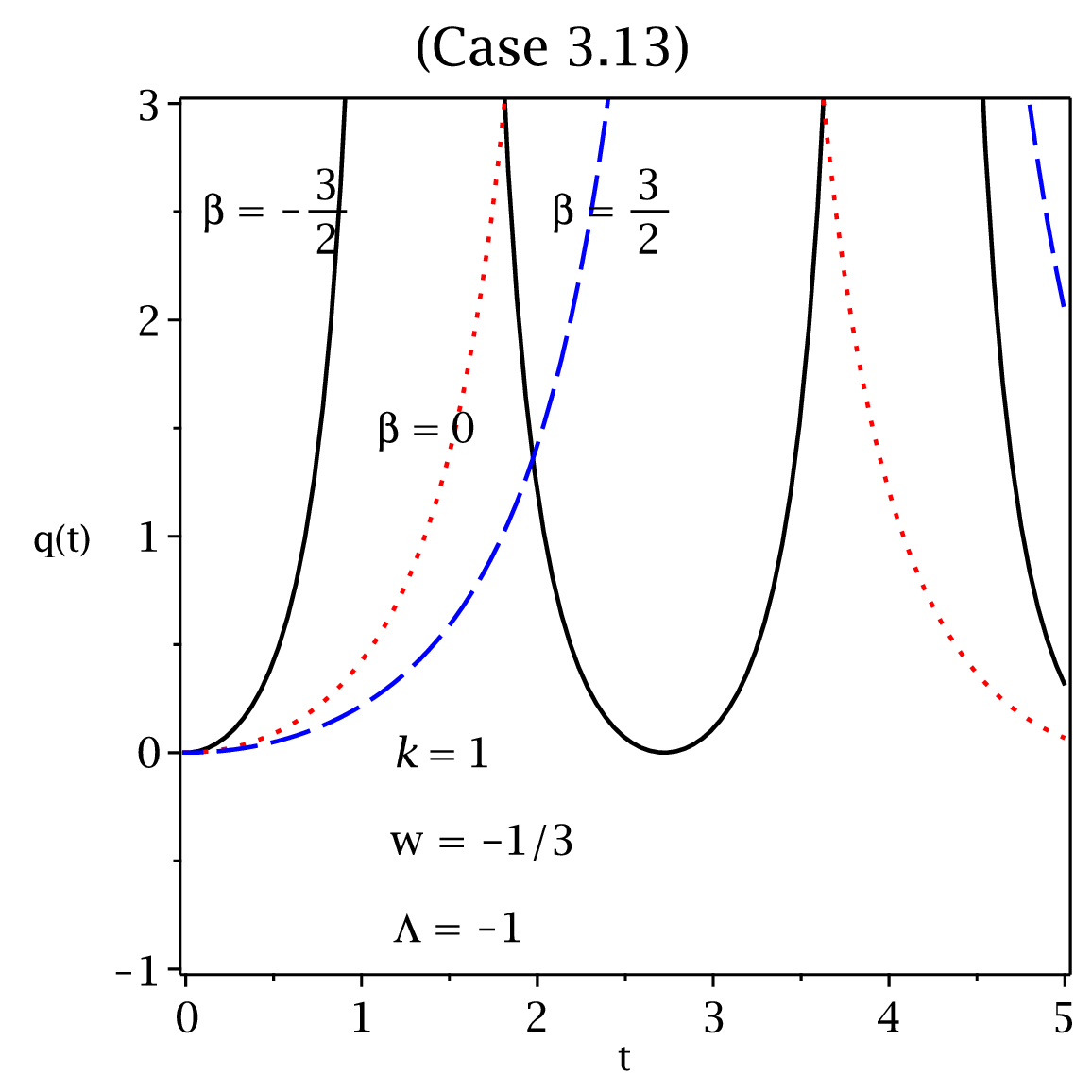}
	\includegraphics[width=3.4cm]{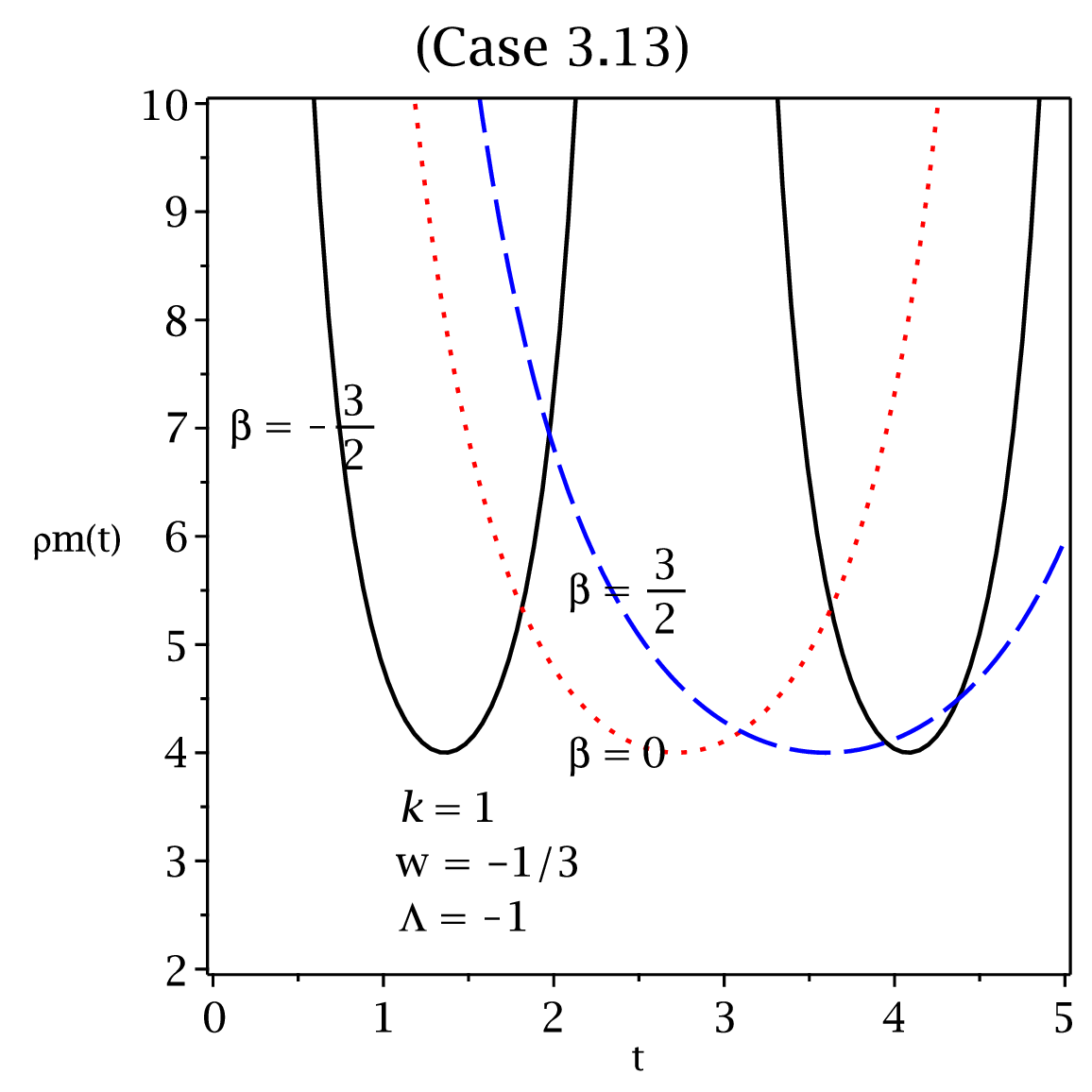}
	\includegraphics[width=3.4cm]{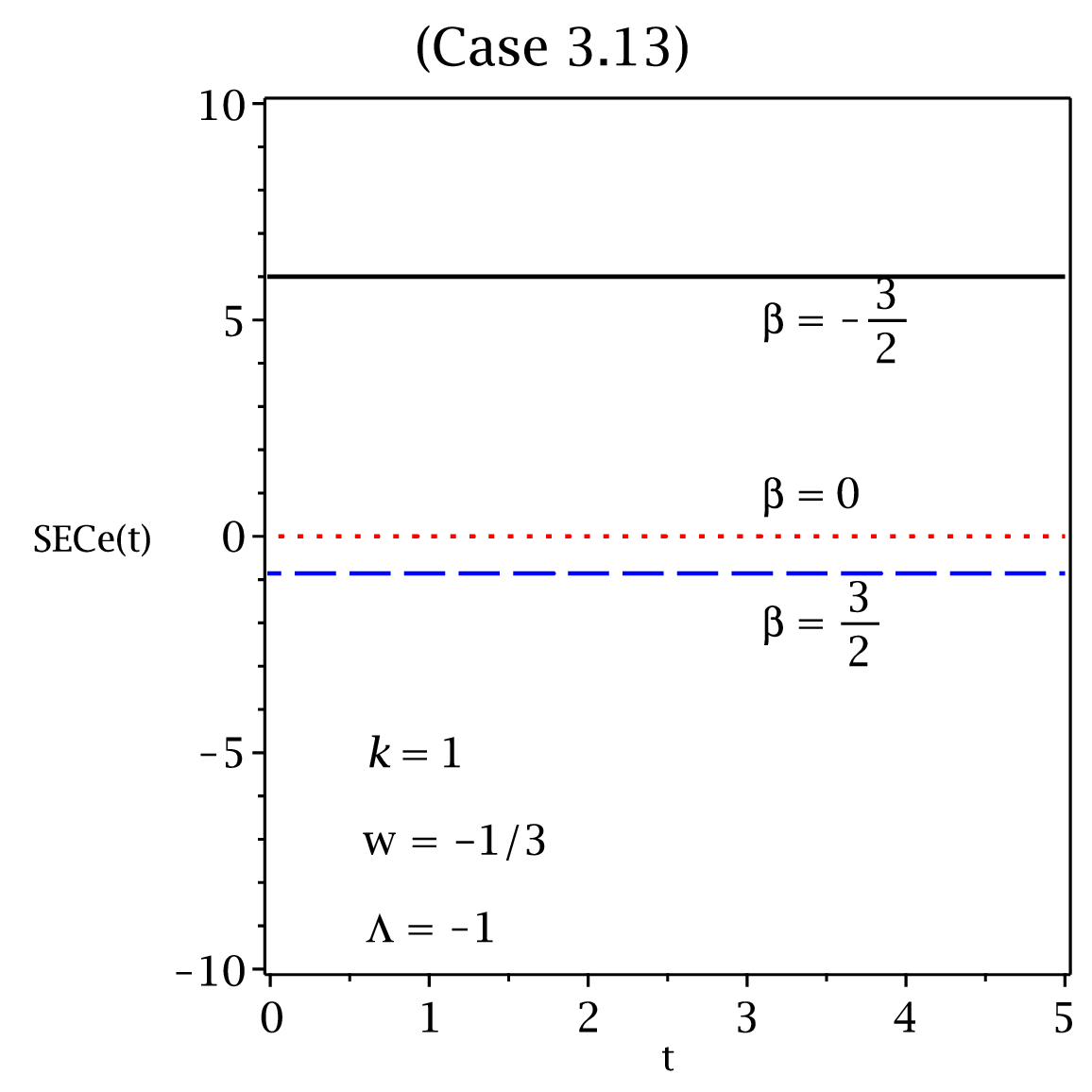}
	\includegraphics[width=3.4cm]{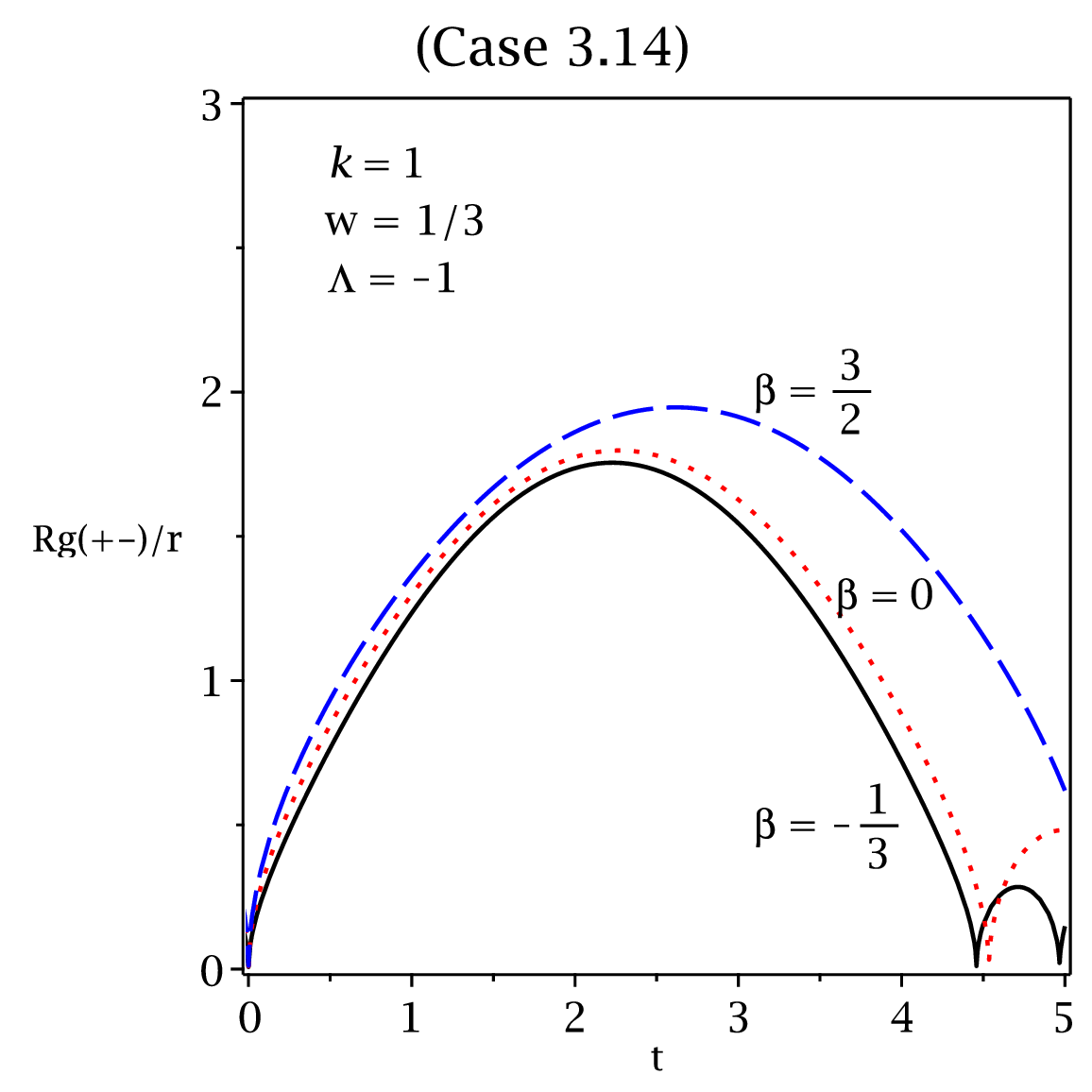}
	\includegraphics[width=3.4cm]{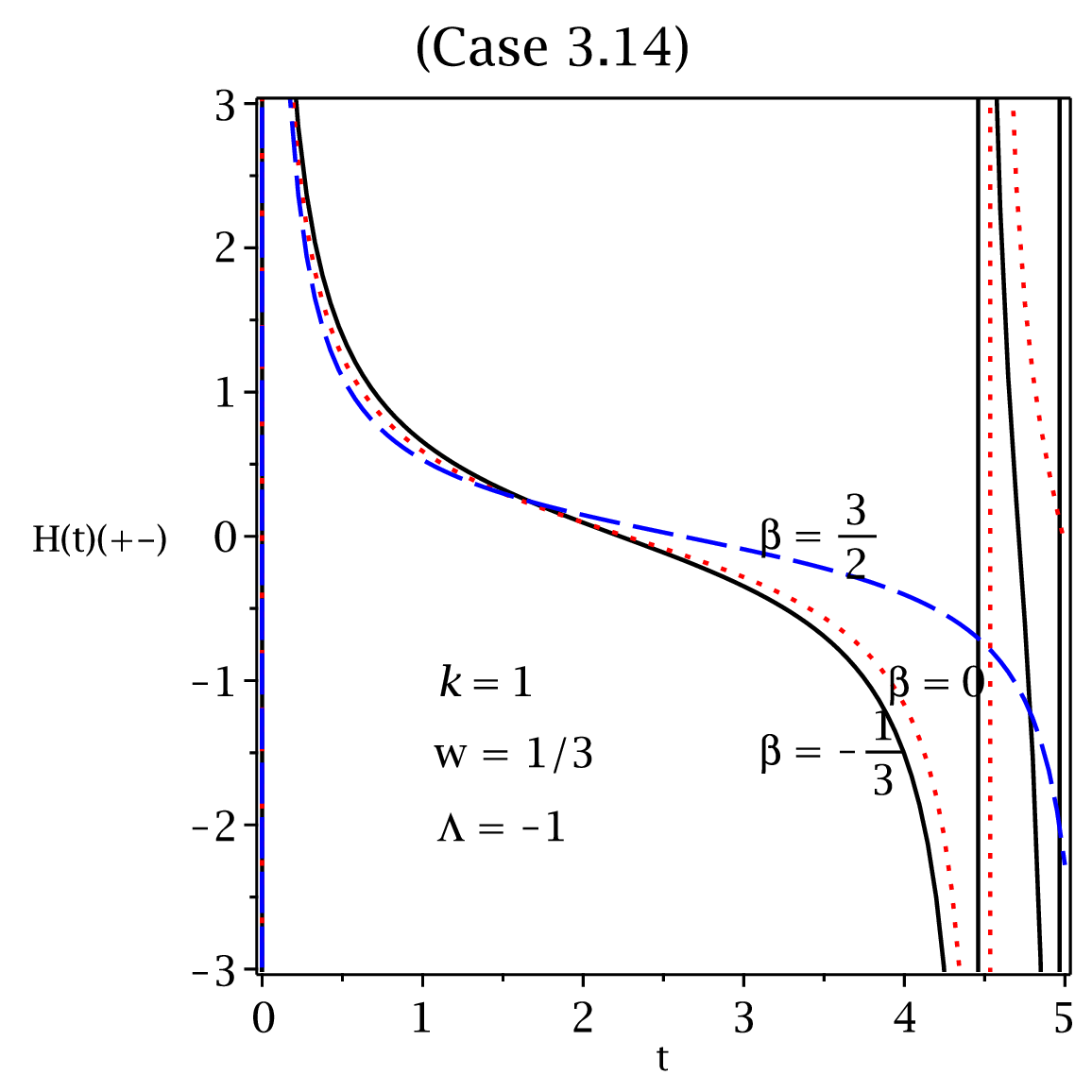}
	\includegraphics[width=3.4cm]{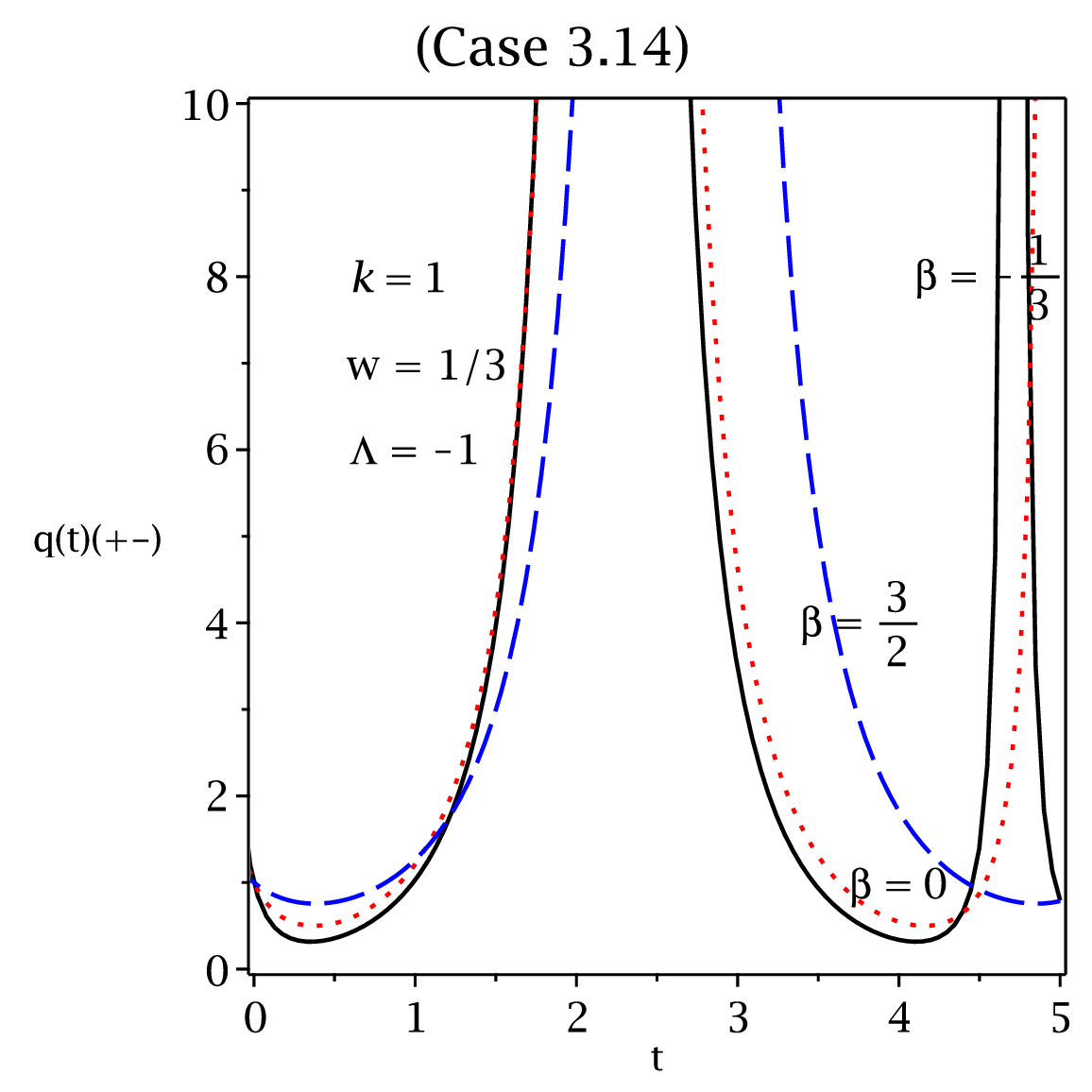}
	\includegraphics[width=3.4cm]{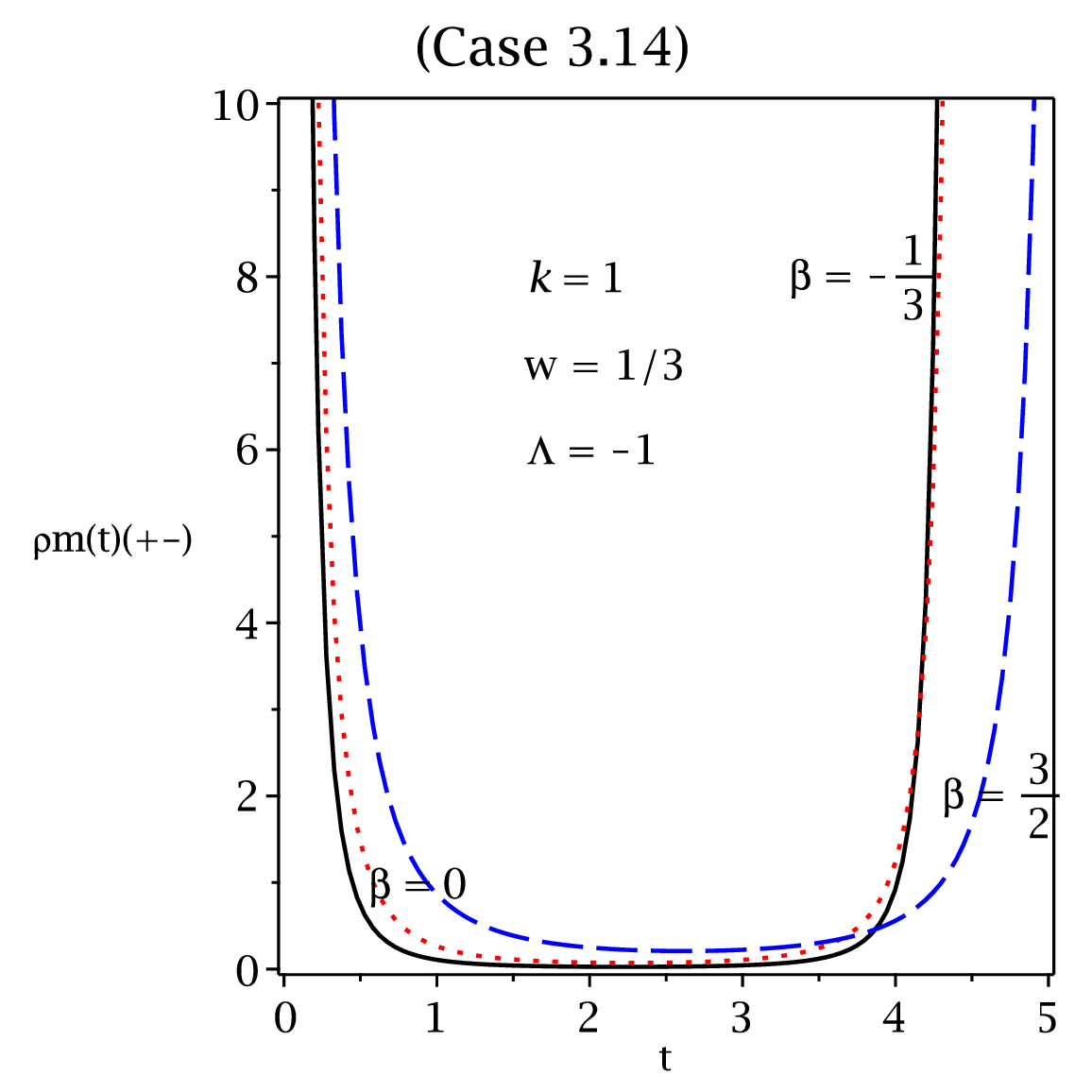}
	\includegraphics[width=3.4cm]{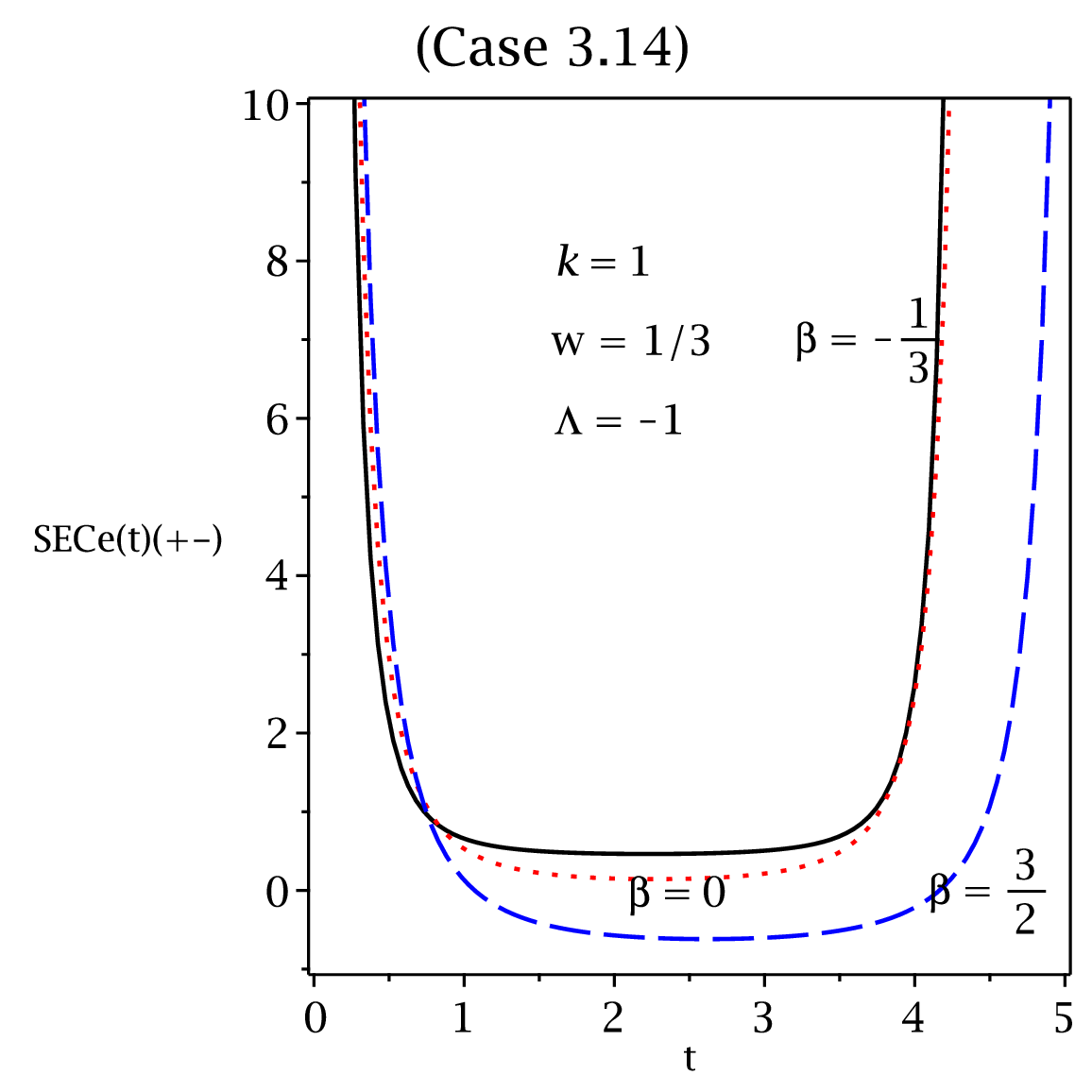}
	\caption{These figures are for $\Lambda<0$ and $k=1$.
		These figures represent the quantities $R_g$ (geometrical radius), 
		$H(t)$ (Hubble parameter) and $q(t)$ (deceleration parameter) $\rho_m(t)$ 
		(energy density of the aether fluid) and $SEC_{e} \equiv SEC_{\rm eff}$ 
		(strong energy condition for the effective fluid) for the different
		values of $\beta=-3/2$ (black solid line), $\beta=0$ (red dotted line), 
		$\beta=3/2$ (blue dashed line). Assuming that $8 \pi G=1$ and
		$R_g(t=0)=0$. Assuming also that $C_1=1$, $C_2=0$ (Cases 3.11, 3.12, 3.13 and 3.14);
		$C_1=1$, $C_2=0$ (Cases 3.08 and 3.09).}
	\label{Figure-311-314}
\end{minipage}	
\end{figure}

These analyses are shown in the Figures \ref{Figure-101-104}-\ref{Figure-311-314}.  
These figures represent the quantities $R_g$ (geometrical radius), 
$H(t)$ (Hubble parameter) and $q(t)$ (deceleration parameter) 
calculated for different values of $\beta$.

We can see from the geometrical radius and the density matter of these figures that we have
models for:

\begin{enumerate} 
\item {\bf Big Bang} for the Cases 1.01-1.04 (FIG. 1.), 1.08-1.11 (FIG. 2.), 1.12, 1.14-1.15 (FIG. 3.), 2.01-2.05 (FIG. 4.),  2.08-2.11 (FIG. 5.),
2.12, 2.14-2.16 (FIG. 6.), 3.01-3.04 (FIG. 7.), 3.07-3.06, 3.08-3.09 (FIG. 8.), 3.11, 3.13-3.14 (FIG. 9.),
\item {\bf Big Crunch} for the Cases 3.02-3.04 (FIG. 7.), 3.07-3.06, 3.09 (FIG. 8.), 3.13-3.14 (FIG. 9.),
\item {\bf Initial Finite Universe} for the Cases  1.07-1.06 and 1.05 (FIG. 2.), 1.13 (FIG. 3.), 
2.07 and 2.06 (FIG. 5.),
2.13 (FIG. 6.), 3.05 and 3.10 (FIG. 8.), 3.12 (FIG. 9.).
\item {\bf Big Rip} \cite{Barrow2009} for $w=-2$ in the  Cases 1.05 (FIG. 2.), 2.06 (FIG. 5.) and 3.10 (FIG. 8.). In the analysis presented in Ha {\it et.al.} (2012) \cite{Ha2012} this case should represent a Big Rip model.
\end{enumerate}

\section{Analysis of the effective energy Conditions}

In this section, similar to what is usually done when we have the Einstein equations with a cosmological constant,  in which the term with $\Lambda$ is treated as a component of the energy-momentum tensor on the right-hand side, let us consider the aether term on the right side of the field equations and treat it as an extra component of the energy-momentum tensor. It is important to note that, as far as we know, there is no rigorous formulation for the energy conditions in the EA theory, but there exists a proof for the positivity of the energy density, $\rho_m \ge 0$ \cite{Garfinkle1}.

Before analyzing the energy conditions, let us discuss the differences between the GR and EA theories, at least for the strong energy condition. This condition is usually used to define the existence of a dark energy in GR. Let us recall that the weak, null, and dominant energy conditions \cite{Hawking1973} are obtained by imposing conditions of physical reasonability on the matter fluid, i.e., for any observer, i) the density energy must be positive; ii) the pressure must not exceed the energy density; iii) the sound velocity in the fluid must not be greater than the vacuum light speed (this condition is not fulfilled in EA theory due to the non-validity of LI). However, the strong energy condition ($\rho_{Einstein}+3p_{Einstein} \ge 0$, for an isotropic fluid in the GR ) comes from a geometrical condition on the Riemann tensor. The term $R_{ab} v^a v^b$ ($v^a$ is any timelike vector) in the analogous Raychaudhuri equation for timelike geodesics \cite{Hawking1973} must contribute negatively to ensure the convergence of a congruence of these geodesics, which defines the attractivity of the gravitation. Thus, this condition must also be valid in the EA theory considering $T^{aether}_{ab}$ as part of an effective energy-momentum tensor $T^{\rm eff}_{ab}$, where
\bq
	T^{\rm eff}_{ab}=T^{aether}_{ab}+8\pi G T^{matter}_{ab}.
	\lb{Tabe}
	\eq
	We assume that the effective energy-momentum tensor corresponds to an isotropic fluid
	with an energy density $\rho_{\rm eff}$, pressure $p_{\rm eff}$,  given by
	\bq
	T^{\rm eff}_{ab}=(\rho_{\rm eff}+p_{\rm eff}) v_a v_b + p_{\rm eff} g_{ab},
	\lb{TGR}
	\eq
	and $v^a=\delta^a_t$ is a unit timelike vector representing the fluid velocity. We are also assuming a comoving reference.
	Thus, following \cite{Hawking1973} we must have
	\bqn
    &&	R_{ab} v^a v^b \ge 0 \Rightarrow T^{\rm eff}_{ab} v^a v^b \ge \frac{1}{2} T^e v^a v^b \nb \\
	&&\Rightarrow \rho_{\rm eff} +3p_{\rm eff} \ge 0.
	\lb{Rab}
	\eqn
	The components of the effective energy-momentum tensor are given by
	\bqn
	T^{\rm eff}_{tt}&&=T^{m}_{tt}+T^{aether}_{tt}\nb \\
	&&=8\pi G \rho_{m}-\frac{3\beta}{2} \frac{\dot B^2}{B^2}\nb \\
	&&=\rho_{\rm eff},
	\lb{Te1}
	\eqn
	\bqn
	T^{\rm eff}_{rr}&=&T^{m}_{rr}+T^{aether}_{rr}\nb\\
	&&=\frac{1}{1-k r^2}\left[8\pi G p_m B^2+\frac{\beta}{2}\left( \dot B^2+2 B \ddot B\right)\right]\nb \\
	&&=p_{\rm eff}\frac{B^2}{1-k r^2},
	\lb{Te2}
	\eqn
	\bqn
	T^{\rm eff}_{\theta\theta}&&=T^{m}_{\theta\theta}+T^{aether}_{\theta\theta}\nb \\
	&&=8\pi G p_m r^2 B^2+\frac{r^2 \beta}{2}  \left[\dot B^2+2 B \ddot B\right]\nb \\
	&&=p_{\rm eff} r^2 B^2,
	\eqn
	\bq
	T^{\rm eff}_{\phi\phi}=T^{\rm eff}_{\theta\theta} \sin^2 \theta.
	\eq

Using the equations (\ref{Te1}) and (\ref{Te2}) we get an equation of state {for the effective fluid}
	\bqn
	p_{\rm eff}+\rho_{\rm eff} &=& 8 \pi G(\rho_{m}+p_{m}) + \beta \frac{d}{dt}\left(\frac{\dot B}{B}\right)\nb \\
	&=& 8 \pi G({\rm w} \rho_{m}) + \beta \frac{d}{dt}\left(\frac{\dot B}{B}\right).
	\eqn
	On the other hand, the strong effective energy condition ($SEC_{\rm eff}$) becomes
	\bqn
	\lb{SECef}
	SEC_{\rm eff}&=&\rho_{\rm eff}+3p_{\rm eff}\ge 0 \Rightarrow \nb\\
	SEC_{\rm eff}&=&\rho_{m}+3p_{m}  +\frac{3\beta}{8\pi G}\frac{\ddot B}{B} \ge 0
	\Rightarrow \nb\\
	SEC_{\rm eff}&=&(1+3 {\rm w}) \rho_{m}  +\frac{3\beta}{8\pi G}\frac{\ddot B}{B} \ge 0.
	\eqn
Besides, we must also have that energy density in EA is positive, i.e.,  $\rho_{m} \ge 0$ \cite{Garfinkle1}. Let us assume that $\rho_{\rm eff} \equiv \rho_{Einstein}$, $p_{\rm eff} \equiv p_{Einstein}$. According to the reference \cite{Chan2009}, we can say that when $SEC_{\rm eff}<0$, we have dark energy. Otherwise, it is a normal fluid.

All the energy densities and the strong effective energy conditions are analyzed graphically, and the results are summarized in Tables \ref{table1}, \ref{table2} and \ref{table3}. These analyses are shown in the Figures \ref{Figure-101-104}-\ref{Figure-311-314}. The quantities $\rho_m(t)$ (energy density of the aether fluid) and $SEC_{e} \equiv SEC_{\rm eff}$ (strong energy condition for the effective fluid) are calculated for different values of $\beta$.

We can see from the Tables \ref{table1}, \ref{table2} and \ref{table3} that most of the models
have positive energy density of the matter except for Cases 1.06, 2.02, 2.02 and 3.07.
We can consider these last cases as unphysical models.
However, when we analyze the effective energy condition we can observe that most of the models 
have an epoch of effective dark energy is prevalent, except the Cases 2.05, 2.10, 2.16, 2.11, 
3.06, 3.04 where we have always an effective normal energy fluid.

\section{Conclusions}

In this paper, we have studied all possible analytical solutions of the FLRW metric in EA theory for all values of the cosmological constant $\Lambda$ and spatial curvature $k$ with many reasonable values of the equation-of-state parameter ${\rm w}$. We have presented the cosmological models under three different groups based on the value of the cosmological constant. We have studied the effective energy conditions for each of the models. We have also graphically analyzed the cosmological dynamics of the models by plotting the geometric radius, Hubble and deceleration parameters along with the effective energy conditions. In order to compare our results with GR, each graph includes the case $\beta=0$. Notice that some of these GR models obtained in this work are not presented yet in the literature so far. 

One of the fundamental results is that the aether does not qualitatively change the dynamics of the cosmological models but merely scales the geometric radius, Hubble, and deceleration parameters depending on the value of $\beta$. Another surprising result we found is that eight models are entirely void of any aether meaning in such a universe (a particular choice of Cosmological constant, spatial curvature, and equation-of-state parameter) aether does not play any role cosmologically, although it affects the solar system dynamics.

\section {Acknowledgments}

The financial assistance from FAPERJ/UERJ (MFAdaS) is gratefully acknowledged. The author (RC) acknowledges the financial support from FAPERJ (no.E-26/171.754/2000, E-26/171.533/2002 and E-26/170.95006). MFAdaS and RC also acknowledge the financial support from Conselho Nacional de Desenvolvimento Cient\'{\i}fico e Tecnol\'ogico - CNPq - Brazil.  The author (MFAdaS) also acknowledges the financial support from Financiadora de Estudos  Projetos - FINEP - Brazil (Ref. 2399/03).

\appendix
\section{Summary of the cosmological models}
\begin{table*}
	\centering
	\begin{minipage}{175 mm}
		\caption{Summary of the Cosmological Solutions for $\Lambda>0$}
		\label{table0a}
		\begin{tabular}{@{}|c|c|c|c|}
			\hline
			Case & $k$ & w & $B(t)$ \\
\hline
			\multirow{1}{*}{$(1.01)$} & -1 & -1 & $-\frac{1}{2\sqrt { \left( \beta+2 \right) {
						C_1}}}
			\left( 2\,\beta+4- {{\rm e}
				^{\pm 2{\frac {\sqrt {{C_1}\,\beta+2\,{C_1}} \left( t+{C_2}
							\right) }{\beta+2}}}} \right)  \left( {{\rm e}^{\pm {\frac {
							\sqrt {{C_1}\,\beta+2\,{C_1}} \left( t+{C_2} \right) }{
							\beta+2}}}} \right) ^{-1}$ \\
			(1a) & & &  \\
\hline
			\multirow{1}{*}{$(1.02)$} & -1 & -2/3 & ${C_1}+\frac{1}{4{\Lambda}{{C_2}}}\, \left( -3+\Lambda\,{{
					C_1}}^{2} \right) {{\rm e}^{{\frac {\sqrt {6}\sqrt {\Lambda}
							t}{3\sqrt {\beta+2}}}}}+{C_2}\,{
				{\rm e}^{-{\frac {\sqrt {6}\sqrt {\Lambda}t}{3\sqrt {\beta+2}}}}}$ \\
\hline
			\multirow{1}{*}{$(1.03)$} & -1 & -1/3 & ${C_1}\,{
				{\rm e}^{{\frac {\sqrt {6}\sqrt {\Lambda}t}{3\sqrt {\beta+2}}}}}+{
				C_2}\,{{\rm e}^{-{\frac {\sqrt {6}\sqrt {\Lambda}t}{3\sqrt {
								\beta+2}}}}}$ \\
\hline
			\multirow{1}{*}{$(1.04)$} & -1 & 1/3 & $\pm \frac{1}{2{\Lambda}}\,\sqrt{-{{\rm e}^{{\frac {4t
								\sqrt {\Lambda}}{\sqrt {6\,\beta+12}}}}}\Lambda\, \left( {{\rm e}^{
						{\frac {8t\sqrt {\Lambda}}{\sqrt {6\,\beta+12}}}}}{C_1}\,\sqrt {
					\Lambda}\sqrt {6\,\beta+12}-
				{C_2}\,\sqrt {\Lambda}\sqrt {6\,\beta
					+12}+6\,{{\rm e}^{{\frac {4t\sqrt {\Lambda}}{\sqrt {6\,\beta+12}}}}}
				\right) } \times$ 
			$\left( {{\rm e}^{{\frac {4t\sqrt {\Lambda}}{\sqrt {6\,
								\beta+12}}}}} \right) ^{-1}$ \\
\hline
\multirow{1}{*}{$(1.05)$} & 0 & -2 & $\frac{1}{{9}^{1/3}} \left( {\frac {1}{\sqrt {6\,\beta+12}}{{\rm e}^{{
				\frac {3t\sqrt {\Lambda}}{\sqrt {6\,\beta+12}}}}} \left( \sqrt {6}
	\sqrt {{\frac {\Lambda}{\beta+2}}}\sqrt {6\,\beta+12}+6\,\sqrt {
		\Lambda} \right)  \left( {{\rm e}^{{\frac {6t\sqrt {\Lambda}}{\sqrt 
					{6\,\beta+12}}}}}{C_1}-{C_2} \right) ^{2}} \right) ^{\frac{2}{3}} \times$
				\\ 
				(1b)	& & &
$ \left( {{\rm e}^{{\frac {6t\sqrt {\Lambda}}{\sqrt {6\,\beta+12}}}}}
{C_1}-{C_2} \right) ^{-2}$ \\
\hline
$(1.06)$ & & & \\
(1c) & 0 & $=-1$ & $C_2 e^{C_1 t}$ \\
\hline
\multirow{1}{*}{$(1.07)$} & 0 & $\neq -1$ & ${C_2}\, \left[ - \left( \cosh \left( \frac{1}{2}\,{\frac { \left( {\rm w}+1
		\right)  \left( t+{C_1} \right) \sqrt {6}\sqrt { \left( \beta+2
			\right) \Lambda}}{\beta+2}} \right)  \right) ^{2} \right] ^{\frac{1}{ \left( 3
		\,{\rm w}+3 \right)}}.$ \\
\hline
			\multirow{1}{*}{$(1.08)$} & 0 & -2/3 & ${C_1}+\frac{1}{4}\,{\frac {{{C_1}}^{2}}{{C_2}}{{\rm e}^{{
				\frac {2t\sqrt {\Lambda}}{\sqrt {6\,\beta+12}}}}}}+{C_2}\,{{\rm e}
	^{-{\frac {2t\sqrt {\Lambda}}{\sqrt {6\,\beta+12}}}}}
$ \\
\hline
\multirow{1}{*}{$(1.09)$} & 0 & -1/3 & ${C_1}\,{{\rm e}^{{\frac {\sqrt {6}\sqrt {\Lambda}t}{3\sqrt {
					\beta+2}}}}}+{C_2}\,{{\rm e}^{-{\frac {\sqrt {6}\sqrt {
					\Lambda}t}{3\sqrt {\beta+2}}}}}
$ \\
\hline
\multirow{1}{*}{$(1.10)$} & 0 & 0 & $\frac{\beta+2}{6\Lambda}  \left[ \frac{3}{2}(e^{ \sqrt{\frac{6\Lambda}{\beta+2}}t} C_1-C_2) e^{ \sqrt{\frac{6\Lambda}{\beta+2}}t} \left( \frac{6\Lambda}{\beta+2}\right) \right] ^{\frac{2}{3}} e^{- \sqrt{\frac{6\Lambda}{\beta+2}}t}$ \\
(1d) & & &  \\
\hline
\multirow{1}{*}{$(1.11)$} & 0 & 1/3 & $\pm \frac{1}{2}\,{\frac {1}{\sqrt {\Lambda}}\sqrt {-{{\rm e}^{{\frac {4t\sqrt {
							\Lambda}}{\sqrt {6\,\beta+12}}}}}\sqrt {\Lambda}\sqrt {6\,\beta+12}
		\left( -{C_2}+{{\rm e}^{{\frac {8t\sqrt {\Lambda}}{\sqrt {6\,
							\beta+12}}}}}{C_1} \right) } \left( {{\rm e}^{{\frac {4t\sqrt {
						\Lambda}}{\sqrt {6\,\beta+12}}}}} \right) ^{-1}}
$ \\
\hline
			\multirow{1}{*}{$(1.12)$} & 1 & -1 & $\frac{1}{2{\sqrt { \left( \beta+2 \right) {
				C_1}}}}\, \left( 2\,\beta+4+  {{\rm e}^
	{\pm {2\frac {\sqrt {{C_1}\,\beta+2\,{C_1}} \left( t+{C_2}
				\right) }{2\left( \beta+2\right)}}}} \right)  \left( {{\rm e}^{\pm {\frac {
				\sqrt {{C_1}\,\beta+2\,{C_1}} \left( t+{C_2} \right) }{
				\beta+2}}}} \right) ^{-1}$ \\
\hline
			\multirow{1}{*}{$(1.13)$} & 1 & -2/3 & ${C_1}+\frac{ \left( 3+\Lambda\,{{
						C_1}}^{2}\right)}{4{\Lambda}{{C_2}}}  {{\rm e}^{ {\frac {\sqrt {6}\sqrt {\Lambda}t}{
							3\sqrt {\beta+2}}}}}+{C_2}\,{{\rm e}^{-{\frac {\sqrt {6}\sqrt {\Lambda}t}{3\sqrt {\beta+2}}}}}$ \\
\hline
			\multirow{1}{*}{$(1.14)$} & 1 & -1/3 & ${C_1}\,{
				{e}^{{\frac {\sqrt {6}\sqrt {\Lambda}t}{3\sqrt {\beta+2}}}}}+{
				C_2}\,{{e}^{-{\frac {\sqrt {6}\sqrt {\Lambda}t}{3\sqrt {
								\beta+2}}}}}$ \\
\hline
			\multirow{1}{*}{$(1.15)$} & 1 & 1/3 & $\pm \frac{1}{2{
					\Lambda}}\,\sqrt{ {{\rm e}^{{\frac {4\,t\sqrt 
								{\Lambda}}{\sqrt {6\,\beta+12}}}}}\Lambda\, \left( -{{\rm e}^{{
							\frac {8\,t\sqrt {\Lambda}}{\sqrt {6\,\beta+12}}}}}{C_1}\,\sqrt {
					\Lambda}\sqrt {6\,\beta+12}+
				6\,{{\rm e}^{{\frac {4\,t\sqrt {\Lambda}}{
								\sqrt {6\,\beta+12}}}}}+{C_2}\,\sqrt {\Lambda}\sqrt {6\,\beta+12}
				\right) } \times$
			 $\left( {{\rm e}^{{\frac {4\,t\sqrt {\Lambda}}{\sqrt {6\,
								\beta+12}}}}} \right) ^{-1}$ \\
\hline
		\end{tabular}
		\medskip
		\medskip
		\\ \begin{justify}
			Notes:  
			(1a) Calculated for any value of $\Lambda>0$. 
			(1b) The singular time of the Case 1.05 is $t_s=\frac{1}{6}\,{\frac {\sqrt {6\,\beta+12}}{\sqrt {\Lambda}}\ln  \left( {\frac {{C_2}}{{C_1}}} \right) }$.
			(1c) Calculated for any value of $\Lambda>0$ and $\beta$. For $\beta=0$ this solution is analogous to the one presented in Ha {\it et.al.} (2012) \cite{Ha2012} (Equation 2.21). For $\rm w=-1$ and $\beta=0$ this solution is analogous to the one presented in Ha {\it et.al.} (2012) \cite{Ha2012} (Equation 2.20). 
			(1d) In order to compare to GR theory let us put $\beta=0$. Then we must choose $C_2=C_1$. In order to get the same metric function $B(t)$ as in  d'Inverno textbook \cite{d'Inverno1992},  we must choose $C_1^2=C=\frac{8}{3} \pi G B(t)^3 \rho_{m}(t)$. 
		\end{justify}
	\end{minipage}
\end{table*}


\begin{table*}
	\centering
	\begin{minipage}{175 mm}
		\caption{Summary of the Cosmological Solutions for $\Lambda=0$}
		\label{table0b}
		\begin{tabular}{@{}|c|c|c|c|}
\hline
			Case & $k$ & w & $B(t)$ \\
\hline
\multirow{1}{*}{$(2.01)$} & -1 & -1 & $-\frac{1}{2\sqrt { \left( \beta+2 \right) {
						C_1}}}
			\left( 2\,\beta+4- {{\rm e}
				^{\pm 2{\frac {\sqrt {{C_1}\,\beta+2\,{C_1}} \left( t+{C_2}
							\right) }{\beta+2}}}} \right)  \left( {{\rm e}^{\pm {\frac {
							\sqrt {{C_1}\,\beta+2\,{C_1}} \left( t+{C_2} \right) }{
							\beta+2}}}} \right) ^{-1}$ \\
\hline
\multirow{1}{*}{$(2.02)$} & -1 & -2/3 & ${\frac { \left( -2+2\,{{C_1}}^{
						2}+\beta\,{{C_1}}^{2} \right) {t}^{2}}{4{C_2}\, \left( \beta+
					2 \right) }}+{C_1}\,t+{C_2}$ \\
\hline
\multirow{1}{*}{$(2.03)$} & -1 & -1/3 & ${C_1}\,t+{C_2}$ \\
			(2a) & & &  \\
\hline
\multirow{1}{*}{$(2.04)$} & -1 & 0 & $-\frac{1}{8\sqrt{-(\beta+2)}} \left\{ \sqrt{2}\epsilon (\beta+2) C_1  
\tan^{-1}\left[ {\frac{\sqrt{2}}{4}} \frac{ \sqrt{-(\beta+2)} [4 B(t)+C_1]}{\sqrt{B(t) (\beta+2) [2 B(t)+C_1]}} \right]-\right.$ \\
(2b) & & & $\left. 4 \epsilon \sqrt{B(t) (\beta+2) [2 B(t)+C_1]} \sqrt{-(\beta+2)}+\right.\left.8 t \sqrt{-(\beta+2)}+8 C_2 \sqrt{-(\beta+2)} \right\}=0$\\
\hline
\multirow{1}{*}{$(2.05)$} & -1 & 1/3 & $\pm {\frac {\sqrt {- \left( 4+2\,\beta
						\right)  \left( t{C_1}\,\beta-{C_2}\,\beta-2\,{C_2}-{t
						}^{2}+2\,{C_1}\,t \right) }}{\beta+2}}$ \\
\hline
\multirow{1}{*}{$(2.06)$} & 0 & -2 & ${\frac {{18}^{2/3} }{ 9\left( {C_1}\,t+{C_2}
		\right) ^{\frac{2}{3}}}}$ \\
(2c) & & & \\
\hline
$(2.07)$ (2d) & 0 & $-1$ & $C_2 e^{C_1 t}$ \\
\hline
			\multirow{1}{*}{$(2.08)$} & 0 & -2/3 & $\frac{1}{4}\,{{C_1}}^{2}{t}^{2}+\frac{1}{2}\,{C_1}{C_2}\,t+\frac{1}{4}\,{{C_2}}^{2}$ \\
(2a) & & &  \\
\hline
\multirow{1}{*}{$(2.09)$} & 0 & -1/3 & ${C_1}\,t+{C_2}$ \\
(2a) & & &  \\
\hline
\multirow{1}{*}{$(2.10)$} & 0 & $\neq -1$ & $\left[ \frac{3}{2}\, \left( {C_1}\,t+{C_2} \right)  \left( 1+{\rm w} \right) \right]^{ \frac{2}{3(1+{\rm w})}}$ \\
\hline
\multirow{1}{*}{$(2.11)$} & 0 & 0 & $\frac{1}{4}\, \left( 12\,{C_1}\,t+12\,{C_2} \right) ^{\frac{2}{3}}$ \\
(2a) & & &  \\
\hline
\multirow{1}{*}{$(2.12)$} & 1 & -1 & $\frac{1}{2{\sqrt {{
							C_1}\, \left( \beta+2 \right) }}}\,
			\left( 2\,\beta\,{{C_1}}^{2}+4
			\,{{C_1}}^{2}+  {{\rm e}^{\pm {2\frac {\sqrt {{C_1}\,\beta+
								2\,{C_1}} \left( t+{C_2} \right) }{{C_1}\, \left( \beta
							+2 \right) }}}} \right)
			\left( {{\rm e}^{\pm {\frac {\sqrt {
								{C_1}\,\beta+2\,{C_1}} \left( t+{C_2} \right) }{{
								C_1}\, \left( \beta+2 \right) }}}} \right) ^{-1}$ \\
\hline
\multirow{1}{*}{$(2.13)$} & 1 & -2/3 & ${\frac { \left( 2+2\,{{C_1}}^{2
					}+\beta\,{{C_1}}^{2} \right) {t}^{2}}{4{C_2}\, \left( \beta+2
					\right) }}+{C_1}\,t+{C_2}$ \\
\hline
\multirow{1}{*}{$(2.14)$} & 1 & -1/3 & ${C_1}\,t+{C_2}$ \\
			(2a) & & &  \\
\hline
$(2.15) (2b)$& 1 & 0 & $-\frac{1}{8\sqrt{\beta+2}} \left\{ -\sqrt{2}\epsilon (\beta+2) C_1  
\tan^{-1}\left[ {\frac{\sqrt{2}}{4}} \frac{ \sqrt{(\beta+2)} [4 B(t)-C_1]}{\sqrt{-B(t) (\beta+2) [2 B(t)-C_1]}} \right]+\right.$ \\
& & & $\left.4 \epsilon \sqrt{-B(t) (\beta+2) [2 B(t)-C_1]} \sqrt{\beta+2}+8 t \sqrt{\beta+2}+8 C_2 \sqrt{\beta+2} \right\}=0$\\
\hline
\multirow{1}{*}{$(2.16)$} & 1 & 1/3 & $\pm {\frac {\sqrt {- \left( 4+2\,\beta
						\right)  \left( t{C_1}\,\beta-{C_2}\,\beta-2\,{C_2}+{t
						}^{2}+2\,{C_1}\,t \right) }}{\beta+2}}$ \\
\hline
		\end{tabular}
		\medskip
		\medskip
		\\
		\begin{justify}
			Notes: 
			(2a) Calculated for any value of $\beta$. 
			(2b) where $\epsilon=\pm 1$. For $\beta=0$ these solutions are analogous to the ones presented in d'Inverno textbook \cite{d'Inverno1992}. 
			(2c) The singular time of the Case 2.06 is $t_s=-{\frac {{C_2}}{{C_1}}}$.
			(2d) For any $\beta$ even $\beta=0$ this solution is analogous to the one presented in Ha {\it et.al.} (2012) \cite{Ha2012} (Equation 2.25). 
		\end{justify}
	\end{minipage}
\end{table*}


\begin{table*}
	\centering
	\begin{minipage}{175 mm}
		\caption{Summary of the Cosmological Solutions for $\Lambda<0$}
		\label{table0c}
		\begin{tabular}{@{}|c|c|c|c|}
\hline
			Case & $k$ & w & $B(t)$ \\
\hline
\multirow{1}{*}{$(3.01)$} & -1 & -1 & $-\frac{1}{2\sqrt { \left( \beta+2 \right) {
						C_1}}}
			\left( 2\,\beta+4- {{\rm e}
				^{\pm 2{\frac {\sqrt {{C_1}\,\beta+2\,{C_1}} \left( t+{C_2}
							\right) }{\beta+2}}}} \right)  \left( {{\rm e}^{\pm {\frac {
							\sqrt {{C_1}\,\beta+2\,{C_1}} \left( t+{C_2} \right) }{
							\beta+2}}}} \right) ^{-1}$ \\
\hline
\multirow{1}{*}{$(3.02)$} & -1 & -2/3 & ${\frac {\sqrt {-3+ \left| \Lambda
						\right| {{C_1}}^{2}+ \left| \Lambda \right| {{C_2}}^{2}}}{
					\sqrt { \left| \Lambda \right| }}}+
			{C_1}\,\sin \left( {
				\frac {\sqrt {6}\sqrt { \left| \Lambda \right| }t}{3\sqrt {2+\beta}}}
			\right) +{C_2}\,\cos \left( {\frac {\sqrt {6}\sqrt {
						\left| \Lambda \right| }t}{3\sqrt {2+\beta}}} \right)$ \\
\hline
\multirow{1}{*}{$(3.03)$} & -1 & -1/3 & ${C_1}\,\sin
			\left({\frac {\sqrt {6}\sqrt { \left| \Lambda \right| }t}{
					3\sqrt {2+\beta}}} \right) +{C_2}\,\cos \left( {\frac {\sqrt 
					{6}\sqrt { \left| \Lambda \right| }t}{3\sqrt {2+\beta}}} \right)$ \\
\hline
\multirow{1}{*}{$(3.04)$} & -1 & 1/3 & $\pm \frac{\sqrt {6}}{12 \left| \Lambda \right| }\Bigg\{ \sqrt {6} \left| \Lambda \right|  [ 2\,{C_2}\,\sqrt {2+\beta}\sqrt { \left| \Lambda \right| }\cos \left( {\frac {2\sqrt {6}\sqrt { \left| \Lambda \right| }t}{3\sqrt {2+\beta}}} \right) -$ \\
& & & $2\,{C_1}\,\sqrt {2+\beta}\sqrt { \left| \Lambda \right| }\sin \left( {\frac {
		2\sqrt {6}\sqrt { \left| \Lambda \right| }t}{3\sqrt {2+\beta}}} \right) 
+\sqrt {6} ] \Bigg\} ^{\frac{1}{2}}$ \\ 
\hline
$(3.05)$ & & & \\
(3a) & 0 & $=-1$ & $C_2 e^{C_1 t}$ \\
\hline
\multirow{1}{*}{$(3.06)$} & 0 & $\neq -1$ & $\{
			-\frac{1}{3C_1}\, \left[ -3\,{C_2}\, \left| \sqrt {\Lambda} \left( 1+{\rm w}
			\right)  \right| \cos \left( {\frac {t\sqrt {6} \left| \sqrt {
						\Lambda} \left( 1+{\rm w} \right)  \right| }{2\sqrt {2+\beta}}} \right) + 
			\sqrt {6}\sqrt {2+\beta}\sin \left( {\frac {t\sqrt {6} \left| 
					\sqrt {\Lambda} \left( 1+{\rm w} \right)  \right| }{2\sqrt {2+\beta}}}
			\right)  \right]$ \\
			& & & $\times \left|  \cos \left( {\frac {t\sqrt {6} \left| \sqrt {\Lambda} \left( 1+{\rm w}
					\right)  \right| }{2\sqrt {2+\beta}}} \right)  \right|
			\left(  \left| \sqrt {\Lambda} \left( 1+{\rm w} \right)  \right|  \right)^{-1}
			\left[ \cos \left( {\frac {t\sqrt {6} \left| \sqrt {\Lambda}
					\left( 1+{\rm w} \right)  \right| }{2\sqrt {2+\beta}}} \right)  \right] ^{-1}\}^{\frac{2}{3(1+{\rm w})}}$ \\

\hline
\multirow{1}{*}{$(3.07)$} & 0 & 0 & $\frac{\beta+2}{6|\Lambda|(\beta+2)^{\frac{1}{3\beta+2}}} \left\{\left(\frac{6|\Lambda|}{\beta+2}\right)\left[C_1\sin\left(\frac{1}{2} \sqrt{\frac{6|\Lambda|}{\beta+2}}t\right) + C_2 \cos\left(\frac{1}{2} \sqrt{\frac{6|\Lambda|}{\beta+2}}t\right)\right]\right\}^\frac{2}{3}$ \\
			(3b) & & & \\
\hline
\multirow{1}{*}{$(3.08)$} & 0 & -2/3 & $\sqrt {{{C_1}}^{2}+{{C_2}}^{2}}+{C_1}\,\sin \left( {\frac {t\sqrt {6}\sqrt { \left| \Lambda \right| }}{3\sqrt {2+\beta}}
} \right) +{C_2}\,\cos \left( {\frac {t\sqrt {6}\sqrt {
			\left| \Lambda \right| }}{3\sqrt {2+\beta}}} \right) 
$ \\
\hline
\multirow{1}{*}{$(3.09)$} & 0 & -1/3 & ${C_1}\,\sin
\left( {\frac {\sqrt {6}\sqrt { \left| \Lambda \right| }t}{
		3\sqrt {2+\beta}}} \right) +{C_2}\,\cos \left( {\frac {\sqrt 
		{6}\sqrt { \left| \Lambda \right| }t}{3\sqrt {2+\beta}}} \right)$ \\
\hline
\multirow{1}{*}{$(3.10)$} & 0 & -2 & ${\sqrt [3]{6 \left| \Lambda \right| } \left[ {C_2}\,
	\cos \left( {\frac {\sqrt {6}\sqrt { \left| \Lambda \right| }t}{
			2\sqrt {\beta+2}}} \right) +{C_1}\,\sin \left( {\frac {\sqrt 
			{6}\sqrt { \left| \Lambda \right| }t}{2\sqrt {\beta+2}}} \right) 
	\right] ^{-\frac{2}{3}}}$ \\
(3c) & & & \\
\hline
\multirow{1}{*}{$(3.11)$} & 1 & -1 & $\frac{1}{2{\sqrt { \left( \beta+2 \right) {
							C_1}}}}\, \left( 2\,\beta+4+  {{\rm e}^
				{\pm {2\frac {\sqrt {{C_1}\,\beta+2\,{C_1}} \left( t+{C_2}
							\right) }{2\left( \beta+2\right)}}}} \right)  \left( {{\rm e}^{\pm {\frac {
							\sqrt {{C_1}\,\beta+2\,{C_1}} \left( t+{C_2} \right) }{
							\beta+2}}}} \right) ^{-1}$ \\
\hline
\multirow{1}{*}{$(3.12)$} & 1 & -2/3 & ${\frac {\sqrt {-3+ \left| \Lambda \right| 
						{{C_1}}^{2}+ \left| \Lambda \right| {{C_2}}^{2}}}{\sqrt {
						\left| \Lambda \right| }}}+
			{C_1}\,\sin \left( {\frac {
					\sqrt {6}\sqrt { \left| \Lambda \right| }t}{3\sqrt {2+\beta}}} \right) 
			+{C_2}\,\cos \left( {\frac {\sqrt {6}\sqrt { \left| \Lambda
						\right| }t}{3\sqrt {2+\beta}}} \right)$ \\
\hline
\multirow{1}{*}{$(3.13)$} & 1 & -1/3 & ${C_1}\,\sin
			\left( {\frac {\sqrt {6}\sqrt { \left| \Lambda \right| }t}{
					3\sqrt {2+\beta}}} \right) +{C_2}\,\cos \left( {\frac {\sqrt 
					{6}\sqrt { \left| \Lambda \right| }t}{3\sqrt {2+\beta}}} \right)$ \\
\hline
\multirow{1}{*}{$(3.14)$} & 1 & 1/3 & $\pm \frac{1}{2  \left| \Lambda \right| }\Bigg[ 6\,\sqrt {6}
			\left| \Lambda \right|  \Bigg( 2\,{C_2}\,\sqrt {2+\beta}\sqrt {
				\left| \Lambda \right| }\cos \left( {\frac {2\sqrt {6}\sqrt {
						\left| \Lambda \right| }t}{3\sqrt {2+\beta}}} \right) -$ \\
			& & & $\sqrt {6}-2\,{C_1}\,\sqrt {2+\beta}\sqrt { \left| 
				\Lambda \right| }
			\sin \left( {\frac {2\sqrt {6}\sqrt { \left| \Lambda \right| }t}{
					3\sqrt {2+\beta}}} \right)  \Bigg) \Bigg]^{\frac{1}{2}}$ \\
\hline
		\end{tabular}
		\medskip 
		\medskip
		\\
		\begin{justify}
			Notes: 
		(3a) Calculated for any value of $\Lambda<0$ and $\beta$. For $\beta=0$ and $C_2=0$ this solution is analogous to the one presented in Ha {\it et.al.} (2012) \cite{Ha2012} (Equation 2.28). 
		(3b) where $\beta+2 \geq 0$. In order to compare to GR theory we must choose $C_2=0$ and in order to get the same metric function $B(t)$ as in d'Inverno textbook
\cite{d'Inverno1992} we must choose $C_1^2=18 C$. 
		(3c) The singular time of the Case 3.10 is $t_s=-\frac{1}{3}\,{\frac {\sqrt {\beta+2}\sqrt {6}}{\sqrt { \left| \Lambda \right| }}\arctan \left( {\frac {{C_2}}{{C_1}}} \right) }$.
		\end{justify}
	\end{minipage}
\end{table*}

\section{Summary of the effective energy conditions of the cosmological models}
\begin{table*}
	\centering
	\begin{minipage}{175 mm}
		\caption{Summary of the Effective Energy Conditions for $\Lambda>0$}
		\label{table1}
		\begin{tabular}{@{}|c|c|c|c|}
			\hline
			Case &  $\rho_m$ & $SEC_{\rm eff}$ & Figures \\
			\hline
			\multirow{1}{*}{$(1.01)$} & positive constant for $\forall \beta$ &  negative constant for  $\forall \beta$ & \ref{Figure-101-104} \\
			\hline
			\multirow{2}{*}{$(1.02)$} & positive for $\forall \beta$ & negative for $0 < t < t_1$ for $\forall \beta$ & \ref{Figure-101-104}\\
			&  & positive for $t_1 < t < +\infty$ for $\beta>0$ & (a) \\
			\hline
			\multirow{2}{*}{$(1.03)$} & positive for $\forall \beta$ & positive constant for $\beta \ge 0$ &  \ref{Figure-101-104} \\
			&  & negative constant for $\beta < 0$ & \\
			\hline
			\multirow{3}{*}{$(1.04)$} & positive for $\forall \beta$ &  positive for  $\beta \ge 0$ & \ref{Figure-101-104} \\
			&  & positive for $0 < t < t_1$ \& $\beta<0$ & \\
			&  & negative for $t_1 < t < +\infty$ \& $\beta<0$ &  \\
			\hline
\multirow{1}{*}{$(1.05)$} & positive for $\forall \beta$ & negative for $\forall \beta$ &  \ref{Figure-105-111} \\
			\hline
			\multirow{2}{*}{$(1.06)$} & positive constant for $\beta\ge 0$ & negative for $\forall \beta$ &  \ref{Figure-105-111} (b) \\
			& negative constant for $\beta<0$ &  & \\
			\hline
\multirow{2}{*}{$(1.07)$} & negative for $\forall \beta$ & negative for $0 < t < t_1$ \& $\forall \beta$ &  \ref{Figure-105-111} (b) \\
&  & positive for $t_1 < t < +\infty$ \& $\beta>0$ & \\
			\hline
\multirow{2}{*}{$(1.08)$} & positive for $\forall \beta$  & negative for $\beta< 0$ &  \ref{Figure-105-111} \\
&  & negative $0 <t< t_1$ \& $\beta \ge 0$ & \\
&  & positive for $t_1 <t< +\infty$ \& $\beta \ge 0$ & \\
			\hline
\multirow{2}{*}{$(1.09)$} & positive for $\forall \beta$ & positive for $\beta \ge 0$ &  \ref{Figure-105-111} \\
&  & negative  $\beta < 0$ & \\
			\hline
			\multirow{1}{*}{$(1.10)$} & positive for $\forall \beta$ & positive for $\beta \ge 0$ & \ref{Figure-105-111} \\
&  & positive for $0 < t < t_1$ \& $\beta<0$ & \\
&  & negative for $t_1 < t < +\infty$ \& $\beta<0$ & \\
			\hline
\multirow{3}{*}{$(1.11)$} & positive for $\forall \beta$ &  positive for $\beta \ge 0$ & \ref{Figure-105-111} \\
&  & positive for $0<t<t_1$ \& $\beta<0$ & \\
&  & negative for $t_1 <t< +\infty$ \& $\beta<0$ &  \\
\hline
			\multirow{1}{*}{$(1.12)$} & positive constant for $\forall \beta$ &  negative constant for $\beta \le 0$ & \ref{Figure-112-115} \\
			&  & positive constant for $\beta>0$ &  \\
			\hline
			\multirow{2}{*}{$(1.13)$} & positive for $\forall \beta$  & negative for $\beta \le 0$ &  \ref{Figure-112-115} (b) \\
			&  & negative $0 <t< t_1$ \& $\beta>0$ & \\
			&  & positive for $t_1 <t< +\infty$ \& $\beta>0$ & \\
			\hline
			\multirow{2}{*}{$(1.14)$} & positive for $\forall \beta$ & positive for $\beta \ge 0$ &  \ref{Figure-112-115} \\
			&  & negative  $\beta < 0$ & \\
			\hline
			\multirow{3}{*}{$(1.15)$} & positive for $\forall \beta$ &  positive for $\beta \ge 0$ & \ref{Figure-112-115} \\
			&  & positive for $0<t<t_1$ \& $\beta<0$ & \\
			&  & negative for $t_1 <t< +\infty$ \& $\beta<0$ &  \\
			\hline
		\end{tabular}
		\\
		\medskip
		\begin{justify}
			Notes: notice that $\rho_m >0$ means that it is a physical aether fluid while $\rho_m <0$ it means an unphysical aether fluid. Moreover, $SEC_{\rm eff} > 0$ means it is a normal fluid in GR while $SEC_{\rm eff} < 0$ means it is a dark energy fluid in GR.
		(a) $t_1=t_1(\beta)$ with $SEC_{\rm eff}(t_1)=0$. (b) Means that $R_g(t=0) \neq 0$.
		\end{justify}
	\end{minipage}
\end{table*}


\begin{table*}
	\centering
	\begin{minipage}{175 mm}
		\caption{Summary of the Effective Energy Conditions for $\Lambda=0$}
		\label{table2}
		\begin{tabular}{@{}|c|c|c|c|}
			\hline
			Case &  $\rho_m$ & $SEC_{\rm eff}$ & Figures \\
			\hline
			\multirow{1}{*}{$(2.01)$} & positive constant for $\forall \beta$ & negative constant $\forall \beta$ & \ref{Figure-201-205} \\
			\hline
			\multirow{3}{*}{$(2.02)$}  & negative for $0<t<t_0$ \& $\forall \beta$ & positive for $0<t<t_1$ \& $\forall \beta$ & \ref{Figure-201-205} \\
			& negative for $t_0<t<+\infty$ \& $\beta \ge 0$ & positive for $t_1<t<+\infty$ \& $\beta \ge 0$ & (a) \\
			& positive for $t_0<t<+\infty$ \& $\beta < 0$ & negative for $t_1<t<+\infty$ \& $\beta < 0$ &  (b) \\
			\hline
			\multirow{2}{*}{$(2.03)$} & positive for $\beta \ge 0$ & null for $\forall$ $\beta$ & \ref{Figure-201-205} \\
			& negative for $\beta<0$ &  & \\
			\hline
			\multirow{1}{*}{$(2.05)$} & positive for $\forall \beta$ & positive for $\forall \beta$ & \ref{Figure-201-205} \\
			\hline
			\multirow{1}{*}{$(2.06)$} & positive for $\forall \beta$ & negative for $\forall \beta$ & \ref{Figure-206-211} \\
			\hline
			\multirow{1}{*}{$(2.07)$} & positive $\forall \beta$ &  negative constant for $\forall \beta$ & \ref{Figure-206-211} (c)\\
			\hline
			\multirow{1}{*}{$(2.08)$} & positive $\forall \beta$ & negative $\forall \beta$ & \ref{Figure-206-211}\\
			\hline
			\multirow{1}{*}{$(2.09)$} & positive for $\forall \beta$ &  null for \& $\forall \beta$ & \ref{Figure-206-211} \\
			\hline
			\multirow{1}{*}{$(2.10)$} & positive $\forall \beta$ &  positive for $\forall \beta$ & \ref{Figure-206-211} \\
			\hline
			\multirow{1}{*}{$(2.11)$} & positive for $\forall \beta$ & positive for $\forall \beta$ & \ref{Figure-206-211} \\
			\hline
			\multirow{1}{*}{$(2.12)$} & positive constant $\forall \beta$ & negative constant $\forall \beta$ & \ref{Figure-212-216} \\
			\hline
			\multirow{1}{*}{$(2.13)$} & positive $\forall \beta$ & negative $\forall \beta$ & \ref{Figure-212-216} (c)\\
			\hline
			\multirow{1}{*}{$(2.14)$} & positive for $\forall \beta$ &  null for \& $\forall \beta$ & \ref{Figure-212-216} \\
			\hline
			\multirow{1}{*}{$(2.16)$} & positive for $\forall \beta$ & positive for $\forall \beta$ & \ref{Figure-212-216} \\
						\hline
		\end{tabular}
		\\
		\medskip
		\begin{justify}
			Notes: Notice that $\rho_m >0$ means that it is a physical aether fluid while $\rho_m <0$ it means an unphysical aether fluid. Moreover, $SEC_{\rm eff} > 0$ means it is a normal fluid in GR while $SEC_{\rm eff} < 0$ means it is a dark energy fluid in GR.
		(a) $t_0=t_0(\beta)$ with $\rho_m(t_0)=0$. 
		(b) $t_1=t_1(\beta)$ with $SEC_{\rm eff}(t_1)=0$. 
		(c) Means that $R_g(t=0) \neq 0$.
		\end{justify}
	\end{minipage}
\end{table*}


\begin{table*}
	\centering
	\begin{minipage}{175 mm}
		\caption{Summary of the Effective Energy Conditions for $\Lambda<0$}
		\label{table3}
		\begin{tabular}{@{}|c|c|c|c|}
			\hline
			Case &  $\rho_m$ & $SEC_{\rm eff}$ & Figures \\
			\hline
			\multirow{1}{*}{$(3.01)$} & positive constant $\forall \beta$ & negative constant for $\forall \beta$ &  \ref{Figure-301-304} \\
			\hline
			\multirow{3}{*}{$(3.02)$} & positive for $\forall \beta$ &  negative for $0<t<t_4$ \& $\forall \beta$ & \ref{Figure-301-304}\\
			&  & positive for $t_4<t<t_5$ \& $\beta<0$ & (a) \\
			&  & negative for $t_5<t<+\infty$ \& $\beta < 0$ & \\
			\hline
			\multirow{2}{*}{$(3.03)$} & positive for $\forall \beta$ &  positive constant for $\beta \le 0$ &  \ref{Figure-301-304}\\
			&  & negative constant for $\beta > 0$ & \\
			\hline
			\multirow{1}{*}{$(3.04)$} & positive for $\forall \beta$ &  positive for $\forall \beta$ & \ref{Figure-305-310}\\
			\hline
			\multirow{1}{*}{$(3.05)$} & positive constant for $\forall \beta$ & negative constant for $\forall \beta$ & \ref{Figure-305-310} (b) \\
			\hline
			\multirow{1}{*}{$(3.06)$} & positive for $\forall \beta$ & positive $\forall \beta$ & \ref{Figure-305-310} \\
			\hline
			\multirow{3}{*}{$(3.07)$} & positive for $0<t<t_2$ \& $\forall \beta$ &  positive for $\beta>0$ &  \ref{Figure-305-310} \\
			& positive for $t_3<t<+\infty$ \& $\forall \beta$ & positive for $0<t<t_4$ for $\beta \le  0$ &  (c) \\
			& negative for $t_2<t<t_3$ \& $\forall \beta$ & negative for $t_5<t<+\infty$ for $\beta \le 0$ & (a) \\
			\hline
			\multirow{3}{*}{$(3.08)$} & positive for $\forall \beta$ & negative for $\beta \ge 0$ & \ref{Figure-305-310} \\
			&  & positive for $t_4<t<t_5$ \& $\beta<0$ & (a) \\
			&  & negative for $t_5<t<+\infty$ \& $\beta < 0$ &  \\
			\hline
			\multirow{2}{*}{$(3.09)$} & positive for $\forall \beta$ &  positive constant for $\beta \le 0$ & \ref{Figure-305-310} \\
			&  & negative constant for $\beta > 0$ & \\
			\hline
			\multirow{1}{*}{$(3.10)$} & positive for $\forall \beta$ & negative for $\forall \beta$ & \ref{Figure-305-310} \\
			\hline
			\multirow{1}{*}{$(3.11)$} & positive constant for $\forall \beta$ &  negative  constant for $\forall \beta$ & \ref{Figure-311-314}\\
			\hline
			\multirow{3}{*}{$(3.12)$} & positive for $\forall \beta$ &  negative for $0<t<t_4$ \& $\forall \beta$ & \ref{Figure-311-314} (b) \\
			&  & positive for $t_4<t<t_5$ \& $\beta<0$ & (a) \\
			&  & negative for $t_5<t<+\infty$ \& $\beta < 0$ &  \\
			\hline
			\multirow{2}{*}{$(3.13)$} & positive for $\forall \beta$ &  positive constant for $\beta \le 0$ & \ref{Figure-311-314} \\
			&  & negative constant for $\beta>0$ & \\
			\hline
			\multirow{3}{*}{$(3.14)$} & positive for $\forall \beta$ & positive for $\beta \le 0$ & \ref{Figure-311-314} \\
			&  & negative for $t_4<t<t_5$ \& $\beta>0$ & (a) \\
			&  & positive for $t_5<t<+\infty$ \& $\beta > 0$ &  \\
			\hline
		\end{tabular}
		\\
		\medskip
		\begin{justify}
			Notes: Notice that $\rho_m >0$ means that it is a physical aether fluid while $\rho_m <0$ it means an unphysical aether fluid. Moreover, $SEC_{\rm eff} > 0$ means it is a normal fluid in GR while $SEC_{\rm eff} < 0$ means it is a dark energy fluid in GR.
		(a) $t_4=t_4(\beta)$ with $SEC_{\rm eff}(t_4)=0$ and $t_5=t_5(\beta)$ with $SEC_{\rm eff}(t_5)=0$, where $t_4 < t_5$. 
		(b) Means that $R_g(t=0) \neq 0$.
		(c) $t_2=t_2(\beta)$ with $\rho_m(t_2)=0$ and $t_3=t_3(\beta)$ with $\rho_m(t_3)=0$, where $t_2 < t_3$. 
		\end{justify}
	\end{minipage}
\end{table*}

\section{Graphical analysis of the dynamics and effective energy conditions of the cosmological models}

\section{References}

\end{document}